\documentclass[review]{elsarticle}

\usepackage[colorlinks,citecolor=blue,linktoc=all,linkcolor=cyan]{hyperref}
\usepackage{graphicx}
\usepackage{mathtools}
\usepackage{upgreek}
\usepackage{physics}
\usepackage{multirow}
\usepackage{placeins} % To control floating objects (figures and tables)
\usepackage{pgfplots}
\usepackage{subfigure}

\usepackage[T1]{fontenc}
\usepackage{dsfont}               % use mathds instead of mathbb for outline fonts
\usepackage{mathrsfs}             % provides mathscr without overwriting mathcal
\usepackage{slashed}              % For Dirac slash notation.

\usepackage{amsmath}
\usepackage{amssymb}
\usepackage{amsbsy}
\usepackage{amsfonts}
\usepackage{braket}
\usepackage{tikz}
\usepackage{bbold}
\usetikzlibrary{arrows,shapes,positioning}
\usetikzlibrary{decorations.markings}
\usetikzlibrary{decorations.pathmorphing}
\tikzstyle arrowstyle=[scale=1]
\tikzstyle directed=[postaction={decorate,decoration={markings,
    mark=at position .5 with {\arrow[arrowstyle]{stealth}}}}]
\tikzstyle reverse directed=[postaction={decorate,decoration={markings,
    mark=at position .5 with {\arrowreversed[arrowstyle]{stealth};}}}]    

\newcommand{\diag}{\mathop{}\!\mathrm{diag}}

\numberwithin{equation}{section}
\numberwithin{table}{section}
\numberwithin{figure}{section}

\journal{Progress in Particle and Nuclear Physics}

\allowdisplaybreaks

\usepackage{titlesec}
\usepackage{sectsty}
\titleformat{\section}{\normalfont\Large\bfseries}{\thesection}{1em}{}
\titleformat{\subsection}{\normalfont\large\bfseries}{\thesubsection}{1em}{}
\titleformat{\subsubsection}{\normalfont\normalsize\bfseries}{\thesubsubsection}{1em}{}
\begin{document}
	
    \begin{frontmatter}
		
    \title{Ordinary and exotic mesons in the extended Linear Sigma Model}

		%authors, affiliations, corresponding author mention 

    \author[kielce_univ,frankfurt_univ]{Francesco Giacosa\corref{mycorrespondingauthor}}
    \cortext[mycorrespondingauthor]{Corresponding author}
    \ead{fgiacosa@ujk.edu.pl}
    %%%%%%%%%%%%%%%%%
    \author[wigner_inst]{Péter Kovács}
    %%%%%%%%%%%%%%%%%
    \author[kielce_univ,arizona,baku_ctp,baku_unec]{Shahriyar Jafarzade}
    %%%%%%%%%%%%%%%%%        
    \address[kielce_univ]{Institute of Physics, Jan Kochanowski University, ulica Uniwersytecka 7, P-25-406 Kielce, Poland}
    \address[frankfurt_univ]{Institute for Theoretical Physics, Goethe-University, Max-von-Laue-Straße 1, D-60438 Frankfurt am Main, Germany}
    \address[wigner_inst]{Institute for Particle and Nuclear Physics, HUN-REN Wigner Research Centre for Physics, Konkoly–Thege Miklós út 29-33, 1121 Budapest, Hungary}  
    \address[arizona]{Department of Physics, Arizona State University, Tempe, AZ 85287, USA}
    \address[baku_ctp]{Center for Theoretical Physics, Khazar University, Mehseti 41 Street, AZ1096 Baku, Azerbaijan}
   \address[baku_unec]{Composite Materials Research Center, Azerbaijan State Economic University (UNEC), H. Aliyev 135, AZ1063, Baku, Azerbaijan}
  
    \begin{abstract}
    The extended Linear Sigma Model (eLSM) is a hadronic model based on the global symmetries of QCD and the corresponding explicit, anomalous, and spontaneous breaking patterns. In its basic three-flavor form, its mesonic part contains the dilaton/glueball as well as the nonets of pseudoscalar, scalar, vector, and axial-vector mesons, thus chiral symmetry is linearly realized.  In the chiral limit and neglecting the chiral anomaly, only one term -within the dilaton potential- breaks dilatation invariance, and all terms are chirally symmetric.  Spontaneous symmetry breaking is implemented by a generalization of the Mexican-hat potential, with explicit symmetry breaking responsible for its tilting. The overall mesonic phenomenology up to $\sim 2$~GeV is in agreement with the PDG compilation of masses and partial and total decay widths. The eLSM was enlarged in a straightforward way to include other conventional quark-antiquark nonets (pseudovector and orbitally excited vector mesons, tensor and axial-tensor mesons, radially excited (pseudo)scalar mesons, etc.), as well as two nonets of hybrid mesons, the lightest one with exotic quantum numbers $J^{\mathcal{P}\mathcal{C}} = 1^{-+}$ not allowed for $\bar{q}q$ objects, such as the resonance $\pi_1(1600)$ and the recently discovered $\eta_1(1855)$. In doing so, different types of chiral multiplets are introduced: heterochiral and homochiral multiplets, which differ in the way they transform under chiral transformations. Moreover, besides the scalar glueball that is present from the beginning as dilaton, other glueballs, the tensor, the pseudoscalar and the vector glueballs were coupled to the eLSM: the scalar resonance $f_0(1710)$ turns out to be mostly gluonic, the tensor glueball couples strongly to vector mesons, and the pseudoscalar glueball couples sizably to $\pi \pi \eta'$ and can be assigned to $X(2370)$ or $X(2600)$. In all cases above, masses and decays can be analyzed allowing for a better understanding of both conventional and non-conventional mesons:  whenever data are available, a comparison is performed and, when this is not the case, predictions of decay widths and decay ratios are outlined. The eLSM contains chiral partners on an equal footing and is therefore well suited for studies of chiral symmetry restoration at nonzero temperature and densities: this is done by coupling it to the Polyakov loop. The QCD phase diagram and the location of the critical endpoint were investigated within this framework.  
    \end{abstract}
		
		
    \end{frontmatter}
	
   \newpage
	
    \thispagestyle{empty}
    \tableofcontents
	
	%to begin the line numbers: 
	%\linenumbers	
	%beginning of the core of the manuscript
\newpage
\section{Introduction}
\label{Sec:intro}
Symmetries play an indispensable role in modern particle physics in general and for
quarks and gluons in particular \cite{Martin:1970hmp,Mosel:1989jf}. This fact is well summarized by Wigner's famous article entitled  \textit{The unreasonable effectiveness of mathematics in the natural sciences} \cite{Wigner:1960kfi}.
Together with symmetries, also their breaking is decisive toward the understanding of the physical world and can take place in three different forms: (i) the breaking can be explicit due to some terms that are not symmetry invariant (in simple terms, a geometric figure is not exactly a circle, but is an ellipse with a small but nonzero eccentricity, thus the $O(2)$ rotational symmetry is broken); generally this kind of breaking ought to be small and the symmetry of the system is referred to as an `approximate symmetry'. (ii) The breaking can be anomalous if the symmetry holds at the classical level but is broken by quantum fluctuations (colloquially, the `quantum version' of the circle is an ellipse); this breaking can be in certain cases large, in such a way that `de facto' the symmetry does not appear at all, even not approximately, but that depends case by case. (iii) A symmetry can be broken spontaneously when the ground state of the system is not left invariant by the symmetry transformation, e.g. \cite{Beekman:2019pmi}. A typical example is the Mexican hat potential: a ball placed on the top represents a configuration which is $O(2)$-rotational symmetric, but is unstable: the ball would eventually roll down and stop at a given location along the circle of minima. In this way a direction has been picked up: the ground state is not invariant under rotations. (In matter physics, ferromagnetic material show spontaneous magnetization along a certain direction below a critical temperature, breaking rotational symmetry.) 
But there is more: with minimal kinetic effort, the ball could roll around the circle. In the quantum version of this model, this corresponds to a massless mode, a Goldstone boson \cite{Goldstone:1961eq}, e.g. the pion. On the other hand, oscillating  along the direction orthogonal to the circle costs energy, thus the particle associated to this mode is massive: this is e.g. the Higgs in the Standard Model (SM), or the $\sigma$ particle in chiral models, see below. 
Interestingly, the spontaneous breaking of a symmetry may be traced back to a philosophical discussion initiated by the French philosopher Buridan (and by others before him, among which Aristoteles), who envisaged a donkey perfectly in the middle between two identical piles of hay ($Z_2$ 
 symmetry) and, unable to choose between the two, dies of hunger. The spontaneous breaking of the $Z_2$ symmetry would indeed save the donkey's life.

Symmetries, together with their breaking patterns described above, are also
essential for understanding the states of quarks and gluons, the hadrons.
%Namely, the description of hadrons via models requires a full plethora ofsymmetries together with their anomalous and explicit breaking patterns. 
This is particularly true for the model described in this review: the extended
Linear Sigma Model (eLSM), e.g. \cite{Parganlija:2012fy,Janowski:2014ppa,Kovacs:2016juc,Eshraim:2020ucw,Jafarzade:2022uqo} and refs. therein. In order to see the genesis of it and its main
features as well as its place in the context of high-energy physics (HEP), we
recall basic facts that from quarks and gluons lead to hadronic objects. 

Quarks and gluons enter as fundamental particles the Lagrangian of
Quantum Chromo-Dynamics (QCD) \cite{Muta:2010xua,Ratti:2021ubw}. Quarks appear in six
different flavors ($u$, $d$, $s$, $c$, $b$, $t$) with distinct \textit{bare} masses, taken from the Particle Data Group (PDG) \cite{Workman:2022ynf,ParticleDataGroup:2024cfk}: 
$m_\mathit{u}=2.16 \pm 0.07$~MeV, $m_\mathit{d}=4.70 \pm 0.07$~MeV,
$m_\mathit{s}=93.5 \pm 0.8$~MeV, $m_\mathit{c}=1.2730\pm 0.0046$~GeV, $m_\mathit{b}=4.183 \pm 0.007$~GeV,
$m_\mathit{t}=172.57 \pm 0.29$~GeV.
In view of these values, it is common to distinguish the light
flavor sector ($u$, $d$, $s$) form the heavy one ($c$, $b$, $t$). Denoting with $N_f$ the number of flavors, $N_f=2$ refers to 
the quarks $u$ and $d$, $N_f =3 $ to $u$, $d$ ,and $s$, and finally $N_f =4$ to $u$, $d$, $s$, and $c$.
In this work, we shall
concentrates on the hadrons built within $N_f =3$, but  the cases  $N_f =2$ and $N_f =4$ will be treated as well.

Each quark also carries the color charge, being either red, green, or blue, for a total
of $N_{c}=3$ possibilities, with $N_{c}$ denoting the number of colors. While
fixed in nature, $N_{c}$ can be seen as parameter both in models and in computer (lattice)
simulations of QCD. In particular, the limit of large-$N_{c}$ is extremely
interesting and appealing since certain simplifications do take place \cite{tHooft:1973alw,Witten:1979kh,Lebed:1998st,Lucha:2021mwx,Giacosa:2024scx}.

Gluons also carry color, intuitively corresponding to a color-anticolor
configuration, for a total of $N_{c}^{2}-1 = 8$ of them (in fact, one
combination, the colorless one, must be subtracted).  The QCD Lagrangian is
based on an exact and local `color' gauge symmetry: we may redefine the color
of quarks in a space-time dependent way without changing the Lagrangian,
provided that the gluon fields transform accordingly. Besides the quark masses
listed above, the QCD Lagrangian contains one parameter, the QCD dimensionless strong coupling $g_{\text{QCD}}$. Moreover, QCD (in its basic formulation) is also invariant under parity transformation $\mathcal{P}$, which refers to space-inversion, and under charge-conjugation $\mathcal{C}$, which swaps particles with antiparticles.

In the chiral limit, where all quark masses are assumed to vanish, the classical QCD Lagrangian contains only one dimensionless parameter, the coupling constant $g_{\text{QCD}}$. This means that an additional symmetry is present: dilatation
invariance under the space-time transformation $x^{\mu}\rightarrow\lambda
^{-1}x^{\mu}$. This symmetry is broken explicitly by nonzero quark masses but,
most importantly, it is broken anomalously when quantizing the theory via gluonic loops.
This is the so-called trace or dilatation anomaly, which is one of the most
important properties of QCD \cite{Migdal:1982jp,Thomas:2001kw}. This feature shall play a major role in the hadronic approach described in this work. 

A related quantum property is the running coupling of QCD, according to which the coupling constant $g_{\text{QCD}}$ becomes a function of the energy: 
\begin{equation}
g_{\text{QCD}}\rightarrow g_{\text{QCD}}(\mu) \text{ .}
\end{equation}
Loop calculations within QCD show that $g_{\text{QCD}}(\mu)$ decreases with increasing $\mu$, indicating (i) asymptotic freedom at large energies (the Nobel prize in 2004, see e.g. the Nobel prize lectures \cite{Wilczek:2005az,Politzer:2005kc}), implying that at high energies QCD can be treated perturbatively \cite{Muta:2010xua}, and (ii) an increasing coupling constant at low-energy, eventually leading to a Landau pole at $\Lambda_{\text{QCD}} \sim 250$ MeV within perturbation theory \cite{Deur:2016tte}. While the presence of a pole may be seen as an artifact of perturbation theory \cite{Deur:2016tte,Gies:2001nw}, the emergence of a strong coupling regime for slowly moving quarks and gluons is established.  Even if not yet proven analytically, within this regime confinement takes place: quarks and gluons cannot be observed as independent particles but are confined in hadrons, which are invariant under local color gauge transformations \cite{Greensite:2003bk}. They may be referred as `colorless' or, in analogy to  actual color, an equal combination of red, green, and blue (and their anticolors) generates a `white' object. 
%(iii) In the aforementioned trace anomaly, the divergence of the current is proportional to the $\beta$-function $\beta(\mu) =  \mu d g_{\text{QCD}}/d \mu$, thus a running coupling is necessary for its reali

Hadrons are divided into mesons (hadrons with integer spin) and baryons (hadrons with semi-integer spin). Mesons are further classified into conventional or standard quark-antiquark ($\bar{q} q$) mesons, which form the vast majority of the mesons listed in the Particle Data Group (PDG) compilation \cite{ParticleDataGroup:2024cfk}, as well as exotic or non-conventional mesons , see e.g. \cite{Gross:2022hyw,Amsler:2004ps,Klempt:2007cp,Giacosa:2015wcz}: glueballs (states made of only gluons, one of the earliest predictions of QCD, which are possible because gluons `shine in their own light'), hybrid mesons (a quark-antiquark pair and one gluon), and multiquark objects such as tetraquark states \cite{Jaffe:2004ph,Klempt:2007cp}. The latter can be further understood as diquark-anti-diquark states, meson-meson molecules, dynamically generated companion poles, etc. 
In the hadronic model `eLSM' discussed in this paper, both conventional and non-conventional mesons are of primary significance. In particular, this review concentrates on standard mesons as well as glueballs and hybrid states within the eLSM with a mass up to $2.6$ GeV, but selected heavier states will be discussed as well.

Baryons are also classified into conventional baryons (three-quark states, such as the neutron and proton, and as the vast majority of PDG baryons) and non-conventional ones, such as pentaquark states \cite{Chen:2022asf}.  The inclusion of baryons in the eLSM was performed in the past both for $N_f =2$ \cite{Gallas:2009qp,Gallas:2011qp,Gallas:2013ipa} and $N_f =3$ \cite{Olbrich:2015gln,Olbrich:2017fsd}, but this topic will not be studied in depth here. 

On the experimental side, several reactions have been and are being used to produce and study hadrons, including the light mesons discussed in this paper. Among them, we mention $e^+ e^-$ scattering at different energies (SND, KLOE, BESIII, ALEPH, etc.), meson photoproduction in $\gamma p$ and $\gamma N$ processes (such as GLUEX, CLAS12, COMPASS, in the future EIC), proton-proton and more in general nucleus-nucleus scattering (such as the ongoing LHC experiments, NA61/SHINE, HADES, in the future CMB at GSI/FAIR), as well as proton-antiproton processes (such as Crystal Barrel at LEAR, experiments at Fermilab, in the future PANDA at GSI/FAIR). Although processes like charmonium decays have recently delivered valuable insights, $\pi\pi$ scattering continues to play a vital role in meson spectroscopy The Particle Data Group summarizes the experimental results \cite{ParticleDataGroup:2024cfk} and will be often used as a reference.

The QCD Lagrangian shows, besides an exact and local color symmetry and an approximate and anomalously broken dilatation symmetry, a series of other approximate global symmetries \cite{Mosel:1989jf}:

\begin{itemize}

\item Baryon number conservation $U(1)_{V}$. The simplest symmetry of QCD is a phase transformation of the quark fields, denoted as $U(1)_{V}$. The
consequence is the conservation of baryon number. This symmetry is
automatically fulfilled at the level of mesons, since each meson, containing
an equal number of quarks and antiquarks, is invariant.

\item Flavor symmetry $SU(3)_{V}$. For $N_{f}=3$, in the limit in which the
bare masses of the three light quarks are regarded as equal, the QCD
Lagrangian is invariant under redefinition of the light quarks, resulting in
a special unitary transformation $SU(3)_{V}$. At a fundamental
level, flavor symmetry arises from the fact that gluons interact with any
quark flavor with the same coupling strength $g_{\text{QCD}}$, thus in the limit of
equal masses no difference occurs. In terms of conventional mesons, this
symmetry implies the emergence of mesonic nonets (3 quarks and 3 antiquarks)
for a definite total angular momentum $J$ and for fixed parity $\mathcal{P}$
and charge conjugation $\mathcal{C}$ (the latter is applicable only to certain
multiplet members), commonly expressed as $J^{\mathcal{P}\mathcal{C}}$. The
lowest mass multiplet is realized for $J^{\mathcal{P}\mathcal{C}}=0^{-+}$ and
refers to pseudoscalar mesons, that contains 3 pions, 4 kaons, and the
$\eta(547)$ as well as the $\eta^{\prime}(958)$ mesons. With the exception
of the last one, an octet emerges, realizing the Gell-Mann's `eightfold way'
\cite{GellMann:1961ky,Gell-Mann:1964ewy}. Flavor symmetry $SU(3)_V$ is explicitly broken by unequal quark masses, especially by $m_s -m_{u,d} \sim \Lambda_{QCD}$.

\item Isospin symmetry $SU(2)_{V}$. Flavor symmetry reduces to the
well-known isospin symmetry $SU(2)_{V}$ when only the light quarks $u$ and $d$
are considered. This is the symmetry introduced by Heisenberg to describe the
proton and neutron as manifestations of the same particle
\cite{Heisenberg:1932dw}, the nucleon, and shortly after baptized by Wigner as
isotopic spin, isospin, due to formal mathematical similarities with the spin \cite{Wigner:1936dx}. Later on, the concept of isospin was
extended by Kemmer \cite{Kemmer:1939zz} to the three pions postulated shortly
before by Yukawa \cite{Yukawa:1935xg}, which build an `isotriplet'. The
extension to other particles was then straightforward. Isospin is, even though
not exact, very well conserved in strong interactions, as the nearly equal
masses of the isotriplet pions show, $\left(  m_{\pi^{+}}-m_{\pi^{0}}\right)
/\left(  m_{\pi^{+}}+m_{\pi^{0}}\right)  \simeq0.017$ (for a very recent
puzzling breaking of isospin in kaon productions, see Refs.
\cite{NA61SHINE:2023azp,Brylinski:2023nrb}).

\item The classical group of QCD, denoted as $\mathcal{G}_{cl}$, and chiral symmetry $SU(3)_{L}\times SU(3)_{R}$.  In
the chiral limit, in which the bare masses of the three light quarks are taken as
massless (the chiral limit), the classical Lagrangian of QCD is invariant under the
so-called classical group $\mathcal{G}_{cl}\equiv U(3)_{L}\times U(3)_{R}.$
Namely, gluons do not mix the chirality of quarks, therefore one can transform
independently the right-handed and the left-handed quark components, resulting
in two distinct symmetries $U(3)_{L}$ and $U(3)_{R}$, where $L$ and $R$ stands
for left and right, respectively. The group can be further decomposed as
$\mathcal{G}_{cl}\equiv U(3)_{L}\times U(3)_{R}=U(1)_{L}\times U(1)_{R}\times
SU(3)_{L}\times SU(3)_{R}$, where $\mathcal{G}_{\text{chiral}}\equiv SU(3)_{L}\times
SU(3)_{R}$, is denoted as the chiral symmetry of QCD, and is exact only in the
chiral limit. Note, chiral symmetry $SU(3)_{L}\times SU(3)_{R}$ reduces to the
usual flavor transformation $SU(3)_{V}$ if the parameters of $SU(3)_{L}$ and
$SU(3)_{R}$ are taken as equal: this is expected, since in this case both
quark chiralities are transformed in the same way.

\item Spontaneous symmetry breaking (SSB) of chiral symmetry. At the mesonic
level, chiral transformations mix e.g. the nonet of pseudoscalar mesons
($J^{\mathcal{P}\mathcal{C}}=0^{-+}$ ) with scalar mesons ($J^{\mathcal{P}%
\mathcal{C}}=0^{++}$ ), and vector mesons ($J^{\mathcal{P}\mathcal{C}}=1^{--}%
$) with axial-vector ones ($J^{\mathcal{P}\mathcal{C}}=1^{++}$), usually
referred to as chiral partners. However, experimental data show that these
states are far from being degenerate in mass. The symmetry is broken explicitly by
nonzero and unequal quark masses, which alone cannot explain the
experimental values. Chiral symmetry also undergoes, and most importantly, the
process of spontaneous symmetry breaking, illustrated above by using the
Mexican-hat's example. Indeed, this is more than a simple analogy: in terms of
the mesonic potential within the (pseudo)scalar sector, a typical Mexican-hat
shaped potential is actually realized, being also one of the most outstanding
features of the eLSM discussed in this review. The SSB takes place in QCD
because the ground state (the QCD `vacuum') is not left invariant by chiral
transformations. As a consequence, no degeneracy of chiral partners takes
place and the pseudoscalar mesons emerge as quasi-Goldstone bosons (where
`quasi' means that they are not massless due to the explicit breaking of
chiral symmetry caused by nonvanishing quark masses). The SSB in the chiral limit is expressed as
$SU(3)_{L}\times SU(3)_{R}\rightarrow SU(3)_{V}$.

\item Two $U(1)$ subgroups of the classical group $U(3)_{L}\times U(3)_{R}$
are important. Namely, $U(1)_{L}\times U(1)_{R}$ can be rewritten as
$U(1)_{V}\times U(1)_{A},$ where $U(1)_{V}$ is the already introuced baron
number symmetry and  $U(1)_{A}$ is the so-called axial transformation, that
corresponds to a phase transformation in which the right-handed phase has an
opposite sign w.r.t. the left-handed one. This symmetry is explicitly broken
by nonzero quark masses, spontaneously broken by the QCD vacuum, but also -and
most significantly- anomalously broken by gluonic fluctuations
\cite{tHooft:1976rip,tHooft:1986ooh}: this is the famous axial or chiral $U(1)_A$ anomaly. One famous consequence is that the meson
$\eta^{\prime}(958)$ is much heavier than pions and kaons. The question if
other mesons also feel (and to which extent) this anomaly is interesting,
since novel interaction types were recently discussed in Refs.
\cite{Giacosa:2017pos,Giacosa:2023fdz}.
\end{itemize}

This set of symmetries and their breaking patterns has been used to construct various theoretical models and approaches of the low-energy sector of QCD, dealing with hadrons made of light quarks and gluons. In particular, we recognize the following widely used approaches and methods that make use of quark d.o.f., hadronic d.o.f., and computer simulations.

\bigskip

\textit{Some widely used approaches involving quarks as basic d.o.f.}

(i) The Isgur-Godfrey quark-model \cite{isgur1985} (and later extensions of it \cite{Lucha:1991vn,Lucha:1992rq,Vijande:2004he}) with a funnel-type potential covered a decisive role in the establishment of many resonances listed in the PDG. It is still an important reference for comparison. Note that the quark masses used in this model are not the bare ones contained in the QCD Lagrangian, but constituent quarks with a mass of the order of $\Lambda_{\text{QCD}}$ for the $u$ and $d$ flavors. While chiral symmetry is not explicitly used in this approach (therefore, it is difficult to describe the pion as a Goldstone boson), flavor symmetry is correctly implemented.

(ii) Bag models, in which quarks and gluons freely move inside a bag but cannot escape from it,  were extensively used at the beginning of the QCD era \cite{Chodos:1974je,Chodos:1974pn}. The spectrum of conventional mesons and baryons could be fairly reproduced and news states, such as glueballs, could be  investigated for the first time \cite{Jaffe:1985qp}. 

(iii) A famous and historically important example of a quark-based chiral approach is the so-called Nambu-Jona-Lasinio (NJL) model \cite{Nambu:1961tp,Nambu:1961fr}, that contains chiral symmetry and its SSB via the emergence of a chiral condensate, as well as, in later applications, the chiral anomaly \cite{Klevansky:1992qe,Hatsuda:1994pi,Vogl:1991qt}. This model is not confining but both mesons as quark-antiquark states and baryons as three-quark states can be obtained \cite{Buck:1992wz}. One of the most interesting properties of this model is the emergence of a constituent quark out of a bare quark via the formation of a $\bar{q}q$ condensate. NJL-type models with constituent gluons are also possible, see e.g. \cite{Gounaris:1985uy,Giacosa:2004ug}. 

(iv) The Dyson-Schwinger equations (DSE) enable the calculation of hadronic bound states starting from the QCD Lagrangian by implementing dressed quark and gluon propagators \cite{Alkofer:2000wg,Eichmann:2016yit,Fischer:2006ub}. The infinite class of diagrams can be treated by applying appropriate truncation schemes. One may interpret the DSE as a synthesis of the three approaches above. The resulting description of the hadronic spectrum is very good. 
Recently, the application of the DSE techniques to glueballs delivered masses in agreement with lattice QCD \cite{Huber:2020ngt,Huber:2021yfy,Huber:2023mls}. 

(v) QCD sum rules \cite{Shifman:1978by,Novikov:1983jt,Narison:1996fm,Narison:2022paf,Chen:2022imp} were and are widely applied to connect microscopic quark-antiquark as well as multiquark and gluonic currents to nonzero condensates and to properties of low-energy resonances. 

(vi) More recently, the so-called functional renormalization group techniques (FRG) have been developed to calculate non-perturbatively the flow and other low-energy properties of gauge theories, e.g. \cite{Gies:2001nw,Braun:2007bx,Braun:2014ata,Pawlowski:2005xe}. 
This technique can be also applied to hadronic d.o.f. (see below) allowing to find its quantum behavior at low-energy starting from the classical Lagrangian, see the recent works in Refs. \cite{Koenigstein:2021rxj,Koenigstein:2021syz,Steil:2021cbu} and refs, therein. 

(vii) Holographic approaches can be implemented to study low-energy hadronic properties by exploiting a correspondence between quantum gauge theories and classical (gravitational) field theory in higher dimensions \cite{Sakai:2004cn}.  In particular, this approach can be used to address the glueball phenomenology \cite{Hechenberger:2023ljn,Brunner:2015oqa,Brunner:2015yha,Rinaldi:2022pgh}, thus offering an alternative nonperturbative approach to determine decay properties of these objects.

%\bigskip Next, we move to effective models and theories that deal with hadronic degrees of freedom as fields entering in the defining Lagrangians. 

\bigskip

\textit{Some widely used approaches involving hadronic d.o.f.}

(i) The first version of Linear Sigma Model (LSM) dates back to the pre-QCD era and consisted of a pion triplet, one scalar particle as the chiral partner of the pion (the famous $\sigma$ meson), and the nucleon field \cite{GellMann:1960np,Lee:1967ug,Lee:1968da}. 
The potential along the pion and the $\sigma$ directions takes the typical Mexican-hat form, with the $\sigma$-field taking a nonzero expectation value (v.e.v.) called the chiral condensate. The corresponding particle is the fluctuation along the sigma direction, while the pion is the massless excitation around the circle. %\footnote{A simple visualization is possible by plotting it along the neutral pion and the scalar direction, see later.}. 
A nonzero quark mass generates a tilting of the potential and a nonzero pion mass.
In turn, this model is able to generate a nucleon mass thanks to the chiral condensate without breaking chiral symmetry. Quite interestingly, the Mexican-hat form (without tilting and without the emergence of massless Goldstone bosons) applies also for the quartet of Higgs fields \cite{Pich:2005mk}. 
The LSM, due to the linear realization of chiral symmetry, treats chiral partners on equal footing. This property is useful for establishing their connection and for studies at nonzero temperature and density.

(ii) Chiral perturbation theory (ChPT) has played and plays a major role in the description of low-energy QCD processes, such as pion-pion scattering, see e.g. \cite{Scherer:2002tk,Gasser:1983yg,Gasser:1984gg}. In its simplest form, it consists of only three pions. Formally, it can be obtained by the LSM by integrating out the sigma field, that is by taking it as infinitely heavy \cite{Pich:1993uq}. The resulting Lagrangian is expressed in terms of derivatives of the pion fields, whose power increases order by order. Namely, an expansion in the pion momentum (via derivatives of the pion field) is understood. This is an example of a non-linear realization of chiral symmetry, since only the pion triplet is retained, but not its chiral partner (the $\sigma$ in this case).  
Generalization to the whole mesonic octet as well as to other fields have been performed \cite{Bijnens:1997ni,Cirigliano:2003yq,Ecker:1988te,Ecker:1989yg}.

(iii) There are also models that contain both mesonic and quark d.o.f. Some of them are denoted as quark-level Linear Sigma Model(s), see e.g. Refs. \cite{Scadron:2013vba,Tripolt:2013jra}. 
In general, one may link the NJL model to the LSM by applying bosonization techniques \cite{Vogl:1991qt,Klevansky:1992qe,Eguchi:1976iz}, where the quark-level LSM appear as an intermediate stage.

\bigskip
\textit{Lattice QCD simulations}

QCD can be successfully simulated on a lattice with a discrete and finite number of space-time points using Monte Carlo integration techniques \cite{Hoelbling:2014uea}. While chiral fermions are  notoriously difficult to include in lattice studies, steady improvements allowed for decreasing the quark masses. Nowadays the physical pion is correctly reproduced, thus its Goldstone boson nature is captured by lattice studies as well. In general, the whole hadronic spectrum agrees very well with the PDG \cite{Dudek:2011tt,Dudek:2013yja,Johnson:2020ilc}. When considering only gluons, the so-called Yang-Mills (YM) part of the QCD Lagrangian, lattice simulations predict a whole tower of glueball states, providing strong evidence for their existence \cite{Morningstar:1999rf,Chen:2005mg,Gui:2012gx,Gregory:2012hu,Athenodorou:2020ani}. The lightest gluonic state is the scalar one with a mass of about 1.7 GeV, the second lightest is the tensor (2.2 GeV), and the third is the pseudoscalar (2.6 GeV). All of them will be discussed in this review. 
Hybrid mesons have also been determined in lattice studies, according to which the lightest state agrees quite well with the exotic resonance $\pi_1(1600)$  \cite{Woss:2020ayi}.

\bigskip
Next, we describe in more detail how the LSM-type models evolved into the eLSM by incorporating vector d.o.f. and dilatation invariance.

Important steps toward the development of the LSM were achieved by the Syracuse group in the '80 \cite{Salomone:1980sp,Gomm:1984zq,Gomm:1985ut,Rosenzweig:1981cu,Rosenzweig:1982cb}. In these works, a LSM with a chiral multiplet involving a pseudoscalar nonet and a scalar nonet were developed. Moreover, the dilaton field was introduced in order to describe at the hadronic composite level the aforementioned anomalous breaking of dilatation invariance, resulting in a logarithmic potential. The fluctuations of the latter are identified with the scalar glueball, which is then naturally coupled to the conventional (pseudo)scalar mesons. Besides, also the first model including the pseudoscalar glueball was put forward.
Modern extensions of that approach consider additional chiral nonets made of tetraquark states \cite{Fariborz:2007ai,Fariborz:2018xxq,Napsuciale:2004au,Giacosa:2006tf}.

Next, a relevant improvement has been the inclusion of vector mesons into the LSM. 
A class of approaches makes use of a local flavor gauge invariance, according to which vector mesons -besides a purely mass term- are seen as gauge bosons. This approach was carried out in nonlinear ChPT-like approaches \cite{Bando:1987br,Harada:2003jx,Gasiorowicz:1969kn} involving pions and vector fields, and also in LSM-type approaches, in which pseudoscalar and scalar d.o.f. (in short, (pseudo)scalar) and vector and axial-vector d.o.f. (in short, (axial-)vector) are taken into account. 
The rationale behind these flavor-local (or isospin-local, since $N_f=2$ was considered) approaches is that a certain realization of vector meson dominance (VMD) is achieved \cite{OConnell:1995nse}. However, as noted in Ref. \cite{Urban:2001ru} there is no fundamental reason to consider local chiral invariance; VMD can be also obtained through a different realization \cite{OConnell:1995nse,Urban:2001ru}. Most importantly, the phenomenology of (axial-)vector mesons cannot be correctly described by imposing local chiral invariance.

All the arguments and attempts above led to the development of the eLSM, which consists in merging the main elements of the previous approaches, in particular:

\begin{enumerate}
\item The mesonic eLSM consists of (at least) three flavors, $N_{f}=3$ \cite{Parganlija:2012fy,Janowski:2014ppa}.
Previous versions with two flavors \cite{Parganlija:2010fz,Janowski:2011gt} were important intermediate steps toward
the development of the eLSM, but the model is intended to include mesons
containing strange quark(s). Of course, a smaller number of flavors can always be recovered as a subset of the $N_{f}=3$ case. The extension to $N_{f}=4$ has also been carried out \cite{Eshraim:2014eka}: on the one hand, caution is required because the purely low-energy sector is left and there is a significant breaking of chiral symmetry due to the charm mass; on the other hand, the results show that certain decays still fulfill the predictions of chiral symmetry.

\item The dilaton field is an important part of the eLSM \cite{Parganlija:2012fy,Janowski:2011gt,Janowski:2014ppa}. In the chiral limit
and neglecting the chiral anomaly, the eLSM contains only one dimensionful
parameter, an effective energy scale denoted as $\Lambda_{G}$, with $\Lambda_{G}\sim\Lambda_{\text{QCD}}$. This energy scale appears in the (logarithmic) 
dilaton potential. All other terms carry dimension 4, and the corresponding
coupling constants are dimensionless, a fact that strongly constrains the
model. As in \cite{Rosenzweig:1981cu,Rosenzweig:1982cb}, the dilaton field condenses and mimics the
generation of the gluon condensate. In turn, the scalar glueball is present in
the eLSM from its very beginning.

\item The interaction terms fulfill chiral symmetry ($SU(3)_L\times
SU(3)_R$) as well as parity and charge conjugation symmetries. Moreover, large-$N_c$ arguments can be applied to isolate dominating terms from subdominant ones \cite{Giacosa:2024scx}. At leading order, one may consider only the former, thus further reducing the list of implemented terms.

\item The
(pseudo)scalar potential of the eLSM is a generalization of the Mexican-hat
potential, thus also nonstrange and strange condensates $\phi_{N}$ and
$\phi_{S}$ emerge via SSB \cite{Parganlija:2012fy}. As a consequence, $\phi_{N,S}\sim\Lambda_{G},$
showing that the trace anomaly is of fundamental relevance. 

\item The mesonic eLSM contains, besides (pseudo)scalar mesons, vector and
axial-vector nonets, thus generalizing the previous $N_{f}=2$ studies of Refs.
\cite{Ko:1994en,Urban:2001ru}. A consequence of SSB,  the  $a_{1}$-$\pi$ mixing of pseudoscalar and
axial-vector d.o.f, emerges. This is an important aspect, since, although quite technical, it affects the phenomenology.

\item The nonzero quark masses are taken into account by proper terms that
break chiral invariance $SU(3)_L\times SU(3)_R$ and, for unequal masses,
also flavor invariance $SU(3)_V$. In the pseudoscalar sector, the square of
the mass of the pion is proportional to bare quark masses:  $m_{\pi}^{2}\sim
m_{u}+m_{d}$. This very specific relation can be seen as the `smoking gun' for SSB and is confirmed by lattice QCD simulations \cite{Bietenholz:2010az}.

\item The chiral anomaly is described by specific terms involving the determinant of matrices of chiral multiplets, see \cite{Kovacs:2013xca}. 
These interactions fulfill chiral symmetry $SU(3)_L\times SU(3)_R$ but break the axial transformation $U(1)_A.$ As a
consequence, the $\eta$-$\eta^{\prime}$ system can be correctly described. For a recent generalization of these anomalous terms to other nonets than the (pseudo)scalar ones, see \cite{Giacosa:2017pos,Giacosa:2023fdz}, where a  mathematical object denoted as `polydeterminant' has been introduced to allow their description \cite{Giacosa:2025dvu}.

\item The eLSM was enlarged to include more fields than the dilaton plus
(pseudo)scalar and (axial-)vector d.o.f., in particular: pseudoscalar glueball \cite{Eshraim:2012jv}, tensor glueball \cite{Vereijken:2023jor},
vector glueball \cite{Giacosa:2016hrm}, the lightest hybrid meson multiplet \cite{Eshraim:2020ucw}, conventional
(axial-)tensor \cite{Jafarzade:2022uqo}, pseudo/excited vector \cite{Giacosa:2016hrm}, and excited (pseudo)scalar multiplets \cite{Parganlija:2016yxq}. Further extensions to the four-flavor case \cite{Eshraim:2014eka} as well as to isospin breaking phenomenology \cite{Kovacs:2024cdb} were also carried out. 
The main goal is the post-diction of known results and the predictions of novel
processes (mostly masses and decay widths).

\item In the construction of the eLSM and its various generalizations to novel resonances, mesonic nonets are grouped in pairs of chiral partners. Among them, one needs to distinguish between \textit{heterochiral} and \textit{homochiral} multiplets, that transform differently under the chiral group \cite{Giacosa:2017pos}. This general classification shows a returning pattern and is especially important for the chiral anomaly.

\item  The eLSM is in agreement with ChPT, both at a formal and at a numerical level \cite{Divotgey:2016pst}. Formally, one needs to integrate out
all the fields except the pseudoscalar ones to determine the low-energy coupling constants (LECs) within the
eLSM. On the other hand, even if dealing with similar physical processes,
there is a difference between these approaches: ChPT offers a systematic
treatment of the Goldstone bosons and their interactions, while the eLSM
describes (already at tree-level) various resonances up to and in certain
cases above 2 GeV.
In this respect, ChPT is well suited for the precise description of e.g. low-energy pion and kaon (as well as nucleon) scattering processes \cite{Colangelo:2001df}, while the eLSM is useful for an overall phenomenological description of masses and decays up to about 2 GeV \cite{Parganlija:2012fy}. Nevertheless, scattering was also studied in the eLSM \cite{Parganlija:2010fz,Lakaschus:2018rki}.

\item The eLSM is not renormalizable due to the dilaton potential and the
(axial-)vector mesonic fields.  Since this is an effective low-energy
QCD\ hadronic model describing non-point-like and non-fundamental particles,
renormalization is not a compelling requirement. Yet, loops can be calculated
in the eLSM in selected cases (as e.g. the scalar sector and/or broad
resonances \cite{Wolkanowski:2015jtc,Wolkanowski:2015lsa}); the finite dimension of mesons of about 0.5 fm provides a quite natural cutoff. 

\item Baryons are also an important part of the eLSM, both for the two-flavor \cite{Gallas:2009qp,Gallas:2011qp,Gallas:2013ipa} and for the three-flavor \cite{Olbrich:2015gln,Olbrich:2017fsd,Kovacs:2013gxa} cases. The baryonic masses emerge, within the so-called mirror assignment \cite{Detar:1988kn},
as a combination of the dilaton and chiral condensates.

\item The eLSM contains chiral partners in its Lagrangian, making it suitable for studies at nonzero density and temperature \cite{Kovacs:2016juc,Kovacs:2022zcl}. This feature has been stressed from the very beginning as one of the main goals of the approach \cite{Lenaghan:2000ey}. 
\end{enumerate}

In this work, we present the main properties and achievements of the eLSM by following the strategy outlined below.

In Sec.~\ref{Sec:QCD}, after a concise but as much as possible complete recall of QCD with a focus on its exact and approximate symmetries together with their explicit, anomalous, and spontaneous breaking patterns, we introduce composite fields: the dilaton field $G$  and various nonets: (pseudo)scalar, (axial-)vector, excited/pseudovector, (axial-)tensor, and hybrid nonets. The starting point is, for all nonets, the corresponding microscopic quark-antiquark current, which is responsible for transformation properties. Nonets are then grouped into chiral multiplets, which can be of two types: heterochiral and homochiral. Both nonets and chiral multiplets are introduced in great detail since the related properties are general and model-independent. 
In this section we also present scattering reactions and some of the past, planned and ongoing experiments dealing with light mesons. The results of these experiments are summarized in the Particle Data Group \cite{ParticleDataGroup:2024cfk}, which is often used to test the eLSM. 
%For the (pseudo)scalar case, we show explicitly how the mesonic fields are linked to the underlying quark currents.
In Sec.~\ref{Sec:eLSM_Lagr} the basic form of eLSM Lagrangian involving the dilaton and the (pseudo)scalar and (axial-)vector multiplets is introduced for $N_f=3$ and the main aspects of its phenomenology are discussed.
In Sec.~\ref{Sec:other(1,2)} other conventional multiplets with total angular momentum $J=1,2$ are added to the eLSM and their masses and decays are studied, in certain cases by using novel $U(1)_A$ anomalous terms.
In Sec.~\ref{Sec:hybrids} the lightest multiplet involving hybrid mesons, with the nonet
$J^{\mathcal{P}\mathcal{C}}=1^{-+}$ and the nonet $J^{\mathcal{P}\mathcal{C}}=1^{+-}$
are presented.

Sec.~\ref{Sec:glueballs} is devoted to glueballs: first, the phenomenology of the scalar glueball (present in the model as a dilaton) is outlined. Then, further glueballs are coupled to the eLSM: the tensor glueball and the pseudoscalar glueball (the latter via a chiral anomalous coupling). An example of a heavy three-gluon glueball, the vector glueball, is also studied.

Three additional topics are discussed in Sec. ~\ref{Sec:other}: (i) the breaking of the isospin symmetry (due to unequal $u$ and $d$ masses) gives rise to small, but in certain cases well measured, processes; (ii) the extension to $N_f=4$ is briefly discussed, showing that certain interaction terms still fulfill chiral symmetry even though the explicit breaking due to the charm mass term is large; (iii) the inclusion of a radially excited multiplet of (pseudo)scalar states.
In Sec.~\ref{Sec:eLSM_finite_T} we discuss the QCD phase diagram as it emerges from the eLSM, to which the Polyakov loop as well as quarks d.o.f. are added. Finally, in Sec.~\ref{Sec:sum} conclusions and outlooks are presented. Some additional aspects (unitary groups, large-$N_c$ recall, and the non-relativistic limit of $\bar{q}q$ currents) are succinctly described in three appendices. 

 \newpage
 \section{From QCD to composite fields}
 \label{Sec:QCD}

 \subsection{QCD and its symmetries}
 \label{Ssec:symmetries}
The strong interaction is one of the four fundamental interactions in Nature
and describes quarks and gluons as elementary particles. Each quark is a
fermion that may appear in $N_{f}=6$ different flavor forms, as well as in
three different colors (red `R', green `G', blue `B'). Gluons are bosons of the type color-anticolor (for a
total of $9-1$ configurations).

The Lagrangian of QCD \cite{Fritzsch:1973pi} is built under the requirement of being locally
invariant under $SU(3)_C$ color ($C$) transformations, see below. It takes the form
\begin{equation}
\mathcal{L}_{\text{QCD}}=\sum_{i=1}^{N_{f}}\mathrm{Tr}\Big(\bar{q}_{i}(x)(i\gamma
^{\mu}D_{\mu}-m_{i})q_{i}(x)\Big )+\mathcal{L}_{YM}\text{\thinspace
}.\label{QCD-Lag}%
\end{equation}
The first part contains $N_{f}$ fermionic fields
$q_{i}\equiv q_{i}(x)$ (the quarks) with bare mass $m_{i}$ together with their
antiparticles $\overline{q}_{i}$, embedded in a standard Dirac Lagrangian,
where $\gamma^{\mu}$ are the 4 Dirac matrices with $\left\{  \gamma^{\mu
},\gamma^{\nu}\right\}  =2g^{\mu\nu}\mathbb{1}_{4\times 4}$, $\left(  \gamma^{\mu}\right)
^{\dagger}=\gamma^{0}\gamma^{\mu}\gamma^{0}$ and $\bar{q}_{i}=q_{i}^{\dagger
}\gamma^{0}$. In the Dirac representation%
\begin{equation}
\gamma^{0}=\left(
\begin{array}
[c]{cc}%
\mathbb{1} & \mathbb{0}\\
\mathbb{0} & -\mathbb{1}% $\mathbb{1}$ and $\mathbb{0}$ stand for a $2\ times 2$ unit matrix and a zero matrix, respectively.
\end{array}
\right)  \text{ , }\gamma^{k}=\left(
\begin{array}
[c]{cc}%
\mathbb{0} & \sigma_{k}\\
\sigma_{k} & \mathbb{0}%
\end{array}
\right)  \text{ ,}\label{diracm}%
\end{equation}
where $\mathbb{1}=\diag\{1,1\}$, $\mathbb{0}$ is the $2\times2$ null matrix, and
$\sigma_{k}$ for $k=1,2,3$ are the 3 Pauli matrices:%
\begin{equation}
\sigma_{1}=\left(
\begin{array}
[c]{cc}%
0 & 1\\
1 & 0
\end{array}
\right)  \text{ , }\sigma_{2}=\left(
\begin{array}
[c]{cc}%
0 & -i\\
i & 0
\end{array}
\right)  \text{ , }\sigma_{3}=\left(
\begin{array}
[c]{cc}%
1 & 0\\
0 & -1
\end{array}
\right)  \text{ .}%
\label{pauli}
\end{equation}
%FG: up to here check of ch2
For each quark flavor, a quark field is a vector in the color
space:
\begin{equation}
q_{i}(x)=\left(
\begin{array}
[c]{c}%
q_{i,R}(x)\\
q_{i,G}(x)\\
q_{i,B}(x)
\end{array}
\right)  \text{ .}%
\end{equation}
If the quark field
is denoted simply as $q(x)$, it means that it is also vector in flavor space. By
restricting to $N_{f}=3$ (light quarks $u$, $d$, and $s\,$) that means
\begin{equation}
q(x)=\left(
\begin{array}
[c]{c}%
\left(
\begin{array}
[c]{c}%
u_{R}(x)\\
u_{G}(x)\\
u_{B}(x)
\end{array}
\right)  \\
\left(
\begin{array}
[c]{c}%
d_{R}(x)\\
d_{G}(x)\\
d_{B}(x)
\end{array}
\right)  \\
\left(
\begin{array}
[c]{c}%
s_{R}(x)\\
s_{G}(x)\\
s_{B}(x)
\end{array}
\right)
\end{array}
\right)
\text{ ,}
\label{Eq:colored_quarks}
\end{equation}
where, of course, each member above is itself a four-vector in Dirac space, see \ref{Ssec:non-relimit} for an explicit expression.

The covariant derivative $D_{\mu}$ in Eq. (\ref{QCD-Lag}) reads:
\begin{equation}
D_{\mu}:=\mathbb{1}_{3\times 3}\partial_{\mu}-igG_{\mu}\text{ }\,,\label{cov-der}%
\end{equation}
where $g_{\text{QCD}}=g$ is the QCD\ strong coupling and where the gluonic field
$G_{\mu}\equiv G_{\mu}(x)$ is introduced as an Hermitian and traceless $3\times3$ matrix expressed as:
\begin{equation}
G_{\mu}(x):=\sum_{a=1}^{8}G_{\mu}^{a}(x)t^{a}\, \text{ ,}
\end{equation}
where $t^{a}=\lambda^{a}/2$, with the $\lambda^{a}$ being the Gell-Mann matrices, see \ref{Ssec:UN_group} for a brief recall of the groups $SU(N)$ and $U(N)$ and standard textbooks, e.g. \cite{Halzen:1984mc,Mosel:1989jf,Peskin:1995ev}.
The interaction vertex is similar to the one of QED: for each quark flavor,
there is a quark-antiquark-gluon vertex. Yet, the gluon field itself is colored: namely, in view of the equation above, the field $G_{\mu}^{1}(x)$ may
be seen as a (properly normalized) $R\bar{G}-G\bar{R}$ colored gluon. Note that
$\lambda^{0}=\sqrt{2/3} \mathbb{1}_{3 \times 3}$, which is proportional to the identity matrix, does \textbf{not} appear above, meaning that the colorless configuration $R\bar {R}+G\bar{G}+B\bar{B}$ is \textit{not} realized in nature. This is why 8 gluons (and not 9, as a naive $N_{c}\cdot N_{c}$ counting would suggest) are considered.

The second part of the Lagrangian, called Yang-Mills (YM) Lagrangian,
describes solely the gluonic fields
\begin{equation}
\mathcal{L}_{YM}=-\frac{1}{2}\mathrm{Tr}\left[  G_{\mu\nu}G^{\mu\nu}\right]
\,,
\end{equation}
where the gluon field strength tensor $G_{\mu\nu}\equiv G_{\mu\nu}(x)$ is
given by
\begin{equation}
G_{\mu\nu}(x):=\partial_{\mu}G_{\nu}(x)-\partial_{\nu}G_{\mu}(x)-ig[G_{\mu}(x),G_{\nu
}(x)]\text{ }\,.\label{eq:gluon}%
\end{equation}
This Lagrangian contains three-leg and four-leg gluon interactions: this is an
important and specific property of non-abelian gauge theories (the basic Feynman diagrams of QCD can be found in \ref{App:Large-N}). Indeed, the
self interaction of gluons immediately raises the question: are there purely
gluonic bound states? There is not yet a definite positive answer, but more and more evidence -both from theory and experiment- points toward their physical existence. We shall discuss some specific glueballs
later on.

The $SU(3)_C$ color transformation is
parameterized by a local special unitary matrix
\begin{equation}
U(x)=e^{{-i\sum_{a=1}^{8}\theta^{a}(x)t^{a}}}\text{ ,}%
\end{equation}
where ${\theta^{a}(x)}$ are 8 arbitrary real functions of the space-time
variable $x$. The quark field change as (for each flavor $i=u,d,...$):%
\begin{equation}
q_{i}(x)\rightarrow U(x)q_{i}(x)\,,
\end{equation}
meaning that an arbitrary space-time-dependent reshuffling of the color d.o.f. is
carried out. The QCD Lagrangian (\ref{QCD-Lag}) is invariant under $SU(3)_C$ provided that the
gluon field transforms as
\begin{equation}
G_{\mu}(x)\rightarrow G_{\mu}^{\prime}(x):=U(x)G_{\mu}(x)U^{\dagger}%
(x)-\frac{i}{g}U(x)\left(  \partial_{\mu}U^{\dagger}(x)\right)  \,\text{.}%
\end{equation}
As a consequence, the covariant derivative and gluon field tensor
transform in a simple way:%
\begin{equation}
D^{\mu}q_{i}(x)   \rightarrow U(x)D^{\mu}\Big(q_{i}(x)\Big) \text{ , }
G_{\mu\nu}(x)   \rightarrow G_{\mu\nu}^{\prime}(x)=U(x)G_{\mu\nu
}(x)U^{\dagger}(x)
\text{ ,}
\end{equation}
out of which the local color invariance can be easily proven.
Note, under a global transformation $U(x)=U$, one has
\begin{equation}
q_{i}(x)\rightarrow Uq_{i}\,(x)\text{ , }G_{\mu}(x)\rightarrow UG_{\mu
}(x)U^{\dagger}%
\end{equation}
that shows how the quarks transform under the fundamental representation and the gluons under the adjoint representation of the group $SU(3)_C$.

Above, we expressed the formulas for the physical case of $3$ colors,
$N_{c}=3.$ The extension of QCD to a generic number of colors $N_{c}$ is
rather straightforward: $q_{i}(x)$ becomes a vector with $N_{c}$ color entries and
the gluon field is a $N_{c}\times N_{c}$ matrix with:
\begin{equation}
q_{i}=\left(
\begin{array}
[c]{c}%
q_{i,1}\\
q_{i,2}\\
...\\
q_{i,N_{c}}%
\end{array}
\right)  \text{ , }G_{\mu}:=\sum_{a=1}^{N_{c}^{2}-1}G_{\mu}^{a}t^{a}
\text{ ,}
\end{equation}
where the $N_{c}\times N_{c}$ linearly independent Hermitian matrices $t^{a}$
are chosen to fulfill $Tr[t^{a}t^{b}]=\delta^{ab}/2$ with $a,b=0,\ldots,N_{c}^{2}-1$ and $t^{0}=\mathbb{1}_{N_{c}\times N_c}/\sqrt{2N_{c}}\,$. For $N_{c}=2$ the usual choice is $t^{a}=\sigma^{a}/2$ with the Pauli matrices $\sigma^{a}$, while for $N_{c}=3$ it is $t^{a}=\lambda^{a}/2$ as presented above, see \ref{Ssec:UN_group}.

The QCD Lagrangian is, of course, invariant under proper orthochronous Lorentz
transformations (the so-called restricted Lorentz group as well as space-time
translations \cite{Martin:1970hmp}), just as any other piece of the SM of particle
physics. 
As discussed in general terms in Sec.~\ref{Sec:intro}, there is a set of additional classical global symmetries of QCD that are realized in certain limits, some of which are broken explicitly, spontaneously, and/or anomalously (that is, by quantum fluctuations). Here we need to recall their main features in a more technical language, since all of them are extremely relevant for writing low-energy hadronic models in general and the eLSM in particular.

\bigskip

\begin{itemize}
\item Parity transformations ($\mathcal{P}$ ). It amounts to the reflection
$x=(t,\mathbf{x})\rightarrow(t,-\mathbf{x}),$ which leaves $\mathcal{L}_{\text{QCD}}$ invariant upon performing the replacements:%
\begin{equation}
q_{i}(x)=q_{i}(t,\mathbf{x})\rightarrow\gamma^{0}q_{i}(t,-\mathbf{x})\text{ ;
}G_{\mu}(x)=G_{\mu}(t,\mathbf{x})\rightarrow G^{\mu}(t,-\mathbf{x})\text{
.}\label{parity}%
\end{equation}
In the Dirac notation:
\begin{equation}
\gamma^{0}=\left(
\begin{array}
[c]{cc}%
\mathbb{1} & \mathbb{0}\\
\mathbb{0} & -\mathbb{1}%
\end{array}
\right)
\text{ ,}
\end{equation}
making evident that the intrinsic parity of particles and anti-particles is opposite, see \ref{Ssec:non-relimit} for an explicit calculation.
Parity plays an important role in the construction of models of QCD, since any
interaction term shall fulfill it. 
Moreover, it is also useful for the
classification of both conventional and non-conventional mesons.

\item Charge-conjugation transformation ($\mathcal{C}$ ).  It corresponds to the exchange of particles
with antiparticles. The quark and gluon fields transform as%
\begin{equation}
q_{i}(x)\rightarrow S_{C}\bar{q}_{i}^{T}(x)\text{ with }S_{C}^{\dagger}%
\gamma^{\mu}S_{C}=-\left(\gamma^{\mu}\right)^{T};\text{ }G_{\mu
}(x)\rightarrow-G_{\mu}^{T}(x)\text{ .}%
\end{equation}
Within the Dirac representation one has
\begin{equation}
q_{i}(x)\rightarrow i\gamma^{2}q_{i}^{\ast}(x)\ \text{with }i\gamma
^{2}=\left(
\begin{array}
[c]{cc}%
\mathbb{0} & \varepsilon\\
\varepsilon & \mathbb{0}
\end{array}
\right)  \text{ , }\left(  \varepsilon\right)  _{ij}=\varepsilon_{ij}\text{  ,}%
\end{equation}
which makes the switch between fermion and anti-fermion evident. The QCD Lagrangian
$\mathcal{L}_{\text{QCD}}$ is left invariant by this transformation. Just as parity,
$\mathcal{C}$  is fulfilled by any interaction term and enters in the classification of mesons. In fact, a generic meson (or mesonic nonet) is expressed by the notation%
\begin{equation}
J^{\mathcal{P}\mathcal{C}}
\text{ ,}
\end{equation}
where $J$ is the total angular momentum. As we shall explain in detail in the following, for quark-antiquark bound states the eigenvalues for $\mathcal{P}$ and $\mathcal{C}$ can be obtained by the orbital angular momentum quantum number $L$ and by the spin $S$ as $\mathcal{P}=(-1)^{L+1}$ and $\mathcal{C}=(-1)^{L+S}$. The lightest hadrons are realized
for $L=S=0\rightarrow J^{\mathcal{P}\mathcal{C}}=0^{-+},$ corresponding to the already mentioned
nonet of pseudoscalar mesons: $\pi^{+}\equiv u\bar{d},$ $\pi^{0}\equiv
\sqrt{1/2}(\bar{u}u-\bar{d}d),$ $\pi^{-}\equiv u\bar{d},$ $K^{+}\equiv
u\bar{s},$ $K^{-}\equiv s\bar{u},$ $K^{0}\equiv d\bar{s},$ $\bar{K}^{0}\equiv
s\bar{u}$ , as well as the non-strange and strange objects $\eta_{N}%
\equiv\sqrt{1/2}(\bar{u}u-\bar{d}d)$ and $\eta_{S}\equiv\bar{s}s$ (the
physical states $\eta$ and $\eta^{\prime}$ emerge as a mixing of the latter
two fields).

In terms of quarks, one may schematically write $\mathcal{C}\left\vert u\right\rangle
=\left\vert \bar{u}\right\rangle ,$ etc. Some mesons are eigenstates of $\mathcal{C},$
such as the neutral pion for which $\mathcal{C}\left\vert \pi^{0}\right\rangle
=\left\vert \pi^{0}\right\rangle $, and the two $\eta$-states $\mathcal{C}\left\vert \eta
_{N}\right\rangle =\left\vert \eta_{N}\right\rangle $ and $\mathcal{C}\left\vert
\eta_{S}\right\rangle =\left\vert \eta_{S}\right\rangle $. However, some of
them are not. For instance, for charged pions and for kaons one has
$\mathcal{C}\left\vert \pi^{+}\right\rangle =\left\vert \pi^{-}\right\rangle $ ,
$\mathcal{C}\left\vert K^{+}\right\rangle =\left\vert K^{-}\right\rangle $ ,
$\mathcal{C}\left\vert K^{0}\right\rangle =\left\vert \bar{K}^{0}\right\rangle $. For a given multiplet, the $\mathcal{C}$-eigenvalue refers to the members that are eigenstates of the charge conjugation.

\item Dilatation transformation. This transformation refers to the space-time
dilatation $x^{\mu}\rightarrow\lambda^{-1}x^{\mu}$ by a factor $\lambda^{-1} > 0$
together with the fields transformations%
\begin{equation}
q_{i}(x)\rightarrow q_{i}^{\prime}(x)=\lambda^{3/2}q_{i}(\lambda x)\text{ ,
}G_{\mu}(x)\rightarrow G_{\mu}^{\prime}(x)=\lambda G_{\mu}(\lambda x)\text{ .}%
\end{equation}
It is easy to verify that in the chiral limit (all quark masses $m_{i}=0)$ the
infinitesimal QCD action is invariant:
\begin{equation}
\mathcal{L}_{\text{QCD}}(q_{i}(x),G_{\mu}(x))d^{4}x=\mathcal{L}_{\text{QCD}}(q_{i}^{\prime
}(x^{\prime}),G_{\mu}^{\prime}(x^{\prime}))d^{4}x^{\prime}\text{ .}%
\end{equation}
This is the case because each term of the Lagrangian scales by a factor $\lambda
^{4},$ which is compensated by a factor $\lambda^{-4}$ arising from the
infinitesimal space-time volume.
In turn, this symmetry is physically evident:
in the chiral limit, $\mathcal{L}_{\text{QCD}}$ contains no parameter that carries a dimension.
It is also clear that any mass term breaks this symmetry since $m_{i}\bar
{q}_{i}(x)q_{i}(x)$ scales with $\lambda^{3}.$ 
The symmetry is then explicitly broken by quark masses, but is also anomalously broken by quantum fluctuations \cite{Collins:1976yq,Thomas:2001kw,Greiner:2002ui}. The divergence of the corresponding dilatation current is non-vanishing in the chiral limit as well: 
\begin{equation}
\partial_{\mu}J_{\text{QCD,\,dil}}^{\mu}=\left(  \Theta_{\text{QCD}}\right)  _{\mu}^{\mu
}=\frac{\beta(g)}{2g}G_{\mu\nu}^{a}G^{a\,\mu\nu}+{\sum_{i=1}^{N_{f}}}m_{i}%
\bar{q}_{i}q_{i}\text{ with }J_{\text{QCD,\,dil}}^{\mu}=x_{\nu}\Theta_{\text{QCD}}^{\mu\nu
}\text{ ,}\label{ta}%
\end{equation}
where $\Theta_{\text{QCD}}^{\mu\nu}$ is the energy-momentum tensor of QCD and the beta function $\beta(g)$ is given by (keeping $N_{c}$ and $N_{f}$ general):
\begin{equation}
\beta(g)=\mu\frac{dg(\mu)}{d{\mu}}\overset{\text{one-loop}}{=}-\,\frac
{11N_{c}-2N_{f}}{48\pi^{2}}g^{3}\text{ ,}%
\end{equation}
where  the renormalized running coupling $g_{\text{QCD}}(\mu) = g(\mu)$ has been introduced. 
Through the process of quantizing QCD and applying an appropriate renormalization procedure, the
coupling constant $g$ in Eq. \eqref{cov-der} is dependent on the energy scale
$\mu$ as \cite{Thomas:2001kw}:
\begin{equation}
g^{2}(\mu)=\frac{48\pi^{2}}{\left(  11N_{c}-2N_{f}\right)  \log\frac{\mu
}{\Lambda_{\text{QCD}}}}\, \text{ ,} \label{rcc}%
\end{equation}
where the low-energy QCD scale $\Lambda_{\text{QCD}}\approx250$ MeV
\cite{Deur:2016tte} has been identified with the real part of the Landau Pole. In turn, this
running implies that, at short distances, quarks behave as free particles,
which is another non-trivial feature of QCD (so-called \textquotedblleft
asymptotic freedom\textquotedblright). Note that the presence of a pole is an artifact of the one-loop approximation, but it signals a separation between the high-energy perturbative and the low-energy non-perturbative regions. The latter is where the hadrons live.

Eq. (\ref{rcc}) offers also the basis for studying QCD in the large-$N_{c}$
limit. Following 't Hooft \cite{tHooft:1973alw}, one keeps $\Lambda_{\text{QCD}}$ and $N_{f}$ fixed, and $N_{c}$ is taken as a parameter. It then follows that $g^{2}\sim N_{c}^{-1}$ for large values of $N_c$. In this limit, many simplifications occur: the masses of conventional mesons as well as glueballs and hybrid mesons are $N_{c}$-independent, but their interaction goes to zero. Thus, they became stable for increasing $N_{c},$ see the reviews \cite{Witten:1979kh,Lebed:1998st,Lucha:2021mwx,Giacosa:2024scx} and \ref{App:Large-N}. In
particular, in the lecture of Ref. \cite{Giacosa:2024scx} a simplified connection between large-$N_{c}$ limit and chiral models (such as the eLSM ) is presented.

\item The classical global QCD symmetry group $\mathcal{G}_{cl}\equiv U(3)_{L}\times
U(3)_{R}$ in the chiral limit. We first introduce the left-and right-handed quarks by using the
matrix $\gamma^{5}=i\gamma^{0}\gamma^{1}\gamma^{2}\gamma^{3}$ (with $\left\{
\gamma^{5},\gamma^{\mu}\right\}  =0$, $\left(  \gamma^{5}\right)  ^{\dagger
}=\gamma^{5},\left(  \gamma^{5}\right)  ^{2}= \mathbb{1}$):
\begin{equation}
q_{i,L}=P_{L}q_{i}=\frac{1}{2}\left(  \mathbb{1}-\gamma^{5}\right)  q_{i}\text{ ,
}q_{i,R}=P_{R}q_{i}=\frac{1}{2}\left(  \mathbb{1}+\gamma^{5}\right)  q_{i}\text{ ,}%
\end{equation}
as well as the related transformations:
\begin{equation}
\bar{q}_{i,L}=\bar{q}_{i}P_{R}\text{ , }\bar{q}_{i,R}=\bar{q}_{i}P_{L}\text{ .}%
\end{equation}
In the Dirac representation%
\begin{equation}
\gamma^{5}=\left(
\begin{array}
[c]{cc}%
\mathbb{0} & \mathbb{1}\\
\mathbb{1} & \mathbb{0}%
\end{array}
\right)  \text{ ,}%
\end{equation}
that implies a switch of upper and lower components (see
\ref{Ssec:non-relimit}). Considering $N_{f}=3$ for definiteness from here
down, a transformation under $\mathcal{G}_{cl}$ amounts to:
\begin{equation}
q=q_{L}+q_{R}\xrightarrow{U(3)_{L}\times U(3)_{R}}U_{L}q_{L}+U_{R}q_{R}
\text{ ,}
\end{equation}
where
\begin{equation}
U_{L}=e^{{-i\sum_{k=0}^{8}\theta_{L}^{(k)}}t^{k}}\text{ and  }U_{R}%
=e^{{-i\sum_{k=0}^{8}\theta_{R}^{(k)}}t^{k}}%
\end{equation}
are two independent unitary $U(3)$ matrices (here the zeroth component is retained). This means that the quark flavors for $q_L$ and $q_R$ are -- independently of each other -- mixed. In terms of components, the chiral transformation takes the form:%
\begin{equation}
q_{i}=q_{i,L}+q_{i,R}\xrightarrow{U(3)_{L}\times U(3)_{R}}\left(
U_{L}\right)  _{ij}q_{j,L}+\left(  U_{R}\right)  _{ij}q_{j,R}
\text { .}
\end{equation}
Note that a transformation solely under $U(3)_{L}$ implies
$q=q_{L}+q_{R}\xrightarrow{U(3)_{L}}U_{L}q_{L}+q_{R}$,
where only the left-component is rotated. The Lagrangian $\mathcal{L}_{\text{QCD}}$
is invariant under $\mathcal{G}_{cl}$ if $m_{u}=m_{d}=m_{s}=0.$ In fact, the gluon terms
\begin{equation}
g\bar{q}_{i}\gamma^{\mu}q_{i}G_{\mu}=g\left(  \bar{q}_{i,L}\gamma^{\mu}%
q_{i,L}+\bar{q}_{i,R}\gamma^{\mu}q_{i,R}\right)  G_{\mu}%
\end{equation}
clearly split into right-handed and left-handed separate parts. On the other
hand, the mass terms break it, since:%
\begin{equation}
-m_{i}\bar{q}_{i}q_{i}=-m_{i}\left(  \bar{q}_{i,L}q_{i,R}+\bar{q}_{i,R}%
q_{i,L}\right)  \text{ .}%
\end{equation}

Finally, the $9+9$ (classically) conserved right-handed and left-handed
currents that emerge from the Noether theorem are:%
\begin{equation}
J_{L,k}^{\mu}=\bar{q}_{L}\gamma^{\mu}t^{k}q_{L}\text{ },\text{ }J_{R,k}^{\mu
}=\bar{q}_{R}\gamma^{\mu}t^{k}q_{R}\text{ }%
\end{equation}
with the corresponding conserved charges%
\begin{equation}
Q_{L,k}=\int d^{3}x\bar{q}_{L}\gamma^{0}t^{k}q_{L}\text{ , }Q_{R,k}=\int
d^{3}x\bar{q}_{R}\gamma^{0}t^{k}q_{R}\text{ .}%
\end{equation}
Further breaking patterns (besides nonzero and unequal masses) are described
shortly hereafter.

\item $\mathcal{G}_{cl}\equiv U(3)_{L}\times U(3)_{R}$ is decomposed
as
\begin{equation}
\mathcal{G}_{cl}\equiv U(3)_{L}\times U(3)_{R}=U(1)_{L}\times U(1)_{R}%
\times\underset{\mathcal{G}_{\text{chiral}}}{\underbrace{SU(3)_{L}\times SU(3)_{R}}%
}=\underset{\text{baryon number}}{\underbrace{U(1)_{V}}}\negthickspace\times\quad \underset{\text{axial}}{\underbrace{U(1)_{A}}}\times\underset{\mathcal{G}_{\text{chiral}}}{\underbrace{SU(3)_{L}\times SU(3)_{R}}}%
\end{equation}
where the chiral group $\mathcal{G}_{\text{chiral}}\equiv SU(3)_{L}\times SU(3)_{R}$ has been singled out. 
%Since $U(1)_{L}\times U(1)_{R}$ are simple phases, the equations presented in the previous item  still hold. 

\item Flavor symmetry and baryon number symmetry $U(3)_{V} = U(1)_V \times SU(3)_V$. The group $U(3)_{V}$ is a subgroup of
 $\mathcal{G}_{cl}\equiv U(3)_{L}\times U(3)_{R}$ realized for
the choice $U_{L}=U_{R}=U_{V}$, hence for the parameter choice ${\theta
_{L}^{(k)}=\theta_{R}^{(k)}=\theta_{V}^{(k)}.}$ It amounts to%
\begin{equation}
q=q_{L}+q_{R}\xrightarrow{U(3)_{V}}U_{V}q_{L}+U_{V}q=U_{V}q\text{ with }%
U_{V}=e^{{-i\sum_{k=0}^{8}\theta_{V}^{(k)}}t^{k}}\text{ ,}%
\end{equation}
which is a rotation in flavor space, independent on chirality. 
%Baryon symmetry $U(1)_V$ holds exactly. 
The group $U(3)_{V}$ can be decomposed into $SU(3)_{V}\times U(1)_{V},$ where
$SU(3)_{V}$ corresponds to the flavour symmetry with the parameterization
\begin{equation}
U_{V}=e^{{-i\sum_{k=1}^{8}\theta_{V}^{(k)}}t^{k}} \text{ ,}
\end{equation}
while $U(1)_{V}$ to the phase
\begin{equation}
U_{V}=e^{{-i\theta_{V}^{(0)}}t^{0}}\text{ , }t^{0}=\frac{1}{\sqrt{6}%
}\mathbb{1}_{3\times3}\text{ .}%
\end{equation}

The exact conservation of $U(1)_{V}$ refers to baryon number conservation with
current and charge expressed as%
\begin{equation}
J_{B}^{\mu}=\bar{q}\gamma^{\mu}t^{0}q\text{ , }B=\frac{1}{3}\int
d^{3}xq^{\dagger}t^{0}q\text{ .}%
\end{equation}
Following the usual convention, each quark has baryon number $1/3$ (with
$B\ket{q}=\frac{1}{3} \ket{q}$), each antiquark $-1/3$ (with
$B\ket{\bar{q}}=-\frac{1}{3} \ket{\bar{q}}$), thus each baryon has $B=1$, each
anti-baryon $B=-1$, and each meson $B=0$ (regardless if conventional or non-conventional).

Flavor symmetry $SU(3)_V$
is not broken by nonzero masses, but by unequal masses. In particular, the
mass differences $m_{s}-m_{u}$ and $m_{s}-m_{d}$ are non-negligible. The mass
difference $m_{d}-m_{u}\ll\Lambda_{\text{QCD}}$ is small, thus considering them equal
is a good approximation. The conserved currents and charges are given by
\begin{equation}
J_{V,k}^{\mu}=\bar{q}\gamma^{\mu}t^{k}q=J_{L,k}^{\mu}+J_{R,k}^{\mu}\text{ ,
}Q_{V,k}=\int d^{3}xq^{\dagger}t^{k}q\text{ for}\; k =1,...,8 \text{ .}
\end{equation}

%When dealing with mesons that contain an equal number of quarks and antiquarks, thus $U(1)_{V}$ generates no change of the mesonic state (it is just as the identity operation). Thus, for mesons referring to the flavortransformations  $SU(3)_{V}$ is perfectly sufficient.

%In fact, for mesons containing an equal number of quarks and antiquarks, $U(1)_{V}$ does not produce any change in the mesonic state (it is the same as the identity operation). Thus, for mesons, it is perfectly sufficient to refer to flavor transformations as $SU(3)_{V}$.
\item Isospin symmetry $SU(2)_{I}.$ This is a subset of flavor transformation;
$SU(2)_{V}$ corresponds to the choice,
\begin{equation}
U_{V}=e^{{-i\sum_{k=1}^{3}\theta_{V}^{(k)}}t^{k}}=\left(
\begin{array}
[c]{cc}%
U_{I} & 0\\
0 & 1
\end{array}
\right)  \text{ , }U_{I}=e^{-{i\sum_{k=1}^{3}\theta_{I}^{(k)}}\frac{\sigma^{k}%
}{2}}\;,%
\end{equation}
where $U_{I}$ is a $2\times2$ unitary matrix parameterized by three angles
${\theta_{I}^{(k)}}$. This is the famous isospin symmetry, which mixes the
flavors $u$ and $d$ and was first introduced by Heisenberg at the level of the nucleon (a rotation in the $u$-$d$ space corresponds to a rotation in the
proton-neutron space) \cite{Heisenberg:1932dw,Wigner:1936dx}. When acting on quarks, the isospin operator reads $I_k = \sigma_k /2$, thus $I_3 \ket{u} = \frac{1}{2} \ket{u}$,  $I_3 \ket{d} = -\frac{1}{2} 
 \ket{d}$, and $\mathbf{I}^2 \ket{u,d} = \frac{3}{4}  \ket{u,d}$ and $\mathbf{I}^2 \ket{s} = 0$.

Isospin symmetry is well fulfilled in low-energy QCD
phenomenology, both for masses and scattering processes \cite{Pennington:2005be}. In fact, the
amplitudes for processes that break isospin are typically proportional to
$\left(  m_{d}-m_{u}\right)  /\Lambda_{\text{QCD}}\ll1.$ It is also interesting to
note that recent experimental findings from heavy ion collisions question the
validity of isospin symmetry \cite{NA61SHINE:2023azp,Brylinski:2023nrb}. For a summary of the quantum numbers of light quarks, see Tab~\ref{Tab:quark_table}.
\begin{table}[ht!]
\centering
\renewcommand{\arraystretch}{1.5}
\begin{tabular}{|c|c|c|c|c|c|c|c|}
\hline
 & $Q$ & $I$ & $I_3$ & $Y$ & $S$ & $B$ & $m_i$ (MeV)  \\ \hline
u & $\frac{2}{3}$  & $\frac{1}{2}$ & $\frac{1}{2}$ & $\frac{1}{3}$ & $0$  & $\frac{1}{3}$ & $ 2.16\pm 0.07$  \\ \hline
d & $-\frac{1}{3}$  & $\frac{1}{2}$ & $-\frac{1}{2}$ & $\frac{1}{3}$ & $0$  & $\frac{1}{3}$ & $ 4.70\pm 0.07$  \\ \hline
s & $-\frac{1}{3}$  & $0$  & $0$  & $-\frac{2}{3}$  & $1$ & $\frac{1}{3}$ & $ 92.5\pm 0.8$  \\ \hline
$\overline{u}$ & $-\frac{2}{3}$  & $\frac{1}{2}$ & $-\frac{1}{2}$ & $-\frac{1}{3}$ & $0$  & $-\frac{1}{3}$ & $ 2.16\pm 0.07$  \\ \hline
$\overline{d}$ & $\frac{1}{3}$  & $\frac{1}{2}$ & $\frac{1}{2}$ & $-\frac{1}{3}$ & $0$  & $-\frac{1}{3}$  & $ 4.70\pm 0.07$ \\ \hline
$\overline{s}$ & $\frac{1}{3}$  & $0$  & $0$  & $\frac{2}{3}$  & $-1$ & $-\frac{1}{3}$ & $ 92.5\pm 0.8$  \\\hline
\end{tabular}
\caption{Quantum numbers and masses for the light (anti)quarks.}
\label{Tab:quark_table}
\end{table}

\item Charge-symmetry transformation. An important isospin transformation, called
charge-symmetry transformation and denoted by $\mathcal{C}_{I}$ (where $I$ stands for isospin in order to distinguish it from the previously introduced charge conjugation $\mathcal{C}$ )\footnote{Caution is needed, because the same name `charge' may refer to different operations, but charge-conjugation $\mathcal{C}$ and charge transformation $\mathcal{C}_I$ are in general different objects.} forms a discrete subgroup of
isospin: it corresponds to a $180^{\circ}$ rotation around
the second isospin axis (${\theta_{I}^{(1)}=\theta_{I}^{(3)}=0,}$ ${\theta
_{I}^{(2)}=\pi),}$ leading to (see e.g. Ref. \cite{Gazdzicki:1991ih} and also \ref{Ssec:UN_group}):
\begin{equation}
\mathcal{C}_{I}=e^{-{i\pi}\frac{\sigma^{2}}{2}}=\varepsilon=\left(
\begin{array}
[c]{cc}%
0 & -1\\
1 & 0
\end{array}
\right)  \text{ .}%
\end{equation}
This transformation is an inversion of the third component of isospin:
$\mathcal{C}_{I}\left\vert u\right\rangle =\left\vert d\right\rangle $ and
$\mathcal{C}_{I}\left\vert d\right\rangle =-\left\vert u\right\rangle $, schematically
$u\leftrightarrow d$. For antiquarks $\mathcal{C}_{I}\left\vert \bar{u}\right\rangle
=\left\vert \bar{d}\right\rangle $ and $\mathcal{C}_{I}\left\vert \bar{d}\right\rangle
=-\left\vert \bar{u}\right\rangle $. As a consequence, for pions and kaons one
has $\mathcal{C}_{I}\left\vert \pi^{+}\right\rangle =-\left\vert \pi^{-}\right\rangle $
, $\mathcal{C}_{I}\left\vert \pi^{0}\right\rangle =-\left\vert \pi^{0}\right\rangle ,$
$\mathcal{C}_{I}\left\vert K^{+}\right\rangle =-\left\vert K^{0}\right\rangle $ ,
$\mathcal{C}_{I}\left\vert K^{-}\right\rangle =-\left\vert \bar{K}^{0}\right\rangle ,$
while for the two isoscalar $\eta$ states one gets $\mathcal{C}_{I}\left\vert \eta\right\rangle
=\left\vert \eta\right\rangle $ and $\mathcal{C}_{I}\left\vert \eta^{\prime}\right\rangle
=\left\vert \eta^{\prime}\right\rangle $.

\item $\mathcal{G}$-parity. A $\mathcal{G}$-transformation combines the charge conjugation
operator $\mathcal{C}$  and the isospin charge-symmetry transformation $\mathcal{C}_{I}$ as
\begin{equation}
\mathcal{G}=\mathcal{C}\cdot \mathcal{C}_{I}\text{ .}%
\end{equation}
The basic idea behind it is simple: while $\left\vert \pi^{+}\right\rangle $
is neither an eigenstate of $\mathcal{C}$ nor of $\mathcal{C}_{I},$ but it is an eigenstate of $\mathcal{G}$
with $\mathcal{G}\left\vert \pi^{+}\right\rangle =-\left\vert \pi^{+}\right\rangle $. It
turn out that isospin multiplets with integer isospin are eigenstates of $\mathcal{G}$
parity with eigenvalue $\mathcal{C}(-1)^{I}$, thus for quark-antiquark states
$\mathcal{G}=(-1)^{L+S+I}.$ $\mathcal{G}$-parity is reported in the PDG for all resonances for which it is a good quantum number (for integer values of $I$): it is conserved with a very good degree of accuracy in
strong processes (including decays) because $\mathcal{P}$ is exactly conserved and $\mathcal{C}_{I}$ is almost conserved.

\item Strangeness and other discrete internal quantum numbers. Strangeness transformation can also be seen as a specific $U(3)_V$
transformation corresponding to the choice $\theta_{V,k\neq0,8}=0$ and
$\theta_{S}=\theta_{V,8}=-\sqrt{2}\theta_{V,0}$, leading to
\begin{equation}
U_{S}=\exp\left[  \frac{\theta_{S}}{6\sqrt{3}}\left(
\begin{array}
[c]{ccc}%
0 & 0 & 0\\
0 & 0 & 0\\
0 & 0 & 1
\end{array}
\right)  \right]  \text{ ,}%
\end{equation}
which implies that $S\left\vert u\right\rangle =0,$ $S\left\vert
d\right\rangle =0.$ By convention, we assign $S\left\vert s\right\rangle
=-\left\vert s\right\rangle$, implying that $S\left\vert \bar{s}\right\rangle
=\left\vert \bar{s}\right\rangle$. For the kaons, one has $S\left\vert
K^{+}\right\rangle =\left\vert K^{+}\right\rangle ,$ $S\left\vert
K^{0}\right\rangle =\left\vert K^{0}\right\rangle $ (positive strangeness),
and $S\left\vert \bar{K}^{0}\right\rangle =-\left\vert \bar{K}^{0}%
\right\rangle ,$ $S\left\vert K^{-}\right\rangle =-\left\vert K^{-}%
\right\rangle $ (negative strangeness). The pions and the $\eta,\eta^{\prime}$
have zero strangeness. In principle, similar quantum numbers may be assigned
for the light quarks $u$ and $d$ (and denoted as $U$ and $D$), but this is
superfluous if isospin is considered\footnote{Note that the symmetries under $U$, $D$, and $S$, which are phase transformations for a chosen quark flavor, are exact regardless of the quark masses. Thus, the actual symmetry of QCD for unequal and non-zero masses is given by $U(1)_U \times U(1)_D \times U(1)_S$.}. On the other hand, for heavy quarks analogous quantum numbers are introduced for the charmness $C,$ bottomness
$B^{\prime},$ and topness $T^{\prime},$ but they will be not used in this work. 
In the light sector, two quantum numbers are relevant: the strong hypercharge
$Y=B+S$ and the electric charge $Q=I_{3}+Y/2,$ see Table \ref{Tab:quark_table} for a summary.

\item (General) axial transformations. Another important but utterly different
subset of the classical group $\mathcal{G}_{cl}$ is obtained for ${\theta
_{L}^{(k)}=-\theta_{R}^{(k)}=\theta_{A}^{(k)},}$ for which
\begin{equation}
U_{L}=U_{R}^{\dagger}=U_{A}\text{ .}%
\end{equation}
The quark field transforms as:
\begin{equation}
q=q_{L}+q_{R}\xrightarrow{U_{A}}U_{A}q_{L}+U_{A}^{\dagger}q_{R}%
=e^{{i\sum_{k=0}^{8}\theta_{A}^{(k)}}t^{k}\gamma^{5}}q\text{ .}%
\end{equation}
One may be tempted to refer to these transformations with $U(3)_{A}$ (and that
is sometimes done), but care is needed because this set of transformations
does \textit{not} form a group. In fact, if we perform this transformation
first with $U_{A}=U_{1}$ and subsequently with $U_{A}=U_{2},$ the quark field
changes as%
\begin{equation}
q=q_{L}+q_{R}\rightarrow U_{2}U_{1}q_{L}+U_{2}^{\dagger}U_{1}^{\dagger}%
q_{R}=U_{2}U_{1}q_{L}+\left(  U_{1}U_{2}\right)  ^{\dagger}q_{R}\text{ ,}%
\end{equation}
which is \textit{not} an axial transformation, since in general $U_{1}%
U_{2}\neq U_{2}U_{1}.$ Nevertheless, there are 9 classically conserved
currents and charges (in the chiral limit):%
\begin{equation}
J_{A,k}^{\mu}=\bar{q}\gamma^{\mu}\gamma^{5}t^{k}q=-J_{L,k}^{\mu}+J_{R,k}^{\mu
}\text{ , }Q_{A,k}=\int d^{3}xq^{\dagger}\gamma^{5}t^{k}q\text{ ,
}k=0,1,...8\text{ .}%
\end{equation}

\item Axial $U(1)_{A}$ transformation. For the specific choice of $\mathcal{G}_{cl}$ given by
\begin{equation}
U_{L}=U_{R}^{\dagger}=U_{A}=e^{-i{\theta_{A}^{(0)}t}^{0}}\text{ ,}%
\end{equation}
the axial transformation is a group, denoted as $U(1)_{A}$. The corresponding
current and charge are $J_{A,0}^{\mu}=\bar{q}\gamma^{\mu}\gamma^{5}t^{0}q$ and
$Q_{A,0}=\int d^{3}xq^{\dagger}\gamma^{5}t^{0}q.$ This symmetry is broken by
nonzero quark masses and, in addition and most significantly, by quantum
fluctuations \cite{tHooft:1976rip,Jackiw:2008,Bell:1969ts,Shifman:1988zk},
leading to
\begin{equation}
\partial_{\mu}J_{A,0}^{\mu}=\sum_{i=1}^{N_{f}}2m_{i}\bar{q}\gamma^{\mu}%
\gamma^{5}t^{0}q+\sqrt{6}\frac{g^{2}N_{f}}{8\pi^{2}}\epsilon_{\mu\nu\rho
\sigma}G^{\mu\nu\,,a}G^{\rho\sigma\,,a}\,,
\end{equation}
where the second term is anomalous, realizing the QCD chiral or axial anomaly. As explained by Zee \cite{Zee:2003mt}, this
arises form the fact that in the quantum version of the theory one cannot
fulfill at the same time $U(1)_{V}$ and $U(1)_{A}.$ Another explanation uses
the path integral approach: even if the QCD Lagrangian is invariant in the chiral
limit, the integration measure is not \cite{Fujikawa:1980eg,Fujikawa:2022cee}.

This anomalous breaking is particularly important in the low-energy\ QCD
phenomenology, especially for the correct understanding of the resonances
$\eta$ and $\eta^{\prime}$ \cite{Feldmann:1998vh}. The effect of this anomaly
on the phenomenology is driven by the so-called instantons, which are
non-perturbative solutions of the Euclidean equations of motions
\cite{Diakonov:2002fq,tHooft:1986ooh,Coleman:1978ae}. The chiral anomaly is taken
into account at the Lagrangian level in effective models. This is also the
case of the eLSM, where different terms with the correct symmetry breaking
patterns are introduced. Recently, novel anomalous mesonic terms were
investigated in Refs. \cite{Giacosa:2017pos,Giacosa:2023fdz}, see Sec.~\ref{Sec:other(1,2)}. In summary, because of this anomaly, the QCD symmetry group at the quantum level in the chiral limit reduces to
\begin{equation}
\mathcal{G}_{\text{QCD}} \equiv SU(3)_{L}\times SU(3)_{Q}\times U(1)_{V} \equiv \mathcal{G}_{\text{chiral}}\times U(1)_V\text{ .}%
\end{equation}
Since $U(1)_{V}$ is trivial for mesons (it leaves them invariant) one may
consider $\mathcal{G}_{\text{chiral}}\equiv SU(3)_{L}\times SU(3)_{R}$ as the actual symmetry
group that any mesonic Lagrangian should fulfill.

\item Chiral symmetry $\mathcal{G}_{\text{chiral}}\equiv SU(3)_{L}\times SU(3)_{R}$ also undergoes the phenomenon of spontaneous symmetry breaking (SSB). This is due to the
fact that the QCD vacuum $\left\vert 0_{\text{QCD}}\right\rangle $ is \textbf{not}
left invariant by the axial charges: $Q_{A,k}\left\vert 0_{\text{QCD}}\right\rangle
\neq0,$ for $k=1,,...,8$. However, in the chiral limit the QCD Hamiltonian
$H_{\text{QCD}}$ commutes with the charges $\left[  H_{\text{QCD}},Q_{A,k}\right]  =0$ for
$k=1,...,8.$ Then, the states $Q_{A,k}\left\vert 0_{\text{QCD}}\right\rangle $ for
$k=1,...,8$ correspond to massless particles:%
\begin{equation}
H_{\text{QCD}}Q_{A,k}\left\vert 0_{\text{QCD}}\right\rangle =Q_{A,k}H_{\text{QCD}}\left\vert
0_{\text{QCD}}\right\rangle =0\text{ .}%
\end{equation}
These are the 8 Goldstone bosons that emerge as a consequence of SSB. As will become clear later, these 8 Goldstone bosons $Q_{A,k}\left\vert 0_{\text{QCD}}%
\right\rangle $ are the already mentioned 8 pseudoscalar mesons. In
particular: $Q_{A,k=1,2,3}\left\vert 0_{\text{QCD}}\right\rangle \sim\left\vert
\pi^{k}\right\rangle $ (the pion triplet), $Q_{A,k=4,5,6,7}\left\vert
0_{\text{QCD}}\right\rangle $ corresponds to 4 kaons, and $Q_{A,k=8}\left\vert
0_{\text{QCD}}\right\rangle \sim\left\vert \eta_{8}\right\rangle $ where $\left\vert
\eta_{8}\right\rangle =\sqrt{1/6}\left(  \left\vert \bar{u}u\right\rangle
+\left\vert \bar{d}d\right\rangle -2\left\vert \bar{s}s\right\rangle \right)
=\sqrt{1/3}\left\vert \eta_{N}\right\rangle -\sqrt{2/3}\left\vert \eta
_{S}\right\rangle $ is the so-called octet configuration (roughly
corresponding to the meson $\eta(547)$). The $U(1)_{A}$-anomalous flavor
singlet configuration
\begin{equation}
Q_{A,k=0}\left\vert 0_{\text{QCD}}\right\rangle \sim\left\vert \eta_{0}\right\rangle
=\sqrt{1/3}\left(  \left\vert \bar{u}u\right\rangle +\left\vert \bar
{d}d\right\rangle +\left\vert \bar{s}s\right\rangle \right)  =\sqrt
{2/3}\left\vert \eta_{N}\right\rangle +\sqrt{1/3}\left\vert \eta
_{S}\right\rangle
\end{equation}
(roughly corresponding to the meson $\eta^{\prime}(958)$) is \textit{not} a
Goldstone boson (but would be such if the chiral anomaly is suppressed, as for
instance in the large-$N_{c}$ limit). For the physical value $N_{c}=3$ the
axial anomaly is strong, and the corresponding meson, the state $\eta^{\prime
}(958)$ is much heavier than the pions, kaons, and the $\eta$ meson. 

In formulas, the SSB implies
\begin{equation}
\mathcal{G}_{\text{chiral}}\equiv SU(3)_{L}\times SU(3)_{R}\xrightarrow{\text{SSB}}SU(3)_{V}%
\text{ .}%
\end{equation}

SSB has an important consequence in the hadronic spectrum: there is no mass
degeneracy between chiral partners.\ This concept will be discussed in depth
later on, but the main idea is simple: a chiral transformation mixes states
with the same $J$ but opposite parity (and, if defined, $\mathcal{G}$-parity). For
instance, chiral symmetry relates pseudoscalar mesons ($J^{\mathcal{P}%
\mathcal{C}}=0^{-+}$) with scalar mesons ($J^{\mathcal{P}\mathcal{C}}=0^{++}%
$).\ Indeed, scalar states are much heavier than pseudoscalar ones, confirming
SSB. The same applies for other chiral partners, such as vector and
axial-vector mesons.

\item
Due to confinement, mesons appear as d.o.f. with zero baryon number. While it is not possible to show how this happens in full QCD, bosonization techniques have been successfully applied to the NJL model mentioned in the introduction: by integrating out the quark fields of the NJL in favor of composite (pseudo)scalar auxiliary fields, the quark-level LSM appears as an intermediate stage; see Ref. \cite{Eguchi:1976iz} and the review articles in Refs. \cite{Vogl:1991qt,Klevansky:1992qe}. Fully integrating out the quark fields leads to LSM-type Lagrangians. This procedure can also be applied to (axial-)vector mesonic fields \cite{Vogl:1991qt}, thus formally linking generalized versions of the NJL models to the eLSM. Interestingly, when the SSB takes place in the NJL model, a potential with a Mexican hat shape is realized at the LSM level, showing that this potential shape emerges from fundamental degrees of freedom. 
\end{itemize}

\bigskip

Next, we need to put all these pieces together in order to use the constrains from
symmetries to set up hadronic models. The first goal is to deal with the trace anomaly
and the dilaton, then we shall move to quark-antiquark nonets and lately to the
so-called chiral multiplets.

\bigskip

\subsection{Dilaton Lagrangian}
\label{Ssec:dilaton}

\begin{figure}[ht!]
    \centering
    \includegraphics[scale=0.7]{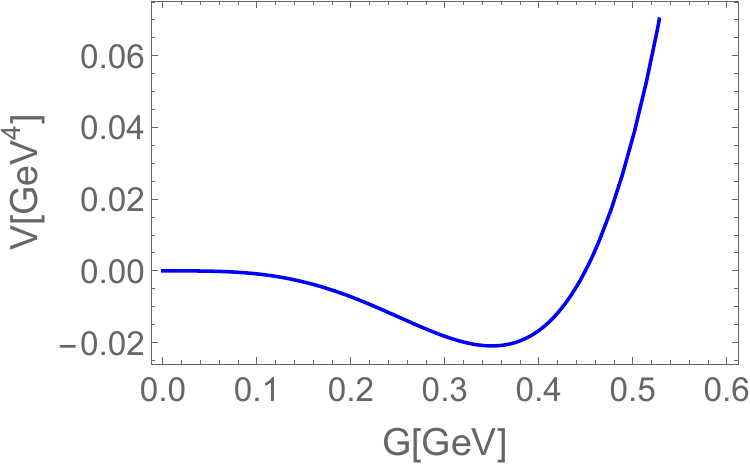}
    \caption{Dilaton potential Eq. \eqref{eq:dilaton} for $m_G=1.65$ GeV and $\Lambda_G=0.35$ GeV.}
    \label{fig:dilaton}
\end{figure}

The first task is to describe the trace anomaly in the low-energy regime.
Namely, gluons are not asymptotic states, but they are expected to form
glueballs, the scalar one with $J^{\mathcal{P}\mathcal{C}}=0^{++}$ being the lightest. Thus, upon
restricting to a unique scalar field $G,$ we write down the model
\begin{equation}
\mathcal{L}_{G}=\frac{1}{2}\left(  \partial_{\mu}G\right)  ^{2}-V_{\text{dil}%
}(G) \text{ ,}
\label{ldil}
\end{equation}
where $V_{\text{dil}}(G)$ is the dilaton potential, that must be chosen to
mimic the trace anomaly. Intuitively, the correspondence between the scalar
glueball/dilaton field and the gluonic fields is set as
\begin{equation}
G^{4}\sim G_{\mu\nu}^{a}G^{a\,\mu\nu}\text{ ,}%
\end{equation}
where both terms carry dimension 4.
The dilatation current for the model of Eq. \eqref{ldil} reads
\begin{equation}
J_{G,dil}^{\mu}=x_{\nu}\left(  \Theta_{G}\right)  ^{\mu\nu}\text{ with
}\left(  \Theta_{G}\right)  ^{\mu\nu}=\partial^{\mu}G\partial^{\nu}G-g^{\mu
\nu}\mathcal{L}_{G}%
\end{equation}
hence its divergence is:%
\begin{equation}
\partial_{\mu}J_{G,dil}^{\mu}=\left(  \Theta_{G}\right)  _{\mu}^{\mu
}=4V_{\text{dil}}-G\frac{\partial V_{\text{dil}}}{\partial G}\text{ }.
\end{equation}
Clearly, a term of the type $V_{\text{dil}}\sim G^{4}$ is such that
$\partial_{\mu}J_{G,dil}^{\mu}=0.$ That would be a dilatation invariant theory
for the scalar field $G$.
On the other hand, following the QCD equation (\ref{ta}), we require that this
divergence is not vanishing, but is negative and is proportional to $G^{4}$:%
\begin{equation}
4V_{\text{dil}}-G\frac{\partial V_{\text{dil}}}{\partial G}=-\frac{\lambda
_{G}}{4}G^{4}\text{ ,}%
\end{equation}
where $\lambda_{G}>0.$ The solution of this first-order differential equation
reads \cite{Migdal:1982jp,Gomm:1984zq,Ellis:1984jv}
\begin{equation}
V_{\text{dil}}(G)=\frac{\lambda_{G}G^{4}}{4}\ln\frac{G}{\Lambda_{G}}%
+\frac{\tilde{\lambda}_{G}G^{4}}{4}
\text{ ,}
\end{equation}
where $\Lambda_{G}$ is an energy scale: this is the energy scale responsible for the trace anomaly at the level of the composite field $G$. The
parameter $\tilde{\lambda}_{G}$ is not fixed by the previous equation. Without
loss of generality, upon requiring that $G_{0}=\Lambda_{G}$ is the minimum of
the potential, the dilatation potential takes the final form:

\begin{equation}\label{eq:dilaton}
V_{\text{dil}}(G)=\frac{\lambda_{G}G^{4}}{4}\left(  \ln\frac{G}{\Lambda_{G}%
}-\frac{1}{4}\right)  \text{ .}%
\end{equation}
The dilaton potential  is plotted in Fig. \ref{fig:dilaton}.
Upon expanding
around the minimum $G_0 = \Lambda_G$, the mass $m_{G}$  of the scalar glueball reads:
\begin{equation}
m_{G}^{2}=\lambda_{G}\Lambda_{G}^{2} \text{ ,}%
\end{equation}
which, according to YM lattice results, lies in the range $1.6$-$1.7$ GeV \cite{Chen:2005,Athenodorou:2020ani} (see also the comparisons in Ref. \cite{Trotti:2022knd}). It
is then also possible to rewrite the dilaton potential by using the
dilaton/glueball mass as%
\begin{equation}
V_{\text{dil}}(G)=\frac{1}{4}\frac{m_{G}^{2}}{\Lambda_{G}^{2}}G^{4}\left(
\ln\frac{G}{\Lambda_{G}}-\frac{1}{4}\right)  \text{ .}%
\end{equation}
Next, the vacuum expectation value of the $\left(  \Theta_{G}\right)  _{\mu
}^{\mu}$ reads
\begin{equation}
\langle\left(  \Theta_{G}\right)  _{\mu}^{\mu}\rangle=-\frac{\lambda_{G}}%
{4}\Lambda_{G}^{4}=-\frac{m_{G}^{2}\Lambda_{G}^{2}}{4}\,.
\end{equation}
Let us recast the QCD result as%
\begin{equation}
\langle\left(  \Theta_{\text{QCD}}\right)  _{\mu}^{\mu}\rangle=-\,\frac
{11N_{c}-2N_{f}}{96\pi^{2}}g^{2}\left\langle G_{\mu\nu}^{a}G^{a\,\mu\nu
}\right\rangle =-\,\frac{11N_{c}-2N_{f}}{24}\left\langle \frac{\alpha_{s}}%
{\pi}G_{\mu\nu}^{a}G^{a\,\mu\nu}\right\rangle =-\,\frac{11N_{c}-2N_{f}}%
{24}C^{4}\text{ }%
\end{equation}
where $C$ refers to the gluon condensate
\begin{equation}
C^{4}=\left\langle \frac{\alpha_{s}}{\pi}G_{\mu\nu}^{a}G^{a\,\mu\nu
}\right\rangle \text{ ,}%
\end{equation}
which in pure YM takes the value $C^{4}=(0.3$-$0.6$ GeV$)^{4}$ \cite{Narison:2014wqa,Ioffe:2005ym,DiGiacomo:2000irz}, thus $C=0.3$-$0.6$ GeV.

Finally, upon setting $N_{f}=0$ (pure YM-case), the assumption that the field $G$ saturates
the trace anomaly leads to%
\begin{equation}
\langle\left(  \Theta_{G}\right)  _{\mu}^{\mu}\rangle=-\frac{m_{G}^{2}%
\Lambda_{G}^{2}}{4}=-\,\frac{11N_{c}}{24}C^{4}%
\end{equation}
hence 
\begin{equation}
    \Lambda_{G}=\sqrt{\frac{11 N_c}{6}} \frac{C^2}{m_G}    \\ \text{  .}
\end{equation}
In the large-$N_{c}$ limit $C^{4}$ scales as $N_{c}$ (because $\alpha_s$ goes as $N_c^{-1}$ and the gluonic loops as $N_c^{2}$), thus $\Lambda
_{G}\sim N_{c}$. In other words, $\Lambda_{G}\sim N_{c}\Lambda_{\text{QCD}}$.
For $N_{c}=3$, the constant $\Lambda_{G}$  ranges between $0.3$ and $1$ GeV (thus a large uncertainty is present).
 
An interesting question is if the fluctuations of the dilaton, the scalar glueballs, can form bound states: as shown in Ref. \cite{Giacosa:2021brl} this is
possible if $\Lambda_{G}$ is small enough (that is, if the attraction is large enough).

In conclusion, the dilaton arises as the field responsible for the trace
anomaly -via a unique dimensionful parameter $\Lambda_{G}$- and plays an
important role in the eLSM, see Section~\ref{Sec:eLSM_Lagr}. The addition of other glueball fields is accomplished as well in Section~\ref{Sec:glueballs}.

\subsection{Quark-antiquark and hybrid nonets}
\label{ssec:qq and hybr}
\subsubsection{General considerations about nonets}

When 3 light quark flavors are considered, there are clearly 9 possibilities to form  a
quark-antiquark object. Here, we describe the general idea of a mesonic
quark-antiquark nonet, that we call $M_{\Gamma}$, which is a $3\times3$ matrix whose
elements are given by the quark-antiquark currents%
\begin{equation}
M_{\Gamma,ij}\overset{\text{transforms as}}{\simeq}\frac{1}{\sqrt{2}}\bar{q}%
_{j}\Gamma q_{i}\label{equiv}%
\end{equation}
with $i,j=u,d,s$ and where the factor $1/\sqrt{2}$ is introduced for later convenience.\ 

The quantity $\Gamma$ is a combination of Lorentz matrices and/or covariant
derivatives, such as $\Gamma=i\gamma^{5},$ $\mathbb{1}_{4\times 4},$ $\gamma^{\mu},$ $D^{\mu},\text{etc}$. The (possibly present) Lorentz indices are suppressed here, but will be retained in actual examples later. Indeed, the value of the total angular
momentum $J$ of a given nonet is contained in the choice of $\Gamma.$ For an
object without Lorentz indices (such as $\Gamma=i\gamma^{5}$) $J=0,$ for one
open Lorentz index (such as $\Gamma=\gamma^{\mu})$ one has $J=1,$ etc. The
matrix $M_{\Gamma}$ needs to be Hermitian since it describes a nonet of physical
mesonic fields, this is why e.g. $\Gamma=i\gamma^{5}$ can be used (but not
$\Gamma=\gamma^{5}$).

In connection with Eq. (\ref{equiv}) a cautionary remark is mandatory: the l.h.s. and r.h.s. cannot be identical, since mesons $M_{ij}$ has dimension [energy], while $\bar{q}_{j}\Gamma q_{i}$ does not (for $\Gamma=i\gamma^{5},$ it carries dimension [energy$^{3}]$ for $\Gamma=D^{\mu}$ it has [energy$^{4}$],$\;\ldots$). Moreover, the r.h.s. is a local quark current, while the l.h.s. represents a nonperturbative (and extended) mesonic object. One should understand that
equivalence in the sense that they behave equally under transformations (in
particular, parity, charge conjugation, flavor and chiral transformations, but
\textit{not }dilatation transformation). With this comment in mind, we will later on use, for simplicity,
the equal sign.

In matrix form, $M_{\Gamma}$ takes the form
\begin{equation}
M_{\Gamma} =\frac{1}{\sqrt{2}}\left(
\begin{array}
[c]{ccc}%
\bar{u}\Gamma u & \bar{d}\Gamma u & \bar{s}\Gamma u\\
\bar{u}\Gamma d & \bar{d}\Gamma d & \bar{s}\Gamma d\\
\bar{u}\Gamma s & \bar{d}\Gamma s & \bar{s}\Gamma s
\end{array}
\right)  \text{ .}\label{MGamma}%
\end{equation}

Parity $\mathcal{P}$  is obtained by applying the parity transformation of Eq.
\eqref{parity} leading to an expression of the type%
\begin{equation}
M_{\Gamma}(t,\mathbf{x})\rightarrow\lambda_{\mathcal{P}}M_{\Gamma}(t,-\mathbf{x})\text{
,}%
\end{equation}
where $\lambda_{\mathcal{P}}$ (often denoted with $\mathcal{P}$  tout-court) is the parity
eigenvalue. Note that for $J>0$ the previous equation refers to the spatial components of the Lorentz vectors (details and examples below).

Under charge conjugation $\mathcal{C}$ , the general expression is
\begin{equation}
M_{\Gamma}\rightarrow\lambda_{\mathcal{C}}M_{\Gamma}^{T}\text{ .}%
\end{equation}
where $\lambda_{\mathcal{C}}$ (often denoted by $\mathcal{C}$  tout-court) is the charge
conjugation eigenvalue (even though, strictly speaking, only the diagonal elements
are eigenstates of $\mathcal{C}$ ). Note that the emergence of the transposed matrix is pretty
intuitive, since under $\mathcal{C}$ one has (up to a sign): $M_{\Gamma,12}\equiv\bar
{d}\Gamma u\longleftrightarrow\bar{u}\Gamma d\equiv M_{\Gamma,21}$.

Under flavor transformations $SU(3)_V$ the matrix $M_{\Gamma}$ changes in a simple way: 
\begin{equation}
M_{\Gamma}\rightarrow U_{V}M_{\Gamma}U_{V}^{\dagger}\text{ ,}%
\end{equation}
regardless of the choice of $\Gamma$. Thus, the matrix $M_{\Gamma}$ transforms
under the adjoint representation of flavor transformations (see \ref{Ssec:UN_group}), as expected for a $\bar{q}q$ object. Chiral transformations are more complicated because they mix different nonets, see Sec. \ref{ssec:chiral-multp}. (The baryon number transformation $U(1)_V$ leaves $M_{\Gamma}$ unchanged.)

Next, the different members of the nonet can be grouped by recognizing
different isospin multiplets within the nonet. To this end, we recall that
the isospin transformation amounts to
\begin{equation}
\left(
\begin{array}
[c]{c}%
u\\
d
\end{array}
\right)  \rightarrow U_{I}\left(
\begin{array}
[c]{c}%
u\\
d
\end{array}
\right)  \text{ , }U_{I}=e^{-{i\sum_{k=1}^{3}\theta_{I}^{(k)}}\frac{\sigma
^{k}}{2}}
\text{ ,}
\end{equation}
with the isospin operators $I_{k} = \sigma^{k}/2$. Then, the quark $\ket{u}$ corresponds to $(1,0)^{T}$ and $\ket{d}$ corresponds to $(0,1)^{T}.$ $I_{3}\left\vert u\right\rangle =\frac{1}{2}\ket{u}$ and $I_{3}\ket{d} =-\frac{1}{2}\ket{d}$, showing that $(u,d)^{T}$ forms an isospin doublet. Of course, $\mathbf{I}^{2}\ket{u} =\frac{1}{2}\left(\frac{1}{2}+1\right)\ket{u}$ and $\mathbf{I}^{2}\ket{d} =\frac{1}{2}\left(\frac{1}{2} + 1 \right) \ket{d}$. Antiparticles transform as
\begin{equation}
\left(
\begin{array}
[c]{cc}%
\bar{u} & \bar{d}%
\end{array}
\right)  \rightarrow\left(
\begin{array}
[c]{cc}%
\bar{u} & \bar{d}%
\end{array}
\right)  U_{I}^{\dagger},
\end{equation}
which can be rearranged in the following way \cite{Halzen:1984mc}:
\begin{equation}
\left(
\begin{array}
[c]{c}%
-\bar{d}\\
\bar{u}%
\end{array}
\right)  \rightarrow U_{I}\left(
\begin{array}
[c]{c}%
-\bar{d}\\
\bar{u}%
\end{array}
\right)  \text{,}%
\end{equation}
according to which $\left\vert -\bar{d}\right\rangle =-\left\vert \bar
{d}\right\rangle $ is the object with $I_{z}=1/2$ (eigenvalue of $I_3$) and $\left\vert
\bar{u}\right\rangle $ with $I_{z}=-1/2$ . In other words, the antiparticle
doublet has been reshaped as a regular particle isodoublet, but in doing so a
sign switch appears. We see explicitly that the isospin charge-symmetry transformation $\mathcal{C}_{I}$
amounts to:
\begin{align}
\mathcal{C}_{I}\left\vert u\right\rangle  &  =\left\vert d\right\rangle \text{ , }%
\mathcal{C}_{I}\left\vert d\right\rangle =-\left\vert u\right\rangle \text{ ,}\\
\mathcal{C}_{I}\left\vert \bar{u}\right\rangle  &  =\left\vert \bar{d}\right\rangle
\text{ , }\mathcal{C}_{I}\left\vert \bar{d}\right\rangle =-\left\vert \bar
{u}\right\rangle \text{ .}%
\end{align}

The usual spin composition can be applied to a quark-antiquark system, hence
the state with $I=1$ and $I_{z}=0$ (analogous to the spin triplet element
$\left\vert \uparrow\downarrow+\downarrow\uparrow\right\rangle /\sqrt{2}$) is
given by
\begin{equation}
\left\vert I=1,I_{z}=0\right\rangle =\frac{1}{\sqrt{2}}\left\vert u\bar
{u}+d\left(  -\bar{d}\right)  \right\rangle =\frac{1}{\sqrt{2}}\left\vert
u\bar{u}-d\bar{d}\right\rangle \text{ ,}%
\end{equation}
while the element with $I=I_{z}=0$ (analogous to the spin triplet element
$\left\vert \uparrow\downarrow-\downarrow\uparrow\right\rangle /\sqrt{2}$) by%
\begin{equation}
\left\vert I=0,I_{z}=0\right\rangle =\frac{1}{\sqrt{2}}\left\vert u\bar
{u}-d\left(  -\bar{d}\right)  \right\rangle =\frac{1}{\sqrt{2}}\left\vert
u\bar{u}+d\bar{d}\right\rangle \text{ .}%
\end{equation}
We then group the nonet into following isospin sub-multiplets.
\begin{itemize}
\item An isospin-triplet state is associated to pion-like states and typically
denoted with the letter $\pi$ (or $\pi_J$, for a generic total angular momentum $J$) for nonets with $\mathcal{P}=-1$ and with the letter $a$ (or $a_J$) for the ones with $\mathcal{P}=+1$. Here we
use the generic notation $\pi_{\Gamma}$. We then have:
\begin{align}
\pi_{\Gamma}^{+} &  \equiv\bar{d}\Gamma u\equiv-\left\vert I=1,I_{z}%
=1\right\rangle \text{ ,}\\
\pi_{\Gamma}^{0} &  \equiv\frac{1}{\sqrt{2}}\left(  \bar{u}\Gamma u-\bar
{d}\Gamma d\right)  \equiv\left\vert I=1,I_{z}=0\right\rangle \text{ ,}\\
\pi_{\Gamma}^{-} &  \equiv\bar{u}\Gamma d\equiv\left\vert I=1,I_{z}%
=-1\right\rangle \text{ .}
\end{align}
The $\mathcal{G}$-parity eigenvalue of this isotriplet is $\lambda_{\mathcal{G}}=$ $-\lambda_{\mathcal{C}},$
or simply $\mathcal{G}=-\mathcal{C}$.
%%%%%%%%%%%%%%%%%%%%
\item Two isospin doublets, corresponding to kaon-like states:%
\begin{align}
K_{\Gamma}^{+}  &  \equiv\bar{s}\Gamma u\equiv\left\vert I=1/2,I_{z}%
=1/2\right\rangle \text{ ,}\\
K_{\Gamma}^{0}  &  \equiv\bar{s}\Gamma d\equiv\left\vert I=1/2,I_{z}%
=-1/2\right\rangle \text{ ,}
\end{align}
and
\begin{align}
\bar{K}_{\Gamma}^{0}  &  \equiv\bar{d}\Gamma s\equiv-\left\vert I=1/2,I_{z}%
=1/2\right\rangle \text{ ,}\\
K_{\Gamma}^{-}  &  \equiv\bar{u}\Gamma s\equiv\left\vert I=1/2,I_{z}%
=-1/2\right\rangle \text{ .}
\end{align}
These states are not G-parity eigenstates. Namely, up to a sign, one gets
$K_{\Gamma}^{+}\overset{\mathcal{C}_{I}}{\rightarrow}K_{\Gamma}^{0}\xrightarrow{
\mathcal{C}}{\rightarrow}\bar{K}_{\Gamma}^{0},$ hence $K_{\Gamma}^{+}\overset
{\mathcal{G}}{\rightarrow}\bar{K}_{\Gamma}^{0}.$
%%%%%%%%%%%%%%%%%%%%
\item Two isoscalar states: the purely nonstrange isospin singlet state
\begin{equation}
\eta_{\Gamma,N}\equiv\frac{1}{\sqrt{2}}\left(  \bar{u}\Gamma u+\bar{d}\Gamma
d\right)  \equiv\left\vert I=0,I_{z}=0\right\rangle \text{ ,}
\end{equation}
and the purely hidden strange object
\begin{equation}
\eta_{\Gamma,S}\equiv\bar{s}\Gamma s\equiv\left\vert I=0,I_{z}=0\right\rangle
\end{equation}
complete the list. The $\mathcal{G}$-parity eigenvalue for these states is $\lambda
_{\mathcal{G}}=$ $\lambda_{\mathcal{C}},$ or just $\mathcal{G}=\mathcal{C}$.
%%%%%%%%%%%%%%%%%%%%
\item The physical isoscalar states of the nonet arise from the mixing of
$\eta_{\Gamma,N}$ and $\eta_{\Gamma,S}$, one of which is light,
$\eta_{\Gamma,L}$ and one of them is heavy $\eta_{\Gamma,H}$.:%
\begin{equation}
\left(
\begin{array}
[c]{c}%
\eta_{\Gamma,L}\\
\eta_{\Gamma,H}%
\end{array}
\right)  =\left(
\begin{array}
[c]{cc}%
\cos\beta_{\Gamma} & \sin\beta_{\Gamma}\\
-\sin\beta_{\Gamma} & \cos\beta_{\Gamma}%
\end{array}
\right)  \left(
\begin{array}
[c]{c}%
\eta_{\Gamma,N}\\
\eta_{\Gamma,S}%
\end{array}
\right)
\text{ .}
\end{equation}
For certain nonets the mixing angle $\beta_{\Gamma}$ is large (as for
pseudoscalar mesons, where instantons are at work), for others it is pretty
small (e.g. vector mesons). The $G$-parity is still $\mathcal{G}=\mathcal{C},$ since it is unaffected
by the isoscalar mixing.

\item The previous relations show that
\begin{equation}
\mathcal{G}=\mathcal{C}(-1)^{I}\text{ ,}
\end{equation}
which requires an integer $I$.

\item Strangeness is simple. For any nonet, the kets $\ket{K^+_{\Gamma}}$ and $\ket{K^0_{\Gamma}}$ carry strangeness $+1$, $\ket{\bar{K}^0_{\Gamma}}$ and $\ket{K^-_{\Gamma}}$ carry $-1$, while all the other states have vanishing strangeness.

\end{itemize}

Finally, the matrix $M$ in terms of the introduced fields with definite
isospin eigenstates read:%
\begin{equation}
M_{\Gamma}=\frac{1}{\sqrt{2}}\left(
\begin{array}
[c]{ccc}%
\frac{\eta_{\Gamma,N}+\pi_{\Gamma}^{0}}{\sqrt{2}} & \pi_{\Gamma}^{+} &
K_{\Gamma}^{+}\\
\pi_{\Gamma}^{-} & \frac{\eta_{\Gamma,N}+\pi_{\Gamma}^{0}}{\sqrt{2}} &
K_{\Gamma}^{0}\\
K_{\Gamma}^{-} & \bar{K}_{\Gamma}^{0} & \eta_{\Gamma,S}%
\end{array}
\right)  \text{ .}%
\end{equation}

The physical fields entering the nonet are also expressed in a compact form as
$\{\pi_{\Gamma},K_{\Gamma},\eta_{\Gamma,L},\eta_{\Gamma,H}\},$ where
$\pi_{\Gamma}$ stands for the isotriplet ($I=1$) states $\pi_{\Gamma}^{+}%
,\pi_{\Gamma}^{-},\pi_{\Gamma}^{0},$ $K_{\Gamma}$ stands for the two isodoublets
$K_{\Gamma}^{+},K_{\Gamma}^{0}$ and $\bar{K}_{\Gamma}^{0},K_{\Gamma}^{-}$, and
$\eta_{\Gamma,L},\eta_{\Gamma,H}$ for the two isoscalars as a mixture of
$\eta_{\Gamma,N},\eta_{\Gamma,S}.$

The factor $1/\sqrt{2}$ in Eq. (\ref{MGamma}) can be now understood as
follows.\ The expansion of $M_{\Gamma}$ in the basis of Hermitian matrices
$t^{a}$ (\ref{Ssec:UN_group}) reads:%
\begin{equation}
M_{\Gamma}=\sum_{k=0}^{8}M^{k}t^{k}=\sum_{k=0}^{8}M^{k}\frac{\lambda^{k}}%
{2}\text{ ,}%
\end{equation}
where the component
\begin{equation}
M^{k}=2\Tr\left[  t^{k}M_{\Gamma}\right]  =\Tr\left[  \lambda^{k}M_{\Gamma
}\right]
\end{equation}
is a normalized physical objects, for instance:
\begin{equation}
M^{3}=\Tr\left[  \lambda^{3}M_{\Gamma}\right]  =\frac{1}{\sqrt{2}}\left(
\bar{u}\Gamma u-\bar{d}\Gamma d\right)  =\pi_{\Gamma}^{0}\text{.}%
\end{equation}

%\bigskip

%\bigskip

\subsubsection{The spectroscopic approach}

Next, we move to the spectroscopic (wave function) approach. Let us consider as an example the  state $\pi_{\Gamma}^{+}$. The corresponding quantum state is
expected to be proportional to
\begin{equation}
\left\vert \pi_{\Gamma}^{+}\right\rangle \sim\left(  \pi_{\Gamma}^{+}\right)
^{\dagger}\left\vert 0_{\text{QCD}}\right\rangle =\left(  \bar{d}\Gamma u\right)
^{\dagger}\left\vert 0_{\text{QCD}}\right\rangle \text{ .}\label{piplusgen}%
\end{equation}
The quantum numbers $J^{\mathcal{P}\mathcal{C}}$ are dictated by the Lorentz structure fixed by $\Gamma$. As a next step, we intend to decompose $\left\vert \pi_{\Gamma}^{+}\right\rangle $ into different parts:%
\begin{equation}
\ket{\pi_{\Gamma}^{+}} = \underset{\text{spacial part}}{\underbrace{\ket{ \text{radial: }n}\ket{\text{angular: }L}}}\ket{\text{spin: }S}
\ket{\text{flavor: }u\bar{d}} \ket{\text{color}}\; ,
\label{spect}
\end{equation}
where the angular momentum $L$ and the spin $S$ enter separately, thus the matching $L$ and $S$ to $J^{\mathcal{P}\mathcal{C}}$ is of particular importance.
We discuss the terms of Eq. (\ref{spect}) one by one, starting from the far right.

\bigskip

(i) The color part is straightforward for any $\bar{q}q$ state, since a
sum over all color d.o.f. is implicitly included in Eq. (\ref{piplusgen}),
leading to:%
\begin{equation}
\left\vert \text{color}\right\rangle =\frac{1}{\sqrt{3}}\left\vert \bar
{R}R+\bar{G}G+\bar{B}B\right\rangle \text{ .}%
\end{equation}

(ii) The flavor part is encoded in the current, in the present case
$\left\vert \text{flavor: }u\bar{d}\right\rangle$, or simply
$\left\vert u\bar{d}\right\rangle$.

(iii) The spin part $\left\vert \text{spin: }S\right\rangle $ can take two
values, either $S=0$ or $S=1.$ In fact, out of two particles with spin $1/2$,
we may construct the singlet state
\begin{equation}
\left\vert \bar{q}q,S=0,S_{z}=0\right\rangle =\frac{1}{\sqrt{2}}\left(
\left\vert \uparrow\downarrow\right\rangle -\left\vert \downarrow
\uparrow\right\rangle \right)
\text{ ,}
\end{equation}
as well as the triplet states
\begin{align}
&\left\vert \bar{q}q,S=1,S_{z}=1\right\rangle  =\left\vert \uparrow
\uparrow\right\rangle \\
&\left\vert \bar{q}q,S=1,S_{z}=0\right\rangle    =\frac{1}{\sqrt{2}}\left(
\left\vert \uparrow\downarrow\right\rangle +\left\vert \downarrow
\uparrow\right\rangle \right) \\
&\left\vert \bar{q}q,S=1,S_{z}=-1\right\rangle   =\left\vert \downarrow
\downarrow\right\rangle
\end{align}
where the first spin refers to the antiquark and the second to the quark (by convention).

The states above are eigenstates of $\mathbf{S}^{2}$ with eigenvalues
$0(0+1)=0$ for the singlet and $1(1+1)=2$ for the triplet, and are eigenstates of $S_{z}$ with eigenvalues indicated in
the kets. For spectroscopic purposes, it is enough to indicate the total spin,
so either $\left\vert S=0\right\rangle $ or $\left\vert S=1\right\rangle $.

(iv) The angular part $\left\vert \text{angular: }L\right\rangle $ refers to the orbital angular momentum $L$, which can take the values $L=0,1,2,\ldots$ implying that $\mathbf{L}^{2}$ $\left\vert\text{angular: }L\right\rangle =L(L+1)\left\vert \text{angular: }L\right\rangle$.

(v) Radial angular momentum $n,$ with $n=1,2,3,\ldots$This is the number of zeros of the radial wave function $\chi(r)$  for the $\bar{q}q$ system with $\int_{0}^{\infty}\left\vert \chi(r)\right\vert ^{2}=1.$ The local current $\bar {d}\Gamma u$ tells us \textit{nothing} about that. One could eventually render the currents nonlocal as to include the (relativistic generalization of the) wave function \cite{Giacosa:2004ug,Faessler:2003yf}, but that is not required for our purposes, since we will work with composite mesonic fields. In the majority of cases, and if not stated otherwise, the radial quantum number $n=1$ is understood.
\bigskip

Various remarks are in order.

\begin{enumerate}
\item Connection between $L,$ $S$ and $J$: the state in Eq.
(\ref{piplusgen}) contains a fixed $J$ that comes from the microscopic current
$\bar{d}\Gamma u.$ On the other hand, the state  in Eq. (\ref{spect}) displays the values $L$ and
\ $S$. Via the composition of angular momenta, $\mathbf{J}%
=\mathbf{L}+\mathbf{S,}$ the possible values for $J$ are integers between $\left\vert L-S\right\vert $ and $L+S$. One way to establish $L$ and $S$ is to study the non-relativistic limit of the current (\ref{piplusgen}) and show that indeed a unique combination of $L$ and $S$ emerges, see \ref{Ssec:non-relimit}.

\item Parity: upon parity transformation, a factor $(-1)^{L}$ emerges from the
angular part $\left\vert \text{angular: }L\right\rangle .$ Yet, an additional
minus sign is due to the intrinsic opposite parity of an anti-fermion w.r.t. a
fermion, leading to:
\begin{equation}
    \mathcal{P}=(-1)^{L+1}\text{ .}%
\end{equation}
Also in this case, the non-relativistic limit shows that the equation above holds. 

\item Charge conjugation $\mathcal{C}.$ Exchanging the quark with the antiquark implies
a factor $(-1)^{L+1}$ from $\left\vert \text{angular: }L\right\rangle$ (just as parity) and a
factor $(-1)^{S+1}$ from $\left\vert \text{spin: }S\right\rangle ,$ leading to
\begin{equation}
\mathcal{C}=(-1)^{L+S}
\text{ .}
\end{equation}
Yet, care is needed: the state $\left\vert \pi_{\Gamma}^{+}\right\rangle$ under consideration is not a charge conjugation
eigenstate, since $u\bar{d}$ is exchanged into $d\bar{u}.$ Explicitly:
\begin{equation}
\mathcal{C}\left\vert \pi_{\Gamma}^{+}\right\rangle  =\mathcal{C}\left\vert \text{radial:
}n\right\rangle \mathcal{C}\left\vert \text{angular: }L\right\rangle \mathcal{C}\left\vert
\text{spin: }S\right\rangle \mathcal{C}\left\vert \text{flavor: }u\bar{d}\right\rangle
\mathcal{C}\left\vert \text{color}\right\rangle 
 = (-1)^{L+S}\left\vert \pi_{\Gamma}^{-}\right\rangle \text{.}%
\end{equation}
The diagonal elements (those wit $I_{z}=0$) are eigenstates of the
$\mathcal{C}$-operation with:%
\begin{equation}
\mathcal{C}\left\vert \pi_{\Gamma}^{0}\right\rangle =(-1)^{L+S}\left\vert \pi_{\Gamma
}^{0}\right\rangle \text{ , }\mathcal{C} \left\vert \eta_{\Gamma,N}\right\rangle
=(-1)^{L+S}\left\vert \eta_{\Gamma,N}\right\rangle \text{ , }\mathcal{C}\left\vert
\eta_{\Gamma,S}\right\rangle =(-1)^{L+S}\left\vert \eta_{\Gamma,S}%
\right\rangle \text{ .}%
\end{equation}
The same applies for the two physical fields $\left\vert \eta_{\Gamma
,L}\right\rangle $ and $\left\vert \eta_{\Gamma,H}\right\rangle .$

\item $\mathcal{G}$-parity amounts to $\mathcal{C}\cdot \mathcal{C}_{I},$ hence one finds (for integer isospin
states) that%
\begin{equation}
\mathcal{G}=\mathcal{C}(-1)^{I}=(-1)^{L+S+I}.
\end{equation}
It follows that
\begin{align*}
G\left\vert \pi_{\Gamma}^{0}\right\rangle  &  =-(-1)^{L+S}\left\vert
\pi_{\Gamma}^{0}\right\rangle \text{ , }G\left\vert \pi_{\Gamma}^{\pm
}\right\rangle =-(-1)^{L+S}\left\vert \pi_{\Gamma}^{\pm}\right\rangle \\
G\left\vert \eta_{\Gamma,N}\right\rangle  &  =(-1)^{L+S}\left\vert
\eta_{\Gamma,N}\right\rangle \text{ , }G\left\vert \eta_{\Gamma,S}%
\right\rangle =(-1)^{L+S}\left\vert \eta_{\Gamma,S}\right\rangle \text{ .}%
\end{align*}

\item Matching. In many cases the constraints imposed by $J^{\mathcal{P}\mathcal{C}}$ are sufficient to unequivocally determine $L$ and $S$. For instance, $J^{\mathcal{P}\mathcal{C}}=0^{-+}$ is only possible for $L=S=0$, while $J^{\mathcal{P}\mathcal{C}}=0^{++}$ for $L=S=1$. The non-relativistic limit of \ref{Ssec:non-relimit} confirms this result. In some cases, however, more choices are available. The vector quantum numbers $J^{\mathcal{P}\mathcal{C}}=1^{--}$ can be obtained for $L=0$, $S=1$ (ground-state vector mesons) but also for $L=2, S=1$ (orbitally excited vector mesons). In this case, even if the nonrelativistic limit would provide an univocal correspondence, in the full relativistic world a mixing of these two configurations is possible and two vector nonets are expected (as well known, $L$ and $S$ are in general not `good quantum numbers' in QFT). Yet, this mixing is expected to be small due to quite large mass differences that $L=0, S=1$ and $L=2, S=1$  choices would generate, the latter much heavier than the former \cite{Godfrey:1985xj}. With this caveat in mind, we shall still assign a pair of $L,S$ to nonets as the `dominant' contribution, see later for examples.

\item Some quantum numbers $J^{\mathcal{P}\mathcal{C}}$  are not accessible to a $\bar{q}q$ pair,
such as:
\begin{align}
0^{+-}\; , \;0^{--},\; 2^{+-},\,\ldots\text{ even}^{+-}\text{ for even}  &
\geq2\nonumber\\
1^{-+}\; , \; 3^{-+},\,\ldots\text{ odd}^{-+}\text{ for odd}  &  \geq1 \; .
\end{align}
Exotic quantum numbers can be realized by glueballs and hybrid mesons. In particular, hybrid mesons with $1^{-+}$ will be studied in this review. In fact, hybrid mesons may form nonets just as quark-antiquark states, but different quantum numbers, such as $1^{-+},$ are possible due to the
additional gluon.

\item The spectroscopic notation for a given $\bar{q}q$ nonet is given by
\begin{equation}
n\text{ }^{2S+1}L_{J}%
\end{equation}
with $n$ being the radial quantum number, $L=S,P,D,F,...=0,1,2,3,...$ the angular
quantum number, and $S=0,1$ the spin quantum number.

\item 
The summary of all available types of $\bar{q}q$ nonets is listed in Table \ref{tab:meson-list}. As one can note, exotic quantum numbers do not appear. 

\begin{table}[ht!]
    \centering    \renewcommand{\arraystretch}{1.3}
    \begin{tabular}
{|c|c|c|c|c|c|c|}\hline
$L$ & $S$ & $\mathcal{P}=(-1)^{L+1}$ & $\mathcal{C}=(-1)^{L+S}$ & $J^{\mathcal{P}\mathcal{C}}$ & $n^{2S+1}L_J$ & Resonances \\\hline\hline
$0$ & $0$ & $-1$ & $+1$ & $0^{-+}$ & $1^{1}S_0$ & $\pi,K,\eta,\eta^{\prime}$\\ \hline
$1$ & $1$ & $+1$ & $+1$ & $0^{++}$ & $1^{3}P_0$ & $a_{0},K_{0}^{\ast},f_{0,L},f_{0,H}%
$\\\hline
$1$ & $0$ & $+1$ & $-1$ & $1^{+-}$ & $1^{1}P_1$ & $b_{1},K_{1B},h_{1L},h_{1H}%
$\\\hline
$0$ & $1$ & $-1$ & $-1$ & $1^{--}$ & $1^{3}S_1$ & $\rho,K^{\ast},\omega_{1L}=\omega
,\omega_{1H}=\phi$\\\hline
$1$ & $1$ & $+1$ & $+1$ & $1^{++}$ & $1^{3}P_1$ & $a_{1},K_{1A},f_{1L},f_{1H}%
$\\\hline
$2$ & $1$ & $-1$ & $-1$ & $1^{--}$ & $1^{3}D_1$ & $\rho_{D},K_{1D}^{\ast},\omega_{D}%
,\phi_{D}$\\\hline
$2$ & $0$ & $-1$ & $+1$ & $2^{-+}$ & $1^{1}D_2$ & $\pi_{2},K_{2},\eta_{2},\eta_{2}^{\prime
}$\\\hline
$1$ & $1$ & $+1$ & $+1$ & $2^{++}$ & $1^{3}P_2$ & $a_{2},K_{2},f_{2L}=f_{2},f_{2H}%
=f_{2}^{\prime}$\\\hline
$2$ & $1$ & $-1$ & $-1$ & $2^{--}$ & $1^{3}D_2$ & $\rho_{2},K_{2}^{\ast},\omega
_{2L}=\omega_{2},\omega_{2H}=\phi_{2}$\\\hline
$3$ & $1$ & $+1$ & $+1$ & $2^{++}$ & $1^{3}F_2$ & $a_{2F},K_{2F},f_{2F,L},f_{2F,H}%
$\\\hline
\end{tabular}
\caption{List of conventional mesons together with their quantum numbers and naming conventions.}
\label{tab:meson-list}
\end{table}
\end{enumerate}

\bigskip

\subsubsection{List of nonets}

Here, we turn to specific nonets of physical resonances and to their properties. We consider $\bar{q}q$ nonets by increasing $J$. Moreover, we present them in pairs of chiral partners, which means that the nonets transform into each other by a chiral transformation. The precise
definition of chiral partners and the construction of chiral multiplet is
described in Sec. \ref{ssec:chiral-multp}. The nonets are also summarized in Table~\ref{tab:meson-list}. The isoscalar mixing angle is listed in Table \ref{tab:mixing}. In Table \ref{tab:mesonlist} the PDG resonances for each nonet are summarized, along with various naming conventions. Finally, all relevant nonets are listed in Table \ref{tab:nonets} with their currents and transformations under $\mathcal{P}$ and $\mathcal{C}$.
\bigskip

\textit{Pseudoscalar meson nonet: }$J^{\mathcal{P}\mathcal{C}}=0^{-+}$ \textit{with} $L=S=0$;
\textit{spectroscopic notation:} $n$ $^{2S+1}L_{J}\equiv$ $1^{\;1}S_{0}$.

The pseudoscalar meson nonet $\{\pi,$ $K,$ $\eta(547),$ $\eta^{\prime}(958)\}$
was already encountered multiplet times, since it corresponds to the lightest physical states in QCD.
In fact, they emerge as quasi-Goldstone bosons due to SSB of chiral symmetry
$SU(3)_{L}\times SU(3)_{R}\rightarrow SU(3)_{V}$. Because of that, any
low-energy model of QCD contains pseudoscalar mesons, both when chiral
symmetry is non-linearly realized, such as in ChPT \cite{Gasser:1983yg,Gasser:1984gg,Leutwyler:1993iq,Pich:1995bw,Jenkins:1995vb,Scherer:2002tk,Bernard:2006gx,Schindler:2005ke,Booth:1996hk,Terschlusen:2016kje}, or linearly realized, such as in LSMs, e.g. \cite{Lee:1967ug,Ko:1994en,Urban:2001ru} (and, of course, in the eLSM).

They are obtained by setting
\begin{equation}
\Gamma\equiv\Gamma_{^{1}S_{0}}=i\gamma^{5}%
\end{equation}
leading to the following elements
\begin{equation}
P_{ij}\equiv\frac{1}{\sqrt{2}}\,\bar{q}_{j}i\gamma^{5}q_{i}\text{ .}%
\end{equation}
The matrix $P$ can be expressed as:
\begin{equation}
P=\frac{1}{\sqrt{2}}\left(
\begin{array}
[c]{ccc}%
\bar{u} i\gamma^5 u & \bar{d}i\gamma^5 u & \bar{s}i\gamma^5 u\\
\bar{u}i\gamma^5 d & \bar{d}i\gamma^5 d & \bar{s}i\gamma^5 d\\
\bar{u}i\gamma^5 s & \bar{d}i\gamma^5 s & \bar{s}i\gamma^5 s
\end{array}
\right) = \frac{1}{\sqrt{2}}%
\begin{pmatrix}
\frac{\eta_{N}+\pi^{0}}{\sqrt{2}} & \pi^{+} & K^{+}\\
\pi^{-} & \frac{\eta_{N}-\pi^{0}}{\sqrt{2}} & K^{0}\\
K^{-} & \bar{K}^{0} & \eta_{S}%
\end{pmatrix}
\,\text{,}%
\end{equation}
where the explicit quark content is retained for a better visualization. Above, $\pi^{\pm},\pi^{0}$ form the pion isotriplet, $K^{+},K^{0}$ and $\bar
{K}^{0},K^{-}$ are pairs of isodoublet kaon states, the isoscalar $\eta_{N}\equiv
(\bar{u}u+\bar{d}d)/\sqrt{2}$ stands for the purely non-strange state, and
the isoscalar $\eta_{S}\equiv\bar{s}s$ stands for the purely strange
state.\footnote{We prefer to work in the strange-nonstrange basis. The mixing
angle in the octet-singlet basis $\beta_{08}$ can be obtained via
$\beta_{08}=\beta_{ps}+35.3^\circ$.} The physical fields emerge as
\begin{equation}
\left(
\begin{array}
[c]{c}%
\eta\equiv\eta(547)\\
\eta^{\prime}\equiv\eta(958)
\end{array}
\right)  =\left(
\begin{array}
[c]{cc}%
\cos\beta_{ps} & \sin\beta_{ps}\\
-\sin\beta_{ps} & \cos\beta_{ps}%
\end{array}
\right)  \left(
\begin{array}
[c]{c}%
\eta_{N}\\
\eta_{S}%
\end{array}
\right)  \text{ },
\end{equation}
where $\beta_{ps}=-43.4^{\circ}$ is the mixing angle obtained in\ Ref.
\cite{Kloe2}. This large mixing angle is a consequence of the axial $U(1)_{A}$
anomaly which, thanks to instanton effects, increases the mass of the singlet
configuration $\eta_{0}=\sqrt{\frac{2}{3}}\eta_{N}+\sqrt{\frac{1}{3}}\eta
_{S}$ (see more details later).

The total $J$ for this state is $J=0,$ since the currents are purely scalar.
Table \ref{tab:meson-list} shows that $L=S=0$ is the only available possibility. Next, we show
in detail how the currents $P_{ij}$ transform under parity and
charge-conjugation transformations. The same procedure can be applied to the
other nonets by following analogous steps. Under parity $\mathcal{P}$ :
\begin{equation}
\sqrt{2}P_{ij} = \bar{q}_{j}(x)i\gamma^{5}q_{i}(x)\rightarrow\bar{q}_{j}(t,-\mathbf{x}) \gamma^{0}i\gamma^{5} \gamma^{0}q_{i}(t,-\mathbf{x}) = -\bar{q}_{j}(t,-\mathbf{x})\gamma^{0}\gamma^{0}i\gamma^{5}q_{i}(t,-\mathbf{x}) = -\bar{q}_{j}(t,-\mathbf{x})i\gamma^{5}q_{i}(t,-\mathbf{x}) \text{ ,}
\end{equation}
thus%
\begin{equation}
P(t,\mathbf{x})\rightarrow-P(t,-\mathbf{x}) \text{ ,}
\end{equation}
implying that parity is given by $\lambda_{\mathcal{P}}=\mathcal{P}=-1$. Under charge conjugation $\mathcal{C}$ :%
\begin{align}
\sqrt{2}P_{ij} &  =\bar{q}_{j}i\gamma^{5}q_{i}\rightarrow(i\gamma^{2}%
q_{j}^{\ast})^{\dagger}\gamma^{0}i\gamma^{5}i\gamma^{2}q_{i}^{\ast}=q_{j}%
^{T}\gamma^{2,\dagger}\gamma^{0}i\gamma^{5}\gamma^{2}q_{i}^{\ast}=-q_{j}%
^{T}\gamma^{2}\gamma^{0}i\gamma^{5}\gamma^{2}q_{i}^{\ast}\nonumber\\
&  =q_{j}^{T}\gamma^{0}i\gamma^{5}q_{i}^{\ast}=-q_{j}^{T}i\gamma^{5}\gamma
^{0}q_{i}^{\ast}=-q_{j}^{T}i\gamma^{5}\gamma^{0,T}q_{i}^{\ast}=-q_{j}%
^{T}i\gamma^{5,T}\left(  q_{i}^{\dagger}\gamma^{0}\right)  ^{T}\nonumber\\
&  =-q_{j}^{T}i\gamma^{5,T}\left(  \bar{q}_{i}\right)  ^{T}=\left(
\bar{q}_{i}i\gamma^{5}q_{j}\right)  ^{T}=\bar{q}_{i}%
i\gamma^{5}q_{j}=\sqrt{2}P_{ji}\text{ ,}%
\end{align}
where the anti-commutation between fermionic fields has been taken into
account. Summarizing:%
\begin{equation}
P\rightarrow P^{T}\text{ ,}%
\end{equation}
implying that $\lambda_{\mathcal{C}}=\mathcal{C}=1$ (where$,$ again, only the diagonal elements
are $\mathcal{C}$ -eigenstates). Indeed, as shown in \ref{Ssec:non-relimit}, these quantum numbers are confirmed when studying the non-relativistic limit of the current.

We conclude this brief survey about pseudoscalars with 3 phenomenological remarks:

(i) The mixing between the neutrally charged kaons $K^{0}$ and $\bar{K}^{0}$ leads to the physical states denoted as $K_{S}$ (weakly decaying into $\pi\pi$) and $K_{L}$ (weakly decaying into $\pi\pi\pi)$, for which $\mathcal{C}\mathcal{P}$ symmetry violation also occurs. However, we will concentrate here on the strong interactions, so we will continue to work with $K^{0}$ and $\bar{K}^{0}$ states.

(ii) The pseudoscalar mesons satisfy the $SU(2)_V$ isospin symmetry at a very
good level of accuracy
\begin{equation}
m_{\pi^{\pm}}-m_{\pi^{0}}\simeq4.6\text{ }\,\text{MeV},\,\,\,m_{K^{0}%
}-m_{K^{\pm}}\simeq3.9\,\text{MeV}\, \text{ .}
\end{equation}

(iii) In general, all pseudoscalar mesons are very narrow states: they may
decay weakly (as the kaons mentioned above or as $\pi^{+}\rightarrow\mu^{+}%
\nu_{\mu}$), and electromagnetically (important decays, also for experimental
detection, are $\pi^{0}\rightarrow\gamma\gamma,$ $\eta\rightarrow\gamma
\gamma,$ $\eta^{\prime}\rightarrow\gamma\gamma$). Besides the strong (but
rather small) decay $\eta^{\prime}\rightarrow\pi^{+}\pi^{-}\eta$ and
$\eta^{\prime}\rightarrow\pi^{0}\pi^{0}\eta$, strong decays are also possible
by violating $\mathcal{G}$-parity (and hence isospin), such as
$\eta\rightarrow\pi^{0}\pi^{0}\pi^{0}$ and $\eta^{\prime}\rightarrow\pi^{0}%
\pi^{0}\pi^{0}.$ We refer to Ref. \cite{Kupsc:2009zzb} for a summarizing discussion of these decays.

\bigskip

\textit{Scalar meson nonet: }$J^{\mathcal{P}\mathcal{C}}=0^{++}$ , \textit{with} $L=S=1$;
spectroscopic notation: $n$ $^{2S+1}L_{J}\equiv$ $1$ $^{3}P_{0}$.

Scalar mesons appear by setting
\begin{equation}
\Gamma\equiv\Gamma_{^{3}P_{0}}=\mathbb{1}_{4\times 4}%
\end{equation}
leading to the following elements
\begin{equation}
S_{ij}=\frac{1}{\sqrt{2}}\,\bar{q}_{j}q_{i}\text{ ,}%
\end{equation}
see \ref{Ssec:non-relimit} for the proof that this scalar current actually corresponds to
$\ L=S=1$ in the nonrelativistic limit.
Formally, the current for scalar mesons can be obtained from the pseudoscalar one upon inserting a matrix $-i\gamma^5$ in it (and by using $(\gamma^5)^2 = \mathbb{1}_{4 \times 4}$): this is the main idea behind chiral partners, see details in \ref{ssec:chiral-multp}. Thus, scalar mesons are the chiral partners of pseudoscalar mesons.

The matrix form for scalar states reads
\begin{equation}
S=\frac{1}{\sqrt{2}}%
\begin{pmatrix}
\frac{\sigma_{N}+a_{0}^{0}}{\sqrt{2}} & a_{0}^{+} & K_{0}^{\star+}\\
a_{0}^{-} & \frac{\sigma_{N}-a_{0}^{0}}{\sqrt{2}} & K_{0}^{\star0}\\
K_{0}^{\star-} & \bar{K}_{0}^{\star0} & \sigma_{S}%
\end{pmatrix}
\,\text{,}%
\end{equation}
where the isovector states are denoted with the letter $a_{0},$ the kaonic states
with $K^{*}_{0},$ and the isoscalar ones with $\sigma_{N}$ and $\sigma_{S}$ (another
viable notation is $f_{0,N}$ and $f_{0,S}$).
Under $\mathcal{P}$  and $\mathcal{C}$, an explicit calculation shows that $S(t,\mathbf{x}%
)\xrightarrow{\mathcal{P}}S(t,-\mathbf{x})$ (thus positive parity, $\mathcal{P}=1$) and
$S\xrightarrow{\mathcal{C}}S^{T}$ (hence, positive charge conjugation, $\mathcal{C}=1$).

Scalar mesons do not appear in ChPT (since formally integrated out), but they are a necessary
ingredient of LSMs. Even in its simplest form, at least one scalar field is
present, the famous $\sigma_N \equiv \sigma$ field, that gives the LSMs their name and is 
necessary for the typical Mexican-hat potential form. Within
the eLSM a full nonet of ground-state scalar mesons is included. Moreover, the
two scalar-isoscalar fields $\sigma_{N}$ and $\sigma_{S}$ acquire a nonzero
vacuum expectation value that will be denoted as $\phi_{N}$ and $\phi_{S}$.
These objects, called the chiral condensates, are directly proportional to the quark-antiquark condensates, as
explicitly encoded in the Gell-Mann-Oakes-Renner relation \cite{Gell-Mann:1968hlm}, which can be measured in lattice QCD\cite{Fukaya:2009fh}. This condensation is at the basis of SSB.

The assignment for scalar states has been a matter of debate for a long time. One may consider the lightest scalar states $\{a_{0}(980)$, $K_{0}^{\star}(700)$, $f_{0}(500)$, $f_{0}(980)\}$, with masses below 1 GeV, but these resonance are typically interpreted as four-quark states rather than conventional mesons \cite{Jaffe:1976ig,Jaffe:2004ph,Oller:1997ti,Oller:1998zr,Pelaez:2015qba}.

The favoured assignment identifies the scalars with { $\{a_{0}(1450),$
$K_{0}^{\ast}(1430),$ $f_{0}(1370),$ $f_{0}(1500)/f_{0}(1710)\}$ }. Note that
scalar-isoscalar mesons $\sigma_{N}$ and $\sigma_{S}$ are expected to mix with
the already encountered scalar glueball $G$. That is why mixing in this sector
differs from the one for pseudoscalar mesons (and other nonets), since it involves at least three
states:
\begin{equation}
\left(
\begin{array}
[c]{c}%
f_{0}(1370)\\
f_{0}(1500)\\
f_{0}(1710)\}
\end{array}
\right)  = \mathcal{O}_B\left(
\begin{array}
[c]{c}%
\sigma_{N}\\
\sigma_{S}\\
G
\end{array}
\right)
\text{ ,}
\label{matrixB}
\end{equation}
where $\mathcal{O}_B$ is an $O(3)$ orthogonal matrix, see e.g. Refs.
\cite{Llanes-Estrada:2021evz,Crede:2008vw,Janowski:2014ppa,Cheng:2006hu,Giacosa:2005zt,Amsler:1995td,Amsler:1995tu,Parganlija:2012fy} and Sec. \ref{Ssec:scalar-glbl}
for more details and for explicit results. 

The scalar mesons have typically a large decay width due to the decays into two
pseudoscalars \cite{Workman:2022ynf,ParticleDataGroup:2024cfk}. The possible two-body strong decays are: $K_0^{\star}\to K\pi$, $a_0 \to K K$, $a_0 \to \eta \pi$, $a_0 \to \eta^{\prime} \pi$, $f_0\to \pi\pi$, $f_0\to K K$, $f_0\to \eta\eta$, $f_0\to \eta\eta^{\prime}$ (for a detailed discussion see e.g. \cite{Parganlija:2012fy}). In certain cases, the simple tree-level result that
implicitly uses Breit-Wigner spectral function is not enough. A possible way
to include loops is to study the spectral function of scalar mesons \cite{Giacosa:2007bn}.
In\ Ref. \cite{Wolkanowski:2015lsa}  it is shown how the four-quark state $a_{0}(980)$ may
emerge as a companion dynamically generated state of the mostly $\bar{q}q$ resonance $a_{0}(1450)$. A similar
pattern holds for the predominantly four-quark state $K_{0}^{\star}(700)$ as a
companion pole of $K_{0}^{\ast}(1430)$ \cite{Wolkanowski:2015jtc}. The full scalar sector up to $1.75$
GeV includes 5 resonances. A full mixing may be studied \cite{Napsuciale:2004au,Fariborz:2007ai} but
is rather difficult to constrain. Within the eLSM, an attempt to include
-within the two-flavor case- a four-quark state $f_{0}(500)$ as well as one
glueball field $G$ and one nonstrange $\bar{q}q$ scalar field $\sigma_{N}$ can be found in Ref. \cite{Lakaschus:2018rki}. However, until now, both nonets of scalars below and above $1$~GeV have not been systematically studied in the eLSM.
%%%%%%%%%%%%%%%%%%%%%%
\bigskip\\
%%%%%%%%%%%%%%%%%%%%%5
\textit{Vector meson nonet: }$J^{\mathcal{P}\mathcal{C}}=1^{--}$ \textit{with} $L=0$ , $S=1$;
\textit{spectroscopic notation:} $n$ $^{2S+1}L_{J}\equiv1$ $^{3}S_{1}$ .
Vector mesons $\{\rho(770)$ , $K^{\ast}(892)$ $,$ $\omega(782),$ $\phi(1024\}$
form the second lightest nonet after pseudoscalar mesons. They are
realized by setting%
\begin{equation}
\Gamma=\Gamma_{^{3}S_{1}}^{\mu}=\gamma^{\mu}%
\end{equation}
leading to the elements
\begin{equation}
V_{ij}^{\mu}\equiv\frac{1}{\sqrt{2}}\,\bar{q}_{j}\gamma^{\mu}q_{i}\text{ ,}%
\end{equation}
out of which the matrix expression reads%

\begin{equation}
V^{\mu}=\frac{1}{\sqrt{2}}%
\begin{pmatrix}
\frac{\omega_{1N}+\rho_{1}^{0}}{\sqrt{2}} & \rho_{1}^{+} & K_{1}^{\ast+}\\
\rho_{1}^{-} & \frac{\omega_{1N}-\rho_{1}^{0}}{\sqrt{2}} & K_{1}^{\ast0}\\
K_{1}^{\ast-} & \bar{K}_{1}^{\ast0} & \omega_{1S}%
\end{pmatrix}
^{\mu}\,=\frac{1}{\sqrt{2}}%
\begin{pmatrix}
\frac{\omega_{N}+\rho^{0}}{\sqrt{2}} & \rho^{+} & K^{\ast+}\\
\rho^{-} & \frac{\omega_{N}-\rho^{0}}{\sqrt{2}} & K^{\ast0}\\
K^{\ast-} & \bar{K}^{\ast0} & \omega_{S}%
\end{pmatrix}
^{\mu},
\label{vectormatrix}
\end{equation}
where on the r.h.s. the suffix $1$ has been omitted.

Under parity $\mathcal{P}$ :
\begin{equation}
V^{\mu}(t,\mathbf{x)}\rightarrow V_{\mu}(t,-\mathbf{x)}%
\end{equation}
thus $V^{0}(t,\mathbf{x)}\rightarrow V^{0}(t,-\mathbf{x)}$ but $V^{k}%
(t,\mathbf{x)}\rightarrow-V^{k}(t,-\mathbf{x)}$ with $k=1,2,3.$ The spacial
coordinates transform with a minus sign. Since the temporal coordinate is seen
as a constraint for vector fields, $\partial_{\mu}V^{\mu}(x)=0$, the vector
mesons are regarded as negative parity states (in short $\mathcal{P}=-1$).
Under charge conjugation $\mathcal{C}$ :
\begin{equation}
V^{\mu}\rightarrow-V^{\mu,T}
\text{ ,}
\end{equation}
where the additional minus sign stands for negative charge conjugation $\mathcal{C}$ (in short $\mathcal{C}=-1$).

In Eq. \eqref{vectormatrix} $\omega_{1N}$ and $\omega_{1S}$ are purely non-strange and
strange states, respectively. The physical fields arise upon mixing
\begin{equation}
\left(
\begin{array}
[c]{c}%
\omega(782)\\
\phi(1020)
\end{array}
\right)  =\left(
\begin{array}
[c]{cc}%
\cos\beta_{v} & \sin\beta_{v}\\
-\sin\beta_{v} & \cos\beta_{v}%
\end{array}
\right)  \left(
\begin{array}
[c]{c}%
\omega_{1N}\\
\omega_{1S}%
\end{array}
\right)  \text{ ,}%
\label{betav}
\end{equation}
where the small isoscalar-vector mixing angle $\beta_{v}=-3.9^{\circ}$,
taken from the PDG2020 \cite{Zyla:2020zbs}, is obtained by using $SU(3)_V$-inspired relations between the physical masses.

The masses of $\rho$ and $\omega$ are almost degenerate. The main decay of the $\rho$ meson is
$\rho\rightarrow\pi\pi,$ for instance $\rho^{0}\rightarrow\pi^{+}\pi^{-},$
which is of particular importance in the eLSM. Note, $\rho^{0}\rightarrow
\pi^{0}\pi^{0}$ does not take place because it would violate $\mathcal{C}$-parity. On the
other hand, $\mathcal{G}$-symmetry conservation forbids $\omega\rightarrow\pi\pi,$ but
allows for $\omega\rightarrow\pi^{+}\pi^{-}\pi^{0}.$ Other relevant decays
are $K^{\ast}\rightarrow K\pi$ and $\phi\rightarrow \bar{K}K$ (the
latter is very close to the kaon-kaon threshold, implying that $\phi
\rightarrow K^{+}K^{-}$ is visibly larger than $\phi\rightarrow K^{0}\bar
{K}^{0}$: this is a purely kinematic effect because of the slightly smaller
mass of charged kaons; the coupling well fulfills isospin symmetry).

Another important property of vector mesons with $I_z=0$ is their transition
into photons: $\rho^{0}\rightarrow\gamma$, $\omega\rightarrow\gamma$ ,
$\phi\rightarrow\gamma,$ which leads to dilepton decays. 
Moreover the vector meson dominance approach describes interactions of hadrons with
photons as taking part via virtual vector mesons \cite{OConnell:1995nse}.
Vector mesons respect isospin, as the following small mass differences
confirm:
\begin{equation}
m_{\rho^{0}}-m_{\rho^{\pm}}=(-0.7\pm0.8)\,\text{MeV}\,,\,\qquad m_{K^{\star0}%
}-m_{K^{\star\pm}}=(6.7\pm1.2)\,\text{MeV}\,.
\end{equation}
On the other hand, the relations
\begin{equation}
m_{\rho}\simeq m_{\omega}\,,\,\qquad m_{\phi}-m_{K^{\star}}\simeq m_{K^{\star
}}-m_{\rho} \sim 130 \text{ MeV,}
\end{equation}
show that the effect of the heavier $s$-quark is non-negligible.
%%%%%%%%%%%%%%%%%%
\bigskip\\
%%%%%%%%%%%%%%%%%
\textit{Axial-vector meson nonet: }$J^{\mathcal{P}\mathcal{C}}=1^{++}$ \textit{with} $L=$ $S=1$;
\textit{spectroscopic notation:} $n^{\;2S+1}L_{J}\equiv 1^{\;3}P_{1}$.
The nonet of axial-vector mesons is given by \{$a_{1}(1260),$ $K_{1}
(1270)$/$K_{1}(1400),$ $f_{1}(1285),$ $f_{1}^{\prime}(1420)$\}. These states
are the chiral partners of vector mesons, see Sec. \ref{ssec:chiral-multp}. The microscopic current is
obtained by setting 
\begin{equation}
\Gamma= \gamma^{5}\gamma^{\mu} \text{ , thus} \; A_{1,ij}^{\mu
}\equiv\  \frac{1}{\sqrt{2}} \bar{q}_{j}\gamma^{5}\gamma^{\mu}q_{i}, 
\end{equation}
leading to the
nonet:
\begin{equation}
A_{1}^{\mu}=\frac{1}{\sqrt{2}}%
\begin{pmatrix}
\frac{f_{1N}+a_{1}^{0}}{\sqrt{2}} & a_{1}^{+} & K_{1A}^{+}\\
a_{1}^{-} & \frac{f_{1N}-a_{1}^{0}}{\sqrt{2}} & K_{1A}^{0}\\
K_{1A}^{-} & \bar{K}_{1A}^{0} & f_{1S}%
\end{pmatrix}
^{\mu}\text{ }\,.
\end{equation}
The mixing angle $\beta_{av}$ between the isoscalar axial-vector mesons enters
into the usual expression
\begin{equation}
\left(
\begin{array}
[c]{c}%
f_{1}(1285)\\
f_{1}(1420)
\end{array}
\right)  =\left(
\begin{array}
[c]{cc}%
\cos\beta_{av} & \sin\beta_{av}\\
-\sin\beta_{av} & \cos\beta_{av}%
\end{array}
\right)  \left(
\begin{array}
[c]{c}%
f_{1N}\\
f_{1S}%
\end{array}
\right)  \text{ .}%
\end{equation}
The experimental result reads $\beta_{av}=\left(  \pm24.0_{-3.4}^{+3.7}\right)
^{\circ}$ \cite{LHCb:2013ged}, the lattice value is $\beta_{av}=\left(  31\pm2\right)^{\circ}$ \cite{Dudek:2011tt}, and the fit from Ref. \cite{Shastry:2021asu} finds $\beta_{av}=\left(24.9\pm3.2\right)^{\circ}$  (see also Refs. \cite{Jiang:2020eml,Liu:2014doa}), all consistent with each other at the $2 \sigma$-level.
The kaonic axial-vector mesons $K_{1A}$ emerge from another type of mixing that relates these mesons to those of another nonet, that of pseudovector mesons. As a consequence, $K_{1A}$ is contained in the two resonances $K_{1}(1270)$ and $K_{1}(1400)$, see below.
Under $\mathcal{P}$  and $\mathcal{C}$ , one has $A_1^{\mu}(t,\mathbf{x}%
)\xrightarrow{\mathcal{P}} -A_{1,\mu}(t,-\mathbf{x})$ (thus positive parity, $\mathcal{P}=1$) and
$A_1^{\mu}\xrightarrow{\mathcal{C}}{A_1^{\mu}}^{T}$ (positive charge conjugation, $\mathcal{C}=1$). 
In general, these states are quite broad, e.g. the resonance $a_{1}(1230)$ with a width of about $400$ MeV, where the main channel is the $\rho\pi$ mode. Because of this large width, this state cannot be described by a Breit-Wigner (BW) spectral function, but a rather simple modification of it,  the Sill distribution \cite{Giacosa:2021mbz}, is able to capture the effect of the $\rho\pi$ threshold.
%%%%%%%%%%%%%%%%5
\bigskip\\
%%%%%%%%%%%%%%%%!
\textit{Pseudovector meson nonet: }$J^{\mathcal{P}\mathcal{C}}=1^{+-}$ \textit{with} $L=1,$ $S=0$;
\textit{spectroscopic notation:} $n$ $^{2S+1}L_{J}\equiv1$ $^{1}P_{1}$.
The nonet of pseudovector mesons reads \{$b_{1}(1235)$, $K_{1B}$,
$h_{1}(1170)$, $h_{1}(1415)$\}. The corresponding currents are obtained for
\begin{equation}
\Gamma\equiv\Gamma_{^{1}P_{1}}^{\mu}=\gamma^{5}\,\overleftrightarrow{D}^{\mu}
\text{ ,}
\end{equation}
where $\overleftrightarrow{D}^{\mu}:=\overrightarrow{D}^{\mu}-\overleftarrow
{D}^{\mu}$ ($\overrightarrow{D}^{\mu}$ being the covariant derivative $D_{\mu
}=\partial_{\mu}-igG_{\mu}$ acting on the right, and $\overleftarrow{D}^{\mu}$
acting on the left) leading to:
\begin{equation}
B_{ij}^{\mu}=\frac{1}{\sqrt{2}}\bar{q}_{j}\,\gamma^{5}\,\overleftrightarrow
{D}^{\mu}\,q_{i}\text{ .}%
\end{equation}
Intuitively, the reasoning is as follows: the current $\bar{q}_{j}\,\gamma^{5}\,\,q_{i}$ corresponds to $L=S=0,$ so adding a derivative increases $L$ to $1$ but does not change the spin, so $L=1,$ $S=0$.
In matrix form, the nonet reads:
\begin{equation}
B^{\mu}=\tfrac{1}{\sqrt{2}}%
\begin{pmatrix}
\frac{h_{1N}+b_{1}^{0}}{\sqrt{2}} & b_{1}^{+} & K_{1B}^{+}\\
b_{1}^{-} & \frac{h_{1N}-b_{1}^{0}}{\sqrt{2}} & K_{1B}^{0}\\
K_{1B}^{-} & \bar{K}_{1B}^{0} & h_{1S}%
\end{pmatrix}
^{\mu}\,.    
\end{equation}
Under $\mathcal{P}$ and $\mathcal{C}$ , one has $B^{\mu}(t,\mathbf{x}%
)\xrightarrow{\mathcal{P}} -B_{\mu}(t,-\mathbf{x})$ ($\mathcal{P}=1$) and
$B^{\mu}\xrightarrow{\mathcal{C}} -{B^{\mu}}^{T}$ ($\mathcal{C}=-1$).
The isoscalar mixing angle $\beta_{pv}$ is defined by
\begin{equation}
\begin{pmatrix}
h_{1}(1170)\\
h_{1}(1415)
\end{pmatrix}
=%
\begin{pmatrix}
\cos\beta_{pv} & \sin\beta_{pv}\\
-\sin\beta_{pv} & \cos\beta_{pv}%
\end{pmatrix}%
\begin{pmatrix}
h_{1N}\\
h_{1S}%
\end{pmatrix}
\text{ .}   
\label{mixingpv}
\end{equation}
Its value is not yet known; the fit of Ref. \cite{Shastry:2021asu} gives $\beta_{pv}=\left(  25.2\pm3.1\right)  ^{\circ}$, but the mixing angle is found to be compatible with zero in the study of Ref. \cite{BESIII:2018ede}. Since this nonet belongs to a `heterochiral multiplet', the chiral anomaly is likely to affect it, see Ref.
\cite{Giacosa:2023fdz} and Sec. \ref{SSec:heterovectors}. 

The kaonic members of this nonet are denoted by $K_{1B}.$ The physical states $K_{1}(1270)$ and $K_{1}(1400)$ arise from the mixing of $K_{1B}$ and the previously introduced $K_{1A}$ from the axial-vector meson nonet. Namely, a peculiar mixing term of the type $i\left(  K_{1A,\mu}^{-}K_{1B}^{+,\mu}-K_{1A,\mu}^{+}K_{1B}^{-,\mu}\right) $ satisfies both $\mathcal{P}$  and $\mathcal{C}$ symmetry. The outcome of the mixing (in terms of fields) is:
\begin{equation}
\left(
\begin{array}
[c]{c}%
K_{1}(1270)\\
K_{1}(1400)
\end{array}
\right)  ^{\mu}=\left(
\begin{array}
[c]{cc}%
\cos\varphi_{K} & -i\sin\varphi_{K}\\
-i\sin\varphi_{K} & \cos\varphi_{K}%
\end{array}
\right)  \left(
\begin{array}
[c]{c}%
K_{1A}\\
K_{1B}%
\end{array}
\right)  ^{\mu}\text{ .}%
\end{equation}
According to the fit of Ref. \cite{Divotgey:2013jba}, $\varphi_{K}=\left(
56.4\pm4.3\right)  ^{\circ}$, implying that $K_{1}(1270)$ is closer to $K_{1B}$
and $K_{1}(1400)$ to $K_{1A}$, but the mixing is quite large \cite{Divotgey:2013jba}. For the mixing of the quantum states it is
common to write%
\begin{equation}
\left(
\begin{array}
[c]{c}%
\left\vert K_{1}(1270)\right\rangle \\
\left\vert K_{1}(1400)\right\rangle
\end{array}
\right)  =\left(
\begin{array}
[c]{cc}%
\sin\theta_{K} & \cos\theta_{K}\\
-\cos\theta_{K} & \sin\theta_{K}%
\end{array}
\right)  \left(
\begin{array}
[c]{c}%
\left\vert K_{1A}\right\rangle \\
\left\vert K_{1B}\right\rangle
\end{array}
\right)  \text{ }%
\end{equation}
where $\theta_{K}=\left(  -90^{\circ}+\varphi_{K}\right)$, so $\theta
_{K}=\left(  -33.6\pm4.3\right)  ^{\circ}$, in agreement also with the
findings of Ref.~\cite{Hatanaka:2008gu}. Finally, the $b_{1}$ state decays mostly into $\omega\pi$; analogous decays
(into a vector and a pseudoscalar meson) hold for the other nonet members.
%%%%%%%%%%%%%%%%%%
\bigskip\\
%%%%%%%%%%%%%%%%%%
\textit{Orbitally excited vector meson nonet: }$J^{\mathcal{P}\mathcal{C}}=1^{--}$ with $L=2,$ $S=1$; \textit{spectroscopic notation:} $n$ $^{2S+1}L_{J}\equiv1$ $^{3}D_{1}$.
Orbitally excited vector mesons are identified with
\{$\rho(1700)$, $K^{\ast}(1680)$, $\omega(1650)$, $\phi(?)$\}. The
corresponding current is obtained by setting%
\begin{equation}
\Gamma=\Gamma_{^{3}D_{1}}^{\mu}=\overleftrightarrow{D}^{\mu},
\end{equation}
out of which:
\begin{equation}
V_{D,\,ij}^{\mu}=\frac{1}{\sqrt{2}}(\bar{q}_{j}\overleftrightarrow{D}^{\mu
}q_{i})\text{ .}%
\end{equation}
Intuitively, the scalar current $\bar{q}_jq_{i}$ with $L=S=1$ gets an additional
unit of orbital angular momentum when the derivative is introduced, thus $L=2$ and
$S=1.$ In matrix form:
\begin{equation}
V_{D}^{\mu}=\frac{1}{\sqrt{2}}%
\begin{pmatrix}
\frac{\omega_{D,N}+\rho_{D}^{0}}{\sqrt{2}} & \rho_{D}^{+} & K_{1D}^{\ast+}\\
\rho_{D}^{-} & \frac{\omega_{D,N}-\rho_{D}^{0}}{\sqrt{2}} & K_{1D}^{\ast0}\\
K_{1D}^{\ast-} & \bar{K}_{1D}^{\ast0} & \omega_{D,S}%
\end{pmatrix}
^{\mu}\,.
\end{equation}
Under $\mathcal{P}$  and $\mathcal{C}$ , one has $V_D^{\mu}(t,\mathbf{x}%
)\xrightarrow{\mathcal{P}} V_{D,\mu}(t,-\mathbf{x})$ ($\mathcal{P}=-1$) and
$V_D^{\mu}\xrightarrow{\mathcal{C}} -{V_D^{\mu}}^{T}$ ($\mathcal{C}=-1$). 
As already mentioned, these states are also vector mesons.

The predominantly $\omega_{D,S}\equiv\bar{s}s$ state $\phi(?)$ could be
assigned to $\phi(2170)$ (see the quark model review of the PDG 2024 \cite{ParticleDataGroup:2024cfk}), but the mass
seems too large when compared to the hadronic model prediction of Ref.
\cite{Piotrowska:2017rgt} and the quark model prediction of Ref. \cite{isgur1985}, according
to which the mass of this state is about 1.9 GeV. Moreover, the $\phi(2170)$
was interpreted as a non-conventional state (a tetraquark) in Ref.
\cite{Ke:2018evd}.
Besides $\phi(?),$ the phenomenology of the orbitally excited vector mesons is
quite well known experimentally, even if large uncertainties are 
present. The main decay channels are into pseudoscalar-pseudoscalar (just as
the ground state vector mesons) and into vector-pseudoscalar pairs.
\bigskip\\
%%%%%%%%%%%%%%%%%
\textit{Tensor meson nonet: }$J^{\mathcal{P}\mathcal{C}}=2^{++}$ \textit{with} $L=1,$ $S=1$;
\textit{spectroscopic notation:} $n$ $^{2S+1}L_{J}\equiv 1^{\;3}P_{2}$.
The well-known nonet of tensor mesons is described by the resonance
\{$a_{2}(1320)$, $K_{2}^{\ast}(1430),f_{2}$$(1270)$, $f_{2}^{\prime}(1525)$\}.
The tensor current is obtained by choosing%
\begin{equation}
\Gamma\equiv\Gamma_{^{3}P_{2}}^{\mu\nu}=i\left[  \gamma^{\mu}%
\,\overleftrightarrow{D}^{\nu}+\gamma^{\nu}\,\overleftrightarrow{D}^{\mu
}-\frac{2}{3}\left(  g^{\mu\nu}-\frac{k^{\mu}k^{\nu}}{k^{2}}\right)
\overleftrightarrow{D}_{\alpha}\gamma^{\alpha}\right]  =i\gamma^{\mu}%
\partial^{\nu}+i\gamma^{\nu}\,\partial^{\mu}+...
\text{ ,}
\end{equation}
leading to
\begin{equation}
T_{ij}^{\mu\nu}=\bar{q}_{j}\,\Gamma_{^{3}P_{2}}^{\mu\nu}\,q_{i}.
\end{equation}
Intuitively, $\bar{q}_{j}\,\gamma^{\mu}\,q_{i}$ with $L=0,$ $S=1$ gets an
additional unit of orbital angular momentum by inserting the derivative.
The corresponding matrix reads:%
\begin{equation}
T^{\mu\nu}=\frac{1}{\sqrt{2}}%
\begin{pmatrix}
\frac{f_{2N}+a_{2}^{0}}{\sqrt{2}} & a_{2}^{+} & K_{2}^{\ast+}\\
a_{2}^{-} & \frac{f_{2N}-a_{2}^{0}}{\sqrt{2}} & K_{2}^{\ast0}\\
K_{2}^{\ast-} & \bar{K}_{2}^{\ast0} & f_{2S}%
\end{pmatrix}
^{\mu\nu}\,.\label{eq:nonet-t2}
\end{equation}
Under $\mathcal{P}$ and $\mathcal{C}$ , one has $T^{\mu\nu}(t,\mathbf{x}%
)\xrightarrow{\mathcal{P}} T_{\mu \nu}(t,-\mathbf{x})$ ($\mathcal{P}=1$) and
$T^{\mu\nu}\xrightarrow{\mathcal{C}} T^{\mu\nu,T}$ ($\mathcal{C}=1$). 
The physical isoscalar-tensor states are
\begin{equation}
\left(
\begin{array}
[c]{c}%
f_{2}(1270)\\
f_{2}^{\prime}(1525)
\end{array}
\right)  =\left(
\begin{array}
[c]{cc}%
\cos\beta_{t} & \sin\beta_{t}\\
-\sin\beta_{t} & \cos\beta_{t}%
\end{array}
\right)  \left(
\begin{array}
[c]{c}%
f_{2N}\\
f_{2S}%
\end{array}
\right)  \text{  ,}
\end{equation}
where $\beta_{t}=5.7^{\circ}$ is the value of the small mixing angle reported
in the PDG (this is in agreement with their underlying homochiral nature; see
\ref{ssec:chiral-multp}). The decays of tensor mesons are well known experimentally: the
two-pseudoscalar channel dominates, but the vector-pseudoscalar mode is also
relevant. The phenomenology fits very well with an almost ideal nonet of
$\bar{q}q$ states, as shown in detail in Refs.
\cite{Giacosa:2005bw,Burakovsky:1997ci}. The tensor nonet has been also
studied within holographic approaches in Refs. \cite{Katz:2005ir,Mamedov:2023sns},
confirming their standard $\bar{q}q$ interpretation.
%%%%%%%%%%%%%%%%%%%%%5
\bigskip\\
%%%%%%%%%%%%%%%%%%%%%%
\textit{Axial-tensor meson nonet: }$J^{\mathcal{P}\mathcal{C}}=2^{--}$ \textit{with} $L=2,$ $S=1$;
\textit{spectroscopic notation:} $n$ $^{2S+1}L_{J}\equiv 1^{\;3}D_{2}$.
The identification of this nonet is difficult. For the kaonic member, two
states close to the expected mass are reported in the PDG: $K_{2}%
(1770)$ and $K_{2}(1820)$ (a mixing between axial-tensor and pseudotensor kaons, analogous to $K_{1A}$ and $K_{1B}$, is likely
to occur, but its magnitude is unknown). However, the isotriplet state
$\rho_{2}$ and the two isosinglet states $\omega_{2}$ and $\phi_{2}$ are
unknown. Namely, even if some resonances with the correct quantum numbers are
present in the PDG ($\rho_{2}(1940)$, $\rho_{2}(2225)$, $\omega_{2}(1975)$,
and $\omega_{2}(2195)$) they are too heavy to be assigned to the ground-state
axial-tensor meson nonet.
The axial-tensor current reads \cite{Koenigstein:2015asa,Sungu:2020azn}%
\begin{equation}
\Gamma_{2^{--}}^{\mu\nu}=i\left[  \gamma^{5}\gamma^{\mu}\,\overleftrightarrow
{D}^{\nu}+\gamma^{5}\gamma^{\nu}\,\overleftrightarrow{D}^{\mu}-\frac{2}%
{3}\left(  g^{\mu\nu}-\frac{k^{\mu}k^{\nu}}{k^{2}}\right)  \overleftrightarrow
{D}_{\alpha}\gamma^{5}\gamma^{\alpha}\right]
\text{ ,}
\end{equation}
leading to the elements axial-tensor mesons.
\begin{equation}
A_{2,ij}^{\mu\nu}=\bar{q}_{j}\,\Gamma_{^{3}D_{2}}^{\mu\nu}\,q_{i}\,.
\end{equation}
Intuitively, $\bar{q}_{j}\,\gamma^{5}\gamma^{\mu}\,q_{i}$ with $L=S=1$ gets an
additional unit of $L$ by the derivative insertion, leading to $L=2,$ $S=1.$
The axial-tensor matrix reads:%
\begin{equation}
A_{2}^{\mu\nu}=\frac{1}{\sqrt{2}}%
\begin{pmatrix}
\frac{\omega_{2N}+\rho_{2}^{0}}{\sqrt{2}} & \rho_{2}^{+} & K_{2A}^{+}\\
\rho_{2}^{-} & \frac{\omega_{2N}-\rho_{2}^{0}}{\sqrt{2}} & K_{2A}^{0}\\
K_{2A}^{-} & \bar{K}_{2A}^{0} & \omega_{2S}%
\end{pmatrix}
^{\mu\nu}\,.\label{eq:nonet-a2}
\end{equation}
Under $\mathcal{P}$  and $\mathcal{C}$ , one has $A_2^{\mu\nu}(t,\mathbf{x}%
)\xrightarrow{\mathcal{P}} -A_{2,\mu \nu}(t,-\mathbf{x})$ ($\mathcal{P}=-1$) and
$A_2^{\mu\nu}\xrightarrow{\mathcal{C}} - A_2^{\mu\nu,T}$ ($\mathcal{C}=-1$). 
%%%%%%%%%%%%%%%%%%%%
Due to the lack of information and the fact that this nonet belongs to a so-called
homochiral multiplet (see Ref. \cite{Giacosa:2017pos} and Sec. \ref{ssec:chiral-multp}), the mixing in the
isoscalar sector is expected to be small with
\begin{equation}
\omega_{2N}\simeq\omega_{2}\,,\quad\omega_{2S}\simeq\phi_{2}\,.
\end{equation}
One of the possible reasons for the missing states may be their very large decay widths \cite{Jafarzade:2021vhh,Johnson:2020ilc}, see also Refs. \cite{Guo:2019wpx,Abreu:2020wio,Feng:2022jtg}.
\bigskip\\
\textit{Pseudotensor meson nonet: }$J^{\mathcal{P}\mathcal{C}}=2^{-+}$ \textit{with} $L=2,$ $S=0$;
\textit{spectroscopic notation:} $n$ $^{2S+1}L_{J}\equiv 1^{\;1}D_{2}$.
The pseudotensor mesons $\{\pi_{2}(1670)$, $K_{2}(1770)/K_{2}(1820),$
$\eta_{2}(1870),$ $\eta_{2}(1645)\}$ fit rather well as $\bar{q}q$ candidates
for the ground-state pseudotensor meson nonet.
The form of the pseudotensor current is obtained for
\cite{Koenigstein:2015asa}%
\begin{equation}
\Gamma\equiv\Gamma_{^{1}D_{2}}^{\mu\nu}=i\left[  \gamma^{5}%
\,\overleftrightarrow{D}^{\mu}\overleftrightarrow{D}^{\nu}-\left(  \frac{2}%
{3}g^{\mu\nu}-\frac{k^{\mu}k^{\nu}}{k^{2}}\right)  \overleftrightarrow
{D}_{\alpha}\,\gamma^{5}\,\overleftrightarrow{D}^{\alpha}\right]
\end{equation}
leading to the elements
\begin{equation}
P_{2,ij}^{\mu\nu}=\bar{q}_{j}\,\Gamma_{^{1}D_{2}}^{\mu\nu}q_{i}\,.
\end{equation}
Intuitively, $\bar{q}_{j}i\gamma^{5}q_{i}$ with $L=S=0$ jumps to $L=2,$ $S=0$
by inserting two derivatives. In matrix form:
\begin{equation}
P_{2}^{\mu\nu}=\tfrac{1}{\sqrt{2}}%
\begin{pmatrix}
\frac{\eta_{2N}+\pi_{2}^{0}}{\sqrt{2}} & \pi_{2}^{+} & K_{2P}^{+}\\
\pi_{2}^{-} & \frac{\eta_{2N}-\pi_{2}^{0}}{\sqrt{2}} & K_{2P}^{0}\\
K_{2P}^{-} & \bar{K}_{2P}^{0} & \eta_{2S}%
\end{pmatrix}
^{\mu\nu}\,.\label{eq:nonet-pseduotensor}
\end{equation}
Under $\mathcal{P}$ and $\mathcal{C}$, one has $P_2^{\mu\nu}(t,\mathbf{x}%
)\xrightarrow{\mathcal{P}} -P_{2,\mu \nu}(t,-\mathbf{x})$ ($\mathcal{P}=-1$) and
$P_2^{\mu\nu}\xrightarrow{\mathcal{C}} P_2^{\mu\nu,T}$ ($\mathcal{C}=+1$). 
The unknown mixing angle $\beta_{pt}$ within the isoscalar sector is given as
\begin{equation}
\begin{pmatrix}
\eta_{2}(1645)\\
\eta_{2}(1870)
\end{pmatrix}
=%
\begin{pmatrix}
\cos\beta_{pt} & \sin\beta_{pt}\\
-\sin\beta_{pt} & \cos\beta_{pt}%
\end{pmatrix}%
\begin{pmatrix}
\eta_{2N}\\
\eta_{2S}%
\end{pmatrix}
\,.    
\end{equation}
A large mixing angle $\beta_{pt}\approx-40^{\circ}$ is obtained by the fits of
Refs. \cite{Koenigstein:2016tjw,Shastry:2021asu}. While a large mixing is
possible due to the heterochiral nature of the multiplet to which pseudotensor
mesons belong (see Secs. \ref{ssec:chiral-multp} and \ref{Sec:other(1,2)}), a small mixing was found in Ref. \cite{Giacosa:2023fdz} using an
instanton approach  as well as  in
Ref. \cite{Dudek:2013yja} by lattice QCD.
\bigskip\\
\textit{``Excited'' tensor meson nonet: }$J^{\mathcal{P}\mathcal{C}}=2^{++}$ \textit{with} $L=3,$ $S=1$;
\textit{spectroscopic notation:} $n$ $^{2S+1}L_{J}\equiv1$ $^{3}F_{2}$.
These states are the chiral partners of the pseudotensor mesons, but at present the PDG lacks adequate candidates. Their currents are obtained by setting
\begin{equation}
\Gamma\equiv\Gamma_{^{3}F_{2}}^{\mu\nu}=\,\overleftrightarrow{D}^{\mu
}\overleftrightarrow{D}^{\nu}-\left(  \frac{2}{3}g^{\mu\nu}-\frac{k^{\mu
}k^{\nu}}{k^{2}}\right)  \overleftrightarrow{D}_{\alpha}\,\overleftrightarrow
{D}^{\alpha}
\text{ ,}
\end{equation}
leading to the elements
\begin{equation}
T_{2F,ij}^{\mu\nu}=\bar{q}_{j}\,\Gamma_{^{3}F_{2}}^{\mu\nu}q_{i}\,.
\end{equation}
The excited tensor matrix reads:%
\begin{equation}
T_{2F}^{\mu\nu}=\frac{1}{\sqrt{2}}%
\begin{pmatrix}
\frac{f_{2F,N}+a_{2F}^{0}}{\sqrt{2}} & a_{2F}^{+} & K_{2F}^{\ast+}\\
a_{2F}^{-} & \frac{f_{2F,N}-a_{2F}^{0}}{\sqrt{2}} & K_{2F}^{\ast0}\\
K_{2F}^{\ast-} & \bar{K}_{2F}^{\ast0} & f_{2F,S}%
\end{pmatrix}
^{\mu\nu}\,.
\end{equation}
Under $\mathcal{P}$ and $\mathcal{C}$ , one has $T_{2F}^{\mu\nu}(t,\mathbf{x}%
)\xrightarrow{\mathcal{P}} T_{2F,\mu \nu}(t,-\mathbf{x})$ ($\mathcal{P}=1$) and
$T_{2F}^{\mu\nu}\xrightarrow{\mathcal{C}} T_{2F}^{\mu\nu,T}$ ($\mathcal{C}=1$). 
A possible candidate member for this nonet is the isoscalar state $f_{2}(2150)$ \cite{Vereijken:2023jor} (it could be the predominantly non-strange isoscalar member) but
the mixing between the isoscalar members is unknown. % Table \ref{tab:mixing}
%%%%%%%%%%%%%%
\bigskip\\
%%%%%%%%%%%%%
\textit{Hybrid meson nonet I: }$J^{\mathcal{P}\mathcal{C}}=1^{-+}$.
We present the lightest nonet of hybrid states, which are
a bound states of a quark-antiquark pair and one gluon (schematically, $\bar{q}qg$)$.$
Even if the experimental status is not complete yet, some candidates exist.
First, let us discuss the structure. Formally, the currents can be obtained by
choosing%
\begin{equation}
\Gamma = \Gamma_{\text{hyb},1^{-+}}^{\mu}=G^{\mu\nu}\gamma_{\nu}%
\end{equation}
leading to
\begin{equation}
P_{ij}^{\text{hyb},\mu}=\frac{1}{\sqrt{2}}\bar{q}_{j}\Gamma_{\text{hyb}}^{\mu}%
q_{i}\,=\frac{1}{\sqrt{2}}\bar{q}_{j}G^{\mu\nu}\gamma_{\nu}q_{i}\,,
\end{equation}
which contain at least one gluon in the wave function, look at $\left(
P_{ij}^{\text{hyb},\mu}\right)  ^{\dagger}\left\vert 0_{\text{QCD}}\right\rangle$. Since the $\bar{q}q$ spectroscopic notation $^{2S+1}L_{J}$ does not apply to this case (it is a 3-body system), it is omitted. 
Intuitively, the vector current $\bar{q}_{j}\gamma_{\nu}q_{i}$ is complemented by the gluon field tensor $G^{\mu\nu}$, which flips the charge-conjugation but preserves parity. 
In matrix form:%
\begin{equation}
P^{\text{hyb},\mu}=\frac{1}{\sqrt{2}}\left(
\begin{array}
[c]{ccc}%
\frac{\eta_{1N}^{\text{hyb}}+\pi_{1}^{0}}{\sqrt{2}} & \pi_{1}^{\text{hyb}+} & K_{1}%
^{\text{hyb}+}\\
\pi_{1}^{\text{hyb}-} & \frac{\eta_{1N}^{\text{hyb}}-\pi_{1}^{0}}{\sqrt{2}} & K_{1}%
^{\text{hyb}0}\\
K_{1}^{\text{hyb}-} & \bar{K}_{1}^{\text{hyb}0} & \eta_{1S}^{\text{hyb}}%
\end{array}
\right)  ^{\mu}\;\text{ .} \label{hybridmatrix}%
\end{equation}
Under $\mathcal{P}$ and $\mathcal{C}$ , one has $P^{\text{hyb}\,, \mu}(t,\mathbf{x}%
)\xrightarrow{\mathcal{P}} P^{\text{hyb}}_{\mu}(t,-\mathbf{x})$ ($\mathcal{P}=-1$) and
$P^{\text{hyb}}_{\mu}\xrightarrow{\mathcal{C}} {P^{\text{hyb}}}^T_{\mu}$ ($\mathcal{C}=1$). 
The quantum numbers
$J^{\mathcal{P}\mathcal{C}}=1^{-+}$ are not allowed for a purely $\bar{q}q$ system, thus such
hybrid object displays exotic quantum numbers (this is not the case for
the kaonic members, since $\mathcal{C}$  is not defined for them).
%%%%%%%%%%%%%%%%%%%
\bigskip\\
%%%%%%%%%%%%%%%%%%%
\textit{Hybrid meson nonet II: }$J^{\mathcal{P}\mathcal{C}}=1^{+-}$.
The next nonet of hybrid states is built by inserting a $\gamma^{5}$ to the
currents, leading to the chiral partners of the $1^{-+}$-hybrid nonet:
\begin{equation}
\Gamma_{\text{hyb},1^{+-}}^{\mu}=G^{\mu\nu}\gamma^{5}\gamma_{\nu}\text{ ,{}}
\end{equation}
hence:
\begin{equation}
B_{ij}^{\text{hyb},\mu}=\frac{1}{\sqrt{2}}\bar{q}_{j}\Gamma_{\text{hyb},1^{+-}}^{\mu}%
q_{i}=\frac{1}{\sqrt{2}}\bar{q}_{j}G^{\mu\nu}\gamma^{5}\gamma_{\nu}q_{i}\text{
.}%
\end{equation}
These states carry the quantum numbers $J^{\mathcal{P}\mathcal{C}}=1^{+-}$ just as the
pseudovector mesons, thus making them crypto-exotics.
In matrix form:%
\begin{equation}
B^{\text{hyb},\mu}=\frac{1}{\sqrt{2}}\left(
\begin{array}
[c]{ccc}%
\frac{h_{1N,B}^{\text{hyb}}+b_{1}^{\text{hyb},0}}{\sqrt{2}} & b_{1}^{\text{hyb},+} & K_{1B}%
^{\text{hyb}+}\\
b_{1}^{\text{hyb},+} & \frac{h_{1N,B}^{\text{hyb}}-b_{1}^{\text{hyb},0}}{\sqrt{2}} & K_{1B}%
^{\text{hyb}0}\\
K_{1B}^{\text{hyb}-} & \bar{K}_{1B}^{\text{hyb}0} & h_{1S,B}^{\text{hyb}}%
\end{array}
\right)  ^{\mu}\text{ .}%
\end{equation}
Under $\mathcal{P}$  and $\mathcal{C}$ , one has $B^{\text{hyb}\,, \mu}(t,\mathbf{x}%
)\xrightarrow{\mathcal{P}} -B^{\text{hyb}}_{\mu}(t,-\mathbf{x})$ ($\mathcal{P}=1$) and
$B^{\text{hyb}, \mu}\xrightarrow{\mathcal{C}} -B^{\text{hyb},\mu T}$ ($\mathcal{C}=-1$). 
According to lattice QCD results, their masses should be above 2 GeV
\cite{Meyer:2015eta,Dudek:2010wm,Woss:2020ayi}, but at present no experimental candidates can be listed.
The most important properties of the different nonets are summarized in Tables~\ref{tab:mixing}, \ref{tab:mesonlist} and \ref{tab:nonets}.
\begin{table}[ht!]
\centering
\renewcommand{\arraystretch}{1.2}
\begin{tabular}{|c|c|c|c|}
\hline
 $J^{\mathcal{P}\mathcal{C}}$   &  Resonances                                             & Mixing Relation                                                          & Mixing Angle \\ \hline
$0^{++}$ & \begin{tabular}[c]{@{}l@{}}$\eta$\\ $\eta^{\prime}(958)$\end{tabular} & \begin{tabular}[c]{@{}l@{}}$\eta= \eta_N \,\cos{\beta_{ps}}+\eta_S\,\sin{\beta_{ps}}$\\$\eta^\prime= -\eta_N\,\sin{\beta_{ps}}+\eta_S \cos{\beta_{ps}}$\end{tabular} & $\beta_{ps}\simeq -43.4^\circ$ Ref. \cite{AmelinoCamelia:2010me} \\ \hline
$1^{--}$ & \begin{tabular}[c]{@{}l@{}}$\omega(782)$\\ $\phi(1020)$\end{tabular} & \begin{tabular}[c]{@{}l@{}}$\omega= \omega_{1N} \,\cos{\beta_v}+\omega_{1S}\,\sin{\beta_v}$\\$\phi= -\omega_{1N}\,\sin{\beta_v}+\omega_{1S}\cos{\beta_v}$\end{tabular} & $\beta_v\simeq-3.9^\circ$ Ref. \cite{Workman:2022ynf} \\ \hline
$1^{++}$ & \begin{tabular}[c]{@{}l@{}}$f_1(1285)$\\ $f_1(1420)$\end{tabular} & \begin{tabular}[c]{@{}l@{}}$f_1= f_{1N} \,\cos{\beta_{av}}+f_{1S}\,\sin{\beta_{av}}$\\$f_1^{\prime}= -f_{1N}\,\sin{\beta_{av}}+f_{1S}\cos{\beta_{av}}$\end{tabular} & $\beta_{av}^\dagger\simeq 24^{\circ\dagger}$ Ref. \cite{LHCb:2013ged} \\ \hline
$1^{+-}$ & \begin{tabular}[c]{@{}l@{}}$h_1(1170)$\\ $h_1(1415)$\end{tabular} & \begin{tabular}[c]{@{}l@{}}$h_1= h_{1N} \,\cos{\beta_{pv}}+h_{1S}\,\sin{\beta_{pv}}$\\$h_1^{\prime}= -h_{1N}\,\sin{\beta_{pv}}+h_{1S}\cos{\beta_{pv}}$\end{tabular} & $\beta_{pv}\simeq 1^{\circ\dagger}$ Ref. \cite{BESIII:2018ede} \\ \hline
$2^{++}$ & \begin{tabular}[c]{@{}l@{}}$f_2(1270)$\\ $f_2(1525)$\end{tabular} & \begin{tabular}[c]{@{}l@{}}$f_2= f_{2N} \,\cos{\beta_{t}}+f_{2S}\,\sin{\beta_{t}}$\\$f_2^{\prime}= -f_{2N}\,\sin{\beta_{t}}+f_{2S}\cos{\beta_{t}}$\end{tabular} & $\beta_{t}\simeq 3.16^\circ$ Ref. \cite{Workman:2022ynf} \\ \hline
$2^{-+}$ & \begin{tabular}[c]{@{}l@{}}$\eta_2(1645)$\\ $\eta_2^{\prime}(1870)$\end{tabular} & \begin{tabular}[c]{@{}l@{}}$\eta_2= \eta_{2N} \,\cos{\beta_{pt}}+\eta_{2S}\,\sin{\beta_{pt}}$\\$\eta_2^\prime= -\eta_{2N}\,\sin{\beta_{pt}}+\eta_{2S} \cos{\beta_{pt}}$\end{tabular} & $\beta_{pt}\simeq -42.3^{\circ\dagger}$ Ref. \cite{Koenigstein:2015asa,Shastry:2021asu}\\ \hline    
\end{tabular}
\caption{Mixing angles between the isoscalar mesons of the nonets. Numerical values indicated by $\dagger$ are under debate.}
\label{tab:mixing}
\end{table}
%----------------------------------------------------------------------------------------------
%----------------------------------------------------------------------------------------------
\begin{table}[ht!]
\centering
\renewcommand{\arraystretch}{1.2}
\begin{tabular}{|c|c|c|c|c|c|c|c|c|}
\hline
$n^{2S+1}L_J$  &  $J^{\mathcal{P}\mathcal{C}}$ & \begin{tabular}[c]{@{}l@{}}I=1\\ $u\overline{d}$, $d\overline{u}$\\$ \frac{d\overline{d}-u\overline{u}}{\sqrt{2}}$\end{tabular} & \begin{tabular}[c]{@{}l@{}}I=1/2\\ $u\overline{s}$, $d\overline{s}$\\$s\overline{d}$, $s\overline{u}$\end{tabular} & \begin{tabular}[c]{@{}l@{}}I=0\\ $\approx \frac{u\overline{u}+d\overline{d}}{\sqrt{2}}$\end{tabular} & \begin{tabular}[c]{@{}l@{}}I=0\\  $\approx s\overline{s}$\end{tabular} & \begin{tabular}[c]{@{}l@{}} Nonet Name\\
PDG inspired\end{tabular}  & \begin{tabular}[c]{@{}l@{}} Nonet Name\\
Chirally inspired \end{tabular} & \begin{tabular}[c]{@{}l@{}} Chiral\\
Multiplet\end{tabular}   \\ \hline
 $1^1 S_0$ & $0^{-+}$ &    $\pi$ & $K$  & $\eta(547)$  & $\eta^\prime(958)$ &    $P$&    $P$ & \multirow{2}{*}{\begin{tabular}[c]{@{}l@{}} $\Phi=S+iP$ \end{tabular}}            \\ \cline{1-8}                                 
  $1^3 P_0$ & $0^{++}$   &    $a_0(1450)$ & $K_0^\star(1430)$  & $f_0(1370)$ & 
  \begin{tabular}[c]{@{}l@{}} $f_0(1500)$\\
$f_0(1710)$\end{tabular} & $S$  &    $S$  &                                                       \\ \hline \hline
$1^3 S_1$ & $1^{--}$  &   $\rho(770)$  & $K^\star(892)$ & $\omega(782)$ & $\phi(1020)$  & $V^{\mu}$ &    $V^{\mu}$    & \multirow{2}{*}{\begin{tabular}[c]{@{}l@{}} $L^{\mu}=V^{\mu}+A^{\mu}$ \\
$R^{\mu}=V^{\mu}-A^{\mu}$ \end{tabular}}            \\ \cline{1-8}                        %%%%%                                
 $1^3 P_1$ &  $1^{++}$  &  $a_1(1260)$  & $K_{1A}$   & $f_1(1285)$  & $f_1^\prime(1420)$ & $A_1^{\mu}$ &    $A^{\mu}$    & \\ \hline\hline
  $1^1 P_1$ &  $1^{+-}$   & $b_1(1235)$ & $K_{1B}$ &$h_1(1170)$ &   $h_1(1415)$ & $B^{\mu}$ &    $P^{\mu}$   & \multirow{2}{*}{\begin{tabular}[c]{@{}l@{}} $\Phi^{\mu}=S^{\mu}+iP^{\mu}$ \end{tabular}}            \\ \cline{1-8}  
%%%%%%  
   $1^3 D_1$ & $1^{--}$  &   $\rho(1700)$  & $K^\star(1680)$ & $\omega(1650)$ & $\phi(???)$  & $V_D^{\mu}$ &    $S^{\mu}$   &  
     \\ \hline\hline
  $1^3 P_2$ & $2^{++}$  &   $a_2(1320)$ & $K_2^\star(1430)$ & $f_2(1270)$ & $f_2^\prime(1525)$   & $T^{\mu\nu}$ &    $V^{\mu\nu}$    & \multirow{2}{*}{\begin{tabular}[c]{@{}l@{}} $\mathbf{L}^{\mu\nu}=V^{\mu\nu}+A^{\mu\nu}$\\ $\mathbf{R}^{\mu\nu}=V^{\mu\nu}-A^{\mu\nu}$ \end{tabular}}            \\ \cline{1-8}  
%%%%%%  
   $1^3 D_2$ & $2^{--}$  & $\rho_2(???)$  & $K_2(1820)$  & $\omega_2(???)$     & $\phi_2(???)$  &  $A_2^{\mu\nu}$  &    $A^{\mu\nu}$   &  
     \\ \hline\hline
  $1^1 D_2$ & $2^{-+}$  &   $\pi_2(1670)$ & $K_2(1770)$ & $\eta_2(1645)$ & $\eta_2(1870)$   & $P_2^{\mu\nu}$ &    $P^{\mu\nu}$    & \multirow{2}{*}{\begin{tabular}[c]{@{}l@{}} $\Phi^{\mu\nu}= S^{\mu\nu}{}+iP^{\mu\nu}$ \end{tabular}}            \\ \cline{1-8}  
     $1^3 F_2$ & $2^{++}$  & $a_2(???)$  & $K_2^{\ast}(???)$  & $f_2(???)$ & $f_2(???)$ &  $T_{2F}^{\mu\nu}$ &    $S^{\mu\nu}$     & 
     \\ \hline 
\end{tabular}
\caption{List of conventional $q\bar{q}$ mesonic nonets, grouped in pair of chiral partners, together with different naming conventions. On the last column, the chiral multiplets described in Section \ref{ssec:chiral-multp} are displayed. Note, in this work we use the PDG-inspired name, but the chirally inspired name, which makes use solely of the letters $S$, $P$, for the heterochiral nonet $\Phi$ and $V$ and $A$ for the homochiral multiplets $L,R$, is also reported for completeness. } 
\label{tab:mesonlist}
\end{table}
%%%%%%%%%%%%%%%%%%%%%%%%%%%%%%
\begin{table}[ht!]
\centering
\small
\renewcommand{\arraystretch}{1.2}
\begin{tabular}{|c|c|c|c|c|c|c|}
    \hline
    Nonet & $J^{\mathcal{P}\mathcal{C}}$  & Current  & States  & $\mathcal{P}$   & $\mathcal{C}$   \\
    \hline\hline
    $P$ & $0^{-+}$   & $P_{ij}=i\,\overline{q}_j\gamma_5q_i/\sqrt{2}$  &  \begin{tabular}{@{}c@{}}$\pi$, $K$,\\ $\eta(547)$, $\eta^{\prime}(958)$\end{tabular} & $-P(t,-\vec{x})$  & $P^{T}$  \\
    \hline
    $S$ & $0^{++}$ & $S_{ij}=\,\overline{q}_jq_i/\sqrt{2}$  & \begin{tabular}{@{}c@{}}$a_{0}(1450)$, $K_{0}^{\ast}(1430)$,\\ $f_{0}(1370)$, $ f_{0}(1500)/f_{0}(1710)$\end{tabular}  & $S(t,-\vec{x})$  & $S^{T}$ \\
    \hline\hline
    $V^{\mu}$ & $1^{--}$  & $V^{\mu}_{ij}=\,\overline{q}_j\gamma^{\mu}q_i/\sqrt{2}$  & \begin{tabular}{@{}c@{}}$\rho(770)$, $K^{\ast}(892)$,\\ $\omega(782)$, $\phi(1020)$\end{tabular}   & $V_{\mu}(t,-\vec{x})$  & $-V^{\mu\,,T}$ \\
    \hline
    $A_1^{\mu}$ & $1^{++}$  & $A^{\mu}_{1\,,ij}=\,\overline{q}_j\gamma^{5}\gamma^{\mu}q_i/\sqrt{2}$  & \begin{tabular}{@{}c@{}}$a_{1}(1260)$, $K_{1}(1270)/K_{1}(1400)$,\\ $f_{1}(1420)$, $f_{1}(1285)$ \end{tabular} & $-A_{1\mu}(t,-\vec{x})$  & $A_1^{\mu\,,T}$ \\
    \hline\hline
    $B^{\mu}$ & $1^{+-}$  & $B^{\mu}_{ij}=\,\overline{q}_j\gamma^{5}\overleftrightarrow{D}^{\mu}q_i/\sqrt{2}$  & \begin{tabular}{@{}c@{}}$b_{1}(1235)$, $K_{1}(1270)/K_{1}(1400)$,\\ $h_{1}(1415)$, $h_{1}(1170)$\end{tabular}   & $-B_{\mu}(t,-\vec{x})$  & $-B^{\mu\,,T}$ \\
    \hline
    $V_{D}^{\mu}$ & $1^{--}$  & $V^{\mu}_{D\,,ij}=i\,\overline{q}_j\overleftrightarrow{D}^{\mu}q_i/\sqrt{2}$  & \begin{tabular}{@{}c@{}}$\rho(1700)$, $K^{\ast}(1680)$,\\ $\omega(1650)$, $\phi(?)$\end{tabular}  & $V_{D\,,\mu}(t,-\vec{x})$  & $-V_{D}^{\mu\,,T}$ \\
    \hline\hline
    $T^{\mu\nu}$ & $2^{++}$  & \begin{tabular}[c]{@{}l@{}}$T_{ij}^{\mu\nu} = i \bar{q}_j \, \Big[  \gamma^{\mu} \, \overleftrightarrow{D}^\nu + \gamma^{\nu} \, \overleftrightarrow{D}^\mu$ \\
    $-\frac{2}{3}\Big(g^{\mu\nu}-\frac{k^{\mu}k^{\nu}}{k^2}\Big)\overleftrightarrow{D}_{\alpha}\gamma^{\alpha} \Big] \, q_i$\end{tabular} & \begin{tabular}{@{}c@{}}$a_{2}(1320)$, $K_{2}^{\ast}(1430)$,\\ $f_{2}(1270), f_{2}(1525)$\end{tabular}  & $T_{\mu\nu}(t,-\vec{x})$  & $\Big(T^{\mu\nu}\Big)^T$     \\
    \hline
    $A_2^{\mu\nu}$ & $2^{--}$  & \begin{tabular}[c]{@{}l@{}}$A_{2,ij}^{\mu\nu} = i \bar{q}_j \, \Big[ \gamma^5 \gamma^{\mu} \, \overleftrightarrow{D}^\nu + \gamma^5\gamma^{\nu} \, \overleftrightarrow{D}^\mu$\\
    $-\frac{2}{3}\Big(g^{\mu\nu}-\frac{k^{\mu}k^{\nu}}{k^2}\Big)\overleftrightarrow{D}_{\alpha}\gamma^5\gamma^{\alpha} \Big] \, q_i$\end{tabular} & \begin{tabular}{@{}c@{}}$\rho_{2}(?)$, $ K_2(1770)/K_2(1820)$,\\ $\omega_2(?)$, $\phi_2(?)$ \end{tabular}  & $-A_{2,\mu\nu}(t,-\vec{x})$  & $-\Big(A_2^{\mu\nu}\Big)^T$ \\
    \hline\hline
    $P_2^{\mu\nu}$ & $2^{-+}$  & 
        \begin{tabular}[c]{@{}l@{}}$P_{2,ij}^{\mu\nu} = i \bar{q}_j \, \Big[  \gamma^{5} \, \overleftrightarrow{D}^\mu \overleftrightarrow{D}^\nu -$  \\
        $-\frac{2}{3}\Big(g^{\mu\nu}-\frac{k^{\mu}k^{\nu}}{k^2}\Big)\overleftrightarrow{D}_{\alpha}\,\gamma^{5}\,\overleftrightarrow{D}^{\alpha} \Big] \, q_i$\end{tabular} & 
        \begin{tabular}{@{}c@{}}$\pi_2(1670)$, $K_2(1770)/K_2(1820)$,\\
        $\eta_2(1870)$, $\eta_2(1645)$  
        \end{tabular}& $-P_{2,\mu\nu}(t,-\vec{x})$  & $\Big(P_2^{\mu\nu}\Big)^T$  \\
    \hline
    $T_{2F}^{\mu\nu}$ & $2^{++}$  & \begin{tabular}[c]{@{}l@{}}$T_{2F,ij}^{\mu\nu} = \bar{q}_j \, \Big[  \, \overleftrightarrow{D}^\mu \overleftrightarrow{D}^\nu -$  \\
        $-\frac{2}{3}\Big(g^{\mu\nu}-\frac{k^{\mu}k^{\nu}}{k^2}\Big)\overleftrightarrow{D}_{\alpha}\,\overleftrightarrow{D}^{\alpha} \Big] \, q_i$\end{tabular} & \begin{tabular}{@{}c@{}}$a_{2}(?)$, $ K_2(?)$,\\     $f_2(?)$, $f_2^\prime(?)$ \end{tabular}  & $T_{2F\,,\mu\nu}(t,-\vec{x})$  & $\Big(T_{2F}^{\mu\nu}\Big)^T$   \\
    \hline\hline
    $P^{\mu,\text{hyb}}$ & $1^{-+}$  & $P^{\mu,\text{hyb}}_{ij}=\,\overline{q}_jG^{\mu\nu}\gamma_{\nu}q_i/\sqrt{2}$  & \begin{tabular}{@{}c@{}}$\pi_1(1600)$, $K_1(?)$,\\ $\eta_1(?)$, $\eta_1(??)$\end{tabular}  & $P_{\mu}^{\text{hyb}}(t,-\vec{x})$  & $\Big(P^{\mu\,,\text{hyb}}\Big)^T$ \\
    \hline
    $B^{\mu,\text{hyb}}$ & $1^{+-}$  &     $B^{\mu,\text{hyb}}_{ij}=\,\overline{q}_jG^{\mu\nu}\gamma_{\nu}\gamma^5q_i/\sqrt{2}$  & 
        \begin{tabular}{@{}c@{}}$b_1(1200?)$, $K_1(?)$,\\ 
        $h_1(?)$, $h_1(??)$ 
        \end{tabular} & $-B_{\mu}^{\text{hyb}}(t,-\vec{x})$  & $-\Big(B^{\mu\,,\text{hyb}}\Big)^T$ \\
    \hline
\end{tabular}
\caption{Particle content and transformation properties under $\mathcal{P}$  parity and $\mathcal{C}$  charge conjugation of the different nonets. Under flavor transformation, each nonet transforms as $M_{\Gamma} \rightarrow U_V M_{\Gamma} U_V^{\dagger} $.}
\label{tab:nonets}
\end{table}

%\bigskip

\subsection{Chiral multiplets}
\label{ssec:chiral-multp}

\subsubsection{General ideas and hetero/homochirality}

Let us consider a certain nonet $M_{1}\equiv 1/\sqrt{2}\,\bar{q}_{j}\Gamma q_{i}$
for a given choice of $\Gamma.$ Then, the nonet of chiral partners is obtained (up to a phase) by inserting a matrix $\gamma^5$, giving $M_{2}\equiv 1/\sqrt{2}\,\bar{q}_{j}\Gamma \gamma^5 q_{i}$, hence:
\begin{equation}
M_{1} \equiv \frac{1}{\sqrt{2}}\,\bar{q}_{j}\Gamma q_{i}\xLeftrightarrow{\text{chiral
partners}}\frac{1}{\sqrt{2}}\,\bar{q}_{j} \Gamma \gamma^{5}q_{i}=M_{2} \; .
\end{equation}
This is so because a chiral transformation $SU(3)_{L}%
\times SU(3)_{R}$ mixes $M_{1}$ with $M_{2}.$ Indeed, the specific chiral transformation denoted as `axial' with $q \rightarrow e^{{i\sum_{k=1}^{8}\theta_{A}^{(k)}}t^{k}\gamma^{5}}q$ clearly does the job due to the explicit presence of $\gamma^{5}.$ It is evident that
$M_{1}$ with $M_{2}$ carry the same total angular momentum $J$ but have
opposite parity: namely, $\gamma^{5}$ does not introduce any Lorentz index but
switches $\mathcal{P}$  due to $\{\gamma^{0},\gamma^{5}\}=0.$ The charge conjugation $\mathcal{C}$ is more
subtle: in some cases (homochiral multiplet) it switches, in other (heterochiral partners) it does not.

We also recall that, in the $N_{f}=2$ case, chiral transformation link states with the
same $J,$ but opposite parity $\mathcal{P}$  and $G$-parity, in particular \cite{Ko:1994en,Koch:1997ei}:
\begin{align}
\sigma_{N}  & \Longleftrightarrow\mathbf{\pi} \text{ , }\eta_{N}%
\Longleftrightarrow\mathbf{a}_{0}\text{ ,}\\
\mathbf{\rho}  & \Longleftrightarrow\mathbf{a}_{1}\text{ , }\omega
_{N}\Longleftrightarrow f_{1N}\text{ .}%
\end{align}
In fact, for the (pseudo)scalars: $\sigma_{N}$ has $\mathcal{G}=\mathcal{C}=1,$ but the pion
triplet $\mathbf{\pi}$ has $\mathcal{G}=-\mathcal{C}=-1,$ and $\eta_{N}$ has $\mathcal{G}=\mathcal{C}=1$ but
$\mathbf{a}_{0}$ has $\mathcal{G}=-\mathcal{C}=-1$.  
For the (axial-)vector sector: $\mathbf{\rho}$ has $\mathcal{G}=-\mathcal{C}=1$ and $\mathbf{a}_{1}$ has
$\mathcal{G}=-\mathcal{C}=-1,$ while $\omega_{N}$ has $\mathcal{G}=\mathcal{C}=-1$ and $f_{1N}$ has $\mathcal{G}=\mathcal{C}=1.$
For the general $N_{f}=3$ case, kaons are present, for which $\mathcal{G}$-parity is not a good quantum number, but we shall recover the above results as a special case.

Next, we group the nonets $M_{1}$ and $M_{2}$ into appropriate chiral
multiplets, for which two possibilities exist: heterochiral multiplets and
homochiral multiplets \cite{Giacosa:2017pos}, see also the summarizing tables \ref{tab:nonets} and \ref{tab:chiral-transformations} listing the main multiplets and their properties. 
%%%%%%%%%%
\bigskip\\
%%%%%%%%%%
\textbf{(1) Heterochiral multiplets.}
The first possibility refers to the case 
\begin{equation}
\left[  \Gamma,\gamma^{5}\right]  =0\text{ , }
\end{equation}
implying that
\begin{equation}
\overline{q}_{j,R}\,\Gamma q_{i,R}=\overline{q}_{j,L}\,\Gamma q_{i,L}=0
\text{ ,}
\end{equation}
and
\begin{equation}
\overline{q}_{j,R}\,\Gamma q_{i,L}=\overline{q}_{j}\,P_{L}\Gamma P_{L}%
q_{i}=\overline{q}_{j}\,P_{L}^{2}\Gamma q_{i}=\overline{q}_{j}\,P_{L}\Gamma
q_{i}=\frac{1}{2}\overline{q}_{j}\,\Gamma q_{i}-\frac{1}{2}\overline{q}%
_{j}\,\gamma^{5}\Gamma q_{i}\neq0\text{ .}%
\end{equation}
The corresponding chiral multiplets are defined as:%
\begin{equation}
\Phi_{\Gamma,ij}=\sqrt{2}\overline{q}_{j,R}\,\Gamma q_{i,L}\text{
.}\label{phi}%
\end{equation}
It is then clear that the object $\Phi_{\Gamma}$ contains both chiral partners
$M_{1}$ and $M_{2}$ in a linear combination. In general one has
\begin{equation}
\Phi_{\Gamma}=M_{1}+iM_{2}%
\end{equation}
where the phase $i$ takes into account that the currents must be Hermitian.
The transformation of $\Phi_{\Gamma}$ under chiral transformations
$SU(3)_{L}\times SU(3)_{R}$ is easily obtained form Eq. \eqref{phi} as:%
\begin{equation}
\Phi_{\Gamma}\rightarrow U_{L}\Phi U_{R}^{\dagger}\text{ .}%
\end{equation}
This is also why this chiral multiplet is called \textit{heterochiral}: both $U_{L}$ and $U_{R}$ appear in
its chirally transformed expression. 
For heterochiral multiplets, the chiral partners $M_{1}$ and $M_{2}$ have the
same $\mathcal{C}$, see below for examples.
%%%%%%%%%%%%%%%
\bigskip\\
%%%%%%%%%%%%%%%
\textbf{(2) Homochiral multiplets}
The second possibility refers to the case
\begin{equation}
\left\{  \Gamma,\gamma^{5}\right\}  =0\text{ ,}%
\end{equation}
implying that
\begin{equation}
\overline{q}_{j,R}\,\Gamma q_{i,L}=0
\end{equation}
and
\begin{equation}
\overline{q}_{j,L}\,\Gamma q_{i,L}\neq0\text{ , }\overline{q}_{j,R}\,\Gamma
q_{i,R}\neq0\text{ .}%
\end{equation}
In this case, one introduces the left-handed and the right-handed currents
\begin{equation}
L_{\Gamma,ij}=\sqrt{2}\overline{q}_{j,L}\,\Gamma q_{i,L}\text{ , }%
R_{\Gamma,ij}=\sqrt{2}\overline{q}_{j,R}\,\Gamma q_{i,R}%
\end{equation}
that can be expressed as:%
\begin{equation}
L_{\Gamma}=M_{1}+M_{2}\text{ , }R_{\Gamma}=M_{1}-M_{2}\text{ .}%
\end{equation}
Under chiral transformations, the left-handed and right-handed currents
transform as:%
\begin{equation}
L_{\Gamma}\rightarrow U_{L}L_{\Gamma}U_{L}^{\dagger}\text{ , }R_{\Gamma
}\rightarrow U_{R}R_{\Gamma}U_{R}^{\dagger}\text{ ,}%
\end{equation}
hence they are called \textit{homochiral}. 
For these multiplets, the chiral partners $M_{1}$ and $M_{2}$ have opposite
$\mathcal{C}$, see Table~\ref{tab:nonets}. 

\subsubsection{List of chiral multiplets}

Below we list various quark-antiquark chiral multiplets by increasing values of $J$ and by following the grouping of the previous subsection where chiral partners were presented one after the other. In the end, we also present one chiral multiplet for vectorial hybrid mesons. Tables \ref{tab:nonets} and \ref{tab:chiral-transformations} summarize the multiplets and their most important properties.
%%%%%%%%%%%%
\bigskip\\
%%%%%%%%%%%%
1) $J^{\mathcal{P}\mathcal{C}}=0^{\pm+},$ \textit{heterochiral multiplet of (pseudo)scalar states,
with pseudoscalar nonet }$P$ ($J^{\mathcal{P}\mathcal{C}}=0^{-+}$; $L=S=0$; $n^{\; 2S+1}%
L_{J}\equiv$ $1^{\; 1}S_{0})$ \textit{and the scalar meson nonet }$S$
($J^{\mathcal{P}\mathcal{C}}=0^{++}$; $L=S=1$; $n$ $^{2S+1}L_{J}\equiv$ $1$ $^{3}P_{0}).$ \textit{In short: heteroscalars.}
The most important multiplet for chiral models is the one that involves
(pseudo)scalar states. It is obtained from the general equations above by
setting $\Gamma=1,$ hence:%
\begin{align}
\Phi_{ij}  &  \equiv\Phi_{ij}=\sqrt{2}\overline{q}_{j,R}\,\Gamma
q_{i,L}=S_{ij}+iP_{ij}\text{ ,}\\
\Phi &  =S+iP
\end{align}
where $S$ and $P$ refer to the scalar and pseudoscalar ground state nonets.
This is clearly a heterochiral multiplet with $\Phi\rightarrow U_{L}\Phi
U_{R}^{\dagger}$ under chiral transformations. Thanks to this simple
transformation, it is very easy to construct $SU(3)_{L}\times SU(3)_{R}$
chirally invariant objects by using the property of the trace, such as
$Tr\left[  \Phi\Phi^{\dagger}\right]  $ or $Tr\left[  \Phi\Phi^{\dagger}%
\Phi\Phi^{\dagger}\right]  $. Another possibility is to use the determinant,
$\det\Phi,$ which however is invariant only under $SU(3)_{L}\times SU(3)_{R}$
but not $U(1)_{A}.$ Thanks to the latter, this multiplet, as well as other heterochiral
ones, contain interaction terms that explicitly break the axial symmetry
$U(1)_{A},$ in agreement with the axial anomaly. In particular, a mass
contribution and mixing among the isoscalar members is possible (as for the
renowned $\eta$-$\eta^{\prime}$ system of the pseudoscalar nonet). In the next
section, we will make use of both the trace and the determinant, as well as an
extension of the latter.
Moreover, using the previous results for parity and charge conjugation, we
obtain
\begin{equation}
\Phi(t,\mathbf{x})\xrightarrow{\mathcal{P}}\Phi^{\dagger
}(t,-\mathbf{x})\text{ , }\Phi\xrightarrow{\mathcal{C}}
\Phi^{T}\text{ .}%
\end{equation}
It is also useful to express the matrix $\Phi$ in explicit form:
\begin{equation}
\Phi=\sum_{k=0}^{8}(S_{k}+iP_{k})t^{k}=\frac{1}{\sqrt{2}}%
\begin{pmatrix}
\frac{(\sigma_{N}+a_{0}^{0})+i(\eta_{N}+\pi^{0})}{\sqrt{2}} & a_{0}^{+}%
+i\pi^{+} & K_{0}^{\star+}+iK^{+}\\
a_{0}^{-}+i\pi^{-} & \frac{(\sigma_{N}-a_{0}^{0})+i(\eta_{N}-\pi^{0})}%
{\sqrt{2}} & K_{0}^{\star0}+iK^{0}\\
K_{0}^{\star-}+iK^{-} & \bar{K}_{0}^{\star}+i\bar{K}^{0} & \sigma_{S}%
+i\eta_{S}%
\end{pmatrix}
\,\,.\label{Eq:J0_sc_ps_nonet}
\end{equation}
The matrix $\Phi$ develops nonzero v.e.v. because of the condensation of the isoscalar-scalar fields $\sigma_N$ and $\sigma_S$. Any LSM needs to contain (at least some of) the fields above.
%%%%%%%%%%%%%%
\bigskip\\
%%%%%%%%%%%%%%
2) $J^{\mathcal{P}\mathcal{C}}=1^{\pm\pm},$ \textit{homochiral multiplet of (axial-)vector states,
with the vector meson nonet }$V^{\mu}$ ($J^{\mathcal{P}\mathcal{C}}=1^{--}$; $L=0,$ $S=1$; $n$
$^{2S+1}L_{J}\equiv$ $1$ $^{3}S_{1})$ \textit{and the axial-vector meson nonet
}$A_{1}^{\mu}$ ($J^{\mathcal{P}\mathcal{C}}=1^{++}$; $L=S=1$; $n$ $^{2S+1}L_{J}\equiv$ $1$
$^{3}P_{1}).$ \textit{In short: homovectors.}

By setting $\Gamma =\gamma^{\mu}$ we get the homochiral
left-handed and right-handed currents (referred to as $L^{\mu}$ and $R^{\mu}$),
which are a combination of vector and axial-vector nonets:
\begin{align*}
\left(  L_{\gamma^{\mu}}\right)  _{ij}^{\mu}  & \equiv L_{ij}^{\mu}=\sqrt
{2}\overline{q}_{j,L}\gamma^{\mu}q_{i,L}\,\text{ , }\left(  R_{\gamma^{\mu}%
}\right)  _{ij}^{\mu}\equiv R_{ij}^{\mu}=\sqrt{2}\overline{q}_{j,R}\gamma
^{\mu}q_{i,R}\,,\\
L^{\mu}  & =V^{\mu}+A_{1\,}^{\mu}\text{ , }R^{\mu}=V^{\mu}-A_{1\,}^{\mu}\text{
.}%
\end{align*}
Under $U(3)_{L}\times U(3)_{R}$ they change as
\begin{equation}
L^{\mu}\rightarrow U_{L}L^{\mu}U_{L}^{\dagger}\text{ , }R^{\mu}\rightarrow U_{R}R^{\mu}%
U_{R}^{\dagger}\text{ .}%
\end{equation}
Interestingly, these objects are invariant under the axial group $U(1)_{A}$, so interaction terms involving $L^{\mu}$ and $R^{\mu}$ do not result in axial anomalous terms. In general homochiral nonets share this property, hence the
isoscalar members are expected to be closer to nonstrange and strange members,
as it is indeed for the well-known vector meson nonet, see Eq. (\ref{betav}). On the other hand, it is quite easy to write chirally invariant terms, e.g. $Tr[L_{\mu}]Tr[L^{\mu}]$, but these terms are suppressed in the large-$N_c$ limit (the coupling constant in front of them scales as $N_c^{-1}$ instead of $N_c^0$ present for a standard mass term).
Under parity and charge conjugation one gets: 
\begin{align}
& L^{\mu}(t,\mathbf{x})\xrightarrow{\mathcal{P}}R_{\mu
}(t,-\mathbf{x})\text{ , }R^{\mu}(t,\mathbf{x})\xrightarrow{\mathcal{P}%
}-L_{\mu}(t,-\mathbf{x})\text{ ;}\\
& L^{\mu}\xrightarrow{\mathcal{C}}-R^{\mu,T}\text{ ,
}R^{\mu}\xrightarrow{\mathcal{C}}-L^{\mu,T}\text{ .}%
\end{align}
The explicit expressions read:%
\begin{equation}
L^{\mu}:=\sum_{k=1}^{8}(V_{k}^{\mu}+A_{1\,k}^{\mu})t^{k}=\frac{1}{\sqrt{2}}%
\begin{pmatrix}
\frac{(\omega_{N}+\rho^{0})}{\sqrt{2}}+\frac{(f_{1N}+a_{1}^{0})}{\sqrt{2}} &
\rho^{+}+a_{1}^{+} & K^{\star+}+K_{1A}^{+}\\
\rho^{-}+a_{1}^{-} & \frac{(\omega_{N}-\rho^{0})}{\sqrt{2}}+\frac
{(f_{1N}-a_{1}^{0})}{\sqrt{2}} & K^{\star0}+K_{1A}^{0}\\
K^{\star-}+K_{1A}^{-} & \bar{K}^{\star0}+\bar{K}_{1A}^{0} & \omega_{S}+f_{1S}%
\end{pmatrix}
^{\mu}\,\,,\label{Eq:J1_nonet_L}
\end{equation}
\begin{equation}
R^{\mu}:=\sum_{k=0}^{8}(V_{k}^{\mu}-A_{1\,k}^{\mu})t^{k}=\frac{1}{\sqrt{2}}%
\begin{pmatrix}
\frac{(\omega_{N}+\rho^{0})}{\sqrt{2}}-\frac{(f_{1N}+a_{1}^{0})}{\sqrt{2}} &
\rho^{+}-a_{1}^{+} & K^{\star+}-K_{1A}^{+}\\
\rho^{-}-a_{1}^{-} & \frac{(\omega_{N}-\rho^{0})}{\sqrt{2}}-\frac
{(f_{1N}-a_{1}^{0})}{\sqrt{2}} & K^{\star0}-K_{1A}^{0}\\
K^{\star-}-K_{1A}^{-} & \bar{K}^{\star0}-\bar{K}_{1A}^{0} & \omega_{S}-f_{1S}%
\end{pmatrix}
^{\mu}\,\,.\label{Eq:J1_nonet_R}
\end{equation}
%%%%%%%%%%%%%
\bigskip\\
%%%%%%%%%%%%%
3) $J^{\mathcal{P}\mathcal{C}}=1^{\pm-},$ \textit{heterochiral multiplet of pseudovector and orbitally excited vector states, with the pseudovector meson nonet }$B^{\mu}$
($J^{\mathcal{P}\mathcal{C}}=1^{+-}$; $L=1,$ $S=0$; $n$ $^{2S+1}L_{J}\equiv$ $1$ $^{1}P_{1})$
\textit{and the orbitally excited vector meson nonet }$V_{D}^{\mu}$
($J^{\mathcal{P}\mathcal{C}}=1^{--}$; $L=2$ , $S=1$; $n$ $^{2S+1}L_{J}\equiv$ $1$ $^{3}D_{1})$.
\textit{In short: heterovectors.}

The next $\bar{q}q$ multiplet deals with (pseudo)vector states. It arises from
the choice $\Sigma=i\overleftrightarrow{D}^{\mu},$ thus%
\begin{align}
\left(  \Phi_{\overleftrightarrow{D}^{\mu}}\right)  _{ij}^{\mu}  & =\Phi
_{ij}^{\mu}=\sqrt{2}\,\overline{q}_{j,R}\,\overleftrightarrow{D}^{\mu
}\,q_{i,L}\,,\\
\Phi^{\mu}  & =B^{\mu}+iV_D^{\mu}\text{ .}%
\end{align}
The structure is heterochiral: $\Phi^{\mu}\rightarrow U_{L}\Phi^{\mu}%
U_{R}^{\dagger}$ under $SU(3)_{L}\times SU(3)_{R},$ just as for the ground-state
(pseudo)scalar multiplet. This fact implies that the axial anomaly may be realized, see later on. Under parity and charge conjugation:%
\begin{equation}
\Phi^{\mu}(t,\mathbf{x})\xrightarrow{\mathcal{P}}\Phi_{\mu
}^{\dagger}(t,-\mathbf{x})\text{ , }\Phi^{\mu}\xrightarrow{\mathcal{C}}-{\Phi^{\mu}}^T\text{ .}%
\end{equation}
%%%%%%%%%%%%%%%
The explicit form is given by:%
%%%%%%%%%%%%%%%
\begin{equation}
\Phi^{\mu}:=\sum_{k=0}^{8}(B_{k}^{\mu}+iV_{D\,,k}^{\mu})t^{k}=\frac{1}%
{\sqrt{2}}%
\begin{pmatrix}
\frac{(\omega_{D}+\rho_{D}^{0})+i(h_{1N}+b_{1}^{0})}{\sqrt{2}} & \rho
_{D}^{+}+ib_{1}^{+} & K_{1D}^{\star+}+iK_{1b}^{+}\\
\rho_{D}^{-}+ib_{1}^{-} & \frac{(\omega_{D}-\rho_{D}^{0})+i(h_{1N}-b_{1}%
^{0})}{\sqrt{2}} & K_{1D}^{\star0}+iK_{1b}^{0}\\
K_{1D}^{\star-}+iK_{1b}^{-} & \bar{K}_{1D}^{\star}+i\bar{K}_{1b}^{0} &
\phi_{D}+ih_{1S}%
\end{pmatrix}
^{\mu}\,\,.    \label{eq:J1-hetero-chiral}
\end{equation}
%%%%%%%%%%%%
\bigskip\\
%%%%%%%%%%%%
4) $J^{\mathcal{P}\mathcal{C}}=2^{\pm\pm},$ \textit{homochiral multiplet of (axial-)tensor mesons,
with the tensor meson nonet }$T^{\mu\nu}$ ($J^{\mathcal{P}\mathcal{C}}=2^{++}$; $L=S=1$; $n$
$^{2S+1}L_{J}\equiv$ $1$ $^{3}P_{2})$ \textit{and the axial-tensor meson nonet
}$A_{2}^{\mu\nu}$ ($J^{\mathcal{P}\mathcal{C}}=2^{--}$; $L=2$ , $S=1$; $n$ $^{2S+1}L_{J}\equiv$
$1$ $^{3}D_{2})$.
\textit{In short: homotensors.}

The choice $\Gamma= \gamma^{\mu}\,\overleftrightarrow{D}^{\nu}+\gamma^{\nu
}\,\overleftrightarrow{D}^{\mu}$ generates tensorial left-handed and
right-handed `homochiral' currents that involve tensor meson nonet $T^{\mu
\nu}$:%
\begin{align}
\left(  L_{\gamma^{\mu}\,\overleftrightarrow{D}^{\nu}+...}\right)  _{ij}%
^{\mu\nu}  & \equiv\mathbf{L}_{ij}^{\mu\nu}=\overline{q}_{j,L}(\gamma^{\mu
}\,\overleftrightarrow{D}^{\nu}+\gamma^{\nu}\,\overleftrightarrow{D}^{\mu
})q_{i,L}\,,\text{ }\left(  R_{\gamma^{\mu}\,\overleftrightarrow{D}^{\nu}%
+...}\right)  _{ij}^{\mu\nu}\equiv\mathbf{R}_{ij}^{\mu\nu}=\overline{q}%
_{j,R}(\gamma^{\mu}\,\overleftrightarrow{D}^{\nu}+\gamma^{\nu}%
\,\overleftrightarrow{D}^{\mu})q_{i,R}\,;\\
\mathbf{L}^{\mu\nu}  & =T^{\mu\nu}+A_2^{\mu\nu}\text{ , }\mathbf{R}%
^{\mu\nu}=T^{\mu\nu}-A_2^{\mu\nu}\text{ .}%
\end{align}
Note, the currents are marked as bold to distinguish them from $L^{\mu\nu
}=\partial^{\mu}L^{\nu}-\partial^{\nu}L^{\mu}$ that appear as kinetic terms of
the right(left)-handed homochiral vector mesons.
Under chiral transformations:
\begin{equation}
\mathbf{L}^{\mu\nu}\rightarrow U_{L}
\mathbf{L}^{\mu\nu}U_{L}^{\dagger} \text{ , }\mathbf{R}^{\mu\nu}\rightarrow
U_{R}\mathbf{R}^{\mu\nu}U_{R}^{\dagger} \text{ ,}
\end{equation}
and under parity and charge
conjugation:
\begin{align}
& \mathbf{L}^{\mu\nu}(t,\mathbf{x})\xrightarrow{\mathcal{P}} \mathbf{R}^{\mu\nu}(t,-\mathbf{x})\text{ , }\mathbf{R}^{\mu\nu}%
(t,\mathbf{x})\xrightarrow{\mathcal{P}}\mathbf{R}^{\mu\nu
}(t,-\mathbf{x})\text{ ;}\\
& \mathbf{L}^{\mu\nu}\xrightarrow{\mathcal{C}}{\rightarrow
}{\mathbf{R}^{\mu\nu}}^T\text{ , }\mathbf{R}^{\mu\nu}\xrightarrow{\mathcal{C}}{\mathbf{L}^{\mu\nu}}^T\text{ .}%
\end{align}
The matrix expressions read:%
\begin{equation}
    \mathbf{L}^{\mu\nu}:=\sum_{k=0}^{8}(T_{k}^{\mu\nu}+A_{2,\,k}^{\mu\nu}%
)t^{k}=\frac{1}{\sqrt{2}}%
\begin{pmatrix}
\frac{(\omega_{2N}+\rho_{2}^{0})}{\sqrt{2}}+\frac{(f_{2N}+a_{2}^{0})}{\sqrt
{2}} & \rho_{2}^{+}+a_{2}^{+} & K_{2}^{\star+}+K_{2A}^{+}\\
\rho_{2}^{-}+a_{2}^{-} & \frac{(\omega_{2N}-\rho_{2}^{0})}{\sqrt{2}}%
+\frac{(f_{2N}-a_{2}^{0})}{\sqrt{2}} & K_{2}^{\star0}+K_{2A}^{0}\\
K_{2}^{\star-}+K_{2A}^{-} & \bar{K}_{2}^{\star0}+\bar{K}_{2A}^{0} &
\omega_{2S}+f_{2S}%
\end{pmatrix}
^{\mu\nu},\,\,
\end{equation}
\begin{equation}
\mathbf{R}^{\mu\nu}:=\sum_{k=0}^{8}(T_{k}^{\mu\nu}-A_{2,\,k}^{\mu\nu}%
)t^{k}=\frac{1}{\sqrt{2}}%
\begin{pmatrix}
\frac{(\omega_{2N}+\rho_{2}^{0})}{\sqrt{2}}-\frac{(f_{2N}+a_{2}^{0})}{\sqrt
{2}} & \rho_{2}^{+}-a_{2}^{+} & K_{2}^{\star+}-K_{2A}^{+}\\
\rho_{2}^{-}-a_{2}^{-} & \frac{(\omega_{2N}-\rho_{2}^{0})}{\sqrt{2}}%
-\frac{(f_{2N}-a_{2}^{0})}{\sqrt{2}} & K_{2}^{\star0}-K_{2A}^{0}\\
K_{2}^{\star-}-K_{2A}^{-} & \bar{K}_{2}^{\star0}-\bar{K}_{2A}^{0} &
\omega_{2S}-f_{2S}%
\end{pmatrix}
^{\mu\nu}\,\,.    
\end{equation}
%%%%%%%%%%%%%%%%%
\bigskip\\
%%%%%%%%%%%%%%%%%
5) $J^{\mathcal{P}\mathcal{C}}=2^{\pm+},$ \textit{heterochiral multiplet of (pseudo)tensor states
and their chiral partners, with the pseudotensor meson nonet }$P_2^{\mu\nu}$
($J^{\mathcal{P}\mathcal{C}}=2^{-+}$; $L=2$,$S=0$; $n^{\;2S+1}L_{J}\equiv$ $1^{\;1}D_{2})$
\textit{and the  excited tensor meson nonet }$T_{2F}^{\mu\nu}$ ($J^{\mathcal{P}\mathcal{C}}=2^{++}$;
$L=3$ , $S=1$; $n^{\;2S+1}L_{J}\equiv$ $1^{\; 3}F_{2})$.
\textit{In short: heterotensors.}

This heterochiral nonet is obtained for $\Gamma =\frac{g^{\mu\nu}}%
{4}\overleftrightarrow{D}^{\alpha}\overleftrightarrow{D}_{\alpha
}-\overleftrightarrow{D}^{\mu}\overleftrightarrow{D}^{\nu}$ , out of which: %
\begin{align}
\left(  \Phi_{\overleftrightarrow{D}^{\mu}\,\overleftrightarrow{D}^{\nu}+...}\right)  _{ij}%
^{\mu\nu}\equiv\Phi_{ij}^{\mu\nu}  & =\overline{q}_{j,R}\,(\frac{g^{\mu\nu}}{4}%
\overleftrightarrow{D}^{\alpha}\overleftrightarrow{D}_{\alpha}%
-\overleftrightarrow{D}^{\mu}\overleftrightarrow{D}^{\nu})\,q_{i,L}\,,\\
\Phi^{\mu\nu}  & =T^{\mu\nu}_{2F} + iP^{\mu\nu}_2\text{ .}%
\end{align}
Under the chiral group $\Phi^{\mu\nu}\rightarrow U_{L}\Phi^{\mu\nu}%
U_{R}^{\dagger},$ while under parity and charge conjugation:%
\begin{equation}
\Phi^{\mu\nu}(t,\mathbf{x})\xrightarrow{\mathcal{P}}\Phi^{\mu
\nu,\dagger}(t,-\mathbf{x})\text{ , }\Phi^{\mu\nu}\xrightarrow{\mathcal{C}} \Phi^{\mu\nu,T}\text{ .}%
\end{equation}
In matrix form:%
\begin{equation}
    \Phi^{\mu\nu}=\sum_{k=0}^{8}(T_{2F,\,k}^{\mu\nu}{}+iP_{2,\,k}^{\mu\nu})t^{k}%
=\frac{1}{\sqrt{2}}%
\begin{pmatrix}
\frac{(f_{2F,N}+a_{2F}^{0})+i(\eta_{2N}+\pi_{2}^{0})}{\sqrt{2}} & a_{2F}%
^{+}+i\pi_{2}^{+} & K_{2F}^{\star+}+iK_{2P}^{+}\\
a_{2F}^{-}+i\pi_{2}^{-} & \frac{(f_{2F,N}-a_{2F}^{0})+i(\eta_{2N}-\pi_{2}^{0}%
)}{\sqrt{2}} & K_{2F}^{\star0}+iK_{2P}^{0}\\
K_{2F}^{\star-}+iK_{2P}^{-} & \bar{K}_{2F}^{\star}+i\bar{K}_{2P}^{0} &
f_{2F,S}+i\eta_{2S}%
\end{pmatrix}
^{\mu\nu}\,\,.
\end{equation}
%%%%%%%%%%%%55
\bigskip\\
%%%%%%%%%%%%%%%
6) $J^{\mathcal{P}\mathcal{C}}=1^{\pm\mp},$ \textit{homochiral multiplet of vectorial hybrid mesons,
with the exotic hybrid nonet }$\Pi^{\text{hyb},\mu}$ ($J^{\mathcal{P}\mathcal{C}}=1^{-+})$ \textit{and
the cryptoexotic hybrid nonet }$B^{\text{hyb},\mu}$ ($J^{\mathcal{P}\mathcal{C}}=1^{+-}$).
\textit{In short: hybrid homovectors.}

Upon choosing the hybrid current $\Gamma =G_{\mu\nu}\gamma^{\nu},$ we obtain
the right-handed and left-handed hybrid currents:
\begin{align}
L_{\mu}^{\text{hyb}}  & =\overline{q}_{L}G_{\mu\nu}\gamma^{\nu}q_{L}\,,\text{
}R_{\mu}^{\text{hyb}}=\overline{q}_{R}G_{\mu\nu}\gamma^{\nu}q_{R}\,,\qquad\\
L_{\mu}^{\text{hyb}}  & =\Pi_{\mu}^{\text{hyb}}+B_{\mu}^{\text{hyb}}\text{ ,
}R_{\mu}^{\text{hyb}}=\Pi_{\mu}^{\text{hyb}}-B_{\mu}^{\text{hyb}}\text{ . }%
\end{align}
Under chiral transformations $L_{\mu}^{\text{hyb}}\rightarrow U_{L}L_{\mu
}^{\text{hyb}}U_{L}^{\dagger}$ , $R_{\mu}^{\text{hyb}}\rightarrow U_{R}R_{\mu
}^{\text{hyb}}U_{R}^{\dagger},$ while under parity and charge conjugation:%
\begin{align}
& L_{\mu}^{\text{hyb}}(t,\mathbf{x})\xrightarrow{\mathcal{P}}R^{\text{hyb,}\mu}(t,-\mathbf{x})\text{ , }R_{\mu}^{\text{hyb}}%
(t,\mathbf{x})\xrightarrow{\mathcal{P}}L^{\text{hyb,}\mu
}(t,-\mathbf{x})\text{ ;}\\
& L_{\mu}^{\text{hyb}}\xrightarrow{\mathcal{C}}R_{\mu
}^{\text{hyb,}T}\text{ , }R_{\mu}^{\text{hyb}}\xrightarrow{\mathcal{C}}L_{\mu}^{\text{hyb,}T}\text{ .}%
\end{align}
The hybrid multiplets can be written via matrices just as regular
quark-antiquark nonets:%
\begin{equation}
L_{\mu}^{\text{hyb}}=P_{\mu}^{\text{hyb}}+B_{\mu}^{\text{hyb}}=\frac
{1}{\sqrt{2}}\left(
\begin{array}
[c]{ccc}%
\frac{\eta_{1,N}^{\text{hyb}}+\pi_{1}^{\text{hyb}\,,0}}{\sqrt{2}}+\frac{h_{1N,B}^{\text{hyb}}%
+b_{1}^{\text{hyb}\,,0}}{\sqrt{2}} & \pi_{1}^{\text{hyb}\,,+}+b_{1}^{\text{hyb}\,,+} &
K_{1}^{\text{hyb}\,,+}+K_{1B}^{\text{hyb}\,,+}\\
\pi_{1}^{\text{hyb}\,,-}+b_{1}^{\text{hyb}\,,-} & \frac{\eta_{1N}^{\text{hyb}}-\pi_{1}^{\text{hyb}\,,0}%
}{\sqrt{2}}+\frac{h_{1N,B}^{\text{hyb}}-b_{1}^{\text{hyb}\,,0}}{\sqrt{2}} & K_{1}%
^{\text{hyb}\,,0}+K_{1B}^{\text{hyb}\,,0}\\
K_{1}^{\text{hyb}\,,-}+K_{1B}^{\text{hyb}\,,-} & \bar{K}_{1}^{\text{hyb}\,,0}+\bar{K}%
_{1B}^{\text{hyb}\,,0} & \eta_{1S}^{\text{hyb}}+h_{1S,B}^{\text{hyb}}%
\end{array}
\right)  _{\mu}\text{ ,}%
\end{equation}
\begin{equation}
R_{\mu}^{\text{hyb}}=P_{\mu}^{\text{hyb}}-B_{\mu}^{\text{hyb}}=\frac
{1}{\sqrt{2}}\left(
\begin{array}
[c]{ccc}%
\frac{\eta_{1N}^{\text{hyb}}+\pi_{1}^{\text{hyb}\,,0}}{\sqrt{2}}-\frac{h_{1N,B}^{\text{hyb}}%
+b_{1}^{\text{hyb}\,,0}}{\sqrt{2}} & \pi_{1}^{\text{hyb}\,,+}-b_{1}^{\text{hyb}\,,+} &
K_{1}^{\text{hyb}\,,+}-K_{1B}^{\text{hyb}\,,+}\\
\pi_{1}^{\text{hyb}\,,-}-b_{1}^{\text{hyb}\,,-} & \frac{\eta_{1N}^{\text{hyb}}-\pi_{1}^{\text{hyb}\,,0}%
}{\sqrt{2}}-\frac{h_{1N,B}^{\text{hyb}}-b_{1}^{\text{hyb}\,,0}}{\sqrt{2}} & K_{1}%
^{\text{hyb}\,,0}-K_{1B}^{\text{hyb}\,,0}\\
K_{1}^{\text{hyb}\,,-}-K_{1B}^{\text{hyb}\,,-} & \bar{K}_{1}^{\text{hyb}\,,0}-\bar{K}%
_{1B}^{\text{hyb}\,,0} & \eta_{1S}^{\text{hyb}}-h_{1S,B}^{\text{hyb}}%
\end{array}
\right)  _{\mu}\text{ .}%
\end{equation}
%%%%%%%%%%%%%%
\bigskip\\
%%%%%%%%%%%%%%
\begin{table}[ht!]
\centering
\renewcommand{\arraystretch}{1.2}
\begin{tabular} {|c|c|c|c|c|}
    \hline
    Chiral Multiplet & Parity $(\mathcal{P})$ & Charge conjugation $(\mathcal{C})$ & $U(3)_L \times U(3)_R$ & Type \\
    \hline \hline
    $\Phi=S+iP$ & $\Phi^{\dagger}(t,-\vec{x})$ & $\Phi^{T}$ & $U_L\Phi U^{\dagger}_R$ & Heterochiral \\
    \hline \hline
    $R^{\mu}=V^{\mu}-A_1^{\mu}$ & $L_{\mu}(t,-\vec{x})$ & $-(L^{\mu })^{T}$ & $U_R R^{\mu}U_R^{\dagger}$  & \multirow{2}{*}{Homochiral} \\ \cline{1-4}
    $L^{\mu}=V^{\mu}+A_1^{\mu}$ & $R_{\mu}(t,-\vec{x})$ & $-(R^{\mu })^{T}$ & $U_L L^{\mu}U_L^{\dagger}$  &  \\
    \hline\hline
    $\Phi^{\mu}=S^{\mu}+iP^{\mu}$ & $\Phi_{\mu}^{\dagger}(t,-\vec{x})$ & $-(\Phi^{\mu})^T$ & $U_L\Phi^{\mu} U^{\dagger}_R$  & Heterochiral \\
    \hline\hline
    $\mathbf{R}^{\mu\nu}=T^{\mu\nu}-A_2^{\mu\nu}$ & $\mathbf{L}_{\mu\nu}(t,-\vec{x})$ & $(\mathbf{L}^{\mu\nu})^{T}$ & $U_{R}\mathbf{R}^{\mu\nu}U^{\dagger}_{R}$    & \multirow{2}{*}{Homochiral} \\ \cline{1-4}
    $\mathbf{L}^{\mu\nu}=T^{\mu\nu}+A_2^{\mu\nu}$ & $\mathbf{R}_{\mu\nu}(t,-\vec{x})$ & $(\mathbf{R}^{\mu\nu})^{T}$ & $U_{L}\mathbf{L}^{\mu\nu}U^{\dagger}_{L}$  &   \\
    \hline\hline
    $\Phi^{\mu\nu}=T_{2F}^{\mu\nu}+iP_2^{\mu\nu}$ & $\Phi_{\mu\nu}^{\dagger}(t,-\vec{x})$ & $(\Phi^{\mu\nu})^T$ & $U_L\Phi^{\mu\nu} U^{\dagger}_R$  & Heterochiral\\
    \hline\hline
    $R^{\text{hyb}\,,\mu}=P^{\text{hyb}\,,\mu}-B^{\text{hyb}\,,\mu}$ & $L_{\mu}^{\text{hyb}}(t,-\vec{x})$ & $-(L^{\text{hyb}\,,\mu })^{T}$ & $U_R R^{\text{hyb}\,,\mu}U_R^{\dagger}$   & \multirow{2}{*}{Homochiral} \\ \cline{1-4}
    $L^{\text{hyb}\,,\mu}=P^{\text{hyb}\,,\mu}+B^{\text{hyb}\,,\mu}$ & $R_{\mu}^{\text{hyb}}(t,-\vec{x})$ & $-(R^{\text{hyb}\,\mu })^{T}$ & $U_L L^{\text{hyb}\,,\mu}U_L^{\dagger}$  & \\
    \hline
\end{tabular}
\caption{Heterochiral and homochiral multiplets and their chiral + axial, $\mathcal{C}$ and $\mathcal{P}$  transformation properties.}
\label{tab:chiral-transformations}
\end{table}

%\FloatBarrier	

 \subsection{Experimental study of light mesons: brief survey}

 The standard $\bar{q}q$ fields listed in this section have been and are
investigated, in conjunction with their exotic counterparts, in a variety of
experiments. The experimental findings are summarized in the PDG compilation \cite{ParticleDataGroup:2024cfk}, whose averages/fits are often used for comparison with theoretical calculations, such as the model results discussed in the next sections. The main quantities that can be investigated in the eLSM are the meson masses, partial decay widths and, in certain cases scattering lengths. It is compelling to present a short overview of some of the main reactions and related experiments with which these quantities can be measured. We divide them according to the reactions used to study them.

\begin{enumerate}
\item  The reaction
\begin{equation}
e^{+}e^{-}\rightarrow\gamma\rightarrow V\rightarrow...
\text{ ,}
\end{equation}
where $V$ stands for a quark-antiquark vector meson that decays further into
lighter mesons and photons, has been extensively used experimentally. The
produced $V$ depends on the center of mass of the $e^{+}e^{-}$ system. In CMD-3 and SND at VEPP-2000 (Novosibirsk) the reactions $e^{+}e^{-}\rightarrow\rho^{0}%
,\omega$ (involving $\bar{u}u$ and $\bar{d}d$ pairs) are studied together with the further decays of the type $\rho^{0}\rightarrow\pi^{+}\pi^{-}$ and $\omega\rightarrow\pi^{+}\pi^{0} \pi^{-}$ \cite{SND:2023gan,Gribanov:2019qgw}. 
In the KLOE and KOLE-2 experiments at Da$\phi$Ne \cite{Kloe2} the state $V$ coincides with the $\phi\simeq\bar{s}s$ meson, which has different decay rates, the most abundant being $\phi\rightarrow KK$. 
Other decays such as $\phi\rightarrow\gamma\pi^{0}\pi^{0}$ and $\phi\rightarrow\eta\gamma$ are
important for the analysis of light mesons (below 1 GeV). At higher energies, in the BESIII (and earlier BES) experiment in Beijing, the produced $V$ is the $J/\psi\equiv\bar{c}c$ meson (or excited charmonia vector states), see e.g. \cite{BESIII:2018ede,BESIII:2010gmv}. The study of the decays of the $J/\psi$ into light mesons (both strong and radiative) has been crucial for the determination of their properties. Belle \cite{Shwartz:2015efa} and Babar \cite{BaBar:2015kii,BaBar:2009wpw}  concentrated on the production of bottomonia $\bar{b}b$ states, which also decay into light hadrons. Currently, the BelleII update is underway (data taken since December 2024) \cite{Belle-II:2018jsg} and also addresses the study of low-energy resonances \cite{Belle-II:2024msd}. Finally, we also
mention the ALEPH experiment at LEP (CERN) where $e^{+}e^{-}\rightarrow Z^{0}$ has
been extensively studied. For light mesons, the further decay $Z^{0}%
\rightarrow\tau^{+}\tau^{-}$ with $\tau^{-}\rightarrow\nu_{\tau}\rho$ and
$\tau^{-}\rightarrow\nu_{\tau}a_{1}(1260)$ allow to measure the spectral
functions of $\rho$ and $a_{1}(1260),$ whose difference is controlled by
spontaneous chiral symmetry breaking.
%%%%%%%%%%%%%%%
\item Another relevant reaction is the photoproduction process of the type 
\begin{equation}
\gamma N\rightarrow XN^{\prime} 
\text{ ,}
\end{equation} 
where $N$ stands for a proton or a light nucleus target in the initial state and $N^{\prime}$ for the recoiled one, while $X$ stands for the produced meson in the final state. The exchanged particles between the photon and the target are mesons, typically pions. In this way, mesonic resonances have been studied at CLAS12 \cite{Rizzo:2016idq} and GLUEX \cite{Ghoul:2015ifw} at Jefferson Lab (USA) as well as at the COMPASS experiment at CERN \cite{Ryabchikov:2019rgx,Ryabchikov:2019rgx,COMPASS:2021ogp}. At CLAS12, for example, the potentially hybrid meson $\pi_{1}(1600)$, with special attention to $\pi_{1}(1600)\rightarrow \eta^{\prime}\pi$, was investigated. The COMPASS experiment also implemented pions instead of photons, allowing for a detailed study of $\pi_{1}(1600)$.
%(A previous experiment with pion beams is WA79 at CERN-SPS \cite{}) 
Previous photoproduction experiments include LEPS (Spring-8, Japan) and MAMI \& MESA (Mainz, Germany) \cite{Kashevarov:2015bfa}. In the near future, updates at the Jefferson Lab \cite{Accardi:2023chb} are planned. Next, the EIC (Electron Ion Collider) at BNL is expected to continue research (including hadron spectroscopy) using photoproduction on proton targets \cite{Abir:2023fpo}.

\item Another class of processes deals with nucleus-nucleus scattering, in particular proton-proton scattering:
\begin{equation}
NN\rightarrow\text{hadrons, }pp\rightarrow\text{hadrons}
\text{ .}
\end{equation}
The final state may contain many particles, especially for high-energy collisions. This type of process has been studied in fixed target SPS experiments at WA102 \cite{WA102:1998nzq} (CERN SPS, 1990s), with emphasis on glueball candidates and scalar/tensor mesons below 2.6 GeV. Currently SPS experiments are ongoing, such as NA61/SHINE, which has recently announced an anomalous isospin symmetry breaking in kaon production \cite{NA61SHINE:2023azp}. Another experiment of this type is HADES at GSI, which focused on light vector mesons in vacuum and in the medium, e.g. \cite{HADES:2012sir}.  The currently running major LHC experiments (ALICE, CMS, LHCb, ATLAS) also produce a large number of hadrons. The LHCb is extremely important for the spectroscopy of heavy quarks (including pentaquark states) \cite{LHCb:2015yax}, but also light mesons are studied, e.g. the mixing of the kaonic states $K_1(1270)$ and $K_1(1400)$ \cite{LHCb:2014osj}. We mention the study of the light resonances $a_0(980)$ and $f_0(980)$ in CMS \cite{CMS:2023rev} and ALICE\cite{Humanic:2022hpq}. The development of femtoscopy techniques also allows for novel studies of scattering and decays \cite{Fabbietti:2020bfg}. In the future, $pp$ and $pN$ scattering will be studied at the GSI/FAIR facility, with the CBM experiment being among the first to be realized \cite{Friman:2011zz}.

%%%%%%%%%%%%%%
\item As a last reaction we mention the proton-antiproton annihilation 
\begin{equation} 
p\bar{p}\rightarrow X\rightarrow...
\text{ ,}
\end{equation} 
in which the isoscalar meson $X$ can have arbitrary non-exotic quantum numbers $J^{PC}$. In this respect, the fusion is similar to $e^{+}e^{-}$, but enriched by a much larger number of mesons. Their mass depends on the energy range scanned by the specific experiment. In this context we mention LEAR (CERN), which studied $X$ up to $2.5$ GeV, thus in the low energy domain and therefore useful for low-energy spectroscopy (a notable result of the Crystal Barrel Collaboration is related to the scalar state $f_0(1500)$ \cite{CrystalBarrel:1996sqr}). The Fermilab experiments E760/E835 worked at the charmonium production energies and produced charmonium states with different quantum numbers, such as the scalar state $\chi_{c0} (1 ^3 P_0)$, which further decays into light mesons, e.g. two pions \cite{FermilabE835:2003gwi}. In the future, the PANDA experiment at GSI/FAIR is expected to produce particles in the mass range between $2.2$-$5$ GeV, opening up the possibility of discovering glueballs in fusion processes \cite{Lutz:2009ff}.

\end{enumerate}
 
\section{The Lagrangian of the eLSM and its consequences}
\label{Sec:eLSM_Lagr}

\subsection{General considerations}
\label{Ssec:eLSM_Lagr_gen_cons}

Effective approaches of QCD, such as the eLSM, are constructed to possess the global symmetries of QCD, i.e. the $SU(3)_L\times SU(3)_R$ chiral symmetry, and the $\mathcal{C}$ and $\mathcal{P}$  discrete symmetries. Since the chiral symmetry is not exact, violating terms are also needed. 

It is worth noting that in earlier versions of linear sigma models, when vector and axial-vector fields were first added to the Lagrangian, local chiral invariance was usually considered. This local invariance was broken to a global one with the introduction of (axial-)vector mass terms, see \cite{Gasiorowicz:1969kn,Ko:1994en} and refs. therein. This approach was used to reflect the vector meson dominance hypothesis \cite{Sakurai:1960ju} and to be consistent with the current-field identity \cite{Arnowitt:1969cc}. It also had the advantage of requiring fewer coupling constants, since local symmetry forced certain coefficients to coincide. However, it turned out that in order to correctly describe the meson decay widths and the pion-pion scattering length(s), the introduction of six-dimensional terms was also necessary \cite{Ko:1994en}. Since this effective model can be regarded as a low-energy theory, valid only up to a certain scale, this is not a problem in itself, but in this way it is somehow hard to explain why one does not include power eight, ten, etc. terms in the Lagrangian. On the other hand, there is no deep reason to make the chiral symmetry local, since it is global in QCD. It has been shown in \cite{Urban:2001ru} in a two-flavor LSM that experimental data can be well reproduced by using global chiral symmetry. Consequently, we follow this approach for the eLSM, i.e. we consider the chiral symmetry as a global symmetry as in 
QCD. Moreover, we request the validity of dilatation invariance, meaning that -in the chiral limit and in absence of the chiral anomaly- the allowed terms have \textit{exactly}
order 4 (dimensionless coupling, besides the parameter $\Lambda_G$ in the dilaton sector, see Sec. \ref{Ssec:dilaton}). This is the condition (and not renormalizability) that imposes that only a limited number of terms is retained in the eLSM \cite{Giacosa:2009bj}.
The eLSM constructed under these requirements is able to reproduce experimental masses and decay widths up to $\sim 2$~GeV with good precision \cite{Parganlija:2012fy}.

In the next subsection, we shall present the three-flavor form of the eLSM
with (pseudo)scalar and  (axial-)vector mesons. Here we merely discuss some general
arguments and its building blocks.
\begin{itemize}
\item An example of a chirally symmetric and dilatation (as well as $\mathcal{C}$  and
$\mathcal{P}$ ) invariant Lagrangian term is given by the quartic term:%
\begin{equation}
-\lambda_{2}Tr\left[  \Phi\Phi^{\dag}\Phi\Phi^{\dag}\right]
\end{equation}
Namely, $\lambda_{2}$ is dimensionless (dilatation invariant). Moreover, it
should be $\lambda_{2}>0$ so that the effective potential is bounded from
below. Chiral symmetry can be easily proved by applying $\Phi\rightarrow
U_{L}\Phi U_{R}^{\dagger},$ leading to
\begin{equation}
-\lambda_{2}Tr\left[  \Phi\Phi^{\dag}\Phi\Phi^{\dag}\right]  \rightarrow
-\lambda_{2}Tr\left[  U_{L}\Phi U_{R}^{\dagger}U_{R}\Phi^{\dag}U_{L}^{\dagger
}U_{L}\Phi U_{R}^{\dagger}\Phi^{\dag}U_{R}\Phi^{\dag}U_{L}^{\dagger}\right]
=-\lambda_{2}Tr\left[  \Phi\Phi^{\dag}\Phi\Phi^{\dag}\right]  
\text{ .}
\end{equation}
Invariance under parity and charge conjugation can be shown in a similar way.
Moreover, it turns out that in the large-$N_{c}$ limit $\lambda_{2}\propto
N_{c}^{-1}$ is the dominant term of this type of quartic interaction \cite{Giacosa:2024scx}.
\item Another dilatation, $\mathcal{P}$, $\mathcal{C}$, and chirally invariant term is given by $-\lambda_{1}\left( Tr\left[  \Phi\Phi^{\dag}\right] \right) ^{2},$ where $\lambda_{1}\propto N_{c}^{-2}$ in the
large-$N_{c}$ limit, thus the $\lambda_1$-term is suppressed
w.r.t. the  $\lambda_2$-term. Even if subleading, this type of terms may be
important for some specific processes, such as suppressed decays and mixing patterns.
%%%%%%%%%%%%%%%%%%%
\item The \textquotedblleft mass\textquotedblright\ term for (pseudo)scalar
mesons is given by the dilatation and chirally (as well as $\mathcal{C}$ and $\mathcal{P}$ )
invariant term
\begin{equation}
-\lambda_{G\Phi}G^{2}Tr\left[  \Phi\Phi^{\dag}\right]
\text{ ,}
\end{equation}
where $\lambda_{G\Phi}\propto N_{c}^{-2}$ is the dominant term for this type
of glueball-glueball-meson-meson interaction. The important point is that
$\lambda_{G\Phi}<0,$ implying the `wrong' mass sign for (pseudo)scalar fields
when the dilaton/glueball field $G$ condenses, $G=G_0$. This feature is at the basis
of the Mexican-hat form of the (pseudo)scalar potential. We shall discuss the form of the potential in Sec. \ref{Ssec:mexican_hat}. 
Note that since $G_0 \propto N_c$, the contribution to the squared masses arising from this term goes as $N_c^0$, as expected (see \ref{App:Large-N}).
%%%%%%%%%%%%%%%%%%
\item We stress that terms of the type $G^{-2}Tr\left[  \Phi\Phi^{\dag
}\Phi\Phi^{\dag}\Phi\Phi^{\dag}\right]  $ or $G^{4}Tr\left[  \Phi\Phi^{\dag
}\right]  ^{-2}$ fulfill all the symmetries above, but are nonanalytic in $G=0.$. These terms are not considered in the eLSM, since a smoothness of the potential for
any finite value of the fields is a natural requirement for effective models.
Moreover, a vanishing $G$ corresponds to restoration of dilatation invariance,
while a vanishing $\Phi$ a restoration of chiral invariance, both of them
expected to take place at high $T$ and $\mu$. 
%%%%%%%%%%%%%%%%%%
\item When right-handed and left-handed homovector fields are introduced,
corresponding invariant terms can be introduced, e.g.:%
\begin{equation}
2h_{3}Tr\left[  \Phi R_{\mu}\Phi^{\dag}L^{\mu}\right]
\text{ ,}
\end{equation}
with $h_{3}$ being dimensionless and $\propto N_{c}^{-1}$ (dominant).
%%%%%%%%%%%%%%%%%%
\item The mass term for (axial-)vector homochiral fields fulfilling the required
symmetries is given by
\begin{equation}
\lambda_{GLR}G^{2}Tr\left[  R_{\mu}R^{\mu}+L_{\mu}L^{\mu}\right]
\text{ .}
\end{equation}
Here, $\lambda_{GLR}>0$ guarantees the standard (non tachyonic) sign for the
masses of (axial-)vector fields.
%%%%%%%%%%%%%%%%%
\item Besides traces, one may generate chirally, $\mathcal{P}$, and $\mathcal{C}$ invariant but $U(1)_{A}$
anomalous terms, such as
\begin{equation}
c_{2}\left(  \det\Phi - \det\Phi^{\dagger}\right)^2  \text{ .}%
\end{equation}
In fact, $\det\Phi$ transforms as%
\begin{equation}
\det\Phi\rightarrow\det U_{L}\Phi U_{R}^{\dagger}=\det U_{L}\det\Phi\det
U_{R}^{\dagger}\text{ ,}%
\end{equation}
which is invariant under $SU(3)_{L}\times SU(3)_{R}$ transformations since
$\det U_{L}=\det U_{R}=1,$ but not under $U(1)_{A}$.
Namely, the latter
amounts to $U_{L}=U_{R}^{\dagger}=e^{i\alpha}\mathbb{1}_{3\times 3},$ implying $\det
\Phi\rightarrow e^{6i\alpha}\det\Phi\neq\det\Phi.$ There are important
phenomenological consequence of such terms for the isoscalar-pseudoscalar mesons $\eta$ and $\eta^{\prime}.$ New anomalous
terms that break the axial anomaly and also involve fields with $J>0$ were
recently considered \cite{Giacosa:2017pos,Giacosa:2023fdz}. In general, anomalous terms also break dilatation invariance and involve dimensional coupling, which can be understood as an average over instanton ensembles \cite{Giacosa:2023fdz}. 
%%%%%%%%%%%%%%%%
\item There are terms that break explicitly $SU(3)_L\times SU(3)_R$ due to
nonzero and unequal quark masses. An example is provided by
\begin{equation}
Tr\left[  H\left(  \Phi+\Phi^{\dagger}\right)  \right]\;
\text{ ,}
\end{equation}
with $H=\mathrm{diag}\{h_{0N}/2,h_{0N}/2,h_{0S}/\sqrt{2}\}$, $h_{0N}\propto m_{n}=(m_u + m_d)/2$, and $h_{0S}\propto
m_{s}$. 
In fact,
\begin{equation}
Tr\left[  H\left(  \Phi+\Phi^{\dagger}\right)  \right]  \rightarrow Tr\left[
H\left(  U_{L}\Phi U_{R}^{\dagger}+U_{R}\Phi^{\dag}U_{L}^{\dagger}\right)
\right]
\end{equation}
is not chirally invariant. In the limit in which $H$ is proportional
to the identity, $SU(3)_{V}$ flavor symmetry is
still fulfilled. Yet, the breaking due to $m_{s}>m_{n}$ is non negligible. For
the choice of $H$ above, isospin symmetry $SU(2)_{V}$ is still retained (the isospin violating case is discussed in Section~\ref{Ssec:isospin_breaking}). 
\end{itemize}

%\bigskip

\subsection{The explicit form of the eLSM  Lagrangian}
\label{Ssec:elsm_Lagr}
%\textbf{generator with small t ; caption of Tables and Figures; G-parity; other tables masses and resonances PDG (tensor, etc.); matrix notation for identity matrix and null matrix}
The eLSM Lagrangian with the dilaton/glueball field and with scalar, pseudoscalar, vector, and axial-vector $\bar{q}q$ mesonic nonets (or, more shortly, with heteroscalars and homovectors), is constructed under the requirement of being --in the chiral limit-- invariant under global chiral transformations $SU(3)_L \times SU(3)_R$ and being -besides the dilatation potential and axial anomalous terms- invariant under dilatation transformations. Additional terms that break chiral symmetry and dilatation invariance due to the nonzero quark masses, and the special case of the axial anomaly $U(1)_A$ terms that fulfill $SU(3)_L \times SU(3)_R$ but break $U(1)_A$, are also introduced. The explicit form of the eLSM Lagrangian reads:
\begin{align}
\mathcal{L}  &  = \mathcal{L}_{\text{dil}} + \mathcal{L}_{\Phi} + \mathcal{L}_{U(1)_A} + \mathcal{L}_{LR} + \mathcal{L}_{\Phi LR}\text{ ,}\label{Eq:Lagr_tot}\\
& \text{with} \nonumber\\
\mathcal{L}_{\text{dil}} & = \frac{1}{2}(\partial_{\mu}G)^{2}-\frac{1}{4}\frac{m_{G}^{2}}{\Lambda_G^{2}}\left(  G^{4}\ln\frac{G^{2}}{\Lambda_G^{2}}-\frac{G^{4}}{4}\right)\text{ ,} \label{Eq:ELSM_Lagr_dil}\\ 
 \mathcal{L}_{\Phi} & = \mathop{\mathrm{Tr}}[(D_{\mu}\Phi)^{\dagger}(D_{\mu}\Phi)]-m_{0}^{2}\left(\frac{G}{G_{0}}\right)^{2}\mathop{\mathrm{Tr}}(\Phi^{\dagger}\Phi)-\lambda_{1}[\mathop{\mathrm{Tr}}(\Phi^{\dagger}\Phi)]^{2}-\lambda
_{2}\mathop{\mathrm{Tr}}(\Phi^{\dagger}\Phi)^{2} + \mathop{\mathrm{Tr}}[H(\Phi+\Phi^{\dagger})] \text{ ,} \label{Eq:ELSM_Lagr_Phi}\\
\mathcal{L}_{U(1)_A} &= c_2(\det\Phi-\det\Phi^{\dagger})^2 \text{ ,}
\label{Eq:ELSM_Lagr_U1A}\\
%\mathcal{L}_{U(1)_A} &= c_{1}(\det\Phi+\det\Phi^{\dagger}) + c_2(\det\Phi-\det\Phi^{\dagger})^2 + c_m(\det\Phi+\det\Phi^{\dagger})\mathop{\mathrm{Tr}}(\Phi^{\dagger} \Phi)  \text{ ,} \label{Eq:ELSM_Lagr_U1A}\\
\mathcal{L}_{LR} &= -\frac{1}{4}\mathop{\mathrm{Tr}}(L_{\mu\nu}^{2}+R_{\mu\nu}^{2})+\mathop{\mathrm{Tr}}\left[ \left(\left(\frac{G}{G_{0}}\right)^{2} + \Delta \right) \frac{m_{1}^{2}}{2}  (L_{\mu}^{2}+R_{\mu}^{2})\right] +i\frac{g_{2}}{2}(\mathop{\mathrm{Tr}}\{L_{\mu\nu}[L^{\mu},L^{\nu}]\}+\mathop{\mathrm{Tr}}\{R_{\mu\nu}[R^{\mu},R^{\nu}]\}) \nonumber\\
& + g_{3}[\mathop{\mathrm{Tr}}(L_{\mu}L_{\nu}L^{\mu}L^{\nu})+\mathop{\mathrm{Tr}}(R_{\mu}R_{\nu}R^{\mu}R^{\nu})]+g_{4}[\mathop{\mathrm{Tr}}\left(  L_{\mu}L^{\mu}L_{\nu}L^{\nu}\right)
+\mathop{\mathrm{Tr}}\left(  R_{\mu}R^{\mu}R_{\nu}R^{\nu}\right)]{\nonumber}\\
& + g_{5}\mathop{\mathrm{Tr}}\left(  L_{\mu}L^{\mu}\right)\mathop{\mathrm{Tr}}\left(  R_{\nu}R^{\nu}\right) + g_{6} [\mathop{\mathrm{Tr}}(L_{\mu}L^{\mu})\,\mathop{\mathrm{Tr}}(L_{\nu}L^{\nu})+\mathop{\mathrm{Tr}}(R_{\mu}R^{\mu})\,\mathop{\mathrm{Tr}}(R_{\nu}R^{\nu})]\text{ ,} \label{Eq:ELSM_Lagr_LR}\\
\mathcal{L}_{\Phi LR} & = \frac{h_{1}}{2}\mathop{\mathrm{Tr}}(\Phi^{\dagger}\Phi)\mathop{\mathrm{Tr}}(L_{\mu}^{2} + R_{\mu}^{2})+h_{2}\mathop{\mathrm{Tr}}[\vert L_{\mu}\Phi \vert ^{2}+\vert \Phi R_{\mu} \vert ^{2}]+2h_{3} \mathop{\mathrm{Tr}}(L_{\mu}\Phi R^{\mu}\Phi^{\dagger})\text{ ,}\label{Eq:ELSM_Lagr_PhiLR}%
\end{align}
where
\begin{align}
D^{\mu}\Phi &  \equiv\partial^{\mu}\Phi-ig_{1}(L^{\mu}\Phi-\Phi R^{\mu
})-ieA^{\mu}[t_{3},\Phi]\;,\nonumber\\
L^{\mu\nu}  &  \equiv\partial^{\mu}L^{\nu}-ieA^{\mu}[t_{3},L^{\nu}]-\left\{
\partial^{\nu}L^{\mu}-ieA^{\nu}[t_{3},L^{\mu}]\right\}  \;\text{,}\nonumber\\
R^{\mu\nu}  &  \equiv\partial^{\mu}R^{\nu}-ieA^{\mu}[t_{3},R^{\nu}]-\left\{
\partial^{\nu}R^{\mu}-ieA^{\nu}[t_{3},R^{\mu}]\right\}  \;\text{,}\nonumber
\end{align}
and
\begin{align}
\label{eq:expl_sym_br_epsilon}
H  &= H_{0}t_{0}+H_{8}t_{8}=\left(
\begin{array}
[c]{ccc}%
\frac{h_{0N}}{2} & 0 & 0\\
0 & \frac{h_{0N}}{2} & 0\\
0 & 0 & \frac{h_{0S}}{\sqrt{2}}%
\end{array}
\right)  \; \text{ , } \\
\Delta &= \Delta_{0}t_{0}+\Delta_{8}t_{8}=\left(
\begin{array}
[c]{ccc}%
\frac{\tilde{\delta}_{N}}{2} & 0 & 0\\
0 & \frac{\tilde{\delta}_{N}}{2} & 0\\
0 & 0 & \frac{\tilde{\delta}_{S}}{\sqrt{2}}%
\end{array}
\right)  \equiv\left(
\begin{array}
[c]{ccc}%
\delta_{N} & 0 & 0\\
0 & \delta_{N} & 0\\
0 & 0 & \delta_{S}%
\end{array}
\right)  \text{ .} \label{eq:expl_sym_br_delta}%
\end{align}
The latter are the constant external fields for the (pseudo)scalar and (axial)vector fields that describe the effect of the nonzero quark masses and thus break chiral symmetry and also flavor symmetry (yet, in the form above  they still satisfy isospin symmetry).
%\footnote{The isospin violating case is discussed in Section~\ref{Ssec:isospin_breaking}}. 
The $\Phi$, $L^{\mu}$ and $R^{\mu}$ nonets are defined in Eqs.~\eqref{Eq:J0_sc_ps_nonet}, \eqref{Eq:J1_nonet_L}, \eqref{Eq:J1_nonet_R}, respectively, $A^{\mu}$ is the electromagnetic field and $t^3$ is the third generator of $U(3)$ (see \ref{Ssec:UN_group}).  

Let us go thorough the different parts of the Lagrangian and their most important properties. 

(i) Eq.~\eqref{Eq:ELSM_Lagr_dil} is the dilaton Lagrangian, which was already discussed in Section~\ref{Ssec:dilaton} and can be related to the scalar glueball.

(ii) Eq.~\eqref{Eq:ELSM_Lagr_Phi} is the chirally symmetric (pseudo)scalar part, that is all terms, except the last one, are invariant under chiral transformations given in Table \ref{tab:chiral-transformations}. The first term is the kinetic term, with a derivative that also includes the $L^{\mu}/R^{\mu}$ vector and the $A^{\mu}$ electromagnetic fields; it is similar to the covariant derivatives in local gauge theories, except that here there are only global transformations. The second term reduces to a mass term through the condensation of the dilaton field $G=G_0$, the third and fourth terms are (pseudo)scalar self-interactions, while the last term is the explicit symmetry breaking term with $H \propto \text{diag}(m_n,m_n,m_s)$. Note that upon setting $G=G_0$, the dilaton field decouples from the rest of the Lagrangian.

(iii) Eq. \eqref{Eq:ELSM_Lagr_U1A} is an $U(1)_A$ anomalous term, which represents one possible implementation, e.g. Ref. \cite{Parganlija:2012fy}. However, other anomalous terms are also possible, see Section~\ref{Ssec:U1A_breaking}.

(iv) Eq.~\eqref{Eq:ELSM_Lagr_LR} includes the kinetic, the mass, the self-interaction and explicit breaking terms for the left- and right-handed vector fields $L^{\mu}$ and $R^{\mu}$. 

(v) Finally, the term of  Eq.~\eqref{Eq:ELSM_Lagr_PhiLR} consists of the interaction terms between the (pseudo)scalars and (axial)vectors. 

After fixing the Lagrangian, the standard procedure is to assume non-zero condensates -- that is spontaneous symmetry breaking (SSB) -- for certain fields (besides, of course, the field $G$, that condenses due to the dilaton potential). In the next Section we discuss which conditions the parameters must fulfill to have SSB. Once it takes place, the fields that condense are those with the quantum numbers of the vacuum, i.e. the scalar-isoscalar states $\sigma_N$, $\sigma_S$ (or equivalently $\sigma_0$, $\sigma_8$) and $\sigma_3$. The last one, $\sigma_3$, is responsible for isospin breaking and is usually neglected since its effect is much smaller compared to the others (for more details on isospin breaking\footnote{It worth noting that other condensates are also considered in the literature, like the pion \cite{Baym:1973zk} and kaon condensates \cite{Politzer:1991ev}, or vector condensates \cite{Sannino:2002wp}.} see Sec.~\ref{Ssec:isospin_breaking}). Next, the $\sigma_{N/S}$ fields are shifted by their vacuum expectation values:
\begin{equation}\label{eq:scalar-shift}
    \sigma_{N/S}\to \sigma_{N/S} + \phi_{N/S} \text{ , } \quad \text{with} \quad\phi_{N/S} \equiv \langle \sigma_{N/S}\rangle \text{ .}
\end{equation}
 After the shifts, the quadratic terms generally give the tree-level masses of all particles, i.e. the spectrum, and the decay widths can be calculated. However, due to the inclusion of the (axial-)vector meson fields in the covariant derivative of the (pseudo)scalar fields, a complication arises, namely there will be quadratic mixing terms between the members of the (pseudo)scalar and (axial)vector nonets. For example, one such possible term is of the form
\begin{equation}
    \sim g_1 a_1^{\mu \pm, 0}\partial_{\mu}\pi^{\mp, 0}
\end{equation}
This mixing can be eliminated by redefining the axial vector field,
\begin{equation}
    a_1^{\mu \pm, 0} \to a_1^{\mu \pm, 0} + Z_{\pi}w_{a_1}\partial^{\mu} \pi^{\pm, 0}.
\end{equation}   
Requiring the disappearance of the mixing term and also the canonical form of the kinetic term for the pion field -- i.e. $1/2\partial_{\mu}\pi^{\pm,0} \partial^{\mu}\pi^{\mp,0}$ -- determines the  coefficients $w_{a_1}$ and $Z_{\pi}$. This procedure with explicit formulas for the masses and decay widths can be found in Ref. \cite{Parganlija:2010fz} for the two flavor case, and in Ref. \cite{Parganlija:2012fy} for the three flavor case. 

%In each othe $N$-$S$ (or $0$-$8$) sector, which was already discussed in the previous sections. It is worth noting that the situation in the scalar sector is the most problematic, where the identification of the physical states -- due to the broad nature and various mixing of these states -- with the fields in the model is far from trivial \cite{Parganlija:2012fy}. 

\subsection{The Mexican hat potential}
\label{Ssec:mexican_hat}

To discuss the shape of the potential for the (pseudo)scalar sector and the emergence of chiral condensates, it is convenient to rewrite part of the eLSM Lagrangian as follows
\begin{equation}
\mathcal{L}_{dil} + \mathcal{L}_{\Phi} = \mathcal{L}_{G \Phi} =\mathcal{L}_{G \Phi}^{\text{(kin)}}-V_{G\Phi} \text{ ,}
\end{equation}
where $V_{G\Phi}\equiv V_{G\Phi}(G,\sigma_{N},\sigma_{S},...,\mathbf{\pi
},...)$ is the potential that depends on all the 19 (pseudo)scalar fields.
Among all of them, only those  having the same quantum numbers of the vacuum
$J^{\mathcal{P}\mathcal{C}}=0^{++}$ can condensate: $G,\sigma_{N},\sigma_{S}.$ The explicit form
of the potential by retaining only these fields is given by:
\begin{equation}
V_{G\Phi}(G,\sigma_{N},\sigma_{S},0,...,0)=\frac{\lambda_{G}}{4}G^{4}\left(
\ln\frac{G}{\Lambda_{G}}-\frac{1}{4}\right)  +\frac{\lambda_{G\Phi}G^{2}}{2}\left(
\sigma_{N}^{2}+\sigma_{S}^{2}\right)  +\frac{\lambda_{2}}{8}\left(  \sigma_{N}%
^{4}+2\sigma_{S}^{4}\right) +\frac{\lambda_{1}}{4}\left(  \sigma_{N}^{2}+\sigma
_{S}^{2}\right)  ^{2}-h_{N}\sigma_{N}-h_{S}\sigma_{S}%
\label{vgphi}
\end{equation}
The minimum is found by setting the derivatives to zero:
\begin{align}
\partial_{G}V_{G\Phi}  & =\lambda_{G}G^{3}\ln\frac{G}{\Lambda_{G}}%
+\lambda_{G\Phi}G\left(  \sigma_{N}^{2}+\sigma_{S}^{2}\right)  =0\text{ , }\\
\partial_{\sigma_{N}}V_{G\Phi}  & =\lambda_1\sigma_N(\sigma_N^2+\sigma_S^2)+\frac{\lambda_2\sigma_N^3}{2}+\lambda_{G\phi} G^2\sigma_N-h_N=0\text{ , } \label{phin}\\
\partial_{\sigma_{S}}V_{G\Phi}  & =\lambda_1\sigma_S(\sigma_N^2+\sigma_S^2)+\lambda_2\sigma_S^3+\lambda_{G\phi} G^2\sigma_S-h_S=0\text{ .} \label{phis}%
\end{align}
We also refer to Refs. \cite{Giacosa:2015wcz,Giacosa:2024scx}
for a pedagogical introduction on the subject, including the search for minima (with and without dilaton) and the simplified one-flavor case. 

As already mentioned, the field $G$ condenses because of the dilaton potential, see Sec. \ref{Ssec:dilaton}. Fixing the dilaton as s $G=G_0$, it is instructive to write down (up to a constant) the potential along the $\sigma_{N}$ and $\pi^{0}$ directions:
\begin{equation}
V_{G\Phi}(G=G_{0}=\Lambda_G,\sigma_{N},0,0,.,\pi^{0},0,...)=\frac{m_{0}^{2}}{2}\left(
\sigma_{N}^{2}+(\pi^{0})^{2}\right)  +\frac{\lambda_{2}}{2}\left(  \sigma_{N}%
^{2}+(\pi^{0})^{2}\right)  -h_{N}\sigma_{N} \label{vsigmapi}
\text{ .}
\end{equation}

We distinguish different scenarios. 
If $\lambda_{G\Phi}>0$ and in the limit $h_{N,S}\rightarrow0$, only the dilaton field $G$
condenses to $G = G_0 = \Lambda_G$, while the v.e.v. of $\sigma_{N}$ and $\sigma_{S}$ vanish:
\begin{equation}
G_{0}=\Lambda_{G}\text{ , }\sigma_{N}=\text{ }\phi_{N}=0\text{ , }\sigma
_{S}=\text{ }\phi_{S}=0 \text{ .}
\end{equation}

In this case, the masses read:
\begin{equation}
m_{\sigma_{N}}^{2}=m_{\pi}^{2}=m_{0}^{2}=\lambda_{G\Phi}G_{0}^{2}%
=\lambda_{G\Phi}\Lambda_{G}^{2}>0
\text{ ,}
\end{equation}
thus the scalar-isoscalar field $\sigma_{N}$ and its chiral partner,
the pseudoscalar $\pi^{0},$ carry the same mass: chiral symmetry is evident.
The schematic form of the potential is depicted in Fig.~\ref{fig:parabola}. As well known, this case is not realized in Nature, since chiral partners are \textit{not} degenerate in mass. Note that allowing for small but nonzero $h_{N,S}$ modifies slightly the potential, and $\phi_{N},\phi_{S}$ are small but nonzero with $\phi_{N,S} \propto m_{n,s}$, yet the investigation of this case is not relevant for our purposes.
\begin{figure}[ht!]
    \centering
    \begin{tikzpicture}
\draw[black, line width = 0.50mm]   plot[smooth,domain=-2:2] (\x, {4+\x*\x});
\draw [thick,red,thick] (-1.2,5.5) to [out=-90,in=-90] (1.2,5.5);
\draw [dashed,red,thick] (-1.2,5.5) to [out=90,in=90] (1.2,5.5);
\draw [thick,red,thick] (-1.5,6.5) to [out=-90,in=-90] (1.5,6.5);
\draw [dashed,red,thick] (-1.5,6.5) to [out=90,in=90] (1.5,6.5);
\draw [thick,red,thick] (-1.8,7.5) to [out=-90,in=-90] (1.8,7.5);
\draw [dashed,red,thick] (-1.8,7.5) to [out=90,in=90] (1.8,7.5);
 \draw[->,black,thick] (0,4) -- (0,9) node[above] {$V(G_0,\sigma_N,\pi^0)$};
 \draw[->,black,thick] (0,4) -- (4,4)  node[below] {$\sigma_N$};
 % \draw[->,black,thick] (0,4) -- (4,2) node[below] {$\sigma_N$};

 \draw[->,black,thick] (-1.3,3) -- (4.5,7.5) node[below] {$\pi^0$};
  
  %\draw[->,black,thick] (0.,4) -- (2.5,5.5) node[below] {$\pi^0$};
%\draw[->,black,thick] (-3,-0.5) -- (3,-0.5) node[below] {$\sigma_N$};
  %\draw[->,black,thick] (0,4) -- (-1.5,3) node[below] { $\pi^0$ };

   \draw[black,thick] (0,4) -- (-1.5,4);
   %\draw[black,thick] (0,4) -- (-1.,4.5);
  \draw[black,thick] (0,4) -- (0,3);
\end{tikzpicture}
        \caption{The potential in Eq. \eqref{vsigmapi} is depicted for $\lambda_{G\Phi}>0$ in the chiral limit ($h_N = 0$). A unique minimum in the origin is present. %\tcred{PK: I've changed the $\sigma_N$ axis, I think it is more natural this way ...} 
        }
    \label{fig:parabola}
\end{figure}
%\bigskip

Next, we turn to the physically relevant case $\lambda_{G\Phi}<0,$ for which all
condensates $G_{0},\phi_{N},\phi_{S}$ do not vanish, even not in the limit
$h_{N,S}\rightarrow0^{+}$. The potential along $\sigma_{N}$ and $\pi^{0}$ has
the same form as in\ Eq. (\ref{vsigmapi}), but now with
\begin{equation}
m_{0}^{2}=\lambda_{G\Phi}G_{0}^{2}<0 \text{ .}
\end{equation}
Upon expanding around the minimum, the masses of the chiral partner of the pion reads (neglecting the large-$N_c$ suppressed $\lambda_1$-term):
$m_{\sigma_{N}}^{2}=m_0^2+3\lambda_2 \sigma_N^2 /3 = -2m_0^2+3m_{\pi}^{2}$ with
$m_{\pi}^{2}=h_{N}/\phi_{N}\rightarrow0^{+}$ for $h_{N}\rightarrow0^{+}$
(Goldstone boson). 
Note, $m_{\sigma_{N}}^{2}-m_{\pi}^{2}=-2m_0^2 +2m_{\pi}^2 = \lambda_2  \phi_N^2 >0$  shows that the mass difference between chiral partners is proportional to the chiral condensate $\phi_N$, which in turn is proportional to the gluon condensate $G_0 \sim \Lambda_G$. This is a recurring pattern in the eLSM that appears also in much more lengthy mass expressions. 

The form of the potential in the $\{\sigma_N,\pi^0 \}$-subspace is depicted in Fig \ref{Mexican-hat}
for the case $h_{N}\rightarrow0^{+}$ (left panel), in which the form of the potential
is the well-known Mexican-hat shape with a circle of
minima, and for $h_{N}>0$ (right panel), where an additional titling due to nonzero quark light masses takes place: a unique minimum for $\sigma_N=\phi_N>0$ and $\pi^0 = 0$ is present.
Note that as long as $h_{N}>0$, no matter how small, the
potential has a unique minimum, since it is always tilted, even if very
little. In this case the pion field can never condense. Only for $h_{N}=0$
(strictly zero) multiple minima are actually present.
\begin{figure}[ht!]
\begin{center}
\begin{tikzpicture}[domain=-1.3101:1.31,samples=150,scale=1]

\draw[->,black,thick] (0,-3) -- (0,2) node[above] {$V(G_0,\sigma_N,\pi^0)$};
\draw[color=black,scale=1.9pt]
plot (\x,{(\x)^4-(\x)^2});
\node at (-1.8,-0.9) { };
\node at (1.6,-0.9) {};
\draw [black,thick] (-1.4,-0.5) to [out=-90,in=-90] (1.4,-0.5);
\draw [dashed,black,thick] (-1.4,-0.5) to [out=90,in=90] (1.4,-0.5);
\draw[->,black,thick] (0.,-0.5) -- (2.5,1.) node[below] {$\pi^0$};
\draw[->,black,thick] (-3,-0.5) -- (3,-0.5) node[below] {$\sigma_N$};

\draw[black,thick] (-2.5,-2.) -- (0.,-0.5);
\end{tikzpicture}
\begin{tikzpicture}[domain=-1.3101:1.31,samples=100,scale=1]
\draw [black,thick] (-1.4,-0.2) to [out=-90,in=-90] (1.4,-0.7);
\draw [dashed,black,thick] (-1.4,-0.2) to [out=90,in=90] (1.4,-0.7);
\draw[->] (-3,-0.5) -- (3,-0.5) node[below] {$\sigma_N$};
\draw[->] (0,-3) -- (0,2) node[above] {$V(G_0,\sigma_N,\pi^0)$};
\draw[color=black,scale=1.9pt]
plot (\x,{(\x)^4-(\x+0.08)^2});
\node at (1.6,-0.9) { };
\draw[->,black,thick] (0.0,-0.5) -- (2.5,1.) node[below] {$\pi^0$};
\draw[black,thick] (-2.5,-2.) -- (0.,-0.5);
\end{tikzpicture}
\end{center}
\caption{The potential in Eq. \eqref{vsigmapi} is depicted for $\lambda_{G\Phi} < 0$. Left: in the chiral limit $h_N=0$, the Mexican-hat potential with minima along the chiral circle  is realized. Right: the Mexican-hat potential in the presence of nonzero light quark mass, $h_N>0$, is tilted and has a unique minimum.}
\label{Mexican-hat}
 \end{figure}

\subsection{Parameterizations and tree-level masses}
\label{Ssec:parameterization}

To compute predictions from the model, the parameters of the eLSM Lagrangian of Sec. \ref{Ssec:elsm_Lagr}  must be determined. This can be done with the following procedure, called parameterization, which consists of calculating physical quantities, such as tree-level masses and decay widths, from the model and comparing them with the corresponding experimental values taken from the PDG \cite{Workman:2022ynf} using a $\chi^2$ minimization. The $\chi^2$ is defined as
\begin{equation}
     \chi^2(x_1,\dots,x_N)=\sum_{i=1}^{M}\left[\frac{Q_i(x_1,\dots,x_N)-Q_i^{\text{exp}}}{\delta
      Q_i}\right]^2,
\end{equation}
where $x_1,\dots,x_N$ are the parameters of the Lagrangian, $Q_1(x_1,\dots,x_N), \ldots, Q_M(x_1,\dots,x_N)$ are the physical quantities, such as masses, decay widths, etc., calculated from the model, $Q_1^{\text{exp}}, \ldots, Q_M^{\text{exp}}$ are the corresponding experimental values, while $\delta Q_1, \ldots, \delta Q_M$ are the corresponding uncertainties. In principle, these uncertainties can be chosen as the experimental errors listed in the PDG, but this would usually lead to fits that are too biased towards quantities known very precisely (e.g. the $\eta(547)$ mass). Thus, the following prescription is used: $\delta Q_i = \max\{5\%, Q_i^{\text{exp}}\}$, according to which a small experimental error is replaced by $5 \%$ of the experimental value. 
%and in the case of very uncertain scalars sometimes $10\%$ or even $20\%$ is used instead of $5\%$. 
This procedure is also motivated by the fact that isospin breaking processes (typically a few percent effect) are here neglected.

Next, one searches for the absolute minimum of the $\chi^2$ in this multidimensional parameter space, which is in general, not an easy task. Usually, due to the highly nonlinear nature of the equations, there are many local minima (hundreds). For this task a MINUIT code can be used \cite{James:1975dr}. One of the advantages of this parameterization is that it works for both $M < N$ and $M \geq N$, but if the number of conditions $M$ is smaller than the number of unknown parameters $N$, then the resulting fit will not be restrictive. In practice, this would mean finding many very different parameter sets with very similar $\chi^2$ values and with large variances of the fitted values. Note that this procedure is not necessary in simpler versions of the model (e.g. without the vector mesons), or in cases where vacuum fluctuations are neglected. In these cases it may be even possible to solve the parameterization equations analytically.      

In the following we make an important simplification, namely we decouple the dilaton field by setting $G=G_{0}$ in Eq.~\eqref{Eq:Lagr_tot}. Accordingly, we show the tree-level masses and parameterization results in the eLSM presented in Refs.~\cite{Parganlija:2012fy,Parganlija:2011hvn}. The Table~\ref{Tab:s_ps_masses_PDG} shows the assignments of the fields entering the model (Eq.~\eqref{Eq:J0_sc_ps_nonet}) and their PDG counterparts, along with their quark contents and experimental central values for their masses.
\begin{table}[!ht]
\centering
\begin{tabular}
[c]{|c|c|c|c|c|c|}\hline
Field & PDG & Quark content & $I^{\mathcal{G}}$ & $J^{\mathcal{P}\mathcal{C}}$ & Mass (MeV)\\\hline
$\pi^{+},\pi^{-},\pi^{0}$ & $\pi$ & $u\bar{d},d\bar{u},\frac{u\bar{u}-d\bar
{d}}{\sqrt{2}}$ & $1^{-}$ & $0^{-+}$ & $138.04$ \\\hline
$K^{+},K^{-},K^{0},\bar{K}^{0}$ & $K$ & $u\bar{s},s\bar{u},d\bar{s},s\bar{d}$
& $1/2$ & $0^{-}$ & $495.64$\\\hline
$\text{Eigenv}^{-}(\eta_{N}, \eta_{S})$ & $\eta$ &  $\frac{u\bar{u}+d\bar{d}}{\sqrt{2}}$, $s\bar{s}$ & $0^{+}$ & $0^{-+}$ & $547.86$\\\hline
%$\eta_{N}a+\eta_{S}b$ & $\eta$ & $\frac{u\bar{u}+d\bar{d}}{\sqrt{2}}a+s\bar
%{s}b$ & $0$ & $0^{-+}$ & $547.86$\\\hline
 $\text{Eigenv}^{+}(\eta_{N}, \eta_{S})$  & $\eta^{\prime}(958)$ &  $\frac{u\bar{u}+d\bar{d}}{\sqrt{2}}$, $s\bar{s}$ & $0^{+}$ & $0^{-+}$ & $957.78$\\\hline
%$-\eta_{N}a+\eta_{S}b$ & $\eta^{\prime}(958)$ & $\frac{u\bar{u}+d\bar{d}%
%}{\sqrt{2}}a+s\bar{s}b$ & $0$ & $0^{-+}$ & $957.78$\\\hline
$a_{0}^{+},a_{0}^{-},a_{0}^{0}$ & $a_{0}(1450)$ & $u\bar{d},d\bar{u}%
,\frac{u\bar{u}-d\bar{d}}{\sqrt{2}}$ & $1^-$ & $0^{++}$ & $1474$\\\hline
$K_{S}^{+},K_{S}^{-},K_{S}^{0},\bar{K}_{S}^{0}$ & $K_{0}^{\ast}(1430)$ &
$u\bar{s},s\bar{u},d\bar{s},s\bar{d}$ & $1/2$ & $0^{+}$ & $1425$\\\hline
 $\text{Eigenv}^{-}(\sigma_{N}, \sigma_{S})$ & $f_{0}(1370)$ &  $\frac{u\bar{u}+d\bar{d}}{\sqrt{2}}$, $s\bar{s}$ & $0^+$ & $0^{++}$ & $1350$\\\hline
%$\sigma_{N}$ & $f_{0}(1370)$ & $\frac{u\bar{u}+d\bar{d}}{\sqrt{2}}$ & $0$ &
%$0^{++}$ & $1350$\\\hline
 $\text{Eigenv}^{+}(\sigma_{N}, \sigma_{S})$ & $f_{0}(1500)$ & $\frac{u\bar{u}+d\bar{d}}{\sqrt{2}}$, $s\bar{s}$ & $0^+$ & $0^{++}$ & $1505$\\\hline
%$\sigma_{S}$ & $f_{0}(1500)$ & $s\bar{s}$ & $0$ & $0^{++}$ & $1504$\\\hline
$\rho^{+},\rho^{-},\rho^{0}$ & $\rho(770)$ & $u\bar{d},d\bar{u},\frac{u\bar
{u}-d\bar{d}}{\sqrt{2}}$ & $1^+$ & $1^{--}$ & $775.16$\\\hline
$K^{\ast+},K^{\ast-},K^{\ast0},\bar{K}^{\ast0}$ & $K^{\ast}(892)$ & $u\bar
{s},s\bar{u},d\bar{s},s\bar{d}$ & $1/2$ & $1^{-}$ & $895.51$\\\hline
$\omega_{N}$ & $\omega(782)$ & $\frac{u\bar{u}+d\bar{d}}{\sqrt{2}}$ & $0^-$ &
$1^{--}$ & $782.66$\\\hline
$\omega_{S}$ & $\phi(1020)$ & $s\bar{s}$ & $0^-$ & $1^{--}$ & $1019.46$\\\hline
$a_{1}^{+},a_{1}^{-},a_{1}^{0}$ & $a_{1}(1230)$ & $u\bar{d},d\bar{u}%
,\frac{u\bar{u}-d\bar{d}}{\sqrt{2}}$ & $1^-$ & $1^{++}$ & $1230$\\\hline
$K_{1}^{+},K_{1}^{-},K_{1}^{0},\bar{K}_{1}^{0}$ & $K_{1}(1270)$ & $u\bar
{s},s\bar{u},d\bar{s},s\bar{d}$ & $1/2$ & $1^{+}$ & $1253$\\\hline
$f_{1N}$ & $f_{1}(1285)$ & $\frac{u\bar{u}+d\bar{d}}{\sqrt{2}}$ & $0^+$ &
$1^{++}$ & $1281.9$\\\hline
$f_{1S}$ & $f_{1}(1420)$ & $s\bar{s}$ & $0^+$ & $1^{++}$ & $1426.3$\\\hline
\end{tabular}
\caption{eLSM fields, PDG correspondence, quark content, quantum numbers and PDG central mass values from \cite{Workman:2022ynf}. In the case of charged particles isospin-averaged values are used. Here $\text{Eigenv}^{\pm}$ stands for the larger/smaller eigenvalue of the non-strange and strange states.}
\label{Tab:s_ps_masses_PDG}
\end{table}
In both the scalar and pseudoscalar sectors there is a mixing between the non-strange and strange (or similarly the $0$ and $8$) states, which can be resolved by $2 \times 2$ orthogonal transformations. The tree-level mass expressions calculated in the model are listed in Table~\ref{Tab:mass-spin0_eLSM} for the scalar and pseudoscalar (spin-0) mesons and in Table~\ref{Tab:mass-spin1_eLSM} for the (axial-)vector mesons. 
\begin{table}[ht!]
%[ptb]
\centering
\renewcommand{\arraystretch}{1.3} 
\begin{tabular}{|c|c|}
\hline
Mass squares  & Analytical expressions  \\ \hline\hline	
$m_{\pi}^{2}$ &  $Z_{\pi}^{2}\left[  m_{0}^{2}+\left(  \lambda_{1}%
+\frac{\lambda_{2}}{2}\right)  \phi_{N}^{2}+\lambda_{1}\phi_{S}^{2}\right]
\equiv\frac{Z_{\pi}^{2}h_{0N}}{\phi_{N}}$ \\
\hline
$m_{K}^{2}$ &  $Z_{K}^{2}\left[  m_{0}^{2}+\left(  \lambda_{1}+\frac
{\lambda_{2}}{2}\right)  \phi_{N}^{2}-\frac{\lambda_{2}}{\sqrt{2}}\phi_{N}%
\phi_{S}+\left(  \lambda_{1}+\lambda_{2}\right)  \phi_{S}^{2}\right]$ \\
\hline
$m_{\eta_{N}}^{2}$ &  $Z_{\pi}^{2}\left[  m_{0}^{2}+\left(  \lambda_{1}%
+\frac{\lambda_{2}}{2}\right)  \phi_{N}^{2}+\lambda_{1}\phi_{S}^{2}%
+c_{2}\,\phi_{N}^{2}\phi_{S}^{2}\right]\equiv Z_{\pi}^{2}\left(  \frac{h_{0N}}{\phi_{N}}+c_{2}\,\phi_{N}^{2}%
\phi_{S}^{2}\right)$\\
\hline
$m_{\eta_{S}}^{2}$ &  $Z_{\eta_{S}}^{2}\left[  m_{0}^{2}+\lambda_{1}\phi
_{N}^{2}+\left(  \lambda_{1}+\lambda_{2}\right)  \phi_{S}^{2}+\frac{c_{2}}%
{4}\phi_{N}^{4}\right]\equiv Z_{\eta_{S}}^{2}\left(  \frac{h_{0S}}{\phi_{S}}+\frac{c_{2}}{4}%
\phi_{N}^{4}\right)$\\
\hline
$m_{\eta_{NS}}^2$ &  $Z_{\eta_N} Z_{\eta_S}\frac{c_2}{2} \phi_{N}^3\phi_{S}$\\
\hline
$m_{a_{0}}^{2}$ &  $m_{0}^{2}+\left(  \lambda_{1}+\frac{3}{2}\lambda
_{2}\right)  \phi_{N}^{2}+\lambda_{1}\phi_{S}^{2}$  \\
\hline
$m_{K_{0}^{\star}}^{2}$  & $Z_{K_{0}^{\star}}^{2}\left[  m_{0}^{2}+\left(
\lambda_{1}+\frac{\lambda_{2}}{2}\right)  \phi_{N}^{2}+\frac{\lambda_{2}%
}{\sqrt{2}}\phi_{N}\phi_{S}+\left(  \lambda_{1}+\lambda_{2}\right)  \phi
_{S}^{2}\right]$ \\
\hline
$m_{\sigma_{N}}^{2}$  &  $m_{0}^{2}+3\left(  \lambda_{1}+\frac{\lambda_{2}}%
{2}\right)  \phi_{N}^{2}+\lambda_{1}\phi_{S}^{2}$  \\\hline
$m_{\sigma_{S}}^{2}$  & $m_{0}^{2}+\lambda_{1}\phi_{N}^{2}+3\left(  \lambda
_{1}+\lambda_{2}\right)  \phi_{S}^{2}$ \\\hline
 $m^2_{\sigma_{NS}}$ & $2\lambda_1 \phi_{N} \phi_{S}$\\\hline
\end{tabular}
\caption{Mass expressions of spin-0 mesons (scalars and pseudoscalars) within the eLSM.}
\label{Tab:mass-spin0_eLSM}
\end{table}
\begin{table}[ht!]
\centering
\renewcommand{\arraystretch}{1.3} 
\begin{tabular}{|c|c|}
\hline
Mass squares  & Analytical expressions \\ \hline\hline
 $m_{\rho}^{2}$  &  $m_{1}^{2}+\frac{1}{2}(h_{1}+h_{2}+h_{3})\phi_{N}^{2}%
+\frac{h_{1}}{2}\phi_{S}^{2}+2\delta_{N}$ \\ \hline
$m_{K^{\star}}^{2}$  &  $m_{1}^{2}+\frac{1}{4}\left(  g_{1}^{2}+2h_{1}%
+h_{2}\right)  \phi_{N}^{2}  +\frac{1}{\sqrt{2}}\phi_{N}\phi_{S}(h_{3}-g_{1}^{2})+\frac{1}{2}(g_{1}%
^{2}+h_{1}+h_{2})\phi_{S}^{2}+\delta_{N}+\delta_{S}$ \\\hline
 $m_{\omega_{N}}^{2}$  &  $m_{1}^{2}+\frac{1}{2}(h_{1}+h_{2}+h_{3})\phi_{N}^{2}%
+\frac{h_{1}}{2}\phi_{S}^{2}+2\delta_{N}=m_{\rho}^{2}$ \\ \hline
$m_{\omega_{S}}^{2}$  &  $m_{1}^{2}+\frac{h_{1}}{2}\phi_{N}^{2}+\left(
\frac{h_{1}}{2}+h_{2}+h_{3}\right)  \phi_{S}^{2}+2\delta_{S}$\\\hline
$m_{a_{1}}^{2}$  &  $m_{1}^{2}+\frac{1}{2}(2g_{1}^{2}+h_{1}+h_{2}-h_{3})\phi
_{N}^{2}+\frac{h_{1}}{2}\phi_{S}^{2}+2\delta_{N}$\\\hline
$m_{K_{1}}^{2}$  &  $m_{1}^{2}+\frac{1}{4}\left(  g_{1}^{2}+2h_{1}+h_{2}\right)
\phi_{N}^{2} -\frac{1}{\sqrt{2}}\phi_{N}\phi_{S}(h_{3}-g_{1}^{2})+\frac{1}{2}\left(
g_{1}^{2}+h_{1}+h_{2}\right)  \phi_{S}^{2}+\delta_{N}+\delta_{S}$ \\\hline
$m_{f_{1N}}^{2}$  &  $m_{1}^{2}+\frac{1}{2}(2g_{1}^{2}+h_{1}+h_{2}-h_{3})\phi
_{N}^{2}+\frac{h_{1}}{2}\phi_{S}^{2}+2\delta_{N}=m_{a_{1}}^{2}$\\\hline
$m_{f_{1S}}^{2}$  &  $m_{1}^{2}+\frac{h_{1}}{2}\phi_{N}^{2}+\left(  2g_{1}%
^{2}+\frac{h_{1}}{2}+h_{2}-h_{3}\right)  \phi_{S}^{2}+2\delta_{S}$ \\\hline
	\end{tabular}%
\caption{Mass expressions of spin-1 mesons (vectors and axial vectors) within the eLSM.}\label{Tab:mass-spin1_eLSM}
\end{table}%
%----------------------------------------------------------------------------------
In the case of (pseudo)scalars, as mentioned before, due to the mixing with (axial-)vectors, so-called $Z$-factors appear in the mass expressions (for more details see Ref. \cite{Parganlija:2012fy}), which are given by 
\begin{align}
Z_{\pi}  &  =Z_{\eta_{N}}=\frac{m_{a_{1}}}{\sqrt{m_{a_{1}}^{2}-g_{1}^{2}%
\phi_{N}^{2}}}\text{ ,} & Z_{K}  &  =\frac{2m_{K_{1}}}{\sqrt{4m_{K_{1}}%
^{2}-g_{1}^{2}(\phi_{N}+\sqrt{2}\phi_{S})^{2}}}\text{ ,}\label{Z_pi}\\
Z_{\eta_{S}}  &  =\frac{m_{f_{1S}}}{\sqrt{m_{f_{1S}}^{2}-2g_{1}^{2}\phi
_{S}^{2}}}\text{ ,} & Z_{K_{0}^{\star}}  &  =\frac{2m_{K^{\star}}}%
{\sqrt{4m_{K^{\star}}^{2}-g_{1}^{2}(\phi_{N}-\sqrt{2}\phi_{S})^{2}}}\;.
\label{Z_K_S}%
\end{align}
Moreover, in the N-S sectors of the scalars and pseudoscalars there are $2\times 2$ mixings with off-diagonal mass squares $m^2_{\sigma_{NS}}$ and $m^2_{\eta_{NS}}$ respectively. In these sectors the mass eigenvalues are given by
\begin{align}
m_{f_0^H/f_0^L}^{2} &= \frac{1}{2}\left[  m_{\sigma_{N}}^{2}+m_{\sigma_{S}%
}^{2}\pm\sqrt{(m_{\sigma_{N}}^{2}-m_{\sigma_{S}}^{2})^{2}+4m_{\sigma_{NS}}^{4}%
}\right] \;, \label{Eq:f0L_f0H} \\
m_{\eta^{\prime}/\eta}^{2} &= \frac{1}{2}\left[  m_{\eta_{N}}^{2}+m_{\eta_{S}%
}^{2}\pm\sqrt{(m_{\eta_{N}}^{2}-m_{\eta_{S}}^{2})^{2}+4m_{\eta_{NS}}^{4}%
}\right] \;. \label{Eq:eta_eta_pr}%
\end{align}
It is worth noting that, in contrast to the (pseudo)scalar sector, the eLSM Lagrangian of Sec. \ref{Ssec:elsm_Lagr} does not contain a  mixing of vectors and axial-vectors isoscalar states at tree-level. As discussed in Sec. \ref{Sec:QCD}, this mixing is rather small, so it can be in first approximation safely neglected (or included as a phenomenological corrections to some specific decay channels). Yet, one may add large-$N_c$ suppressed terms of the type $Tr[L_{\mu}]Tr[L^{\mu}]+...$, that would generate an isoscalar mixing in the (axial-)vector sectors as well.

\begin{figure}[ht!]
    \centering
    \includegraphics[scale=0.65]{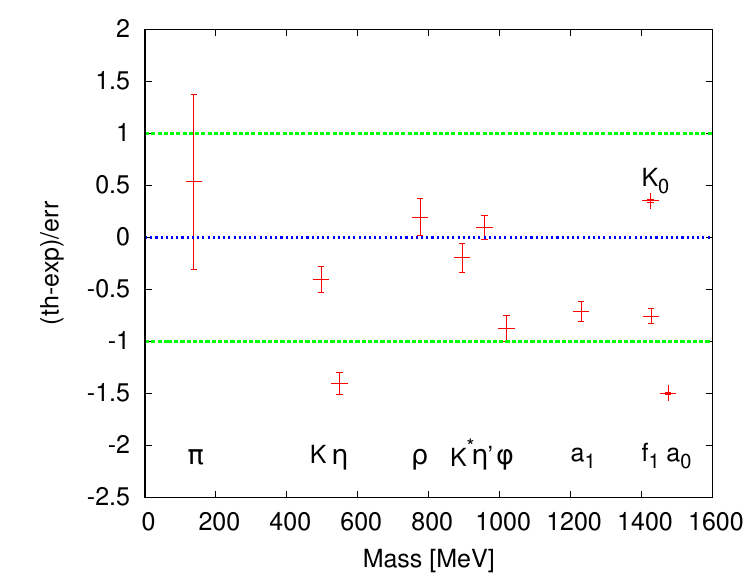}\includegraphics[scale=0.65]{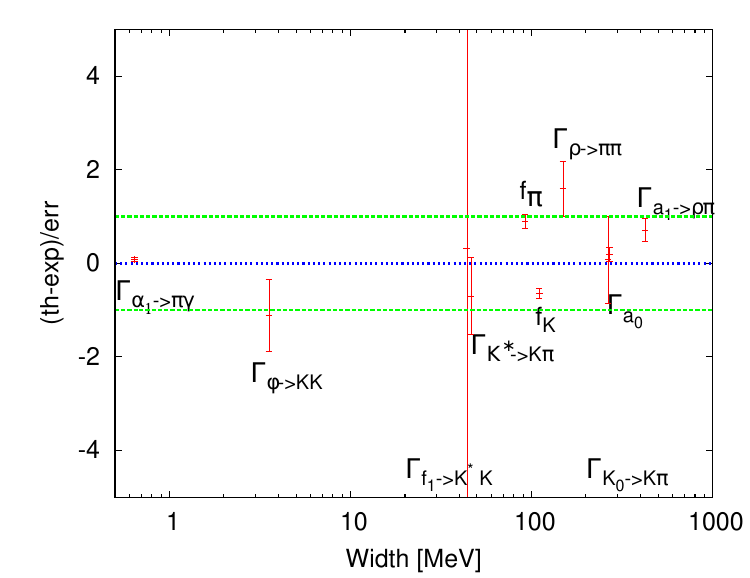}
    \caption{eLSM results compared to the experiment from Ref.~\cite{Parganlija:2012fy}. The figure on the left represents the masses, while the one on the right represents the decay rates. The $x$-axis lists the masses (decay widths) of the resonance, while the $y$-axis represents the deviations from the experimental value as compared to the experimental uncertainty by using $\delta Q_i = \max\{5\%, Q_i^{\text{exp}}\}$. Vertical lines on the data points represent the fit errors. Those decay channels located between the green dashed lines are described well in the model.}
    \label{fig:elsm-summary}
\end{figure}

As an example, a fit result is shown in Table~\ref{Tab:fit_res_eLSM_2013}, while the corresponding fit parameters are given in Table~\ref{Tab:params_eLSM_2013} (both taken from \cite{Parganlija:2012fy}). The comparison of model results for certain observables with experimental data is presented in Fig.~\ref{fig:elsm-summary}.
\begin{table}[ht!]
\centering
\begin{tabular}{|c|c|c|c|c|c|}\hline
Observable & Fit $\left[  \text{MeV}\right]  $ & Experiment $\left[
\text{MeV}\right]  $ &Observable & Fit $\left[  \text{MeV}\right]  $ & Experiment $\left[
\text{MeV}\right]  $\\\hline
$f_{\pi}$ & $96.3\pm0.7$ & $92.2\pm4.6$ &$f_{K}$ & $106.9\pm0.6$ & $110.4\pm5.5$ \\\hline
$m_{\pi}$ & $141.0\pm5.8$ & $138\pm6.9$& $m_{K}$ & $485.6\pm3.0$ & $495.6\pm24.8$ \\\hline
$m_{\eta}$ & $509.4\pm3.0$ & $547.9\pm27.4$ &$m_{\eta^{\prime}}$ & $962.5\pm5.6$ & $957.8\pm47.9$\\\hline
$m_{\rho}$ & $783.1\pm7.0$ & $775.5\pm38.8$ & $m_{K^{\star}}$ & $885.1\pm6.3$ & $893.8\pm44.7$\\\hline
$m_{\phi}$ & $975.1\pm6.4$ & $1019.5\pm51.0$ &$m_{a_{1}}$ & $1186\pm6.0$ & $1230\pm62$\\\hline
  $m_{f_{1}\left(  1420\right)  }$ & $1372.4\pm5.3$ & $1426\pm71$ &$m_{a_{0}}$ & $1363\pm1$ & $1474\pm74$\\\hline
$m_{K_{0}^{\star}}$ & $1450\pm1$ & $1425\pm71$ &$\Gamma_{\rho\rightarrow\pi\pi}$ & $160.9\pm4.4$ & $149.1\pm7.4$ \\\hline
$\Gamma_{K^{\star}\rightarrow K\pi}$ & $44.6\pm1.9$ & $46.2\pm2.3$ &$\Gamma_{\phi\rightarrow\bar{K}K}$ & $3.34\pm0.14$ & $3.54\pm0.18$\\\hline
$\Gamma_{a_{1}\rightarrow\rho\pi}$ & $549\pm43$ & $425\pm175$ &$\Gamma_{a_{1}\rightarrow\pi\gamma}$ & $0.66\pm0.01$ & $0.64\pm0.25$\\\hline
$\Gamma_{f_{1}\left(  1420\right)  \rightarrow K^{\star}K}$ & $44.6\pm39.9$ &
$43.9\pm2.2$ &$\Gamma_{a_{0}}$ & $266\pm12$ & $265\pm13$\\\hline
$\Gamma_{K_{0}^{\star}\rightarrow K\pi}$ & $285\pm12$ & $270\pm80$ &&&\\\hline
\end{tabular}
\caption{An example of fit results from \cite{Parganlija:2012fy}, together with the experimental values taken from \cite{Workman:2022ynf}. The uncertainties correspond to $\delta Q_i = \max\{5\%, Q_i^{\text{exp}}\}$, see text.}
\label{Tab:fit_res_eLSM_2013}
\end{table}

%%%%%%%%%%%%%%%%%%%%%%%%%%%%%%%%
\begin{table}[ht!]
    \centering
   \begin{tabular}
[c]{|c|c|c|}\hline
Parameter & Value & $N_c$-scaling\\\hline
$C_1$ $\left[  \text{GeV}^{2}\right]  $ & $-0.9183\pm0.0006$ & $N_c^0$ \\\hline
$C_2$ $\left[  \text{GeV}^{2}\right]  $ & $0.4135\pm0.0147$ & $N_c^0$ \\\hline
$c_{2}$ $\left[  \text{GeV}^{-2}\right]  $ & $450.5420\pm7.033$ & $N_c^{-3}$ \\\hline
$\delta_{S}$ $\left[  \text{GeV}^{2}\right]  $ & $0.1511\pm0.0038$ & $N_c^0$ \\\hline
$g_{1}$ & $5.843\pm0.018$ & $N_c^{-1/2}$ \\\hline
$g_{2}$ & $3.0250\pm0.2329$ & $N_c^{-1/2}$ \\\hline
$\phi_{N}$ $\left[  \text{GeV}\right]  $ & $0.1646\pm0.0001$ & $N_c^{\frac{1}{2}}$ \\\hline
$\phi_{S}$ $\left[  \text{GeV}\right]  $ & $0.1262\pm0.0001$ & $N_c^{\frac{1}{2}}$ \\\hline
$h_{2}$ & $9.8796\pm0.6627$ & $N_c^{-1}$  \\\hline
$h_{3}$ & $4.8667\pm0.0864$ & $N_c^{-1}$  \\\hline
$\lambda_{2}$ & $68.2972\pm0.0435$ & $N_c^{-1}$  \\\hline
\end{tabular}
\caption{Parameter values for the fit in Table~\ref{Tab:fit_res_eLSM_2013} and the $N_c$-scaling of the parameters. Note that in the case of $C_1$ and $C_2$ the first terms, i.e. $m_0^2$ and $m_1^2$, scale with $N_c^0$, while the second terms are suppressed by a factor of $N_c$.}
\label{Tab:params_eLSM_2013}
\end{table}
In this particular fit, the masses and decay widths of the two $f_0$ scalar-isoscalar states are not used. Consequently, not all of the parameters can be determined separately. Actually there are two combinations of parameters that appear in every expression, namely $C_1= m_0^2 + \lambda_1(\phi_N^2 + \phi_S^2)$ and $C_2 = m_1^2 + h_1/2(\phi_N^2 + \phi_S^2)$. The number of $x_i$ parameters is $N=11$, namely $C_1$, $C_2$, $c_2$, $\delta_S$, $g_1$, $g_2$, $\phi_N$, $\phi_S$, $h_2$, $h_3$ and $\lambda_2$ and the number of physical $Q_i$ quantities is $M=21$. Note that the determination of the two external fields $h_{0N}$ and $h_{0S}$ is exchanged for the determination of the condensates $\phi_N$ and $\phi_S$ by the field equations \eqref{phin} and \eqref{phis} in vacuum.
Moreover, $\delta_N$ can be merged into $m_1^2$, and thus can be set to zero without loss of generality. The fitted physical quantities include masses, decay widths and the pion and kaon decay constants from the Partially Conserved Axial Current (PCAC) relations. The two-particle decay widths ($A\to BC$) can be calculated as follows
\begin{equation}
\Gamma_{A\rightarrow BC}=\mathcal{I}\frac{|\mathbf{k}|}{8\pi m_{A}^{2}%
}\left\vert \mathcal{M}_{A\rightarrow BC}\right\vert ^{2} \; , \label{Gamma}%
\end{equation}
where $\mathcal{M}_{A\rightarrow BC}$ is the transition matrix element for the decay, while $\mathbf{k}$ is the three-momentum of one of the product particles in the rest frame of particle $A$. All expressions for the two-particle decay widths can be found in \cite{Parganlija:2012fy}. For illustrative purposes, we provide here only the decay width for the process $\rho\to\pi\pi$, which is
\begin{equation}
\Gamma_{\rho\rightarrow\pi\pi}=\frac{m_{\rho}^{5}}{48\pi m_{a_{1}}^{4}}\left[
1-\left(  \frac{2m_{\pi}}{m_{\rho}}\right)  ^{2}\right]  ^{3/2}\left[
g_{1}Z_{\pi}^{2}-\frac{g_{2}}{2}\left(  Z_{\pi}^{2}-1\right)  \right]  ^{2}.
\label{Eq:rho_pi_pi_isosym}
\end{equation}

The value of the reduced chi-square $\chi^2_{\text{red}} \equiv \chi^2/N_{\text{dof}}$, where $N_{\text{dof}}=M-N=10$ is the number of degrees of freedom of the fit, at the minimum is $1.2$. There is an important consequence of this fit, namely that the scalar quark-antiquark mesons ($a_0$ and $K_0^{\star}$) are above $1$~GeV and therefore the light scalar mesons below $1$~GeV ($a_0(980)$ and $K_0^{\star}(800)$) are something else (four-quark states). In particular, they may appear as dynamically generated companion poles of the mostly $\bar{q}q$ resonances $a_0(1450)$ and $K_0^{\star}(1430)$ \cite{Wolkanowski:2015jtc,Wolkanowski:2015lsa}. Another parameterization, with the inclusion of $f_0$'s will be presented in Section~\ref{Sec:eLSM_finite_T}.     

\subsection{General \texorpdfstring{$U(1)_A$}{U1A} anomaly terms}
\label{Ssec:U1A_breaking}

The axial anomaly $U(1)_A$ has been described in Sec. \ref{Ssec:elsm_Lagr} by a specific term, but other terms are possible. A more general Lagrangian can be written as
\begin{equation}
    \mathcal{L}_{U(1)_A}^{\text{gen}} = c_{1}(\det\Phi+\det\Phi^{\dagger}) + c_2(\det\Phi-\det\Phi^{\dagger})^2 + c_m(\det\Phi+\det\Phi^{\dagger})\mathop{\mathrm{Tr}}(\Phi^{\dagger} \Phi)  \text{ .} \label{Eq:ELSM_Lagr_U1A_gen}
\end{equation}
The first term is described in Ref. \cite{tHooft:1976rip} and is also used within the eLSM in Refs. \cite{Kovacs:2016juc, Kovacs:2021kas}, the second term, that first appeared in Refs. \cite{Veneziano:1979ec,Witten:1979vv} and is called Veneziano-Witten term employed in Refs. \cite{Parganlija:2010fz, Parganlija:2012fy, Olbrich:2015gln, Divotgey:2016pst, Parganlija:2016yxq} (and in Sec. \ref{ssec:qq and hybr}), while the last one is a mixed term \cite{Pisarski:2024esv}. 
The first term contains three, the second six and the last five mesonic fields, so $c_1$, $c_2$ and $c_m$ have energy dimensions $1$, $-2$ and $-1$ respectively. Since the parameters are not dimensionless, these terms also break dilatation invariance. Note that instead of the $c_2 (\det\Phi-\det\Phi^{\dagger})^2$ term, a $\bar{c}_2 ((\det\Phi)^2 + (\det\Phi^{\dagger})^2)$ term could also be used, such as in \cite{Pisarski:2024esv}. The difference of the two terms is $\sim \det \Phi \det \Phi^{\dagger}$, which is chirally invariant but is not anomalous; as such, it can be transformed out by redefining the coefficients of some of the chirally invariant terms involving traces, so both $\sim \det\Phi^2$ terms can be used interchangeably.

In \cite{Kovacs:2013xca} it is investigated how these terms are able to reproduce the particle spectrum. It was found that each term ($c_1$, $c_2$ and $c_m$) can give an equally good description of the masses. Moreover, in principle these terms can be used simultaneously and still reproduce the particle spectrum with a good $\chi^2$.  

For the sake of completeness, we list in Table~\ref{Tab:mass-spin0_eLSM_gen_U1A} how these terms modify the tree-level mass expressions of Table~\ref{Tab:mass-spin0_eLSM}.
\begin{table}[ht!]
%[ptb]
\centering
\renewcommand{\arraystretch}{1.3} 
\begin{tabular}{|c|c|}
\hline
Mass squares  & Analytical expressions \\ \hline\hline
$m_{\pi}^{2}$ &  $Z_{\pi}^{2}\left[  m_{0}^{2}+\left(  \lambda_{1}%
+\frac{\lambda_{2}}{2}\right)  \phi_{N}^{2}+\lambda_{1}\phi_{S}^{2} - \textcolor{blue}{(c_1 + c_m (\phi_N^2 + \frac{1}{2}\phi_S^2))\frac{\phi_S}{\sqrt{2}}}\right] $\\
\hline
$m_{K}^{2}$ &  $Z_{K}^{2}\left[  m_{0}^{2}+\left(  \lambda_{1}+\frac
{\lambda_{2}}{2}\right)  \phi_{N}^{2}-\frac{\lambda_{2}}{\sqrt{2}}\phi_{N}%
\phi_{S}+\left(  \lambda_{1}+\lambda_{2}\right)  \phi_{S}^{2} - \textcolor{blue}{(2c_1 + c_m (\phi_N^2 + \phi_S^2 + \sqrt{2}\phi_N\phi_S))\frac{\phi_N}{4}}\right]$ \\
\hline
$m_{\eta_{N}}^{2}$ &  $Z_{\pi}^{2}\left[  m_{0}^{2}+\left(  \lambda_{1}%
+\frac{\lambda_{2}}{2}\right)  \phi_{N}^{2}+\lambda_{1}\phi_{S}^{2}%
+c_{2}\,\phi_{N}^{2}\phi_{S}^{2} - \textcolor{blue}{(c_1 + \sqrt{2} c_2\phi_N^2\phi_S + c_m \frac{1}{2}\phi_S^2)\frac{\phi_S}{\sqrt{2}}} \right]$\\
\hline
$m_{\eta_{S}}^{2}$ &  $Z_{\eta_{S}}^{2}\left[  m_{0}^{2}+\lambda_{1}\phi
_{N}^{2}+\left(  \lambda_{1}+\lambda_{2}\right)  \phi_{S}^{2}+\frac{c_{2}}%
{4}\phi_{N}^{4} + \textcolor{blue}{(c_2 \phi_N^2 - \sqrt{2} c_m \phi_S)\frac{\phi_N^2}{4}}\right]$\\
\hline
$m_{\eta_{NS}}^2$ &  $Z_{\eta_N} Z_{\eta_S}\left[\textcolor{blue}{(2c_1 + \sqrt{2} c_2 \phi_N^2\phi_S + c_m (\phi_N^2 + \phi_S^2))\frac{\phi_N}{2\sqrt{2}}}\right]$\\
\hline
$m_{a_{0}}^{2}$ &  $m_{0}^{2}+\left(  \lambda_{1}+\frac{3}{2}\lambda
_{2}\right)  \phi_{N}^{2}+\lambda_{1}\phi_{S}^{2} + \textcolor{blue}{(c_1 + \frac{1}{2} c_m \phi_S^2)\frac{\phi_S}{\sqrt{2}}}$  \\
\hline
$m_{K_{0}^{\star}}^{2}$  & $Z_{K_{0}^{\star}}^{2}\left[  m_{0}^{2}+\left(
\lambda_{1}+\frac{\lambda_{2}}{2}\right)  \phi_{N}^{2}+\frac{\lambda_{2}%
}{\sqrt{2}}\phi_{N}\phi_{S}+\left(  \lambda_{1}+\lambda_{2}\right)  \phi
_{S}^{2} + \textcolor{blue}{(2c_1 + c_m (\phi_N^2 + \phi_S^2 - \sqrt{2}\phi_N\phi_S))\frac{\phi_N}{4}}\right]$ \\
\hline
$m_{\sigma_{N}}^{2}$  &  $m_{0}^{2}+3\left(  \lambda_{1}+\frac{\lambda_{2}}%
{2}\right)  \phi_{N}^{2}+\lambda_{1}\phi_{S}^{2} - \textcolor{blue}{(c_1 + \frac{1}{2} c_m (6\phi_N^2 + \phi_S^2)\frac{\phi_S}{\sqrt{2}})} $  \\\hline
$m_{\sigma_{S}}^{2}$  & $m_{0}^{2}+\lambda_{1}\phi_{N}^{2}+3\left(  \lambda
_{1}+\lambda_{2}\right)  \phi_{S}^{2} - \textcolor{blue}{\frac{3}{2\sqrt{2}} c_m \phi_N^2 \phi_S}$ \\\hline
$m^2_{\sigma_{NS}}$ & $2\lambda_1 \phi_{N} \phi_{S} - \textcolor{blue}{(c_1 + c_m (\phi_N^2 + \frac{3}{2}\phi_S^2))\frac{\phi_N}{\sqrt{2}}}$\\\hline
\end{tabular}
\caption{Mass expressions of spin-0 mesons (scalars and pseudoscalars) within the eLSM with the general $U(1)_A$ anomaly term. The anomaly contributions are written in blue (online).}
\label{Tab:mass-spin0_eLSM_gen_U1A}
\end{table}
Note that in the case of the scalars there is a non-zero mixing in the nonstrange-strange (N-S) sector even if all $U(1)_A$ anomaly terms are zero (described by the large-$N_c$ suppressed $\lambda_1$-term), while in the pseudoscalar N-S sector there is no mixing if there is no $U(1)_A$ anomaly. Another interesting property is that if we were to use only the $c_2$ type of term, it would only modify the masses in the pseudoscalar sector, but not in the scalar sector. In the recent paper \cite{Pisarski:2024esv} , Pisarski and Rennecke conjecture that the $\sim \det\Phi^2$ term dominates w.r.t. the $\sim \det \Phi$ term, i.e. it is possible that $c_2 >> c_1$. This fact may affect other properties, such as the order of the phase transition in the chiral limit, which changes by using different $U(1)_A$ anomaly terms.
 
% The chiral anomaly is, together with the trace anomaly, an important element of chiral models and deserve a treatment on its own. 

%It is worth mentioning that the main difference between the two anomalous terms would be visible in the decay processes. Since we use the experimental results for masses of $\eta$ and $\eta^\prime$ and the mixing angle, we can choose any of them. Note that these anomaly terms do not influence the spontaneous breaking of the chiral symmetry. The term proportional to $c_1$ is used in \cite{Parganlija:2010fz}  for $N_f=2$ and to $c_2$ in \cite{Parganlija:2012fy}  for $N_f=3$.

 \subsection{The eLSM at large-\texorpdfstring{$N_c$}{Nc}}

 As discussed in detail in Refs. \cite{Witten:1979kh,Lebed:1998st,Lucha:2021mwx,Giacosa:2024scx}, in the limit of a large number of colors, the so-called large-$N_c$ limit, interesting and useful simplifications occur in the 't Hooft limit \cite{tHooft:1973alw}, in which $\Lambda_{\text{QCD}}$ is $N_c$-independent and $g_{\text{QCD}}$ scales as $1/\sqrt{N_c}$: conventional quark-antiquark mesons, hybrids, and glueballs retain their mass and become long-lived. In particular, the decay width of conventional mesons and hybrids scales as $1/N_c$, while that of glueballs as $1/N_c^2$ \cite{Witten:1979kh}.
Moreover, the dominant four-leg interaction between conventional mesons and hybrids involves a single trace and scale as $1/N_c$, while the subdominant ones (product of two traces) goes as $1/N_c^3$. This is an important mean to distinguish between different terms, especially when few data are available (see Sec. \ref{Sec:other(1,2)}). The four-leg glueball scaling behaves as $1/N_c^2$, just as the four-leg interaction with two glueballs and two conventional mesons or hybrids (see Sec. \ref{Sec:hybrids} and \ref{Sec:glueballs} for applications).

The condensates scale as well: the dilaton/glueball as $G_0 \sim \Lambda_G \propto N_c$ , while the quark ones $\phi_{N,S} \propto N_c^{1/2}$.
A special case regards the chiral anomaly, where an additional large-$N_c$ suppression takes place \cite{Witten:1979vv}.
For a detailed description of all the scaling behaviors mentioned above, with special focus on chiral models, we refer to the recent lectures of Ref. \cite{Giacosa:2024scx}.

In Table \ref{Tab:params_eLSM_2013} we report the scaling of the parameters of the eLSM, while in Table \ref{tab:Large-N} we discuss some of its main consequences, that well agree with the general expected rules. For a brief recall of large-$N_c$, see \ref{App:Large-N}.
\begin{table}[ht!]
\centering
\begin{tabular}{|c|c|}\hline
Quantity & Scaling\\\hline
Gluon condensate $G_{0}$ & $N_{c}$\\\hline 
Quark condensates $\phi_{N,S}$ & $N_{c}^{1/2}$\\\hline
Mass parameter $m_{0}^{2}=\lambda_{G\Phi}G_{0}<0$ & $N_{c}^{0}$\\\hline
Masses of mesons ($\bar{q}q,$ glueballs, hybrids) & $N_{c}^{0}$\\\hline 
Exception ($U_{A}(1)$ anomaly): mass of $\eta_{0}$ (split into $\eta
,\eta^{\prime}$) & $N_{c}^{-1}$\\\hline 
Decay of $\bar{q}q$ states into $\bar{q}q,$ e.g. $\Gamma_{\rho\rightarrow
\pi\pi}$ & $N_{c}^{-1}$\\\hline 
Decay of glueballs, e.g. $\Gamma_{G\rightarrow\pi\pi}$ & $N_{c}^{-2}%
$\\\hline 
Exception ($U_{A}(1)$ anomaly): $\tilde{G},$ e.g. $\Gamma_{\tilde
{G}\rightarrow\eta f_{0}}$ & $N_{c}^{-4}$\\\hline 
Four-leg $\bar{q}q$ scattering amplitude, e.g. $A_{\pi\pi\rightarrow\pi\pi}$ &
$N_{c}^{-1}$\\\hline 
Four-leg $GG$ scattering amplitude, e.g. $A_{GG\rightarrow GG}$ & $N_{c}^{-2}%
$\\\hline 
\end{tabular}
\caption{$N_c$-scaling of some phenomenologically relevant eLSM quantities.}
\label{tab:Large-N}
\end{table}
	
\section{Inclusion of other conventional mesons with \texorpdfstring{$J=1,2$}{J12} into the eLSM }
\label{Sec:other(1,2)}

In this section, we discuss the inclusion of additional chiral multiplets into the eLSM: (i) the heterochiral multiplet (heterovectors) including the pseudovector nonet 
($J^{\mathcal{P}\mathcal{C}}=1^{+-}$, $^{2S+1}L_J$ = $^{1}P_1$) and the orbitally excited vector one ($J^{\mathcal{P}\mathcal{C}}=1^{--}$, $^{2S+1}L_J$ =$^{3}D_1$); (ii) the homochiral multiplet (homotensors) with tensor mesons ($J^{\mathcal{P}\mathcal{C}}=2^{++}$,$ ^{2S+1}L_J =^{3}P_2$) and with axial-tensor mesons ($J^{\mathcal{P}\mathcal{C}}=2^{--}$, $^{2S+1}L_J$ =$^{3}D_2$); (iii) the heterochiral multiplet (heterotensors) of pseudotensor mesons ($J^{\mathcal{P}\mathcal{C}}=2^{-+}$, $^{2S+1}L_J$ =$^{1}D_2$) and their chiral partners.
The corresponding resonances are listed in Table \ref{Tab:spin1-2_masses_PDG}.

\begin{table}[ht!]
\centering
\begin{tabular}
[c]{|c|c|c|c|c|c|}\hline
Field & PDG & Quark content & $I^\mathcal{G}$ & $J^{\mathcal{P}\mathcal{C}}$ &  PDG Mass (MeV)\\\hline\hline
$\rho_D^{+},\rho_D^{-},\rho_D^{0}$ & $\rho(1700)$ & $u\bar{d},d\bar{u},\frac{u\bar
{u}-d\bar{d}}{\sqrt{2}}$ & $1^{+}$ & $1^{--}$  & $1720\pm 20$\\\hline
$K_D^{\ast+},K_D^{\ast-},K_D^{\ast0},\bar{K}_D^{\ast0}$ & $K^{\ast}(1680)$ & $u\bar
{s},s\bar{u},d\bar{s},s\bar{d}$ & $1/2$ & $1^{-}$ & $1718\pm 18$\\\hline
$\omega_{DN}$ & $\omega(1650)$ & $\frac{u\bar{u}+d\bar{d}}{\sqrt{2}}$ & $0^{-}$ &
$1^{--}$ & $1670 \pm 30$\\\hline
$\omega_{DS}$ & $\phi(???)$ & $s\bar{s}$ & $0^{-}$ & $1^{--}$ & $-$\\\hline\hline
$b_{1}^{+},b_{1}^{-},b_{1}^{0}$ & $b_{1}(1235)$ & $u\bar{d},d\bar{u}%
,\frac{u\bar{u}-d\bar{d}}{\sqrt{2}}$ & $1^{+}$ & $1^{+-}$ & $1229.5\pm 3.2$\\\hline
$K_{1B}^{+},K_{1B}^{-},K_{1B}^{0},\bar{K}_{1B}^{0}$ &  \begin{tabular}[c]{@{}l@{}} $K_{1}(1400)$\\
$K_{1}(1270)$\end{tabular}  & $u\bar
{s},s\bar{u},d\bar{s},s\bar{d}$ & $1/2$ & $1^{-}$ & \begin{tabular}[c]{@{}l@{}} $1368\pm 38$\\
$1253\pm 7$ \end{tabular}  \\\hline
$h_{1N}$ & $h_{1}(1170)$ & $\frac{u\bar{u}+d\bar{d}}{\sqrt{2}}$ & $0^{-}$ &
$1^{+-}$ & $1166\pm 6$\\\hline
$h_{1S}$ & $h_{1}(1415)$ & $s\bar{s}$ & $0^{-}$ & $1^{+-}$ & $1419^{+9}_{-8}$\\\hline\hline
$\pi_2^{+},\pi_2^{-},\pi_2^{0}$ & $\pi_2(1670)$ & $u\bar{d},d\bar{u},\frac{u\bar{u}-d\bar
{d}}{\sqrt{2}}$ & $1^{-}$ & $2^{-+}$ & $1670.6^{+2.9}_{-1.2}$\\\hline
$K_{2P}^{+},K_{2P}^{-},K_{2P}^{0},\bar{K}_{2P}^{0}$ & \begin{tabular}[c]{@{}l@{}} $K_2(1770)$\\
$K_{2}(1820)$\end{tabular}  & $u\bar{s},s\bar{u},d\bar{s},s\bar{d}$
& $1/2$ & $2^{-}$ & \begin{tabular}[c]{@{}l@{}} $1773\pm 8$ \\
$1819\pm 12 $\end{tabular} \\\hline
$\eta_{2N}$ & $\eta_2(1645)$ & $\frac{u\bar{u}+d\bar{d}}{\sqrt{2}}$ & $0^{+}$ & $2^{-+}$ & $1617\pm 5$\\\hline
%$\eta_{N}a+\eta_{S}b$ & $\eta$ & $\frac{u\bar{u}+d\bar{d}}{\sqrt{2}}a+s\bar
%{s}b$ & $0$ & $0^{-+}$ & $547.86$\\\hline
$\eta_{2S}$ & $\eta_2(1870)$ &  $s\bar
{s}$ & $0^{+}$ & $2^{-+}$ & $1842\pm 8$\\\hline\hline
%$-\eta_{N}a+\eta_{S}b$ & $\eta^{\prime}(958)$ & $\frac{u\bar{u}+d\bar{d}%
%}{\sqrt{2}}a+s\bar{s}b$ & $0$ & $0^{-+}$ & $957.78$\\\hline
$a_{2}^{+},a_{2}^{-},a_{2}^{0}$ & $a_{2}(1320)$ & $u\bar{d},d\bar{u}%
,\frac{u\bar{u}-d\bar{d}}{\sqrt{2}}$ & $1^{-}$ & $2^{++}$ & $1317.7\pm 1.4 $\\\hline
$K_{2}^{+},K_{2}^{-},K_{2}^{0},\bar{K}_{2}^{0}$ & $K_{2}^{\ast}(1430)$ &
$u\bar{s},s\bar{u},d\bar{s},s\bar{d}$ & $1/2$ & $2^{+}$ & $1424\pm 4$\\\hline
$f_{2N}$ & $f_{2}(1270)$ & $\frac{u\bar{u}+d\bar{d}}{\sqrt{2}}$ & $0^{+}$ & $2^{++}$ & $1275.4\pm 0.8$\\\hline
%$\sigma_{N}$ & $f_{0}(1370)$ & $\frac{u\bar{u}+d\bar{d}}{\sqrt{2}}$ & $0$ &
%$0^{++}$ & $1350$\\\hline
$f_{2S}$ & $f_{2}^\prime(1525)$ &$s\bar{s}$ & $0^{+}$ & $2^{++}$ & $1517.3\pm 2.4$\\\hline
%$\sigma_{S}$ & $f_{0}(1500)$ & $s\bar{s}$ & $0$ & $0^{++}$ & $1504$\\\hline
\end{tabular}
\caption{Resonances studied within the eLSM in this Section:  PDG name, quark content, quantum numbers and PDG mass values from \cite{Workman:2022ynf}. One potential candidate for the missing orbitally excited vector meson $\phi(?)$ is the resonance $\phi(2170)$. However, it is disfavored due to its higher mass.}
\label{Tab:spin1-2_masses_PDG}
\end{table}

%eLSM for the excited spin-1 mesons $V_{D}^\mu$ (within heterochiral nonet) and spin-2 heterochiral $P_{2\,\mu\nu}$ and homochiral fields ( $T_{\mu\nu}, A_{2,\mu\nu}$)   mesons. All mesonic fields and their quantum numbers as well as the PDG masses are listed in Table \ref{Tab:spin1-2_masses_PDG}

%For spin-1 heterochiral mesons, we start with the chiral invariant lagrangian and present the corresponding decay widths for the well-established excited mesons $\rho(1700), K^\star(1680), \omega(1650)$. Secondly, we discuss about the lagrangian breaking $U_A(1)$-axial symmetry and present the outcomes for the anomalous decays and isoscalar mixing angle $\beta_{pv}$. 

%For spin-2 heterochiral mesons, we present the results for the decay of the pseudotensor mesons and estimation for the mixing angle in the isoscalar sector.

%The last topic of this section concerns homochiral spin-2 mesons. We introduce the mass and interaction term lagrangians and present the model results compared to the PDG and LQCD. We also present the dominant decay channels for axial-tensor mesons.

\subsection{Heterochiral mesons with \texorpdfstring{$J=1$}{J1}}
\label{SSec:heterovectors}
The heterochiral multiplet $\Phi^{\mu} =B^{\mu}+iV_D^{\mu}$, that includes the pseudovector nonet $B^{\mu}$ with $J^{\mathcal{P}\mathcal{C}}=1^{+-}$ and the orbitally excited vector one with $J^{\mathcal{P}\mathcal{C}}=1^{--}$, can be easily embedded into the eLSM. Namely, the Lagrangian term that is responsible for the mass generation of pseudovector and orbitally vector mesons can be written as:
\begin{align}
\mathcal{L}^{\Phi_\mu}_{\text{mass}}=&\text{Tr}\Big[\Big(\frac{m_{1}^{2}\,G^2}{2\,G_0^2}+\Delta^{\text{pv}}\Big)\Big(\Phi
_{\mu}^{\dagger}\Phi^{\mu}\Big)\Big]+ \frac{\lambda_{\Phi_\mu,1}}{2}\text{Tr}\Big[\Phi^\dagger \Phi\Big]\text{Tr}\Big[\Phi_{\mu}^\dagger \Phi^{\mu}\Big]+\lambda_{\Phi_\mu,2}\text{Tr}\Big[\Phi_{\mu}^{\dagger}\Phi
\Phi^{\mu\dagger}\Phi+\Phi_{\mu}\Phi^{\dagger}
\Phi^{\mu}\Phi^\dagger\Big]\\\nonumber
&+\lambda_{\Phi_\mu,3}\text{Tr}\Big[\Phi
_{\mu}\Phi^{\dagger}\Phi\Phi^{\mu\,\dagger}+\Phi
_{\mu}^{\dagger}\Phi\Phi^\dagger\Phi^{\mu}\Big]
\text{ ,}
\end{align}
where the chiral fields are $\Phi$ \eqref{Eq:J0_sc_ps_nonet} and  $\Phi_{\mu}$ \eqref{eq:J1-hetero-chiral} have been introduced. The explicit chiral symmetry and dilatation breaking terms is the one proportional to $\Delta^{\text{pv}}=\text{diag}(\delta_{N}^{\text{pv}},\delta_{N}^{\text{pv}},\delta_{S}^{\text{pv}})$.
%This Lagrangian enables us to predict the mass of the excited vector isoscalar meson $\omega_{DS}\simeq \phi(1930)$. 
As usual, one may set $\delta_{N}^{\text{pv}}=0$ and one expects that $\delta_{S}^{\text{pv}} \sim \delta_{S}$, see Table \ref{Tab:params_eLSM_2013}.
Note, the field $\Phi^{\mu}$ has been included in the eLSM in Ref. \cite{Giacosa:2016hrm}, but the Lagrangian above is here presented for the first time. Its effect is to introduce a mass splitting between pseudovector and orbitally excited vector mesons, but its detailed study was not yet carried out in detail and is left as an outlook.

Next, a chiral invariant interaction Lagrangian leading to the decay of heterovectors reads:
%for the pseudovector mesons ($B^{\mu}$) and orbitally excited vector mesons ($V^{\mu}_{D}$) within the vector heterochiral multiplet $\Phi^{\mu}$ of Eq. \eqref{eq:J1-hetero-chiral} 
reads: 
\begin{align}  
\label{eq:lag-hetero-chiral-1} \mathcal{L}^{\text{int}}_{\Phi^{\mu}}&=g_{\Phi^{\mu}\Phi\Phi}\text{Tr}\Big[\Phi^{\mu}\,\Phi \partial_{\mu}\Phi+\text{c.c}\Big]+g_{\Phi^{\mu}LR}\text{Tr}\Big[\Phi_{\alpha}^{\dagger}L_{\beta}L^{\alpha}\partial^{\beta}\Phi+R_{\alpha}\Phi_{\beta}^{\dagger}\partial^{\alpha}\Phi R^{\beta}+L_{\alpha}\partial^\beta\Phi\Phi_{\alpha}^{\dagger}L^{\beta}+\partial_\alpha\Phi^\dagger R_{\beta}R^{\alpha}\Phi^{\beta}\Big]
\text{ .}
\end{align}
It reduces to the following two dominant terms which describe the decays of the orbitally excited vector mesons into two pseudoscalar and vector pseudoscalar mesons:
\begin{align}  
\mathcal{L}_{V_D}=ig_{v_dpp}\text{Tr}\Bigg[\Big[\partial_{\mu}P,V_{D}^{\mu}\Big]P\Bigg]+g_{v_dvp}\text{Tr}\Bigg[V_{D}^{\mu\nu}\Big\{V_{\mu\nu},P\Big\}\Bigg] \text{ ,}
\end{align}
see Ref. \cite{Piotrowska:2017rgt} and Table \ref{tab:v_epp} for numerical results. For pseudovector mesons, it reduces to a term proportional to 
$\text{Tr}[ B_{\mu} \{V^{\mu},P \}]$
that has been studied in Ref. \cite{Divotgey:2013jba}. However, it is important to note that a joint study of the decays of both orbitally excited vector mesons and pseudovector mesons has not yet been undertaken. 
\begin{table}[ht!]		
\renewcommand{\arraystretch}{1.1}
  \centering
\begin{tabular}{|c|c|c|c|}
\hline Decay process 			& 	Width (MeV) 	& Decay process 				& 	Width (MeV) 	
		\\
				\hline
    $\rho(1700) \rightarrow \overline{K}K $						&	$ 40\pm 11 $  &	
    $\rho(1700) \rightarrow \pi \pi $						&	$ 140\pm 37 $  	\\ \hline
      $K^\star(1680) \rightarrow K\pi  $						&	$ 82\pm 22 $  	&
      $K^\star(1680) \rightarrow K\eta  $						&	$ 52\pm 14 $  	\\ \hline
       $\omega(1650) \rightarrow \overline{K}K  $						&	$ 37\pm 10 $ &$\rho(1700)\rightarrow\omega\pi$ & $140\pm59$ 	\\ \hline
       $\rho(1700)\rightarrow K^{\ast}(892)K$ & $56\pm23$ &  $\rho(1700)\rightarrow\rho\eta$ & $41\pm17$ \\\hline
       $K^{\ast}(1680)\rightarrow K\rho$ & $64\pm27$ & $K^{\ast}(1680)\rightarrow K\phi$ & $13\pm6$ \\\hline
       $K^{\ast}(1680)\rightarrow K\omega$ & $21\pm9$ & $K^{\ast}(1680)\rightarrow K^{\ast}(892)\pi$ & $81\pm34$\\\hline
       $K^{\ast}(1680)\rightarrow K^{\ast}(892)\eta$ & $0.5\pm0.2$ & $\omega(1650)\rightarrow\rho\pi$ & $370\pm156$ \\\hline
       $\omega(1650)\rightarrow K^{\ast}(892)K$ & $42\pm18$ & $\omega(1650)\rightarrow\omega(782)\eta$ & $32\pm13$\\\hline
       $\phi(1930)\rightarrow K\bar{K}^{\ast}(892)$ & $260\pm109$ & $\phi(1930)\rightarrow\phi(1020)\eta$ & $67\pm28$ \\\hline
			\end{tabular}
		\caption{\label{tab:v_epp} %
		Decays of the radially excited vector mesons \cite{Piotrowska:2017rgt}.		}
	\end{table} 

Very interestingly, the chiral anomaly can be introduced for this chiral multiplet. 
 To this end, a certain interaction Lagrangian type was employed by Giacosa, Pisarski and Koenigstein (GPK) in Ref. \cite{Giacosa:2017pos} and by Giacosa, Pisarski, and Jafarzade (GPJ) in Ref. \cite{Giacosa:2023fdz} for a detailed discussion of its physical properties). It turns out that these Lagrangian terms make use of a mathematical concept known as `mixed discriminant' in mathematics, first introduced in Ref. \cite{Alexandroff1938} and further discussed in Refs. \cite{AAPanov1987,BAPAT1989107}. This object is denoted below as `$\epsilon$-function', because of its connection with the Levi-Civita tensor; it represents an extension of the determinant for distinct matrices and can be called `polydeterminant'. For details, see Ref. \cite{Giacosa:2025dvu}.
%%%%%%%%%%%%
For $3 \times 3$ matrices, it reads as
\begin{equation}
\epsilon[A,B,C]= \frac{1}{3!} \epsilon^{ijk}\epsilon^{i'j'k'}A_{ii'}B_{jj'}C_{kk'} \;.
\label{epsilon}
\end{equation}
It has quite interesting properties such as \begin{align}    \epsilon[A,A,A]=\det A\,,\qquad \epsilon[A,1,1]=\frac{1}{3}\text{Tr} A\ \text{ ,}
\end{align}
and is invariant under exchange of the matrices, $\epsilon[A,B,C]=\epsilon[B,A,C],$ etc.
It can be also easily generalized to the $N$-case with:
\begin{equation}
\epsilon[A_1,A_2,...,A_N]= \frac{1}{N!}  \epsilon^{i_1 i_2 ... i_N}\epsilon^{i_1' i_2'...i_N'}{A_1}_{i_1 i_1'} {A_2}_{j_2 i_2'}...{A_N}_{i_N i_N'} ;.
\end{equation}
Using this object, one may write down a Lagrangian that breaks $U(1)_A$ axial symmetry but retains chiral symmetry (just as the determinant term in the (pseudo)scalar sector, see Sec. \ref{Ssec:elsm_Lagr} and especially Sec. \ref{Ssec:U1A_breaking} and refs. therein). Following Refs. \cite{Giacosa:2017pos,Giacosa:2023fdz}:
 \begin{align}\label{eq:anomal-lag-J1}
    \mathcal{L}^{U(1)_A-\text{anom}}_{\Phi^{\mu}}=a_{\Phi^{\mu}} \Big( \epsilon[\Phi\,, \Phi_\mu \,, \Phi^\mu]
+\text{c.c.}\Big)\ \text{ ,}
 \end{align}
where the coupling is estimated $a_{\Phi^{\mu}}\approx -0.14$  GeV in Ref. \cite{Giacosa:2023fdz} by using the eLSM implemented within dilute instanton gas (DIG) approximation. Anomalous decay channels following the Lagrangian \eqref{eq:anomal-lag-J1} are listed in Table \ref{tab:anom-spin1}. Even though they are quite small, their experimental measurement can still play a significant role in understanding the axial anomaly in spin-1 heterochiral mesons.
\begin{table}[ht!]			\renewcommand{\arraystretch}{1.1}
  \centering
  \begin{tabular}{|c|c|c|c|}
			\hline Decay process 				& 	Width (MeV) &Decay process 				& 	Width (MeV) 						\\
				\hline
    $\rho(1700) \rightarrow h_1(1415) \, \pi$		&	$ 0.027 $  	&     $\phi (2170) \rightarrow b_1(1235) \, \pi$				&	$0.071$   \\\hline
		    $\phi (2170)\rightarrow h_1(1170) \, \eta$							&	$0.012   $&
   $\phi (2170) \rightarrow h_1(1170) \, \eta^\prime(958)$							&	$0.010  $\\\hline
			\end{tabular}				 
		\caption{\label{tab:anom-spin1} %
		Decay rates for the anomalous decay channels of spin-1
mesons estimated in a DIG \cite{Giacosa:2023fdz}.	}
	\end{table}

\begin{figure}[ht!]
       \centering       \includegraphics[scale=0.75]{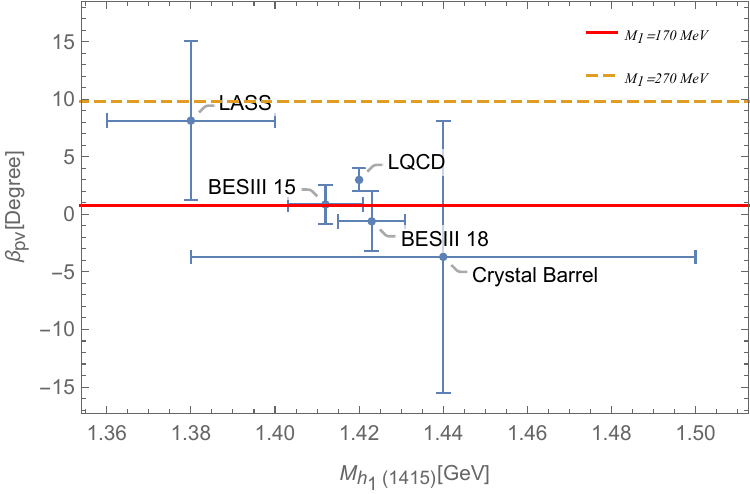}\\
      \caption{The estimate of the mixing angle $\beta_{pv}$ in a eLSM + DIG approach \cite{Giacosa:2023fdz} is compared to the experiments \cite{Aston:1987ak,CrystalBarrel:1997kda,BESIII:2015vfb,BESIII:2018ede}
      and to LQCD \cite{Dudek:2011tt}.  Here $M_1$ is a DIG parameter, which links the mesonic and quark fields.	The theoretical prediction lies between the two horizontal lines (solid orange and dashed yellow).}
        \label{fig:mixing}
 \end{figure}

The mixing angle $\beta_{pv}$ between $h_1(1170)$ and $h_1(1415)$ introduced in \eqref{mixingpv} to be within the range $0$-$10^{\circ}$ in the eLSM+DIG approach of Ref. \cite{Giacosa:2023fdz}, see Fig. \ref{fig:mixing} for a comparison to the experiments in Refs.
\cite{Aston:1987ak,CrystalBarrel:1997kda,BESIII:2015vfb,BESIII:2018ede}
      and to lattice (LQCD) \cite{Dudek:2011tt}.

 The term in Eq. \eqref{eq:anomal-lag-J1} represents an interesting new type of anomalous interactions, which in general are possible for heterochiral multiplets (it affects in particular the masses and the mixing of the isoscalar members). A new plethora of such terms is possible, and a promising outlook is the study of their detailed role in the mesonic phenomenology. 

 \subsection{Homochiral mesons with \texorpdfstring{$J=2$}{J2}}
 \label{Ssec:axial-tensor}
%FG up to here
The tensor homochiral multiplets (homotensors), involving a tensor nonet $T^{\mu\nu}$ \eqref{eq:nonet-t2} (with $J^{\mathcal{P}\mathcal{C}}=2^{++}$) and and (axial-)tensor one $A_2^{\mu\nu}$ 
\eqref{eq:nonet-a2} (with $J^{\mathcal{P}\mathcal{C}}=2^{--}$), were studied within the eLSM in Ref. \cite{Jafarzade:2022uqo}. 
The chiral Lagrangian describing the masses of (axial-)tensor states reads:
\begin{align}\nonumber  
\label{eq:m2-homo-Lag}  \mathcal{L}_{\textbf{L},\textbf{R}}^{\text{mass}}=\text{Tr}\Big[&\Big(\frac{m_{2}^{2}\,G^2}{2\,G_0^2}+\Delta^{\text{ten}}\Big)\Big(\mathbf{L}_{\mu\nu}^2+\mathbf{R}_{\mu\nu}^2\Big)\Big]
   + \frac{h_1^{\text{ten}}}{2}\text{Tr}\Big[\Phi^\dagger \Phi\Big]\text{Tr}\Big[\mathbf{L}^{\mu\nu} \mathbf{L}_{\mu\nu}+ \mathbf{R}^{\mu\nu} \mathbf{R}_{\mu\nu}\Big]  \\
  & +h_2^{\text{ten}}\text{Tr}\Big[\Phi^\dagger \mathbf{L}^{\mu\nu} \mathbf{L}_{\mu\nu} \Phi+\Phi \mathbf{R}^{\mu\nu} \mathbf{R}_{\mu\nu} \Phi^\dagger\Big]+2 h_3^{\text{ten}}\text{Tr}\Big[\Phi \mathbf{R}^{\mu\nu}\Phi^\dagger \mathbf{L}_{\mu\nu} \Big]
   \text{.}
\end{align}
Above, the explicit symmetry breaking term parameterized by the diagonal matrix $\Delta^{\text{ten}}=\text{diag}(\delta_{N}^{\text{ten}},\delta_{N}^{\text{ten}},\delta_{S}^{\text{ten}})$ is the only dimensionful one. By assuming (without loss of generality) $\delta_N^{\text{ten}}\simeq 0$, we obtain $\delta_S^{\text{ten}}$ via the following expression:
\begin{equation}  \delta_S^{\text{ten}}= m_{K_{2}}^2-m_{\mathbb{a}_2}^2 \simeq 0.3\,\text{GeV}^2\,.
\end{equation}
The masses arising from Eq. \eqref{eq:m2-homo-Lag}  are listed in Table \ref{tab:mass-spin2}. In particular, within the nonets we have:
%we observe that the masses of the isovector and nonstrange isoscalar members are identical, since  in the absence of an isoscalar mixing angle within each nonet
\begin{equation}
    m_{\rho_2}^2= m_{\omega_{2N}}^2\,\,\,\text{and}\,\,\, m_{\mathbf{a}_2}^2= m_{f_{2N}}^2
    \text{.}
\end{equation}
The following relations between chiral partners follow: 
\begin{equation}
m_{\mathbf{a}_2}^2-m_{\rho_2}^2=  h_3^{\text{ten}}\phi_N^2 \text{ ,} 
m_{K_{2}}^2-m_{K_{2A}}^2= \sqrt{2} h_3^{\text{ten}}\phi_N\phi_S \text{ , }
m_{f_{2S}}^2-m_{\omega_{2S}}^2=2 h_3^{\text{ten}}\phi_S^2 \text{ ,}
\end{equation}
showing that the mass differences between chiral partners are proportional to the chiral condensate(s), as expected. Using the parameters from Table \ref{Tab:params_eLSM_2013} and the masses of $K_2(1820)$ and $K_2^{*}(1430)$ from Table \ref{Tab:spin1-2_masses_PDG},  we obtain  the parameter $h_3^{\text{ten}}$ as: 
\begin{equation}\label{eq:h3}    h_3^{\text{ten}}=\frac{m_{K_{2A}}^2-m_{K_{2}}^2}{\sqrt{2}\phi_N\phi_S} \simeq - 41\,\text{ .}
\end{equation}

\begin{table}[ht!]
%[ptb]
\centering
\renewcommand{\arraystretch}{1.3} 
\begin{tabular}{|c|c|c|}
\hline
Mass squares  & Analytical expressions & Estimates of masses (MeV) \\ \hline\hline	$m_{\mathbf{a}_2}^2$ &  $\big( m^2_{2}+2\delta_N^{\text{ten}}\big)+\frac{ h_3^{\text{ten}}\phi_N^2}{2}+\frac{h_2^{\text{ten}}\phi_N^2}{2}+\frac{h_1^{\text{ten}}}{2}(\phi_N^2+\phi_S^2)$ & $\textbf{1317}$  \\
\hline
$ m_{K_{2}}^2$ & $\big( m^2_{2}+\delta_N^{\text{ten}}+\delta_S^{\text{ten}} \big) +\frac{h_1^{\text{ten}}}{2}(\phi_N^2+\phi_S^2)+\frac{h_3^{\text{ten}}}{\sqrt{2}} \phi_N\phi_S+\frac{h_2^{\text{ten}}}{4}(\phi_N^2+2\phi_S^2)$ & $\textbf{1427}$  
\\
\hline
$m_{f_{2N}}^2$ & $\big(m^2_{2}+2\delta_N^{\text{ten}}\big)+\frac{ h_3^{\text{ten}}\phi_N^2}{2}+\frac{h_1^{\text{ten}}}{2}(\phi_N^2+\phi_S^2)+\frac{h_2^{\text{ten}}\phi_N^2}{2}=m_{\mathbf{a}_2}^2$ & $1317$  
\\
\hline
$ m_{f_{2S}}^2$ & $\big( m^2_{2}+2\delta_S^{\text{ten}}\big) +\frac{h_1^{\text{ten}}}{2}(\phi_N^2+\phi_S^2)+ h_3^{\text{ten}}\phi_S^2+h_2^{\text{ten}}\phi_S^2$ & $1524$ 
\\
\hline
$ m_{\rho_2}^2$ & $\big( m^2_{2}+2\delta_N^{\text{ten}}\big)-\frac{ h_3^{\text{ten}}\phi_N^2}{2}+\frac{h_2^{\text{ten}}\phi_N^2}{2}+\frac{h_1^{\text{ten}}}{2}(\phi_N^2+\phi_S^2)$ & $1663$  
\\
\hline
$ m_{K_{2A}}^2$ & $\big( m^2_{2}+\delta_N^{\text{ten}}+\delta_S^{\text{ten}} \big)+\frac{h_1^{\text{ten}}}{2}(\phi_N^2+\phi_S^2)-\frac{h_3^{\text{ten}}}{\sqrt{2}} \phi_N\phi_S+\frac{h_2^{\text{ten}}}{4}(\phi_N^2+2\phi_S^2)$ & $\textbf{1819}$ 
\\
\hline
$m_{\omega_{2N}}^2$ & $\big( m^2_{2}+2\delta_N^{\text{ten}}\big)-\frac{ h_3^{\text{ten}}\phi_N^2}{2}+\frac{h_2^{\text{ten}}\phi_N^2}{2}+\frac{h_1^{\text{ten}}}{2}(\phi_N^2+\phi_S^2)=m_{\rho_2}^2$ & $1663$ 
\\
\hline
$m_{\omega_{2S}}^2$ & $\big( m^2_{2}+2\delta_S^{\text{ten}}\big)- h_3^{\text{ten}}\phi_S^2+h_2^{\text{ten}}\phi_S^2+\frac{h_1^{\text{ten}}}{2}(\phi_N^2+\phi_S^2)$ & $1971$
\\
\hline
	\end{tabular}%
\caption{ Masses of $2^{++}$ and $2^{--}$ mesons within the eLSM. Bold masses are taken from PDG.}\label{tab:mass-spin2}
\end{table}%
Our estimate for the (still experimentally missing) isovector member $\rho_2$ of the axial-tensor mesons is $1.67$ GeV, consistent with the prediction of Ref. \cite{isgur1985}. The other members of the axial-tensor nonet have masses below $2$ GeV.

\begin{figure}[ht!]
    \centering    \includegraphics[scale=0.75]{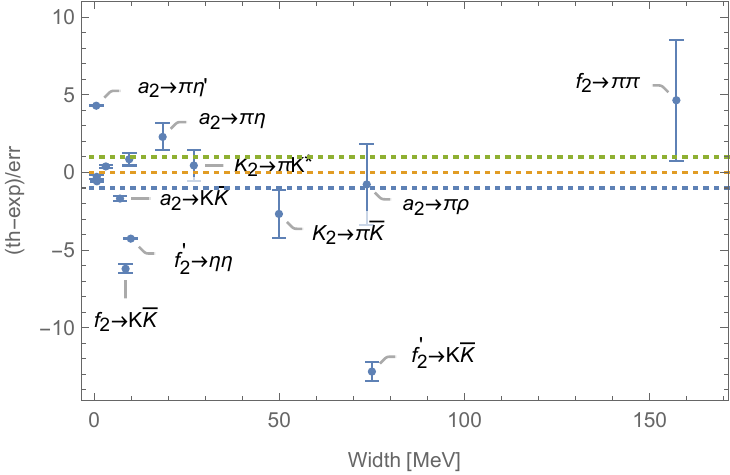}
    \caption{Results for tensor mesons within the eLSM \cite{Jafarzade:2022uqo}. The $x$-axis lists the decay width type, while the $y$-axis represents the deviations from the experimental value as compared to the experimental error. Vertical lines on the data points represent the fit errors. Those decay channels located between the dashed lines are described well within the model.}
    \label{fig:spin-2}
\end{figure}

The Lagrangian responsible for the decays of (axial-)tensor mesons reads:
\begin{align}\label{eq:int-spin2}    \mathcal{L}^{\text{int}}_{\textbf{L},\textbf{R}}=\frac{g_2^{\text{ten}}}{2}\Big(\text{Tr}\Big[ \mathbf{L}_{\mu\nu}\{ L^{\mu}, L^{\nu}\}\Big]+\text{Tr}\Big[\mathbf{R}_{\mu\nu} \{R^{\mu}, R^{\nu}\} \Big]\Big) +\frac{g_2^{\prime\,\text{ten}}}{6}\text{Tr}\Big[\mathbf{L}_{\mu\nu}+\mathbf{R}_{\mu\nu}\Big]\text{Tr}\Big[\{L^{\mu}, L^{\nu}\}+\{R^{\mu}, R^{\nu}\}\Big]\\\nonumber
+\frac{a^{\text{ten}}}{2}\text{Tr}\Big[\mathbf{L}_{\mu\nu}\{L^{\mu}_{\beta},L^{\nu\beta}\}+\mathbf{R}_{\mu\nu}\{R^{\mu}_{\beta},R^{\nu\beta}\}\Big]+\frac{a^{\prime\,\text{ten}}}{6}\text{Tr}\Big[\mathbf{L}_{\mu\nu}+\mathbf{R}_{\mu\nu}\big]\text{Tr}\Big[\{L^{\mu}_{\beta},L^{\nu\beta}\}+\{R^{\mu}_{\beta},R^{\nu\beta}\}\Big]\\\nonumber
+c_1^{\text{ten}}\,\text{Tr} \Big[ \partial^{\mu}\textbf{L}^{\nu\alpha} \tilde{L}_{\mu\nu}\,\partial_{\alpha}\Phi\,\Phi^\dagger-\partial^{\mu}\textbf{R}^{\nu\alpha} \Phi^\dagger\,\partial_{\alpha}\Phi \tilde{R}_{\mu\nu}- \partial^{\mu}\textbf{R}^{\nu\alpha} \tilde{R}_{\mu\nu}\partial_{\alpha}\Phi^\dagger \Phi +\partial^{\mu}\textbf{L}^{\nu\alpha}\Phi \partial_{\alpha} \Phi^\dagger \tilde{L}_{\mu\nu}\Big]\\\nonumber
   + c_2^{\text{ten}}\,\text{Tr} \Big[ \partial^{\mu}\textbf{L}^{\nu\alpha}\partial_{\alpha}\Phi \tilde{R}_{\mu\nu}\,\,\Phi^\dagger-\partial^{\mu}\textbf{R}^{\nu\alpha} \Phi^\dagger\,\tilde{L}_{\mu\nu}\partial_{\alpha}\Phi - \partial^{\mu}\textbf{R}^{\nu\alpha} \partial_{\alpha}\Phi^\dagger\tilde{L}_{\mu\nu} \Phi +\partial^{\mu}\textbf{L}^{\nu\alpha}\Phi\tilde{R}_{\mu\nu} \partial_{\alpha} \Phi^\dagger \Big]
   \text{.}
\end{align}
This Lagrangian enables us to test eLSM for the tensor mesons and estimate the decay rates for the axial-tensor mesons. Its parameters with their fit values and large-$N_c$ scaling properties are listed in Table \ref{tab:spin2-parameters}.
\begin{table}[ht!]
\centering
		\renewcommand{\arraystretch}{1.}
\begin{tabular}{|c|c|c|}
\hline
Parameters & Numerical values & $N_c$ scaling \\ \hline
   $h_3^{\text{ten}}$        &   $-41$               & $N_c^{-1}$              \\ \hline
         $\delta_S^{\text{ten}}$  &   $0.3\,\text{GeV}^2$               & $N_c^{0}$             \\ \hline
       $g_{2}^{\text{ten}}$    & $(1.392\pm0.024)\cdot10^{4}(\text{MeV})$                 &        $N_c^{-\frac{1}{2}}$     \\ \hline
         $g_{2}^{\prime\,\text{ten}}$  &    $(0.024\pm0.041)\cdot10^{4}(\text{MeV})$              &    $N_c^{-\frac{3}{2}}$         \\ \hline
$c^{\text{ten}}\equiv  c_1^{\text{ten}}+c_2^{\text{ten}}$           &   $(4.8\pm0.9)\cdot10^{-7}(\text{MeV})^{-3}$               & $N_c^{-\frac{1}{2}}$             \\ \hline
$a^{\text{ten}}$           &   $(-2.09\pm 0.06)\cdot 10^{-2}(\text{MeV})^{-1}$               &  $N_c^{-\frac{1}{2}}$          \\ \hline
    $a^{\prime\,\text{ten}}$       &   $(3.5\pm 0.4)\cdot 10^{-3}(\text{MeV})^{-1}$               & $N_c^{-\frac{3}{2}}$          \\ \hline
\end{tabular}
\caption{Parameters of the spin-2 homochiral eLSM Lagrangian(s) describing (axial-)tensor mesons. }
\label{tab:spin2-parameters}
\end{table}
Results for the dominant decay channels of the tensor mesons are compared to the PDG values in Table \ref{tab:spin2-decays}. This comparison is also illustrated in Figure \ref{fig:spin-2}.  Apart from some specific channels like $f_2^\prime(1525)\rightarrow \overline{K}K$ that deserve further clarification, most of the results are in qualitative agreement with the PDG. The deviations can be caused by large-$N_c$ suppressed terms, isospin breaking, and uncertainties in the resonance masses, which are not included in our calculations. Note, the results are also quantitatively in agreement with one obtained in the flavor invariant model of Ref. \cite{Giacosa:2005bw}.

\begin{table}[ht!]
%[ptb]
\centering
\renewcommand{\arraystretch}{1.} 
\begin{tabular}{|c|c|c|c|c|c|}
\hline
Decay process  & eLSM  (MeV)   &  PDG  (MeV)& Decay process  & eLSM  (MeV)   &  PDG  (MeV)\\ \hline\hline
	$\,\;a_2(1320) \rightarrow  \bar{K}\, K  $ &  $4.06\pm 0.14$  & $7.0^{+2.0}_{-1.5}$
&
	$\,\;a_2(1320) \rightarrow \pi\,\eta   $  &  $25.37\pm 0.87$   & $18.5\pm 3.0$ 
\\ \hline
$\,\;a_2(1320) \rightarrow  \pi\,\eta^\prime (958)  $ &  $1.01\pm 0.03$   & $0.58\pm 0.10$
&
	$\,\;K_2^\ast(1430) \rightarrow \pi\, \bar{K}   $ &  $44.82\pm 1.54$    & $49.9\pm 1.9$\\ \hline
	$\,\;f_2(1270) \rightarrow \bar{K}\, K $ &  $3.54\pm 0.29$    & $8.5\pm 0.8 $ 
&
	$\,\;f_2(1270) \rightarrow  \pi\,\pi$  &  $168.82\pm 3.89$   &  $157.2^{+4.0}_{-1.1}$  
\\ \hline
$\,\;f_2(1270) \rightarrow \eta \, \eta$  &  $0.67\pm0.03$ &    $0.75\pm 0.14 $ 
&
$\,\;f_2^\prime(1525) \rightarrow \bar{K}\, K   $  &  $23.72\pm 0.60$   & $75\pm 4$
\\ \hline
	$\,\;f_2^\prime(1525) \rightarrow  \pi\,\pi   $   &  $0.67\pm 0.14$  &  $0.71\pm 0.14$
&
$\,\;f_2^\prime(1525) \rightarrow \eta \, \eta $   &  $1.81\pm 0.05$   & $9.9\pm 1.9 $
\\ \hline
$\,\;a_2(1320) \rightarrow  \rho(770)\,\pi $ &  $71.0\pm 2.6$ &  $73.61\pm 3.35$ &
$\,\;K_2^\ast(1430) \rightarrow \bar{K}^{\ast}(892)\,\pi   $ &  $27.9\pm 1.0$  &  $26.92 \pm 2.14$ \\ \hline
	$\,\;K_2^\ast(1430) \rightarrow \rho(770)\,K   $ &  $10.3\pm 0.4$  &  $9.48\pm 0.97$ &
	$\,\;K_2^\ast(1430) \rightarrow \omega(782)\,\bar{K}   $ &  $3.5\pm 0.1$  &  $3.16\pm 0.88$   \\\hline
$\,\;f_2^\prime(1525) \rightarrow \bar{K}^{\ast}(892)\,K+\mathrm{c}%
.\mathrm{c}. $ & $19.89\pm0.73$  &     &&&
\\ \hline
\end{tabular}%
\caption{Decay rates of the ground-state $2^{++}$ resonances.
}\label{tab:spin2-decays}
\end{table}
Chiral symmetry relates chiral partners, in this specific case the well-known tensor mesons to the poorly known axial-tensor ones.
Here, within the eLSM we use the parameters in Table \ref{tab:spin2-parameters} to estimate the decays of the axial-tensor mesons. The results are presented in Table \ref{tab:axial-tensor decays} and are compared to the lattice QCD ones of Ref. \cite{Johnson:2020ilc} in Table \ref{tab:axial-Lattice}, showing that they are qualitatively in agreement with each other. 
We observe that, even when taking the lower limit for the decay rates in Table \ref{tab:axial-tensor decays}, the ground-state axial-tensor mesons are very broad. This can be a viable and rather simple explanation for the missing status of $\rho_2$, $\omega_{2N}$ and $\omega_{2S} \sim \phi_2$ in the list of Table \ref{Tab:spin1-2_masses_PDG}: they might be too broad to be measured in experiments. 
 \begin{table}[ht!] %[ptb]
		\centering
		\renewcommand{\arraystretch}{1.}
		\begin{tabular}[c]{|c|c|c|c|}
			\hline
			Decay process  & eLSM (MeV) & Decay process  & eLSM (MeV)  \\
			\hline \hline
			$\,\;\rho_2(?) \rightarrow \rho(770)\, \eta$ & $ \approx 99\pm 50$ & 
			$\,\;\rho_2(?) \rightarrow \bar{K}^\ast(892)\, K + \mathrm{c}.\mathrm{c}.$ & $  \approx 85\pm 43$   \\
			\hline
			$\,\;\rho_2(?) \rightarrow \omega(782)\, \pi$ & $ \approx 419\pm 210 $ &
			$\,\;\rho_2(?) \rightarrow \phi(1020)\, \pi$ & $ \approx 0.8 $ \\
			\hline 
			$\,\;K_{2A} \rightarrow \rho(770)\, K$ & $ \approx 195\pm 98$ &
			$\,\;K_{2A} \rightarrow \bar{K}^\ast(892)\, \pi $ & $ \approx 316\pm 158 $  \\
			\hline
			$\,\;K_{2A} \rightarrow \bar{K}^\ast(892)\, \eta$ & $ \approx0.01$ 
		&			$\,\;K_{2A} \rightarrow \omega(782)\, \bar{K}$ & $ \approx 51\pm 26$ \\
			\hline
			$\,\;K_{2A} \rightarrow \phi(1020)\, \bar{K}$ & $  \approx 50\pm 25$ &
			$\,\;\omega_{2N} \rightarrow \rho(770)\, \pi$ & $ \approx 1314\pm 657 $  \\
			\hline
			$\,\;\omega_{2N} \rightarrow \bar{K}^\ast(892)\, K + \mathrm{c}.\mathrm{c}.$ & $ \approx 85\pm 43 $ & 
			$\,\;\omega_{2N} \rightarrow \omega(782)\, \eta$ & $ \approx 93\pm 47$  \\
			\hline
			$\,\;\omega_{2N} \rightarrow \phi(1020)\, \eta$ & $ \approx 0.06 $  &$\,\;\omega_{2S} \rightarrow \omega(782)\, \eta^\prime(958)$ & $ \approx 0.3 $  \\
			 \hline
			$\,\;\omega_{2S} \rightarrow \bar{K}^\ast(892)\, K + \mathrm{c}.\mathrm{c}.$ & $ \approx 510\pm 255$  & 
			$\,\;\omega_{2S} \rightarrow \omega(782)\, \eta$ & $ \approx 1.0\pm0.5 $  \\
			\hline
			$\,\;\omega_{2S} \rightarrow \phi(1020)\, \eta$ & $ \approx 101\pm 51 $ & & \\
			\hline
		\end{tabular}
		\caption{Esimates of the axial-tensor meson decays.}\label{tab:axial-tensor decays}
		\end{table}
%%%%%%%%%%%%%%%%%%%%%%%%%%%%%%%%%
%We refer to Ref. \cite{Jafarzade:2022bnf} for the decays of other axial mesons.
\begin{table}[ht!] %[ptb]
		\centering
		\renewcommand{\arraystretch}{1.1}
		\begin{tabular}[c]{|c|c|c|c|c|c|}
			\hline
			Decay process  & eLSM  (MeV) & LQCD (MeV) & Decay process  & eLSM  (MeV) & LQCD (MeV)  \\
			\hline \hline
			$\,\;\rho_2(?) \rightarrow \rho(770)\, \eta$ & $ \approx 30$ & $ $   &
			$\,\;\rho_2(?) \rightarrow \bar{K}^\ast(892)\, K + \mathrm{c}.\mathrm{c}.$ & $  \approx 27$ & $36$  \\
			\hline
			$\,\;\rho_2(?) \rightarrow \omega(782)\, \pi$ & $ \approx 122 $ & $125$ &
			$\,\;\rho_2(?) \rightarrow \phi(1020)\, \pi$ & $ \approx 0.3 $ & $ $ \\
			\hline 
			$\,\;K_{2A} \rightarrow \rho(770)\, K$ & $ \approx 53$ & $ $ &
			$\,\;K_{2A} \rightarrow \bar{K}^\ast(892)\, \pi $ & $ \approx 87 $ & $ $ \\
			\hline
			$\,\;K_{2A} \rightarrow \bar{K}^\ast(892)\, \eta$ & $ \approx0.004$ & $ $ &
			$\,\;K_{2A} \rightarrow \omega(782)\, \bar{K}$ & $ \approx 13.8$ & $ $ \\
			\hline
			$\,\;K_{2A} \rightarrow \phi(1020)\, \bar{K}$ & $  \approx 13.7$ & $ $ &
			$\,\;\omega_{2N} \rightarrow \rho(770)\, \pi$ & $ \approx 363 $  & $365$ \\
			\hline
			$\,\;\omega_{2N} \rightarrow \bar{K}^\ast(892)\, K + \mathrm{c}.\mathrm{c}.$ & $ \approx 25 $ & $36$  &
			$\,\;\omega_{2N} \rightarrow \omega(782)\, \eta$ & $ \approx 27$ & $17$ \\
			\hline
			$\,\;\omega_{2N} \rightarrow \phi(1020)\, \eta$ & $\approx 0.02 $ & $  $ &
					$\,\;\omega_{2S} \rightarrow \bar{K}^\ast(892)\, K + \mathrm{c}.\mathrm{c}.$ & $ \approx 100$  & $148$  \\
			\hline
			$\,\;\omega_{2S} \rightarrow \omega(782)\, \eta$ & $ \approx 0.2 $ & $ $ &
			$\,\;\omega_{2S} \rightarrow \omega(782)\, \eta^\prime(958)$ & $ \approx 0.02 $ & $ $ \\
			\hline
			$\,\;\omega_{2S} \rightarrow \phi(1020)\, \eta$ & $ \approx 17 $ & $44$ & & &  \\
			\hline
		\end{tabular}
		\caption{ Decay rates of axial-tensor mesons fitted to LQCD Ref. \cite{Johnson:2020ilc}. }\label{tab:axial-Lattice}
		\end{table}			
%\bigskip

\subsection{Heterochiral mesons with \texorpdfstring{$J=2$}{J2}}

The spin-2 heterochiral mesons (heterotensors) contain the rather well-established pseudotensor mesons described by the nonet $P_2^{\mu\nu}$ \eqref{eq:nonet-pseduotensor} and their, up to now missing, chiral partners, the orbitally excited tensor mesons. For this reason, a joint detailed study of masses and decays of both nonets could not yet be undertaken.

The mass term for the heterochiral spin-2 mesons can be constructed by following the standard procedure: 
\begin{align}
\mathcal{L}_{\Phi_{\mu\nu}}^{\text{mass}}=&\text{Tr}\Big[\Big(\frac{m_{2^\ast}^{ 2}\,G^2}{2\,G_0^2}+\Delta^{\text{pt}}\Big)\Big(\Phi
_{\mu\nu}^{\dagger}\Phi^{\mu\nu}\Big)\Big]+\frac{\lambda_{\Phi_{\mu\nu}\,,1}}{2}\text{Tr}\Big[\Phi^\dagger \Phi\Big]\text{Tr}\Big[\Phi_{\mu\nu}^\dagger \Phi^{\mu\nu}\Big]+\lambda_{\Phi_{\mu\nu}\,,2}\text{Tr}\Big[\Phi_{\mu\nu}^{\dagger}\Phi
\Phi^{\mu\nu\dagger}\Phi+\Phi_{\mu\nu}\Phi^{\dagger}
\Phi^{\mu\nu}\Phi^\dagger\Big]\\\nonumber
&+\lambda_{\Phi_{\mu\nu}\,,3}\text{Tr}\Big[\Phi
_{\mu\nu}\Phi^{\dagger}\Phi\Phi^{\mu\nu\,\dagger}+\Phi
_{\mu\nu}^{\dagger}\Phi\Phi^\dagger\Phi^{\mu\nu}\Big] \text{ ,}
\end{align}
where the parameter $\Delta^{\text{pt}}$ describes, as usual, a mass-driven contribution. 
Due to the poor knowledge of the orbitally excited tensor mesons, this Lagrangian has been not yet used to calculate masses. 

Next, we turn to additional interactions that lead to decays of heterotensors via chirally invariant under $U(3)_L \times U(3)_R$, giving:
 \begin{align}   \label{eq:lag-hetero-chiral-2} \mathcal{L}^{\text{int}}_{\Phi^{\mu\nu}}=g^{(1)}_{\Phi^{\mu\nu}}\text{Tr}\Big[\Phi^{\mu\nu}\,L_{\mu} \Phi^\dagger R_{\nu}+\text{c.c}\Big]+g^{(2)}_{\Phi^{\mu\nu}}\text{Tr}\Big[\Phi^{\mu\nu}\,\mathbf{L}_{\mu\nu} \Phi^\dagger +\text{c.c}\Big]\, \text { .}
\end{align}
For pseudotensor mesons only, it reduces to the following Lagrangian
\begin{equation}    \mathcal{L}_{P_2}=g_{p_2vp}\text{Tr}\Bigg[P_{2}^{\mu\nu}\Big[V_{\mu},\partial_{\nu}P\Big]\Bigg]+g_{p_2t_2p}\text{Tr}\Bigg[P_{2}^{\mu\nu}\Big\{T_{\mu\nu},P\Big\}\Bigg]\, \text{ ,}
\end{equation}
whose decay properties were studied in Ref. \cite{Koenigstein:2016tjw},  see also Table \ref{tab:pseudotensor}. The coupling constants $g_{p_2vp}$ and $g_{p_2t_2p}$ were fixed using the PDG value for  $\pi_2(1670) \rightarrow \rho(770)\, \pi$ and $K_2(1770) \rightarrow \bar{K}_2^\ast(1430) \pi$, respectively. The other entries of table \ref{ab:pseudotensor} were not yet reported as branching rations in the PDG, see however Ref. \cite{Koenigstein:2016tjw} for details.  
\begin{table}[ht!] %[ptb]
		\centering		\renewcommand{\arraystretch}{1.}
		\begin{tabular}[c]{|c|c|c|c|}
			\hline
			Decay process & theory (MeV) & Decay process & theory (MeV)   
   \\
			\hline 
   $\pi_2(1670) \rightarrow \rho(770)\, \pi$ &  $80.6\pm10.8$ 
  &
    $\pi_2(1670) \rightarrow \bar{K}^\ast(892)\, K +\textbf{c}.\mathrm{c}$ &  $11.7\pm 1.6$ 
   \\\hline    
   $K_2(1770) \rightarrow \rho(770) K$ &  $22.2\pm 3.0$ 
  & 
    $K_2(1770) \rightarrow \bar{K}^\ast(892) \pi$ &  $25.5\pm 3.4$ 
   \\\hline 
   $K_2(1770) \rightarrow \bar{K}^\ast(892) \eta$ &  $20.5\pm 1.4$ 
   & 
   $K_2(1770) \rightarrow \omega(782) K$ &  $8.3\pm 1.1$ 
   \\\hline 
    $K_2(1770) \rightarrow \phi(1020) K$ &  $4.2\pm 0.6$ & $\pi_2(1670) \rightarrow f_{2}(1270)\pi$ &  $146.4 \pm 9.7$ 
   \\\hline 
   $K_2(1770) \rightarrow \bar{K}_2^\ast(1430) \pi$ &  $84.5 \pm 5.6$  &&\\\hline
		\end{tabular}
		\caption{Decay rates of pseudotensor mesons estimated in Ref. \cite{Koenigstein:2016tjw}.}
		\label{tab:pseudotensor}
				\end{table}
%%%%%%%%%%%%%%%%%%
Next, since this chiral multiplet is heterochiral, we can use the `GPKJ' mathematical object defined in Eq. (\ref{epsilon}) to write down
$U(1)_A$ anomalous but chirally invariant terms. Following Refs. \cite{Giacosa:2017pos,Giacosa:2023fdz}, one has: 
\begin{equation}
    \mathcal{L}^{U(1)_A-\text{anom}}_{\Phi^{\mu\nu}}= -b_{\Phi^{\mu\nu}} \Big(\epsilon\Big[ \big(\partial_\mu \Phi\big), \Phi_{ \nu}\,,\Phi^{\mu\nu}\Big]+\text{c.c.}\Big)- c_{\Phi^{\mu\nu}} \Big(\epsilon\Big[\big(\partial_\mu\Phi\big),\big(\partial_\nu \Phi\big),\Phi^{\mu\nu}\Big]+\text{c.c.}\Big)\,.
\end{equation}
The anomalous couplings $b_{\Phi^{\mu\nu}}$ and $c_{\Phi^{\mu\nu}}$ and corresponding decay rates are estimated within DIG in Ref. \cite{Giacosa:2023fdz}. 
The anomalous interaction also allows to estimate the isoscalar mixing angle $\beta_{pt}$, that turns out to be negative and quite small (smaller than $\beta_{pv}$), in agreement with lattice QCD results \cite{Dudek:2013yja}. Note, based on a study of decay ratios, $\beta_{pt}$ was estimated to be \textit{large} in Ref. \cite{Koenigstein:2016tjw} $\beta_{pt}=-42^\circ$ . Such a large mixing has been later on confirmed in the fit of Ref. \cite{Shastry:2021fsk}: for what concerns this important point, a tension between different results is present and novel studies are needed.
%\footnote{We do not present those results since they are much smaller than the results in Table \ref{tab:pseudotensor} due to the unknown parameter $M_2$ linking the spin-2 heterochiral mesons and quark fields. For more details see Ref. \cite{Giacosa:2023fdz}.} 

\section{Hybrid mesons within the eLSM}
\label{Sec:hybrids}

In this section we focus on the ground-state hybrid mesons with the exotic quantum numbers $J^{\mathcal{P}\mathcal{C}}=1^{-+}$ and their chiral partners, the pseudovector hybrid mesons $J^{\mathcal{P}\mathcal{C}}=1^{+-}$, embedded into a homochiral multiplet (hybrid homovectors). Following \cite{Eshraim:2020ucw}, we couple them to the eLSM and study their masses and decay processes. Thus, this chapter shows that the eLSM can be applied to non-conventional hybrid mesons in pretty much the same way as for regular $\bar{q}q$ multiples.

\subsection{Hybrid mesons: basic information}
The existence of hybrid mesons was proposed long ago in the context of the constituent gluon model \cite{Horn:1977rq} as an excitation of the gluon string. The decays of hybrids into conventional mesons was described by the breaking of this string \cite{Isgur:1985vy, Kokoski:1985is}. See also \cite{Gross:2022hyw} for historical and recent phenomenological details \cite{Swanson:2023zlm,Farina:2023oqk}.

The present experimental status of 
the hybrid mesons can be summarized as follows: up to the PDG 2022 (and 2023 update) two isovector states with $J^{\mathcal{P}\mathcal{C}}=1^{-+}$ were reported, the state $\pi_{1}(1400)$ found by the E852 collaboration at Brookhaven \cite{E852:1997gvf,E852:1999xev}, and $\pi_{1}(1600)$ observed also by E852
\cite{E852:2001ikk,E852:1998mbq,Kuhn:2004en,E852:2004rfa} and by the COMPASS
collaboration \cite{COMPASS:2009xrl,COMPASS:2018uzl}. 

However, according to lattice calculations, only one hybrid state should appear close to $1.5$~GeV \cite{Meyer:2015eta,Dudek:2010wm,Woss:2020ayi}. This is also the case for models, such as holographic QCD approaches \cite{Kim:2008qh,Bellantuono:2014lra}.
%which would favor a $\pi_{1}(1400)$ hybrid state.
The analysis of Ref. \cite{JPAC:2018zyd} has shown a way out of this puzzle: the resonances $\pi_{1}(1400)$ and $\pi_{1}(1600)$ correspond to the same pole, and thus to the same physical state, whose mass is closer to $\pi_{1}(1600)$. This view is confirmed by a simple observation (PDG 2022 \cite{Workman:2022ynf}): $\pi_{1}(1400)$ was observed only in $\eta\pi,$ while $\pi_{1}(1600)$ was observed in  $f_1(1285)\pi$ \cite{Kuhn:2004en}, $b_1(1235)\pi$ \cite{E852:2004rfa,Baker:2003jh}, $\eta^\prime \pi$ \cite{Kuhn:2004en}, $\rho\pi$ \cite{COMPASS:2018uzl}, $\eta\pi+ \eta^\prime \pi$ \cite{Kopf:2020yoa}. In the latest version of PDG 2024, the entry $\pi_{1}(1400)$ has been removed from the summary table \cite{ParticleDataGroup:2024cfk}. %%%%%%%%%%%%%%%%
The decay ratios reported by Refs. \cite{Kopf:2020yoa,E852:2004gpn} read:
 \begin{align}
                &\frac{\pi_1(1600\to\pi\eta^{\prime})}{\pi_1(1600\to\pi\eta)}=5.54\pm 1.10^{+1.80}_{-0.27} \text{ ,} \\             &\frac{\pi_1(1600\to\pi f_1(1285))}{\pi_1(1600\to\pi\eta^{\prime})}=3.8\pm 0.78 \text{ ,} 
            \end{align}
where the first decay ratio implies the pseudoscalar mixing angle $\beta_{ps} \approx-15^\circ$ under the assumption that the hybrid state couples to the flavor-singlet pseudoscalar configuration  (according to the relation $\Gamma(\pi_1(1600)\to\pi\eta^{\prime})/\Gamma(\pi_1(1600)\to\pi\eta) = q_{\eta^{\prime}}/q_{\eta}\tan^2(\beta_{ps} +35.3^{\circ} )$).

Recently, in Ref. \cite{BESIII:2022riz} the BESIII experiment reported the existence of an isoscalar hybrid state, denoted $\eta_{1}(1855)$.
%where the analyses are based only on the Breit-Wigner function fit (which can be improved by performing the partial wave analyses on the $\eta\eta^\prime$ peak). 
It is natural to include this state in a nonet of hybrid states, but one needs to keep in mind that an experimental confirmation is required\footnote{This state is not yet included in the PDG summary.}.  Due to its larger mass, it can be considered as a predominantly $\bar{s}sg$ state. In Refs. \cite{Shastry:2022mhk,Shastry:2023ths}, a detailed study of the phenomenology of the whole nonet is presented, including strong and radiative decays, a kaonic hybrid state with a mass of $1.75$~GeV, and a light $\eta_{1}$ (mostly nonstrange) resonance with a mass of about $1.66$ GeV.

\subsection{Masses of hybrid mesons in the eLSM}

The eLSM chiral Lagrangian term describing the mass of the homochiral hybrid  mesons reads:
\begin{align}    \mathcal{L}^{\text{mass}}_{\text{hyb}}=\text{Tr}\Big[&\Big(\frac{m_{\text{hyb}}^{2}\,G^2}{2\,G_0^2}+\Delta^{\text{hyb}}\Big)\Big(L_{\mu}^{\text{hyb}\,2}+R_{\mu}^{\text{hyb}\,2}\Big)\Big]
   + \frac{h_1^{\text{hyb}}}{2}\text{Tr}\Big[\Phi^\dagger \Phi\Big]\text{Tr}\Big[\Big(L_{\mu}^{\text{hyb}\,2}+R_{\mu}^{\text{hyb}\,2}\Big)\Big]  \\\nonumber
  & +h_2^{\text{hyb}}\text{Tr}\Big[\Phi^\dagger L_{\mu}^{\text{hyb}} L^{\mu\,,\text{hyb}} \Phi+\Phi R_{\mu}^{\text{hyb}} R^{\mu\,,\text{hyb}}\Phi^\dagger\Big]+2 h_3^{\text{hyb}}\text{Tr}\Big[\Phi R^{\mu\,,\text{hyb}}\Phi^\dagger L_{\mu}^{\text{hyb}} \Big]
   \text{ ,}
\end{align}
where $\Delta^{\text{hyb}}=\text{diag}(\delta_N^{\text{hyb}},\delta_N^{\text{hyb}},\delta_S^{\text{hyb}})$ is the standard contribution due to bare quark masses with 
$\delta_{N,S}^{\text{hyb}} \propto m_{n,s}^2$. As usual, we set $\delta_N^{\text{hyb}} = 0$. We include $\pi_1(1600)$ in the  $J^{\mathcal{P}\mathcal{C}}=1^{-+}$ nonet and we set the mass of the chiral partner $b_1^{\text{hyb}}$ at about $2\, \text{GeV}$. Then, using the previously determined eLSM parameters from Table \ref{Tab:params_eLSM_2013} (and fixing $\delta^{\text{hyb}}_S \simeq \delta_S = 0.151 \text{GeV}^2$) we obtain the hybrid masses in Table \ref{tab:mass-hybrid}. In particular, the following relations between chiral partners hold:
\begin{equation}
m_{b_{1}^{\text{hyb}}}^{2}  = m_{\pi_{1}^{\text{hyb}}}^{2}-2h_{3}^{\text{hyb}}\phi_{N}%
^{2} \text{ , }
m_{K_{1B}^{\text{hyb}}}^{2} = m_{K_{1}^{\text{hyb}}}^{2}-\sqrt{2}\phi_{N}\phi_{S}%
h_{3}^{\text{hyb}} \text{ , }
m_{h_{1S}^{\text{hyb}}}^{2} =m_{\eta_{1S}^{\text{hyb}}}^{2} -h_{3}^{\text{hyb}}\phi_{S}%
^{2} \text{ .} 
\label{hcp2}
\end{equation}
They are, as usual, proportional to the chiral condensates and the parameter $h_3^{\text{hyb}}$. Note, the heavy $\bar{s}sg$ hybrid $\eta_{1S}^{\text{hyb}}$ with a mass of about 1.75 GeV is part of the lightest hybrid nonet. In 2022, the resonance $\eta_1(1855)$ with exotic quantum numbers $J^{\mathcal{P}\mathcal{C}} =1^{-+}$ has been discovered by the BESIII collaboration in Ref. \cite{BESIII:2022riz}. Just shortly after this discovery, in Ref. \cite{Shastry:2022mhk} it was shown that, when a small but nonnegligible mixing in the isoscalar sector is considered, the resonance $\eta_{1}(1855)$ can be interpreted as predominantly $\eta_{1S}^{\text{hyb}} = \bar{s}sg$. 
\begin{table}[ht!]
%[ptb]
\centering
\renewcommand{\arraystretch}{1.3} 
\begin{tabular}{|c|c|c|}
\hline
Mass squares  & Analytical expressions & Estimates of masses (MeV) \\ \hline\hline	$m_{\pi_1^{\text{hyb}}}^2$ &  $\big( m^2_{\text{hyb}}+2\delta_N^{\text{hyb}}\big)+\frac{ h_3^{\text{hyb}}\phi_N^2}{2}+\frac{h_2^{\text{hyb}}\phi_N^2}{2}+\frac{h_1^{\text{hyb}}}{2}(\phi_N^2+\phi_S^2)$ & $\textbf{1660}$ \\
\hline
$ m_{K_1^\text{hyb}}^2$ & $\big( m^2_{\text{hyb}}+\delta_N^{\text{hyb}}+\delta_S^{\text{hyb}} \big) +\frac{h_1^{\text{hyb}}}{2}(\phi_N^2+\phi_S^2)+\frac{h_3^{\text{hyb}}}{\sqrt{2}} \phi_N\phi_S+\frac{h_2^{\text{hyb}}}{4}(\phi_N^2+2\phi_S^2)$ & $1707$ 
\\
\hline
$m_{\eta_{1N}^{\text{hyb}}}^2$ & $\big(m^2_{\text{hyb}}+2\delta_N^{\text{hyb}}\big)+\frac{ h_3^{\text{hyb}}\phi_N^2}{2}+\frac{h_1^{\text{hyb}}}{2}(\phi_N^2+\phi_S^2)+\frac{h_2^{\text{hyb}}\phi_N^2}{2}=m_{\pi_1^{\text{hyb}}}^2$ & $1660$ 
\\
\hline
$ m_{\eta_{1S}^{\text{hyb}}}^2$ & $\big( m^2_{\text{hyb}}+2\delta_S^{\text{hyb}}\big) +\frac{h_1^{\text{hyb}}}{2}(\phi_N^2+\phi_S^2)+ h_3^{\text{hyb}}\phi_S^2+h_2^{\text{hyb}}\phi_S^2$ & $1751$ 
\\
\hline
$ m_{b_1^{\text{hyb}}}^2$ & $\big( m^2_{\text{hyb}}+2\delta_N^{\text{hyb}}\big)-\frac{ h_3^{\text{hyb}}\phi_N^2}{2}+\frac{h_2^{\text{hyb}}\phi_N^2}{2}+\frac{h_1^{\text{hyb}}}{2}(\phi_N^2+\phi_S^2)$ & $\textbf{2000}$ 
\\
\hline
$ m_{K_{1B}^{\text{hyb}}}^2$ & $\big( m^2_{\text{hyb}}+\delta_N^{\text{hyb}}+\delta_S^{\text{hyb}} \big)+\frac{h_1^{\text{hyb}}}{2}(\phi_N^2+\phi_S^2)-\frac{h_3^{\text{hyb}}}{\sqrt{2}} \phi_N\phi_S+\frac{h_2^{\text{hyb}}}{4}(\phi_N^2+2\phi_S^2)$ & $2063$ 
\\
\hline
$m_{h_{1N}^{\text{hyb}}}^2$ & $\big( m^2_{\text{hyb}}+2\delta_N^{\text{hyb}}\big)-\frac{ h_3^{\text{hyb}}\phi_N^2}{2}+\frac{h_2^{\text{hyb}}\phi_N^2}{2}+\frac{h_1^{\text{hyb}}}{2}(\phi_N^2+\phi_S^2)=m_{b_1^{\text{hyb}}}^2$ & $2000$ 
\\
\hline
$m_{h_{1S}^{\text{hyb}}}^2$ & $\big( m^2_{\text{hyb}}+2\delta_S^{\text{hyb}}\big)- h_3^{\text{hyb}}\phi_S^2+h_2^{\text{hyb}}\phi_S^2+\frac{h_1^{\text{hyb}}}{2}(\phi_N^2+\phi_S^2)$ & $2126$ 
\\
\hline
\end{tabular}%
\caption{Masses of hybrid mesons within the eLSM. The bold entries are input.}
\label{tab:mass-hybrid}
\end{table}%
%%%%%%%%%%%%%%%%%%%%%%%
\begin{table}[ht!]
\centering
\renewcommand{\arraystretch}{1.1}
\begin{tabular}{|c|c|c|c|}
\hline
Parameters & $N_c$-scale & Parameters & $N_c$-scale \\ \hline
   $h_{1,2,3}^{\text{hyb}}$                      & $N_c^{-2}$             &
         $m_{\text{hyb}}$           & $N_c^{0}$             \\ \hline
       $\lambda_{1,2}^{\text{hyb}}$        &        $N_c^{-\frac{1}{2}}$     &
         $\alpha^{\text{hyb}}$         &    $N_c^{-1}$         \\ \hline
$\beta^{\text{hyb}}$                         & $N_c^{-1}$       & &      \\ \hline
\end{tabular}
\caption{Parameters of the hybrid mesons within the eLSM. }
\label{tab:hybrid-parameters}
\end{table}

\subsection{Decays of hybrid mesons in the eLSM}
The resonance $\pi_1(1600)$ was observed in the decay channels  $\pi_1\rightarrow \rho \pi, b_1\pi, f_1\pi, \eta^\prime\pi, \eta\pi $ \cite{ParticleDataGroup:2024cfk}. 
Following the usual steps, an eLSM  chiral Lagrangian that leads to these decay channels can be easily constructed: 
\begin{align}\label{eq:hyb-int}    \mathcal{L}^{\text{int}}_{\text{hyb}}&=i\lambda_1^{\text{hyb}}G\text{Tr}\Big[ L_{\mu}^{\text{hyb}}\Big(\Phi^{\mu}\Phi^\dagger-\Phi\Phi^{\mu\dagger}\Big)+R_{\mu}^{\text{hyb}}\Big(\Phi^{\mu\dagger}\Phi-\Phi^\dagger\Phi^{\mu}\Big)\Big]\\\nonumber
   & +i\lambda_2^{\text{hyb}}\text{Tr}\Big[ [L_{\mu}^{\text{hyb}},L^{\mu}]\Big(\Phi \Phi^\dagger\Big)+[R_{\mu}^{\text{hyb}}R^{\mu}]\Big(\Phi^{\dagger}\Phi\Big)\Big]\\\nonumber    &+\alpha^{\text{hyb}}\text{Tr}\Big[\Tilde{L}_{\mu\nu}^{\text{hyb}}\Phi R^{\mu\nu}\Phi^\dagger-\Tilde{R}_{\mu\nu}^{\text{hyb}}\Phi^\dagger L^{\mu\nu}\Phi\Big]\\\nonumber
   & +\beta^{\text{hyb}}_A\Big(\det\Phi-\det\Phi^\dagger\Big)\text{Tr}\Big[L_{\mu}^{\text{hyb}}\Big(\partial^{\mu}\Phi\cdot\Phi^\dagger-\Phi\cdot\partial^{\mu}\Phi^\dagger\Big)-R_{\mu}^{\text{hyb}}\Big(\partial^{\mu}\Phi^\dagger\cdot\Phi-\Phi^\dagger\cdot\partial^{\mu}\Phi\Big)\Big]\text{ ,}
\end{align}
where the terms proportional to $\lambda_1^{\text{hyb}}$ and $\lambda_2^{\text{hyb}}$ are dilatation invariant, but the one proportional to $\alpha^{\text{hyb}}$ (involving the Levi-Civita tensor) and the axial-symmetry anomalous term proportional to $\beta^{\text{hyb}}_A$ break it (see Table \ref{tab:hybrid-parameters} for the $N_c$ scaling of these parameters). 
The first three terms of the Lagrangian in \eqref{eq:hyb-int} lead to the decay ratios in Table \ref{tab:hybrid-decays}, in which both the ground-state hybrids and their chiral partners are listed. Among these decay channels, $\pi_!(1600)\rightarrow b_1\pi$ is dominant, as well confirmed by data. Another interesting decay channel is the process $\pi_{1}\rightarrow\rho\pi$, which decays further into three pions, was observed at COMPASS \cite{COMPASS:2018uzl}. 
The last term in \eqref{eq:hyb-int} contains the interaction term breaking $U(1)_A$ axial symmetry and gives rise to new channels that  show the effect of the chiral anomaly in the exotic mesons sector. In particular, the eLSM prediction for the decay ratio $(\pi_1\rightarrow \pi \eta^{\prime})/(\pi_1\rightarrow \pi \eta) \sim 12.7$ is roughly comparable with the result $6.8$ of Ref. \cite{Bass:2001zs}.   
We list other anomalous decay ratios in Table \ref{tab:hyb:anom-decays}.
In Ref. \cite{Shastry:2022mhk} a throughout discussion of the masses and strong decays of the lightest ground-state exotic nonet with $J^{\mathcal{PC}}=1^{-+}$ (including the not-yet found light $\eta_1$ and the $K_1$ states). The model, based on flavor symmetry only, can be seen as a special case of the Lagrangian  \eqref{eq:hyb-int}. A rather clear picture emerges, in agreement with both experimental and lattice QCD results of Ref. \cite{Woss:2020ayi}, see Table \ref{tab:hyb-latt}.
\begin{table}[ht!]
    \centering
    \begin{tabular}{|c|c|c|} \hline
        Decay channel & Decay rate (MeV)  \cite{Shastry:2022mhk}  & Decay rate (MeV)  \cite{Woss:2020ayi} \\ \hline
        $\Gamma_{\pi_1^{\text{hyb}}\rightarrow b_1\pi}$ & $220\pm 34$ & $139-529$ \\ \hline
        $\Gamma_{\pi_1^{\text{hyb}}\rightarrow \rho\pi}$ & $7.1\pm 1.8$ & 0-20 \\ \hline
       $\Gamma_{\pi_1^{\text{hyb}}\rightarrow K^{\star}K}$ & $1.2\pm 0.3$ & 0-2 \\ \hline
        $\Gamma_{\pi_1^{\text{hyb}}\rightarrow f_1\pi}$ & $16.2\pm 3.1$ & 0-24 \\ \hline
        $\Gamma_{\pi_1^{\text{hyb}}\rightarrow f_1^\prime\pi}$ & $0.83\pm 0.16$ & 0-2 \\ \hline
       $\Gamma_{\pi_1^{\text{hyb}}\rightarrow \rho\omega}$ & $0.08\pm 0.03$ & $\leq 0.15$ \\ \hline
       $\Gamma_{\pi_1^{\text{hyb}}\rightarrow \eta\pi}$ & $0.37\pm0.03$ & 0-1 \\ \hline
       $\Gamma_{\pi_1^{\text{hyb}}\rightarrow \eta^\prime\pi}$ & $4.6\pm 1.0$ & 0-12 \\ \hline
    \end{tabular}
    \caption{Decay rates of the hybrid mesons from phenomenology (left) and on the lattice (right).}
    \label{tab:hyb-latt}
\end{table}
%In particular, $\pi_1\rightarrow \pi \eta^{\prime} \sim 10$ is in surprisingly good agreement with the eLSM  estimate based on chiral symmetry. 
Finally, a detailed discussion of their radiative production and decays within the same flavor-symmetry based approach was put forward in Ref. \cite{Shastry:2023ths}. 
%%%%%%%%%%%%%%%%%%%%
\begin{table}[ht!]
    \centering   \begin{tabular}
    {|c|c|c|c|}\hline
     Decay Ratios    & Estimates &  Decay Ratios    & Estimates \\\hline  $\Gamma_{K_{1}^{\text{hyb}}\rightarrow Kh_{1}(1170)}/\Gamma_{\pi_{1}^{\text{hyb}}%
\rightarrow\pi b_{1}}$ & 0.050    & $\Gamma_{b_{1}^{\text{hyb}}\rightarrow\pi\omega(1650)}/\Gamma_{\pi_{1}^{\text{hyb}}%
\rightarrow\pi b_{1}}$ & 0.065\\\hline
$\Gamma_{K_{1B}^{\text{hyb}}\rightarrow\pi K^{\ast}(1680)}/\Gamma_{\pi_{1}%
^{\text{hyb}}\rightarrow\pi b_{1}}$ & 0.19 & $\Gamma_{h_{1N}^{\text{hyb}}\rightarrow\pi\rho(1700)}/\Gamma_{\pi_{1}^{\text{hyb}}%
\rightarrow\pi b_{1}}$ & 0.16\\\hline
$\Gamma_{\pi_{1}^{0hyb}\rightarrow\overline{K}^{0}K^{\ast0}}/\Gamma_{\pi
_{1}^{-hyb}\rightarrow\rho^{0}\pi^{-}}$ & $0.61$ & $\Gamma_{\eta_{1N}^{\text{hyb}}\rightarrow\overline{K}^{0}K^{\ast0}}/\Gamma_{\pi
_{1}^{-hyb}\rightarrow\rho^{0}\pi^{-}}$ & $0.61$\\\hline
$\Gamma_{\eta_{1S}^{\text{hyb}}\rightarrow\overline{K}^{0}K^{\ast0}}/\Gamma_{\pi
_{1}^{-hyb}\rightarrow\rho^{0}\pi^{-}}$ & $1.6$ &  $\Gamma_{K_{1}^{0hyb}\rightarrow\overline{K}^{\ast0}\eta}/\Gamma_{\pi
_{1}^{-hyb}\rightarrow\rho^{0}\pi^{-}}$ & $0.0011$\\\hline
$\Gamma_{b_{1}^{0hyb}\rightarrow\pi^{-}a_{1}^{+}}/\Gamma_{\pi_{1}%
^{-hyb}\rightarrow\rho^{0}\pi^{-}}$ & $3.8$ & $\Gamma_{b_{1}^{0hyb}\rightarrow\overline{K}_{1}^{0}K^{0}}/\Gamma_{\pi
_{1}^{-hyb}\rightarrow\rho^{0}\pi^{-}}$ & $0.60$\\\hline
$\Gamma_{h_{1N,B}^{\text{hyb}}\rightarrow\overline{K}_{1}^{0}K^{0}}/\Gamma_{\pi
_{1}^{-hyb}\rightarrow\rho^{0}\pi^{-}}$ & $0.59$ & $\Gamma_{h_{1S,B}^{\text{hyb}}\rightarrow\overline{K}_{1}^{0}K^{0}}/\Gamma_{\pi
_{1}^{-hyb}\rightarrow\rho^{0}\pi^{-}}$ & $1.7801$\\\hline
$\Gamma_{K_{1B}^{0hyb}\rightarrow K_{1}^{0}\eta}/\Gamma_{\pi_{1}%
^{-hyb}\rightarrow\rho^{0}\pi^{-}}$ & $0.010$ & $\Gamma_{K_{1B}^{0hyb}\rightarrow\overline{K}_{1}^{0}\pi^{0}}/\Gamma_{\pi
_{1}^{-hyb}\rightarrow\rho^{0}\pi^{-}}$ & $0.029$ \\\hline
$\Gamma_{K_{1B}^{0hyb}\rightarrow\overline{K}^{0}a_{1}^{0}}/\Gamma_{\pi
_{1}^{-hyb}\rightarrow\rho^{0}\pi^{-}}$ & $0.046$ &$\Gamma_{K_{1}^{0hyb}\rightarrow K^{\ast0}\pi^{0}}/\Gamma_{\pi_{1}%
^{-hyb}\rightarrow\rho^{0}\pi^{-}}$ & $0.00022$\\\hline
    \end{tabular}
    \caption{Decay ratios of the hybrid mesons within the eLSM.}
\label{tab:hybrid-decays}
\end{table}
%%%%%%%%%%%%%%%%%%%%%%
\begin{table}[ht!]
\centering
\renewcommand{\arraystretch}{1.1}
\begin{tabular}{|c|c|c|c|}
\hline
Decay Ratios & Estimates & Decay Ratios & Estimates \\ \hline
   $\Gamma_{\pi_1^{\text{hyb}}\rightarrow \eta^\prime  \pi}/\Gamma_{\pi_1^{\text{hyb}}\rightarrow \eta\pi}$                      & $12.7$              &
    $\Gamma_{K_1^{\text{hyb}}\rightarrow K \eta}/\Gamma_{\pi_1^{\text{hyb}}\rightarrow \eta\pi}$                      & $0.69$              \\ \hline
     $\Gamma_{K_1^{\text{hyb}}\rightarrow K \eta^\prime}/\Gamma_{\pi_1^{\text{hyb}}\rightarrow \eta\pi}$                      & $5.3$            &
     $\Gamma_{\eta_{1N}^{\text{hyb}}\rightarrow \eta^\prime  \eta}/\Gamma_{\pi_1^{\text{hyb}}\rightarrow \eta\pi}$                      & $2.2$              \\ \hline
      $\Gamma_{\eta_{1S}^{\text{hyb}}\rightarrow \eta^\prime  \eta}/\Gamma_{\pi_1^{\text{hyb}}\rightarrow \eta\pi}$                      & $1.57$  & &            \\ \hline
\end{tabular}
\caption{Anomalous decay ratios of hybrid mesons.}
\label{tab:hyb:anom-decays}
\end{table}

\section{Glueballs within the eLSM}\label{Sec:glueballs}
\subsection{General considerations}
According to lattice QCD, glueballs with various quantum numbers are expected to exist, see Refs.~\cite{Chen:2005mg,Athenodorou:2020ani, Morningstar:1999rf,Chen:2021dvn} and Table~\ref{tab:glubmass} for the compilation of the results of Refs.~\cite{Chen:2005mg,Athenodorou:2020ani}. 
These lattice glueball masses have been recently used to calculate the pressure within the Glueball Resonance Gas model in Ref. \cite{Trotti:2022knd}. The pressure has been then compared to the results obtained in independent and purely thermodynamic  lattice QCD simulations \cite{Borsanyi:2012ve}. This comparison shows that  the glueball mass spectrum  obtained in Ref. \cite{Athenodorou:2020ani} is slightly favored.

The existence of glueballs is confirmed by various models, see e.g. Refs. \cite{Chodos:1974pn,Llanes-Estrada:2021evz,Mathieu:2008me,Huber:2021yfy,Huber:2020ngt,Rinaldi:2021dxh,Pawlowski:2022zhh,Chen:2022imp,Huber:2023mls}.
Glueballs can be studied within models by evaluating their decays: a glueball is expected to couple equally strongly to the light quark flavors, resulting in the so-called `flavor blindness'.
Another important source of information about glueballs can be obtained from the radiative decay channels of $J/\psi$, see e.g  the experimental data for pions and kaons in Ref. \cite{BESIII:2015rug,BESIII:2018ubj}. See Refs. \cite{Sarantsev:2021ein,Klempt:2022qjf,Rodas:2021tyb,Guo:2020akt} for phenomenological studies and Refs. \cite{Gui:2012gx,Chen:2014iua,Zou:2024ksc} for lattice ones. 
However, we should stress that, despite intensive searches, up to now no glueball state has been unambiguously identified, even if some, eventually strong, candidates exist. 
%%%%%%%%%%%%%%%%%%%
%In recent theoretical works, the scalar  have been analyzed for the scalar glueball search in \cite{Sarantsev:2021ein}, and for the tensor glueball in \cite{Klempt:2022qjf}. The corresponding peak for the scalar glueball is about 1865 MeV, while for the tensor glueball it is about 2210 MeV. The consideration of the different scalar and tensor states for the analysis of the radiative decays is also carried out in \cite{Rodas:2021tyb}, which favors the most extensive glueball distribution in the state $f_0(1710)$. Some LQCD based calculations of the radiative decays can be found in \cite{Gui:2012gx,Chen:2014iua,Zou:2024ksc}.Alternatively, the eLSM can help elucidate some of the properties of glueballs.
%%%%%%%%%%%%%%%%%%%
Within the eLSM, the first glueball that we present is the already introduced scalar glueball $G$ that appears as a dilaton field in the eLSM, see Sec.  \ref{Ssec:scalar-glbl} and Refs. \cite{Janowski:2011gt,Janowski:2014ppa}. This is the lightest glueball with a mass of about 1.7 GeV.
Next, we move to further glueballs with $\mathcal{C}=1$ (that is, two constituent gluons), that were studied within the eLSM: the tensor glueball in Sec. \ref{Ssec:tensor-glbl} (Ref. \cite{Vereijken:2023jor}), and the pseudoscalar glueball in Sec. \ref{Ssec:pscalar-glbl} (Ref. \cite{Eshraim:2012jv}), which are the second and the third lightest glueball according to lattice with masses of about 2.2 GeV and 2.6 GeV respectively. Finally, we present the results for the vector glueball in Sec. \ref{Ssec:vector-glbl} (Ref. \cite{Giacosa:2016hrm}), which carries $\mathcal{C}=-1$, representing an example of a three-gluon object. 
%----------------------
\begin{figure}[ht!]
    \centering
    \includegraphics[scale=1.2]{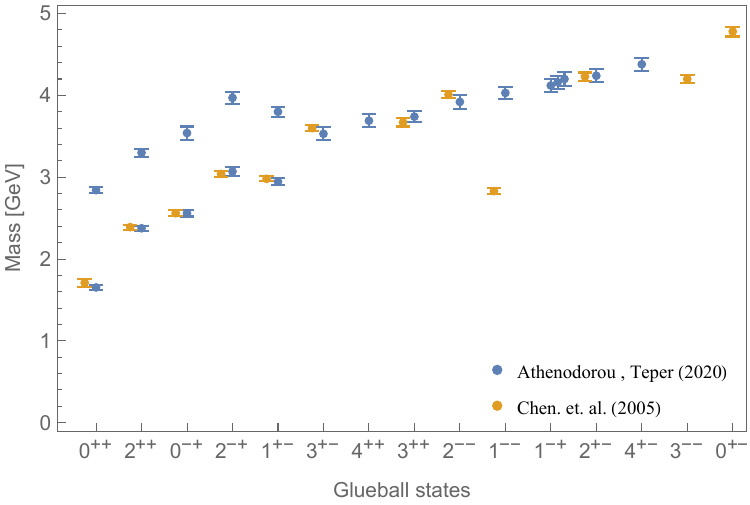}
    \caption{Glueball spectrum obtained in Refs.\cite{Chen:2005,Athenodorou:2020ani}.}
\end{figure}
%----------------------
\begin{table}[ht!]
\begin{center}  
\begin{tabular}{|c|c|c|c|c|c|}
    \hline
    \rule{0pt}{2em} $n\,J^{\mathcal{P}\mathcal{C}}$ & \multicolumn{2}{|c|}{M[MeV]}& \rule{0pt}{2em} $n\,J^{\mathcal{P}\mathcal{C}}$ & \multicolumn{2}{|c|}{M[MeV]}  \\
    \hline
      & Chen et.al. \cite{Chen:2005} &A \& T \cite{Athenodorou:2020ani}  &   & Chen et.al. \cite{Chen:2005} & A \& T \cite{Athenodorou:2020ani}  \\
    \hline
    $\textbf{1\, 0}^{++}$ & 1710(50)(80) & 1653(26) & $\textbf{1\,1}^{--}$ & 3830(40)(190)  & 4030(70) \\
    \hline
    $\textbf{2\,0}^{++}$ &   & 2842(40)  & $\textbf{1\,2}^{--}$ & 4010(45)(200)  & 3920(90)\\
    \hline
    $\textbf{3\,0}^{++}$ &  &  & $\textbf{1\,3}^{--}$& 4200(45)(200)  & \\
    \hline
    $\textbf{1\,2}^{++}$ & 2390(30)(120)     & 2376(32) & $\textbf{1\,0}^{+-}$ & 4780(60)(230)  & \\
    \hline
    $\textbf{2\,2}^{++}$ &    &  3300(50) & $\textbf{1\,1}^{+-}$ & 2980(30)(140) &  2944(42) \\
    \hline
    $\textbf{1\,3}^{++}$ & 3670(50)(180)  & 3740(70) & $\textbf{2\,1}^{+-}$ &  &3800(60)\\
    \hline
    $\textbf{1 4}^{++}$ &    & 3690(80) & $\textbf{1\,2}^{+-}$ & 4230(50)(200)  &4240(80)  \\
    \hline
    $\textbf{1\,0}^{-+}$ & 2560(35)(120)    & 2561(40)& $\textbf{1\,3}^{+-}$ & 3600(40)(170) & 3530(80) \\
    \hline
    $\textbf{2\,0}^{-+}$ &    & 3540(80)  & $\textbf{1\,4}^{+-}$ &  &4380(80)   \\
    \hline
    $\textbf{1\,2}^{-+}$ & 3040(40)(150)    & 3070(60) & $\textbf{1\,1}^{-+}$ &    & 4120(80) \\
    \hline
    $\textbf{2\,2}^{-+}$ &  & 3970(70)  & $\textbf{2\,1}^{-+}$ &    & 4160(80)   \\
    \hline
    $\textbf{3\,2}^{-+}$   & &  &   $\textbf{3\,1}^{-+}$ &    & 4200(90)      \\
    \hline    
    \end{tabular}  
    \caption{Masses of the glueball states taken from the lattice simulations Refs. \cite{Chen:2005,Athenodorou:2020ani}.}
    \label{tab:glubmass}
\end{center}
\end{table}
%------------------
Before introducing the interaction of glueballs with $\bar{q}q$ fields, we mention that a composite model for the YM part of QCD is possible by generalizing the dilaton Lagrangian of Section \ref{Ssec:dilaton} as shown below: 
\begin{align}    \mathcal{L}_{Glueballs}=\mathcal{L}^{\text{kin}} +V_{\text{dil}} +\frac{\alpha\, G^2}{2}T_{G\,,\mu\nu}T_{G}^{\mu\nu}+\frac{\beta G^2P_{G}^2}{2}+\frac{\gamma\, G^2}{2}V_{G\,\mu}V_G^{\mu}+\mathcal{L}_{T_G}^{\text{int}}+\mathcal{L}_{P_G}^{\text{int}}+\mathcal{L}_{V_G}^{\text{int}} \text{ ,}
\end{align}
where $\mathcal{L}^{\text{kin}}$ contains the kinetic terms, $V_{\text{dil}}$ is the usual dilaton potential, and the fields $P_G$, $T_G$, and  $V_G$ refer to the pseudoscalar, tensor, and vector glueballs, respectively. They acquire a mass term by interacting with the scalar glueball that condenses to a nonzero v.e.v. $G_0 = \Lambda_G$; the remaining terms are self-interactions. 
In agreement with dilatation invariance, this model is built containing only dilatation invariant interaction terms with the usual exception of $\Lambda_G$ in $V_{\text{dil}}$. As discussed in Ref. \cite{Trotti:2022knd} this model is suited to study e.g. the scattering of glueballs in a pure YM context.

%Note that the glueball self-interaction term disappears in the large $N_c$ limit. This results in the glueball being a free field with masses proportional to $N_c$, and widths proportional to $N_c^{-1}$ as listed in Table \ref{tab:Large-N}.

\subsection{The scalar glueball}
\label{Ssec:scalar-glbl}

The scalar glueball is the lightest gluonic state predicted by lattice\ QCD (see Table \ref{tab:glubmass})
and is one of the hadronic d.o.f of the eLSM emerging as the excitation of the dilaton
field\ \cite{Parganlija:2012fy,Janowski:2011gt}. The eLSM makes predictions for the lightest
glueball state in a chiral framework, completing previous phenomenological
works on the subject \cite{Amsler:2004ps,Klempt:2007cp,Giacosa:2009bj,Amsler:1995td,Lee:1999kv,Giacosa:2005zt,Cheng:2006hu}. The results of the study of Ref. \cite{Janowski:2014ppa} shows that within a three-state mixing scheme the scalar glueball is predominately contained in the resonance $f_{0}(1710)$. This assignment is in agreement with the pioneering lattice result of Ref. \cite{Lee:1999kv}, with the more recent lattice result of Ref. \cite{Gui:2012gx}, with other hadronic approaches \cite{Amsler:2004ps}, and lately also within the holographic approach of Ref. \cite{Brunner:2015oqa}. %\textbf{Nevertheless analysis of the data \cite{BESIII:2015rug,BESIII:2018ubj} raises the question about the validity of this assumption. Instead, the distribution of the scalar glueball among many other scalar states is discussed in \cite{Klempt:2021wpg}}.

When expanding the eLSM Lagrangian in the scalar-isoscalar sector, it reduces to the following  terms, including the fields ($G,\sigma_N,\sigma_S$) 
\begin{align}    
\mathcal{L}^{\text{mass}}_{G\sigma_N\sigma_S}=\mathcal{L}_{\text{dil}}-\frac{m_0^2}{2}\frac{G^2}{G_0^2}(\sigma_N^2+\sigma_S^2)-\lambda_1(\frac{\sigma_N^2+\sigma_S^2}{2})^2 -\frac{\lambda_2}{4}(\frac{\sigma_N^4}{4}+\sigma_S^4)+h_{\sigma_N}\sigma_N+h_{\sigma_S}\sigma_S-\frac{\epsilon_S\sigma_S^2}{2} \text{ .}
\end{align}
An important remark concerns the $\epsilon_S$-term (the one on the very right), that arises as an additional contribution to the eLSM  of the type $-2\text{Tr}[\hat{\epsilon}\Phi \Phi^{\dagger}]$, with $\hat{\epsilon} = \text{diag}(\epsilon_N,\epsilon_N,\epsilon_S)$, and $\epsilon_{N,S} \propto m_{n,s}^2$. This standard mass term for (pseudo)scalar mesons is subleading w.r.t. the term proportional to the matrix $H$, but is nevertheless important in the scalar-isoscalar sector. As usual, without loss of generality, we set $\epsilon_N = 0$.
Also, the large-$N_c$ suppressed parameter $\lambda_1$, that could not be univocally fixed in Sec. \ref{Ssec:parameterization}, is here taken into account.
%The explicit expression for the bare masses of $\sigma_N$ and $\sigma_S$ (with $\epsilon_S=0$ ) are presented in Table \ref{tab:mass-spin0}. 
Upon minimizing the potential as in Sec. \ref{Ssec:mexican_hat}, the bare glueball mass  reads
\begin{align}
    M_G^2=\Big(\frac{m_0}{G_0}\Big)^2\Big(\phi_N^2+\phi_S^2\Big)+\frac{m_G^2G_0^2}{\Lambda^2}\Big(1+3\ln{\Big|\frac{G_0}{\Lambda_G}\Big|}\Big)\, \text{ ,}
    \label{MG}
\end{align}
where $M_G$ is the bare mass of the glueball in the eLSM ($m_G$ is the dilaton mass, or the bare mass in the YM limit). 

In total, 5 parameters ($\Lambda$, $m_{G}$, $\lambda_{1}$, $h_{1}$ and $\epsilon_{S}$) are relevant in the scalar
isoscalar sector. They were fitted to the experimental quantities of Table \ref{tab:scalar-fit}.
\begin{table}[ht!]
\centering
\begin{tabular}{|c|c|c|c|c|c|c|}\hline
Resonances & PDG Mass  & eLSM Mass  & $\Gamma^{\text{PDG}}_{f_{0}\rightarrow \pi\pi}$   & $\Gamma^{\text{PDG}}_{f_{0}\rightarrow K\overline{K}}$ & $\Gamma^{\text{eLSM}}_{f_{0}\rightarrow \pi\pi}$   & $\Gamma^{\text{eLSM}}_{f_{0}\rightarrow K\overline{K}}$  \\ \hline
$f_0(1370)$ & 1200-1500 & 1444 & $-$ & $-$ & $423.6 $ & $117.5$ \\ \hline
$f_0(1500)$ & $1505\pm 6$& 1534 & $38.04\pm 4.95$ & $9.37\pm 1.69$& 39.2 & 9.1 \\ \hline
$f_0(1710)$ & $1720\pm 6$ & 1750 & $29.3\pm 6.5$ & $ 71.4\pm 29.1 $& 28.3& 73.4 \\ \hline
\end{tabular}
\caption{Fitting data for conventional scalar mesons and glueball.   }
\label{Fitting data for ``scalar glueball'' within eLSM.}
\label{tab:scalar-fit}
\end{table}

The admixtures as calculated in Ref. \cite{Janowski:2014ppa} read (see also Eq. (\ref{matrixB})):
\begin{equation}
\left(
\begin{array}
[c]{c}%
f_{0}(1370)\\
f_{0}(1500)\\
f_{0}(1710)
\end{array}
\right)  =\left(
\begin{array}
[c]{ccc}%
-0.91 & 0.24 & -0.33\\
0.30 & 0.94 & -0.17\\
-0.27 & 0.26 & 0.93
\end{array}
\right)  \left(
\begin{array}
[c]{c}%
\sigma_{N}\\
\sigma_{S}\\
G
\end{array}
\right)  \text{  ,} 
\label{mixmat}%
\end{equation}
that leads to the probabilities glueball contents of $f_0$ resonances in Table \ref{tab:f0-mixtures}. These outcomes show that the scalar glueball is mostly located in the resonance $f_0(1710)$. The corresponding model parameters are shown in Table \ref{tab:scalar-parameters}. Interestingly, the parameter $m_G$ is close to $1.5$ GeV, but this quantity refers to the bare dilaton mass without quark-antiquark field. The inclusion of the latter fields increases the glueball mass via a v.e.v. contribution (see Eq.~\eqref{MG}) and via mixing, thus lifting up the predominantly glueball state to about 1.7 GeV.
Moreover, the parameter $\Lambda _G$ turns out to be large, of the order of 3 GeV. As in Sec. \ref{Ssec:parameterization}, various local minima exist, yet all of those that generate a satisfactory scalar phenomenology are such that $\Lambda _G \gtrsim 1$ GeV, thus larger than the right edge of the interval quoted in Sec. \ref{Ssec:dilaton}. This outcome is in agreement with the study in Ref. \cite{Ellis:1984jv}: a small scale $\Lambda_G$ generates a very broad scalar glueball. Of course, a viable explanation would be to admit that the scalar glueball is too broad to be detected experimentally (similarly to the axial-tensor mesons in Sec. \ref{Ssec:axial-tensor}). Yet such a broad glueball would be also in disagreement with large-$N_c$ expectations. In conclusion, we note that this specific point of the dilaton potential, the value of the dimensionful parameter $\Lambda_G$, is not yet fully clarified. A possible way to determine it would be to measure the scattering of two scalar glueballs in lattice YM theory, see the discussion in Ref. \cite{Giacosa:2021brl}.\\
\begin{table}[ht!]
    \centering
    \begin{tabular}
    {|c|c|c|c|}\hline
     Resonances & $\sigma_N$ & $\sigma_S$    & G  \\\hline
     $f_0(1370)$ & 83\% &6\% & 11\%\\\hline
     $f_0(1500)$ & 9\% & 88\% & 3\%\\\hline         $f_0(1710)$ & 8\% &6\% & 86\%\\\hline
    \end{tabular}
    \caption{Results for the $\bar{q}q$-meson and glueball mixtures of the $f_0$-states. }
    \label{tab:f0-mixtures}
\end{table}
%%%%%%%%%%%%%%%%%%%
\begin{table}[ht!]
\centering
  \begin{tabular}
[c]{|c|c|c|c|}\hline
Parameter & Value & Parameter & Value\\\hline
$\Lambda_G$ & $3297\,[$MeV$]$ &
$m_{G}$ & $1525\,[$MeV$]$\\\hline
$\lambda_{1}$ & $6.25$ &
$h_{1}$ & $-3.22$ \\\hline
$\epsilon_{S}$ & $0.4212\times10^{6}\,[$MeV$^{2}]$ &&\\\hline
\end{tabular}
    \label{tab:scalar-parameters}
    \caption{Parameter values for the $J^{\mathcal{P}\mathcal{C}}=0^{++}$ sector.}
\end{table}
Further predictions of the approach are reported in Table \ref{tab:scalar-predictions}.
\begin{table}[ht!]
\renewcommand{\arraystretch}{1.} 
\centering
\begin{tabular}
[c]{|c|c|c|c|c|c|}\hline
Decay Channel & eLSM [MeV] & PDG [MeV]&Decay Channel & eLSM [MeV] & PDG [MeV]\\\hline
$f_{0}(1370)\rightarrow K\bar{K}$ & $117.5$ & - &$f_{0}(1370)\rightarrow\eta\eta$ & $43.3$ & -\\\hline
$f_{0}(1370)\rightarrow\rho\rho\rightarrow4\pi$ & $13.8$ & -& $f_{0}(1500)\rightarrow\eta\eta$ & $4.7$ & $5.56\pm1.34$\\\hline
$f_{0}(1500)\rightarrow\rho\rho\rightarrow4\pi$ & $0.2$ & $>54.0\pm
7.1$& $f_{0}(1710)\rightarrow\eta\eta$ & $57.9$ & $34.3\pm17.6$\\\hline
$f_{0}(1710)\rightarrow\rho\rho\rightarrow4\pi$ & $0.5$ & -& & &\\\hline
\end{tabular}
\caption{Results of the eLSM compared to the PDG.}
\label{tab:scalar-predictions}
\end{table}
The eLSM implies that $f_{0}(1370)$ decays predominantly into two pions with a
decay width of about $400$ MeV, in agreement with its interpretation as
predominantly nonstrange $\bar{q}q$ state\ \cite{Parganlija:2010fz,Janowski:2011gt}. This is in
qualitative agreement with the experimental analysis of Ref.\ \cite{Bugg:2007ja},
where $\Gamma_{f_{0}(1370)\rightarrow\pi\pi}=325$ MeV, $\Gamma_{f_{0}%
(1370)\rightarrow4\pi}\approx50$ MeV, and $\Gamma_{f_{0}(1370)\rightarrow
\eta\eta}/\Gamma_{f_{0}(1370)\rightarrow\pi\pi}=0.19\pm0.07$. Moreover, the
decay channel $f_{0}(1500)\rightarrow\eta\eta$ is in good agreement with the PDG,
while the decay channel $f_{0}(1710)\rightarrow\eta\eta$ is slightly larger than but compatible with the PDG value, see Table \ref{tab:scalar-predictions}.
%%%%%%%%%%%%%%%%%%%%%%%%%
Recently, it has been argued that a scheme with three scalar states is not sufficient to describe the scalar glueball \cite{Klempt:2021wpg}. According to this study, based on $J/\psi$ decays \cite{Rodas:2021tyb} and on the partial wave analyses %\footnote{Partial wave analyses are also used in Refs. \cite{BESIII:2022iwi,BESIII:2022zel,BESIII:2013qqz,BESIII:2012rtd} for the insvestigation of the glueball states.} 
of the data from BESIII \cite{BESIII:2015rug,BESIII:2018ubj,BESIII:2022iwi,BESIII:2022zel,BESIII:2013qqz,BESIII:2012rtd}, more states (such as $f_0(2020)$, $f_0(2100$)) are considered. The investigation of such an enlarged scenario in the eSLM is possible by including the excited (pseudo)scalar states \cite{Parganlija:2016yxq}. Such a study represents an interesting outlook for the eLSM. 
%In conclusion, the present results show that $f_{0}(1710)$ is the best scalar glueball candidate.

%\bigskip

\subsection{The tensor glueball}\label{Ssec:tensor-glbl}
The second lightest  glueball is the tensor glueball ($J^{\mathcal{P}\mathcal{C}}=2^{++}$), see Table \ref{tab:glubmass}. As in the scalar sector, there are various isoscalar $J^{\mathcal{P}\mathcal{C}}=2^{++}$ resonances listed in PDG. 
We show them in Table \ref{tab:tensor-data}.
\begin{table}[ht!]
\centering
\renewcommand{\arraystretch}{1.} 
\begin{tabular}{|c|c|c|c|}
\hline
Resonances & Masses (MeV) & Decay Widths (MeV) & Decay Channels   \\ 
\hline
$f_2(1910)$   & $1900\pm 9$       &  $167\pm 21$            &   $\pi\pi$, $KK$, $\eta\eta$, $\omega\omega$, $\eta\eta^\prime$, $\eta^\prime\eta^\prime$, $\rho\rho$, $a_{2}(1320)\pi$, $f_{2}(1270)\eta$ \\ 
\hline
$f_2(1950)$   &  $1936\pm 12$      &  $464\pm 24$            &   $\pi\pi$,   $KK$, $\eta\eta$, $K^{*}K^{*}$, $4\pi$ \\ 
\hline
$f_2(2010) $  & $2011^{+60}_{-80}$       &  $202\pm 60$            & $KK$, $\phi\phi$            \\ 
\hline
$f_2(2150)$   & $2157\pm 12$        &  $152\pm 30$            &  $\pi\pi$, $\eta\eta$, $KK$, $a_{2}(1320)\pi$, $f_{2}(1270)\eta$   \\ 
\hline
$f_J(2220)$   &   $2231.1\pm 3.5 $     &   $23^{+8}_{-7}$           & $\eta\eta^\prime$               \\ 
\hline
$f_2(2300)$   &  $2297\pm 28$        &    $149\pm 41$           &     $KK$,    $ \phi\phi$        \\ 
\hline
$f_2(2340)$   &  $2345^{+50}_{-40}$      &  $322^{+70}_{-60}$            & $\eta\eta$, $\phi\phi$  \\ 
\hline
\end{tabular}
\caption{Spin-2 resonances heavier than 1.9 GeV listed in PDG \cite{Workman:2022ynf}.}
\label{tab:tensor-data}
\end{table}
The chirally invariant eLSM interaction Lagrangian describing the decay of the tensor glueball reads:
\begin{align}
  \mathcal{L}_{T_{G}}^{\text{int}}=  \lambda_{T_G}^{(1)}T_{G\,,\mu\nu}\Big(\text{Tr}\Big[ \{ L^{\mu}, L^{\nu}\}\Big]+\text{Tr}\Big[ \{R^{\mu}, R^{\nu}\} \Big]\Big)+\lambda_{T_G}^{(2)}T_{G\,,\mu\nu}\Big(\text{Tr}\Big[ \Phi \textbf{R}^{\mu\nu}\Phi^\dagger\Big]+\text{Tr}\Big[ \Phi^\dagger\textbf{L}^{\mu\nu}\Phi\Big]\Big)\,.  
\end{align}
The corresponding decay ratios are listed in Table \ref{tab:tenglu-results}, where the mass of the tensor glueball was set to 2.2 GeV, in agreement with the analysis of Ref. \cite{Klempt:2022qjf}.  
We observe that the dominant decay
channels are the vector-vector decay products, especially $\rho\rho$ and $K^\star\overline{K}^\star$. Large-$N_c$ analyses of the tensor glueball decaying into $\pi\pi$ ($\Gamma_{G_{2}\rightarrow\pi\pi}\simeq15$ MeV) is comparable to the one obtained within the so-called
Witten-Sakai-Sugimoto model \cite{Brunner:2015oqa} (
$\Gamma_{G_{2}\rightarrow\pi\pi}\simeq28$ MeV).
\begin{table}[ht!]
\centering
%\small
\renewcommand{\arraystretch}{1.5} 
\begin{tabular}{|c|c|c|c|c|c|}
\hline
Decay Ratio   & eLSM & Decay Ratio  & eLSM & Decay Ratio   & eLSM \\ 
\hline\hline
$ \frac{T_G \rightarrow \bar{K}\, K}{T_G \rightarrow  \pi\,\pi} $ & $0.4$ & $ \frac{T_G \rightarrow \rho\, \rho}{T_G \rightarrow  \pi\,\pi} $ & $55$ & $ \frac{T_G\rightarrow a_1(1260)\,\pi}{T_G \rightarrow  \pi\, \pi} $ & $0.24$ 
\\ 
\hline
 $ \frac{T_G \rightarrow \eta \, \eta}{T_G \rightarrow  \pi\,\pi}$  & $0.1$ & $ \frac{T_G \rightarrow \bar{K}^\ast \, \bar{K}^\ast }{T_G \rightarrow  \pi\,\pi} $ & $46$ & $ \frac{T_G\rightarrow K_{1A}\,K}{T_G\rightarrow \pi\,\pi} $ & $0.08$
\\ 
\hline
$ \frac{T_G \rightarrow \eta \, \eta^\prime}{T_G \rightarrow  \pi\,\pi}$  & $0.004$ & $ \frac{T_G \rightarrow \omega \, \omega}{T_G \rightarrow  \pi\,\pi}$  & $18$ & $ \frac{T_G\rightarrow f_1(1285)\,\eta}{T_G\rightarrow \pi\,\pi}$  & $0.02$
\\ 
\hline
$ \frac{T_G \rightarrow \eta^\prime \, \eta^\prime}{T_G \rightarrow  \pi\,\pi}$  & $0.006$ &
$ \frac{T_G \rightarrow \phi  \,\phi }{T_G \rightarrow  \pi\,\pi}$  & $6$ &	
$ \frac{T_G\rightarrow f_1(1285)\,\eta}{T_G\rightarrow \pi\,\pi}$  & $0.02$ 
\\ 
\hline
 & & & & $ \frac{T_G\rightarrow f_1(1420)\,\eta}{T_G\rightarrow \pi\,\pi}$  & $0.01$ \\ \hline
\end{tabular}%
\caption{Decay ratios of the tensor glueball within the eLSM. }\label{tab:tenglu-results}
\end{table}
In Table \ref{tab:spin-2 ratios} we also compared the theoretical predictions to the experimental data for some tensor $2^{++}$ resonances listed in PDG. It shows that the relatively broad resonance $f_2(1950)$ is the favored tensor glueball candidate despite having a mass slightly smaller than the quenched LQCD predictions in Table \ref{tab:glubmass}. Namely, unquenching effects may lead to lowering the masses of the tensor glueball. 
\begin{table}[ht!]
\renewcommand{\arraystretch}{1.} 
\centering
\begin{tabular}{|c|c|c|c|}
\hline
Resonances & Branching Ratios       & PDG                   & eLSM 
\\ \hline \hline
$f_2(1910)$   & $\rho(770)\rho(770) / \omega(782)\omega(782)$ &   $2.6\pm 0.4$ &    $3.1$        \\ 
 \hline
$f_2(1910)$   & $f_2(1270)\eta/a_2(1320)\pi$ &   $0.09\pm 0.05$ &    $0.07$       \\ 
 \hline
$f_2(1910)$   & $\eta \eta / \eta\eta^\prime(958)$ &   $<0.05$ &    $\sim 8$       \\ 
 \hline
$f_2(1910)$   & $\omega(782) \omega(782) /\eta \eta\prime(958)$ &   $2.6 \pm 0.6$  &    $\sim 200$       \\ 
\hline \hline
$f_2(1950)$   & $\eta\eta/\pi\pi$ &     $0.14 \pm 0.05$                  & $0.081$                       \\ 
\hline
$f_2(1950)$   & $K\overline{K}/\pi\pi$ &     $\sim 0.8$                  & $0.32$                       \\ 
\hline
$f_2(1950)$   & $4\pi/\eta\eta$ &     $>200$                  & $>700$                       \\ 
\hline 
\hline
$f_2(2150)$   & $f_2(1270)\eta/a_2(1320)\pi$ &     $0.79\pm 0.11$                  & $0.1$                       \\ 
\cline{1-4}
\hline
$f_2(2150)$   & $K\overline{K} / \eta \eta$ &     $1.28\pm 0.23$                 & $\sim 4$                       \\ 
\hline
$f_2(2150)$   & $\pi \pi / \eta \eta$ &     $<0.33$                 & $\sim 10$                       \\ 
\hline \hline
$f_J(2220)$   & $\pi \pi / K\overline{K}$ &     $1.0 \pm 0.5$                 & $\sim 2.5$                       \\ 
\hline
\end{tabular}
\caption{Test of spin-2 resonances as the tensor glueball.}
\label{tab:spin-2 ratios}
\end{table}

Also for the tensor glueball, a split into a multitude of states is possible, as Ref. \cite{Klempt:2022qjf} shows by studying $J/\psi$ decays. Such a study within the eLSM would imply adding excited (axial-)tensor nonets; see the discussion in Ref. \cite{Vereijken:2023jor}.

\subsection{The pseudoscalar glueball}
\label{Ssec:pscalar-glbl}

The pseudoscalar glueball is the third-lightest glueball \cite{Masoni:2006rz}. Recently, BESIII measured the pseudoscalar ($0^{-+}$) resonances $X(2370)$ \cite{BESIII:2010gmv,BESIII:2023wfi} and $X(2600)$ \cite{BESIIICollaboration:2022kwh}. 
The pseudoscalar glueball field $P_G$ couples in a chirally invariant but $U(1)_A$-breaking way to light mesons \cite{Eshraim:2012jv}:
\begin{equation}
\mathcal{L}_{P_G}^{\text{int}}=ic_{P_G}P_G\left(
\text{\textrm{det}}\Phi-\text{\textrm{det}}\Phi^{\dag}\right)  \text{ .}
\label{intlag}%
\end{equation}
%FG up to here
The branching ratios of $P_G$ relative to the total decay width of the pseudoscalar glueball $\Gamma
_{P_G}^{tot}$ are reported in Table \ref{tab:ps-glueball} for two choices
of the pseudoscalar masses: $M_{P_G}=2.6$ GeV predicted by lattice QCD
\cite{Chen:2005, Athenodorou:2020ani} and in agreement with $X(2600)$, and $M_{P_G}=2.37$ GeV, in agreement with the
$X(2370)$. 
%%%%%%%%%%%%%
Quite remarkably, the eLSM + DIG approach of Ref. \cite{Giacosa:2023fdz} could estimate the coupling constant entering in Eq.~\eqref{intlag}:  $c_{P_G} \approx 11$.
Namely, while chiral symmetry and its anomalous breaking dictate the form of the Lagrangian that was already presented in Ref. \cite{Eshraim:2012jv} (and in a similar form even before in Ref. \cite{Rosenzweig:1981cu}), the microscopic DIG method allows us to estimate the actual value of the coupling constants. This is  an important advance. 

As a consequence, one gets $\Gamma(P_G\rightarrow K\bar{K}\pi)\approx 0.24$ GeV and $\Gamma(P_G\rightarrow \pi \pi \eta^\prime) \approx
0.05$ GeV.  
Notably, both resonances $X(2370)$ and $X(2600)$ have been seen by BESIII in the $\pi \pi \eta^\prime$ channel.
Our results support the interpretation that these resonances may contain a large gluonic amount, with a decay controlled (and in some channels enhanced) by the chiral
anomaly. We note that the mass found in lattice QCD \cite{Athenodorou:2020ani} agrees with the 
identification of $X(2600)$ as a predominantly gluonic, but further studies are needed. 
\begin{table}[ht!]
\centering
\begin{tabular}
[c]{|c|c|c|}\hline
Quantity & $M_{P_G}=2.6$ GeV & $M_{P_G}=2.37$
GeV\\\hline
$\Gamma_{P_G\rightarrow KK\eta}/\Gamma_{P_G}^{tot}$ & $0.049$ &
$0.043$\\\hline
$\Gamma_{P_G\rightarrow KK\eta^{\prime}}/\Gamma_{P_G}^{tot}$ &
$0.019$ & $0.011$\\\hline
$\Gamma_{P_G\rightarrow\eta\eta\eta}/\Gamma_{P_G}^{tot}$ & $0.016$
& $0.013$\\\hline
$\Gamma_{P_G\rightarrow\eta\eta\eta^{\prime}}/\Gamma_{P_G}^{tot}$
& $0.0017$ & $0.00082$\\\hline
$\Gamma_{P_G\rightarrow\eta\eta^{\prime}\eta^{\prime}}/\Gamma_{\tilde
{G}}^{tot}$ & $0.00013$ & $0$\\\hline
$\Gamma_{P_G\rightarrow KK\pi}/\Gamma_{P_G}^{tot}$ & $0.47$ &
$0.47$\\\hline
$\Gamma_{P_G\rightarrow\eta\pi\pi}/\Gamma_{P_G}^{tot}$ & $0.16$ &
$0.17$\\\hline
$\Gamma_{P_G\rightarrow\eta^{\prime}\pi\pi}/\Gamma_{P_G}^{tot}$ &
$0.095$ & $0.090$\\\hline
\end{tabular}
\caption{Branching ratios for the pseudoscalar glueball within the eLSM.}
\label{tab:ps-glueball}
\end{table}
In conclusion, we predict that $KK\pi$ is the dominant decay channel ($47 \%$),
followed by sizable $\eta\pi\pi$ and $\eta^{\prime}\pi\pi$ decay channels
(16\% and 10\% respectively). 
These are simple and testable theoretical predictions which can be helpful in
future experimental searches at the PANDA experiment \cite{Lutz:2009ff}, where
glueballs can directly form in the proton-antiproton fusion process. In particular, the expected mass of the pseudoscalar glueball corresponds to the low range for particle formation in the PANDA experiment. 
The pseudoscalar glueball and its radial excitation were further studied within the eLSM in Refs. \cite{Eshraim:2020zrs,Eshraim:2016mds}.

\subsection{The vector glueball}
\label{Ssec:vector-glbl}

The vector glueball is an example of a glueball with $\mathcal{C}=-1$, which therefore contains three constituent gluons. As a consequence, its mass is larger and lies close to 3.8 GeV, see Table \ref{tab:glubmass}. 
The vector glueball has been coupled to the eLSM in Ref. \cite{Giacosa:2016hrm}.
Following that work, here we present the chirally invariant interaction terms describing the coupling of
the vector glueball field $V_G$ to various quark-antiquark
multiplets introduced in Sec. \ref{ssec:chiral-multp}: 
\begin{align}    \mathcal{L}^{int}_{V_G}=\lambda_{V_G}^{(1)}G\,V_{G\,,\mu}\text{Tr}\left[
\Phi^{\dagger}\Phi^{\mu}+\Phi^{\mu\dag}\Phi\right]+\lambda_{V_G}^{(2)}V_{G\,,\mu}\text{Tr}\left[  L^{\mu}%
\Phi\Phi^{\dagger}+R^{\mu}\Phi^{\dagger}\Phi\right]+\lambda_{V_G}^{(3)}\varepsilon_{\mu\nu\rho\sigma}\partial^{\rho}%
V_G^{\sigma}\text{Tr}\left[  L^{\mu}\Phi R^{\nu}\Phi^{\dagger}\right]
\text{ .}
\label{eq:lag-vecglueball}
\end{align}
In the first two interaction terms the main decay modes are
$V_G \rightarrow b_1\pi \rightarrow \omega \pi\pi$, $V_G \rightarrow  \omega \pi\pi$,  and $V_G \rightarrow   \pi KK^\star
(892)$. 
Also the decays involving kaons and $K_{1}(1270)$ and
$K_{1}(1400)$ are sizable.
An interesting small but peculiar decay emerging from the first term involves the scalar glueball:  $G\equiv f_{0}(1710)$, leading to $\frac{\Gamma_{V_G\rightarrow f_{0}(1710)b_{1}\pi}}{\Gamma
_{V_G\rightarrow b_{1}\pi}}=3.9\cdot10^{-6}$:  two glueballs are present in this process.
The third interaction term, which violates dilatation invariance, predicts decays into vector-pseudoscalar pairs.  The dominant decay modes
are into $\rho\pi$, $KK^{\ast}(892)$, and $\rho a_{1}(1230)$ (with increasing
strength). 
Unfortunately, we cannot determine the coupling constants in the present
framework, but we calculate various decay ratios for the three interaction terms. The results are presented in Table \ref{tab:vec-glueball}. 
Extending this analyses to the charmed meson sector we may estimate the following ratio $\frac{\Gamma_{V_G\rightarrow DD}}{\Gamma_{V_G\rightarrow
\omega\pi\pi}}\approx0.029 $
shows that the $\bar{D}D$ mode is also expected to be small. This result is
important because it shows that the vector glueball, even if according to
lattice QCD has a mass above the $\bar{D}D$ threshold, decays quite weakly into
charmed mesons. 
\begin{table}[ht!]
    \centering
    \renewcommand{\arraystretch}{1.5} 
    \begin{tabular}{|c|c|c|c|c|c|}\hline
        Decay Ratio & Value &Decay Ratio & Value &Decay Ratio & Value \\\hline $\frac{_{V_G\rightarrow\eta h_{1}(1170)}}{V_G\rightarrow b_{1}\pi}$ & 0.17 & $\frac{_{V_G\rightarrow KK^{\ast}(892)}}{_{V_G\rightarrow
\rho\pi}}$ & 1.3 &$\frac{_{V_G\rightarrow KK\rho}}{_{V_G\rightarrow\omega\pi\pi}%
}$ & 0.50 \\\hline         $\frac{_{V_G\rightarrow\eta h_{1}(1380)}}{_{V_G\rightarrow b_{1}\pi}}$ & 0.11 &$\frac{_{V_G\rightarrow\eta\omega}}{_{V_G\rightarrow\rho\pi}}$
& 0.16 & $\frac{_{V_G\rightarrow KK\omega}}{_{V_G\rightarrow\omega\pi
\pi}}$ & 0.17 \\ \hline
$\frac{_{V_G\rightarrow\eta^{\prime}h_{1}(1170)}}%
{_{V_G\rightarrow b_{1}\pi}}$ & 0.15 &$\frac{_{V_G\rightarrow\eta^{\prime}\omega}}{_{V_G\rightarrow\rho\pi}}$ & 0.13& $\frac{_{V_G\rightarrow KK\phi}}{_{V_G\rightarrow\omega\pi\pi}%
}$ & 0.21 \\ \hline
$\frac{_{V_G\rightarrow\eta^{\prime}h_{1}(1380)}}%
{_{V_G\rightarrow b_{1}\pi}}$ & 0.10 & $\frac{_{V_G\rightarrow\eta\phi}}{_{V_G\rightarrow\rho\pi}}$ &
0.21 & $\frac{V_G\rightarrow\pi KK^{\ast}(892)}{_{V_G\rightarrow\omega\pi\pi}}$ & 1.2 \\ \hline
$\frac{_{V_G\rightarrow KK_{1}(1270)}}{_{V_G\rightarrow b_{1}%
\pi}}$ & 0.75 & $\frac{_{V_G\rightarrow\eta^{\prime}\phi}}{_{V_G\rightarrow\rho\pi}}$ & 0.18 & $\frac{_{V_G\rightarrow\eta\eta\omega}}{_{V_G\rightarrow%
\omega\pi\pi}}$ & 0.064\\ \hline
$\frac{_{V_G\rightarrow KK_{1}(1400)}}{_{V_G\rightarrow b_{1}%
\pi}}$ & 0.30  & $\frac{_{V_G\rightarrow\rho a_{1}(1230)}}{_{V_G\rightarrow%
\rho\pi}}$ & 1.8 & $\frac{_{V_G\rightarrow\eta\eta^{\prime}\omega}}%
{_{V_G\rightarrow\omega\pi\pi}}$ & 0.019\\ \hline
$\frac{_{V_G\rightarrow K_{0}^{\ast}(1430)K^{\ast}(1680)}%
}{_{V_G\rightarrow b_{1}\pi}}$ & 0.20 &$\frac{_{V_G\rightarrow}\omega f_{1}(1285)}{_{V_G\rightarrow\rho\pi}}$ & 0.55 & $\frac{_{V_G\rightarrow\eta\eta\phi}}{_{V_G\rightarrow%
\omega\pi\pi}}$ & 0.039\\ \hline
$\frac{_{V_G\rightarrow a_{0}(1450)\rho(1700)}}{_{V_G\rightarrow b_{1}\pi}}$ & 0.14 & $\frac{_{V_G\rightarrow}\omega f_{1}(1420)}{_{V_G\rightarrow\rho\pi}}$ & 0.82 & $\frac{_{V_G\rightarrow\eta\eta^{\prime}\phi}}{_{V_G\rightarrow\omega\pi\pi}}$ & 0.011\\ \hline
    \end{tabular}
    \caption{Decay ratios of the vector glueball within the eLSM.}
    \label{tab:vec-glueball}
\end{table}

\section{Other topics}
\label{Sec:other}
In this section, we enlarge/modify the eLSM along 3 different directions: (i) the inclusion of isospin breaking effects; (ii) an example of a radially excited multiplet of (pseudo)scalar mesons; (iii) the extension to the four-flavor case.

\subsection{Isospin breaking}
\label{Ssec:isospin_breaking}

Isospin symmetry is broken in Nature, i.e. the masses of the up and down quarks are different. Since this effect is of the order of a few MeV, it is neglected in many cases, as in the previous sections. As a consequence of the $m_u - m_d$ mass difference, the masses of the charged and neutral mesons of a given isospin multiplet are slightly different. In addition, there is another non-QCD isospin breaking effect, the electromagnetic interaction, which due to the different quark charges also causes a difference between the charged and neutral masses. This has been studied first in a vector meson extended effective model in Ref. \cite{Kapoor:1975rp} and very recently in eLSM in Ref. \cite{Kovacs:2024cdb}. In the latter, besides the modification of the particle masses, the decay widths, the PCAC relation, the role of the electromagnetic mass contributions and the violation of Dashen's theorem \cite{Dashen:1969eg, Gao:1996kr} are also investigated. Technically, the isospin violation can be incorporated into the eLSM by changing the external fields (see \eqref{eq:expl_sym_br_epsilon} and \eqref{eq:expl_sym_br_delta}) to
\begin{align}
    H  &= H_{0}t_{0}+H_3 t_3+H_{8}t_{8} = \frac{1}{2}\diag(h_{0N}+h_{03},h_{0N}- h_{03},\sqrt{2}h_{0S})\;, \label{eq:expl_br_h_isospin}\\
    \Delta &= \Delta_{0}t_{0}+\Delta_{3}t_{3}+\Delta_{8}t_{8} = \diag(\delta_u,\delta_d,\delta_S) \text{ ,} \label{eq:expl_br_delta_isospin}%
\end{align}
and assume that $h_{03}\neq 0$ and $\delta_3\equiv\delta_u-\delta_d\neq 0$.
In this case, besides $\sigma_N$ and $\sigma_S$, $a_0^0$ can also have non-zero vacuum expectation value, denoted as $\phi_3\equiv\braket{a_0^0}$, for a total of 3 scalar condensates. 
The value of $\phi_3$ in the vacuum is determined by minimizing the modified classical potential $V_{G \Phi}(G=G_0,\phi_N,\phi_S,\phi_3,0,...,0)$ after also shifting the $a_0^0$ field by its expectation value, i.e. $a_0^0 \rightarrow \phi_3+ a_0^0$. The classical potential is given by
\begin{align}
    V_{G \Phi}(G=G_0, \sigma_N =\phi_N,\sigma_S =\phi_S,a_0^0=\phi_3,0,...,0) &= \frac{m_0^2}{2} \left[\phi_N^2+\phi_S^2 + (\phi_3)^2\right] + \frac{\lambda_1}{4} \left[\phi_N^2+\phi_S^2 + (\phi_3)^2\right]^2\\ \nonumber  
    &+ \frac{\lambda_2}{4} \left[\frac{\phi_N^4}{2}+3\phi_N^2(\phi_3)^2 + \frac{(\phi_3)^4}{2} + \phi_S^4\right] - h_{0N}\phi_N - h_{0S} \phi_S- h_{03} \phi_3\;,
\end{align}
while the field equations -- given by $\partial V_{G \Phi}/ \partial \phi_{N/S/3} = 0$ -- are
\begin{align}
h_{0N} &= \phi_N \left\{ m_0^2+\lambda_1 \left[\phi_N^2+\phi_S^2+(\phi_3)^2\right]+\frac{\lambda_2}{2} \left[\phi_N^2+3(\phi_3)^2\right] \right\}\;,\label{Eq:h_0N_iso}\\
h_{0S} &= \phi_S \left\{ m_0^2+\lambda_1 \left[\phi_S^2+\phi_N^2+(\phi_3)^2\right]+\lambda_2\phi_S^2 \right\} \;, \label{Eq:h_0S_iso}\\
h_{03} &= \phi_3 \left\{ m_0^2+\lambda_1\left[\phi_N^2+\phi_S^2+ (\phi_3)^2\right]+\frac{\lambda_2}{2}\left[3\phi_N^2+(\phi_3)^2\right]\right\} \;. \label{Eq:h_03_iso}
\end{align}
The tree-level masses and decay widths can be calculated similarly to the isospin-symmetric case, i.e. from the quadratic and triple coupling terms of the Lagrangian written with the shifted fields. In the isospin violating case, the expressions for the masses and decay widths are generally quite long due to the various mixings (see their explicit form in Ref. \cite{Kovacs:2024cdb}). For the sake of illustration, we show here only the mass expressions for the neutral and charged kaons,
\begin{align}
    m_{K^{\pm}}^2 &= Z_{K^{\pm}}^2 \left(m_0^2+\lambda_1(\phi_N^2+\phi_3^2+\phi_S^2)+\lambda_2 \left[-\frac{1}{\sqrt2}(\phi_N + \phi_3)\phi_S +
    \frac{(\phi_N+\phi_3)^2}{2}  + \phi_S^2\right]\right)+\bar{m}^2_{\text{em}}\;,\\
    m_{K^0}^2 &= Z_{K^{0}}^2\left(m_0^2 + \lambda_1(\phi_N^2 + \phi_3^2+\phi_S^2) + \lambda_2\left[-\frac{1}{\sqrt2}(\phi_N-\phi_3) \phi_S +
    \frac{(\phi_N-\phi_3)^2}{2} + \phi_S^2\right]\right)\;,
\end{align}
where
\begin{equation} 
Z_{K^\pm}=\frac{2m_{K_1^\pm}}{\sqrt{4m_{K_1^\pm}^2-g_1^2(\phi_N+\phi_3+\sqrt2\phi_S)^2}} \text{ , } Z_{K^0}=\frac{2m_{K_1^0}}{\sqrt{4m_{K_1^0}^2-g_1^2(\phi_N-\phi_3+\sqrt2\phi_S)^2}} \text{ .}
\end{equation}
The quantity $\bar{m}^2_{\text{em}}$ represents a phenomenological electromagnetic mass contribution to the charged kaon mass. As can be seen in the above expressions, the difference is also due to the differing $Z$ factors. Regarding the decay widths, similar to the isospin symmetric case, we show here the explicit form of the $\rho \to \pi \pi$ decays, 
\begin{equation}
\Gamma_{\rho^{0}\to \pi^{+}\pi^{-}} = \frac{k_{\rho^0}^3}{6\pi M_{\rho^0}^2} \left| B^{\rho}_1  + \frac12 C^{\rho}_1 M_{\rho^{0}}^2\right|^2 \text{ , }
\Gamma_{\rho^{\pm}\to \pi^{\pm}\pi^{0}} = \frac{k_{\rho^\pm}^3}{24\pi M_{\rho^\pm}^2} \left| D^{\rho}_1 - D^{\rho}_2 + F^{\rho}_1 M_{\rho^{\pm}}^2\right|^2 \text{ ,}
\end{equation}
where
\begin{eqnarray}
B^{\rho}_1 &=& Z_{\pi^\pm}^2\left\{\left[g_1 + \phi_N \left(h_3-g_1^2\right) w_{a_1^\pm}\right]\cos\vartheta_V - \frac{2}{3}\phi_3(h_2 + h_3)w_{a_1^\pm}\sin\vartheta_V\right\}\;,\\
C^{\rho}_1 &=& -g_2 Z_{\pi^\pm}^2 w_{a_1^\pm}^2 \cos\vartheta_V \\
D^{\rho}_1 &=& Z_{\pi^\pm}\left\{\left[g_1 + \phi_N \left(h_3-g_1^2\right) w_{a_1^\pm}\right]{\mathds{O}_{P}}_{21} - \phi_3 \big(h_3-g_1^2\big) w_{a_1^\pm} {\mathds{O}_{P}}_{11} \right\}\;,\\
D^{\rho}_2 &=& -Z_{\pi^\pm}\left[g_1 {\mathds{O}_{P}}_{21}+ \phi_N \big(h_3-g_1^2\big) \big(w_{\eta}^a{\mathds{O}_{P}}_{11} + w_{\pi}^a{\mathds{O}_{P}}_{21} \big) - \phi_3\big(h_2 - h_3 + 2g_1^2\big)\big(w_{\eta}^f{\mathds{O}_{P}}_{11} + w_{\pi}^f{\mathds{O}_{P}}_{21}\big)\right]\;,\\
F^{\rho}_1 &=& -g_2 Z_{\pi^\pm} w_{a_1^\pm} \big(w_{\eta}^a{\mathds{O}_{P}}_{11} + w_{\pi}^a{\mathds{O}_{P}}_{21} \big),\quad F^{\rho}_2 = -F^{\rho}_1\;,
\end{eqnarray}
and
\begin{equation}
k_{\rho^0} = \frac12 \sqrt{M_{\rho^0}^2-4M_{\pi^\pm}^2} \text{ , }
k_{\rho^{\pm}} = \frac{\sqrt{(M_{\rho^\pm}^2 - M_{\pi^\pm}^2 - M_{\pi^0}^2)^2-4M_{\pi^\pm}^2 M_{\pi^0}^2}}{2M_{\rho^\pm}} \text{ ,}
\end{equation}
where $\vartheta_V$ is the mixing angle between $\rho^0$ and $\omega_N$ (not to be confused with the previous strange-nonstrange mixing angle $\beta_v$ defined in Eq. \ref{betav}).
The various coefficients $w_X$ refer to the (axial-)vector field shifts given in section II of \cite{Kovacs:2024cdb}, $\mathds{O}_{P}$ is the transformation matrix between the pseudoscalar physical $\pi^0$, $\eta$, $\eta^{\prime}$ fields and the original $\eta_N$, $\pi^0$, $\eta_S$ fields of the Lagrangian, while the various $M_X$ are the physical masses. The expressions for the charged and neutral $\rho\to\pi\pi$ decay widths look more complicated than their isospin symmetric counterparts of Eq.~\eqref{Eq:rho_pi_pi_isosym}, but it can be shown that as $\phi_3 \to 0$ and $\delta_3 \to 0$ the charged and neutral expressions coincide with each other and with Eq.~\eqref{Eq:rho_pi_pi_isosym}. By using the tree-level expressions for the charged and neutral masses, decay widths and pion and kaon decay constants, a global fit can be performed similarly as in the isospin symmetric case. An illustrative fit result is shown in Table~\ref{Tab:isospin_fit}, while the corresponding parameter values are given in Table~\ref{Tab:isospin_params}.
\begin{table}[ht!]
 \label{Tab:isospin_fit}
    \centering
\begin{tabular} [c]{|c||c|c|c|c||c|c|c|}
\hline 
Observable &  Exp. val. [MeV]  &  $\text{Fit}_{\text{DS},\text{no-}\omega}$ [MeV] & $\chi^2$ 
& Observable &  Exp. val. [MeV]  &  $\text{Fit}_{\text{DS},\text{no-}\omega}$ [MeV] & $\chi^2$ \\\hline             
$f_{\pi^+}$  &  $92.06 \pm 4.60$   & $96.72$   & $1.0$  & $f_{K^+}$ &  $110.10 \pm 5.51$   & $110.45$  & $0.0$ \\
\hline	
$\bar M_{\pi}$  &  $138.04 \pm 6.90$   & $140.20$  & $0.1$  & $\Delta M_{\pi}$  &  $-4.59 \pm 0.92$   & $-4.56$   & $0.0$  \\
\hline
$M_{\eta}$   &  $547.86 \pm 27.39$  & $547.39$  & $0.0$  & $M_{\eta^{\prime}}$  &  $957.78 \pm 47.89$    & $952.44$  & $0.0$  \\
\hline
$\bar M_{K}$ &  $495.64 \pm 24.78$   & $482.47$  & $0.3$ & $\Delta M_{K}$      &  $3.93 \pm 0.79$     & $3.93$    & $0.0$  \\
\hline
$\bar M_{\rho}$  &  $775.16 \pm 38.76$  & $761.61$  & $0.1$ & $\Delta M_{\rho}$  &  $0.15 \pm 0.57$    & $0.13$    & $0.0$  \\
\hline
$M_{\omega}$  &  $782.66 \pm 39.13$   & $761.86$  & $0.3$ & $M_{\phi}$  &  $1019.46 \pm 50.97$  & $986.41$  & $0.4$  \\
\hline
$\bar M_{K^{\star}}$   &  $895.50 \pm 44.78$  & $882.52$  & $0.1$ & $\Delta M_{K^{\star}}$   &  $0.08 \pm 0.90$   & $0.19$    & $0.0$ \\
\hline
$\bar M_{a_{1}}$   &  $1230.00 \pm 246.00$ & $1115.71$ & $0.2$   & $M_{f_{1}^L}$   &  $1281.90 \pm 256.38$ & $1222.40$ & $0.1$   \\
\hline  	
$M_{f_{1}^H}$  &  $1426.30 \pm 285.26$ & $1367.72$ & $0.0$ & $\bar M_{K_{1}}$  &  $1253.00 \pm 250.60$  & $1260.84$ & $0.0$  \\
\hline  
$\bar M_{a_{0}}$  &  $1474.00 \pm 294.80$ & $1140.38$ & $1.0$  & $\bar M_{K_{0}^{\star}}$ &  $1425.00 \pm 285.00$  & $1237.23$ & $0.4$  \\
\hline
$M_{f_0^L}$  &  $1350.00 \pm 675.00$ & $1136.72$ & $0.1$ & $M_{f_0^H}$  &  $1733.00 \pm 866.50$ & $1326.58$ & $0.2$  \\
\hline
$\bar \Gamma_{\rho\rightarrow\pi\pi}$  &  $148.53 \pm 7.43$   & $154.85$  & $0.7$ &	$\Delta \Gamma_{\rho\rightarrow\pi\pi}$  &  $-1.70 \pm 1.60$    & $-1.89$   & $0.0$   \\
\hline
$\bar \Gamma_{\phi\rightarrow \bar{K}K}$  &  $1.76 \pm 0.09$    & $1.10$    & $0.3$ &  $\Delta \Gamma_{\phi\rightarrow \bar{K}K}$   &  $-0.65 \pm 0.13$    & $-0.58$  & $0.2$ \\
\hline
$\bar \Gamma_{K^{\star}\rightarrow K\pi}$    &  $46.75 \pm 2.34$    & $46.20$   & $0.0$ & $\Delta \Gamma_{K^{\star}\rightarrow K\pi}$  &  $1.10 \pm 1.80$  & $0.40$  & $0.2$ \\
\hline
$\bar \Gamma_{a_{1}\rightarrow\rho\pi}$  &  $425.00 \pm 175.00$   & $428.34$  & $0.0$ &  $\Gamma_{a_{1}\rightarrow\pi\gamma}$  &  $0.64 \pm 0.25$  & $0.68$  & $0.0$ \\
\hline
$\Gamma_{f_{1}^H\rightarrow K^{\star}K}$ &  $43.60 \pm 8.72$  & $43.84$ & $0.0$ & $\Gamma_{a_{0}}$ &  $265.00 \pm 53.00$ & $239.03$  & $0.2$  \\
\hline    
$\Gamma_{K_{0}^{\star}\rightarrow K\pi}$ &  $270.00 \pm 80.00$   & $333.96$  & $0.6$ & & & & \\
\hline
\end{tabular}
\caption{Detailed fit results in case of isospin violation. Experimental data, fit result in the Dashen case (DS) without fitting the $\omega \to \pi \pi$ decay and the corresponding chi square values.}
\end{table}
\begin{table}[ht!]
    \centering
    \begin{tabular}{|c|c||c|c|}
        \hline 
        Parameter & DS, no-$\omega$ & Parameter & DS, no-$\omega$ \\\hline\hline
        $\phi_{N}$ [MeV]        & $163.93$           &   $\phi_{S}$ [MeV]        & $133.40$   \\
        \hline
        $\phi_{3}$ [MeV]        & $-4.72\times 10^{-3}$   &   $m_{0}^2$ [MeV$^2$]  & $-6.39\times 10^{+5}$\\
        \hline 
        $\tilde{m}_1^2$ [MeV$^2$] & $8.00\times 10^{+5}$  &   $\lambda_{1}$  & $0.09$            \\
        \hline 
        $\lambda_{2}$             & $44.79$          &   $h_{1}$                   & $26.24$           \\
        \hline
        $h_{2}$                   & $23.82$          &   $h_{3}$                   & $5.41$            \\
        \hline
        $g_{1}$                   & $5.53$           &   $g_{2}$                   & $3.01$            \\
        \hline
        $c_{1}$ [$\text{MeV}^{-2}$]  & $2.68\times 10^{-4}$  &   $\tilde{\delta}_{S}$ [MeV$^2$] & $1.46\times 10^{+5}$ \\
        \hline 
        $\delta_{3}$ [MeV$^2$]       & $3.75$        &  $\delta m^2_V$ [MeV$^2$]     & $-1.26\times 10^{+2}$  \\
        \hline
        $\delta m^2_A$ [MeV$^2$]  & $-1.91\times 10^{+5}$ & $m^2_{\text{em},S}$ [MeV$^2$] & $9.93\times 10^{+3}$  \\
        \hline
        $m^2_{\text{em},P}$ [MeV$^2$] & $-3.69\times 10^{+3}$ & $m^2_{\text{em},V}$ [MeV$^2$] & $-3.20\times 10^{+2}$ \\
        \hline
        $m^2_{\text{em},A}$ [MeV$^2$] & $9.91\times 10^{+3}$  & & \\
        \hline
    \end{tabular}
    \caption{Parameter set for the fit result of Table~\ref{Tab:isospin_fit}.}
    \label{Tab:isospin_params}
\end{table}
In the fit, the Dashen case refers to the fulfillment of Dashen's theorem, i.e. the electromagnetic contribution within a nonet is the same for the pion-like and kaon-like particles. In this fit, the $\omega \to \pi \pi$ decay width was not used, which is the only isospin violating decay that cannot be fit within the eLSM at tree-level. The reduced chi-squared of the fit is $\chi^2_{\text{red}}=0.6$, showing a very good agreement with data. 

As a concluding remark, we notice that the eLSM, when properly equipped, is capable of describing also relatively small effects. In the future, further improvements are possible by adding small additional large-$N_c$ suppressed terms, such as the already mentioned nonstrange-strange mixing for homochiral mesons (such as (axial-)vector ones).

\subsection{Radially excited (pseudo)scalar mesons}

Up to now, we solely considered $\bar{q}q$ nonets as states with radial quantum numbers $n=1$. There are, however, quite well-established nonets of radially excited
vector and pseudoscalar mesons \cite{Workman:2022ynf,isgur1985}. For what concerns radially excited vector states, we refer to \cite{Piotrowska:2017rgt} for a detailed treatment in the framework of a flavor-symmetry based approach.
Here, following 
Ref. \cite{Parganlija:2016yxq}, we concentrate on the chiral multiplet of radially excited (pseudo)vector states made out of excited pseudoscalar and scalar nonets.
The nonet of radially excited pseudoscalar states ($J^{\mathcal{P}\mathcal{C}}=0^{-+}$ with
$L=S=0,$ spectroscopic notation $2$ $^{1}S_{0},$ denoted as $P_{E}$) contains
the resonances \{$\pi(1300),$ $K(1460),$ $\eta(1295)$, $\eta(1405)/\eta
(1475)$\} and shares the same transformation properties of the ground-state
pseudoscalar nonet $P$ (note, the two resonances $\eta(1405)$ and $\eta(1475)$
can be interpreted as a single resonances denoted as $\eta(1440)$).
In fact, the radial excitation is formally associated to the same local current.
Similarly, the nonet of radially excited scalar states ($J^{\mathcal{P}\mathcal{C}}=0^{++}$ with
$L=S=1$, spectroscopic notation $2$ $^{1}S_{0},$ denoted as $S_{E}$) contains
the resonances {$a_{0}(1950),$ $K_{0}^{\ast}(1950),$ $f_{0}(1790)$,
$f_{0}(2020)/f_{0}(2100)$}.
This nonet is formally analogous to $S.$ Their
matrix expressions read
\begin{equation}
P_{E}=\frac{1}{\sqrt{2}}%
\begin{pmatrix}
\frac{\eta_{N,E}+\pi_{E}^{0}}{\sqrt{2}} & \pi_{E}^{+} & K_{E}^{+}\\
\pi_{E}^{-} & \frac{\eta_{N,E}-\pi_{E}^{0}}{\sqrt{2}} & K_{E}^{0}\\
K_{E}^{-} & \bar{K}_{E}^{0} & \eta_{S,E}%
\end{pmatrix}
\,\text{, }S_{E}=\frac{1}{\sqrt{2}}%
\begin{pmatrix}
\frac{\sigma_{N,E}+a_{0,E}^{0}}{\sqrt{2}} & a_{0,E}^{+} & K_{0,E}^{\ast+}\\
a_{0,E}^{-} & \frac{\sigma_{N,E}-a_{0,E}^{0}}{\sqrt{2}} & K_{0,E}^{\ast0}\\
K_{0,E}^{\ast-} & \bar{K}_{0,E}^{\ast0} & \sigma_{S,E}
\end{pmatrix}
\text{ ,}
\end{equation}
out of which we construct the heterochiral multiplet of radially excited (pseudo)scalar
states (in short, excited heteroscalars) as $\Phi_{E}=S_{E}+iP_{E}$, which transforms just as $\Phi$ under
chiral, $\mathcal{C}$, and $\mathcal{P}$  transformations.
This is a general feature for radially excited chiral multiplets: for what concerns transformation properties, they are formally identical to the ground-state ones. 
The Lagrangian describing the masses of radially excited (pseudo)scalars is given by 
\begin{align}
\mathcal{L}_{\Phi_E}^{\text{mass}}=\text{Tr}\Big[\Big(\frac{m_{0}^{\ast\,2}\,G^2}{2\,G_0^2}+\Delta_E\Big)\Big(\Phi
_{E}^{\dagger}\Phi_{E}\Big)\Big]+ \frac{\lambda_{\Phi_E}}{2}\text{Tr}\Big(\Phi^\dagger \Phi\Big)\text{Tr}\Big(\Phi_{E}^\dagger \Phi_{E}\Big)+\lambda_{2}^{\ast}\text{Tr}\Big(\Phi_{E}^{\dagger}%
\Phi_{E}\Phi^{\dagger}\Phi\Big)+\xi_{2}\text{Tr}\Big(\Phi
_{E}^{\dagger}\Phi\Phi_{E}^{\dagger}\Phi\Big)
\text{ ,}
\end{align}
where $\Delta_E=\text{diag}(0,0,\epsilon_{S}^{E})$ describes, as usual, the mass contribution arising from nonzero quark masses for this multiplet.
The resulting expressions for the masses are reported in Table \ref{tab:mass-escalar}. 
%----------------------
\begin{table}[ht!]
\centering
\renewcommand{\arraystretch}{1.3} 
\begin{tabular}{|c|c|c|}
\hline
Mass squares  & Analytical expressions & Estimates of masses (MeV) \\ \hline\hline
$m_{\sigma_{N}^{E}}^{2}$  &  $(m_{0}^{\ast})^{2}+\frac{\lambda_{2}^{\ast}%
+\xi_{2}}{2}\phi_{N}^{2}$ & $1790 \pm 35$\\\hline
$m_{a_{0}^{E}}^{2}$  &  $(m_{0}^{\ast})^{2}+\frac{\lambda_{2}^{\ast}+\xi_{2}}%
{2}\phi_{N}^{2}=m_{\sigma_{N}^{E}}^{2}$ & $1790 \pm 35$\\\hline
$m_{\eta_{N}^{E}}^{2}$  &  $(m_{0}^{\ast})^{2}+\frac{\lambda
_{2}^{\ast}-\xi_{2}}{2}\phi_{N}^{2}$ & $1294 \pm 4$\\\hline
$m_{\pi^{E}}^{2}$  &  $(m_{0}^{\ast})^{2}+\frac{\lambda
_{2}^{\ast}-\xi_{2}}{2}\phi_{N}^{2}=m_{\eta_{N}^{E}}^{2}$ & $1294 \pm 4$\\\hline
$m_{\eta_{S}^{E}}^{2}$  &  $(m_{0}^{\ast})^{2}-2\epsilon_{S}^{E}+\left(
\lambda_{2}^{\ast}-\xi_{2}\right)  \phi_{S}^{2} $ & $1432 \pm 10$ \\\hline
$m_{\sigma_{S}^{E}}^{2}$  &  $(m_{0}^{\ast})^{2}-2\epsilon_{S}^{E}+\left(
\lambda_{2}^{\ast}+\xi_{2}\right)  \phi_{S}^{2} $ & $1961 \pm 38$\\\hline
$m_{K^{E}}^{2}$  &  $(m_{0}^{\ast})^{2}-\epsilon_{S}^{E}+\frac{\lambda_{2}%
^{\ast}}{4}\phi_{N}^{2}-\frac{\xi_{2}}{\sqrt{2}}\phi_{N}\phi_{S}+\frac
{\lambda_{2}^{\ast}}{2}\phi_{S}^{2}$ & $1366 \pm 6$\\\hline
$m_{K_{0}^{\star E}}^{2}$  &  $(m_{0}^{\ast})^{2}-\epsilon_{S}^{E}+\frac
{\lambda_{2}^{\ast}}{4}\phi_{N}^{2}+\frac{\xi_{2}}{\sqrt{2}}\phi_{N}\phi
_{S}+\frac{\lambda_{2}^{\ast}}{2}\phi_{S}^{2} $ & $1877 \pm 36$\\\hline
	\end{tabular}%
\caption{ Masses of radially excited (pseudo)scalar mesons within the eLSM (we assume $\sigma_N^E\equiv f_0(1790)$, $\eta_N^{E}\equiv \eta(1295)$ and $\eta_S^{E}\equiv \eta(1440)$).}\label{tab:mass-escalar}
\end{table}%
The chiral interaction Lagrangian for the decays of radially excited (pseudo)scalar mesons reads
\begin{align}\label{eq:excited-int}
    \mathcal{L}_{\Phi_E LR} & = h_{2}^{\ast}\mathop{\mathrm{Tr}}\Big(\Phi_{E}^{\dagger}L_{\mu}L^{\mu}\Phi
+\Phi^{\dagger}L_{\mu}L^{\mu}\Phi_{E}+R_{\mu}\Phi_{E}^{\dagger}\Phi R^{\mu
}+R_{\mu}\Phi^{\dagger}\Phi_{E}R^{\mu}\Big)\\\nonumber
&+2h_{3}^{\ast}\mathop{\mathrm{Tr}}\Big(L_{\mu}\Phi_{E}R^{\mu}\Phi^{\dagger
}+L_{\mu}\Phi R^{\mu}\Phi_{E}^{\dagger}\Big)+\cdots (\text{large-$N_c$ suppressed terms}) \text{ .}
\end{align}
As a result, we obtain the decay channels given in Table \ref{tab:excited-decays} and in Table \ref{tab:excited-decays2}. By using $\Gamma_{f_0(1790)\rightarrow\pi\pi}=270\pm 45\, \text{MeV}$ and $\Gamma_{f_0(1790)\rightarrow KK}=70\pm 40\, \text{MeV}$ in order to fit the parameters of Eq. \eqref{eq:excited-int}, one obtains $h_2^\ast=67\pm 63$ and $h_3^\ast =79\pm 63$.
%%%%%%%%%%%%%%%%%%
\begin{table}[ht!]
\centering
\renewcommand{\arraystretch}{1.1}
\begin{tabular}{|c|c|c|c|}
\hline
Decay Rates & Estimates (MeV) & Decay Rates & Estimates (MeV) \\ \hline
   $\sigma_N^E \rightarrow a_1(1260) \pi$ &  $47 \pm  8$ &  $\sigma_N^E \rightarrow \eta \eta^{\prime}$ &  $ 10 \pm 2$\\ \hline
   $\sigma_N^E \rightarrow \eta \eta$ &  $7 \pm  1$ &  $\sigma_N^E \rightarrow f_1(1285) \eta$ & $ 1 \pm 0$ \\\hline
  $\eta_N^E \rightarrow \eta \pi \pi + \eta^{\prime} \pi \pi + \pi KK$ &  $  7 \pm   3$  & $\eta_{S}^{E} \rightarrow K^{\star} K$ &  $  128 ^{+204}_{-128}$ \\\hline
  $\eta_{S}^{E} \rightarrow \eta \pi \pi$ and $\eta^{\prime} \pi \pi$ & $  156 ^{+245}_{-156}$ &  $\sigma_S^E \rightarrow KK$ &  $  21 ^{+39}_{-21} $\\\hline 
  $\sigma_S^E \rightarrow \eta \eta^{\prime}$ &  $  12 \pm   2$ & $\sigma_S^E \rightarrow \eta \eta$  &   $  6 \pm    1$\\\hline
   $\sigma_S^E \rightarrow K_1 K$ & $  2^{+5}_{-2}$ & $\sigma_S^E \rightarrow \eta^{\prime} \eta^{\prime}$ &   $ 1 \pm 0$ \\\hline
    $a_{0}^{E} \rightarrow \eta \pi$ & $  72 \pm 12$ &  $a_{0}^{E} \rightarrow KK$  &  $  70 \pm 40$ \\\hline
     $a_{0}^{E} \rightarrow \eta^{\prime} \pi$ &  $\; \, 32 \pm \; \, 5$ &  $a_{0}^{E} \rightarrow f_1(1285) \pi$ & $\; \, 16 \pm \; \, 3$\\\hline
      $K_{0}^{\star E}\rightarrow K \pi$ &  $\; \; 51 \pm 35$ & $K_{0}^{\star E} \rightarrow \eta^{\prime} K$ &   $\;  24 \pm \,  4$   \\\hline
       $K_{0}^{\star E} \rightarrow K_1 \pi$ &   $\; \;  \, 6 \pm \, 4$ & $K_{0}^{\star E} \rightarrow \eta K$ &  $\; \; \, 4^{+7}_{-4}$   \\\hline
       $K_{0}^{\star E} \rightarrow a_1(1260) K$  &  $\; \; \; \; 3 \pm \; \, 2$ &
 $K_{0}^{\star E} \rightarrow f_1(1285) K$ & $\; \; \; \; 1 \pm \; \, 1$ \\\hline
\end{tabular}
\caption{Decay rates of excited scalar mesons.}
\label{tab:excited-decays}
\end{table}
%-----------------------
\begin{table}[ht!]
\centering
\renewcommand{\arraystretch}{1.1}
\begin{tabular}{|c|c|c|c|}
\hline
Decay Rates & Estimates (MeV) & Decay Rates & Estimates (MeV) \\ \hline
    $\eta_{S}^{E} \rightarrow KK \pi$ &  $3 \pm  0$ &  $\pi^E \rightarrow \rho \pi$ &  $368 \pm 37$\\ \hline
     $\pi^E \rightarrow 3\pi$ &  $204 \pm 15$ & $\pi^E \rightarrow K K \pi$ &  $\; \; \; \, 2 \pm \;  \, 0$ \\ \hline
      $K^{E} \rightarrow K^{\star} \pi$ &  $112 \pm 11$ & $K^{E} \rightarrow K \pi \pi$ &  $\; \; \, 35 \pm \; \; \, 4$   \\\hline
       $K^{E} \rightarrow \rho K$ & $\; \; \, 20 \pm \; \; \, 2$   &   $K^E \rightarrow \omega K$ &  $\; \; \; \; \, 7 \pm \; \; \, 1$   \\\hline
\end{tabular}
\caption{Decay rates of excited pseudoscalar mesons.}
\label{tab:excited-decays2}
\end{table}
While the assignment of the radially excited pseudoscalars is rather established (ses the quark model review of the PDG \cite{ParticleDataGroup:2024cfk}), this is not the case for the scalars. According to the  model results, the resonances $f_0(1790)$ (with decay width $405\pm 96 \,\text{MeV}$) and $a_0(1950)$ (with decay width $271 \pm 40 $ MeV \cite{BaBar:2015kii}) are predicted to be the nonstrange excited quark-antiquark states (the latter one is the isotriplet member). The state $K_0^\star(1950)$ is a strong candidate for an excited scalar kaon. We also predict an excited strange-antistrange isoscalar state $\sigma_S^E\equiv\overline{s}s$ with a mass $m_{\sigma_S^E}=2038\pm 24 \,\text{MeV}$  positioned between the masses of $f_0(2020)$ and $f_0(2100)$ and with a relatively small decay width ($\Gamma_{\sigma_S^E}\leq 110\, \text{MeV}$).
%%%%%%%%%%%%%%%%
%If we use the excited pseudoscalars during the fit procedure, $\pi(1300)$ and $K(1460)$ are quite well described as excited pseudoscalar quarkonium states. However, in this case, their scalar chiral partners become broad enough (see Ref.\cite{Parganlija:2016yxq} for further details).
%%%%%%%%%%%%%%%
The extension to other radially excited chiral multiplets is possible, e.g. to (axial-)vector mesons forming homochiral fields ($L_{\mu}^E$ and $R_{\mu}^{E}$). Based on recent PDG 2024 data \cite{ParticleDataGroup:2024cfk}, one has the vector states $\rho(1450)$, $K^\ast(1410)$, $\phi(1680)$, $\omega(1420)$ and the axial-vector ones $a_1(1640)$, $K_1(1650)$. One may also extend the procedure to tensor and axial-tensor homochiral multiplets ($\textbf{L}_{\mu\nu}^E$ and $\textbf{R}_{\mu\nu}^E$), using e.g. the resonances $a_2(1700)$, $K^\ast_2(1980)$, $f_2(1640)$, and so on.

\subsection{Extension of the eLSM to \texorpdfstring{$N_f=4$}{Nf4}}
\label{sub:Nf=4}
 
In this subsection, following Ref. \cite{Eshraim:2014eka}, we present the results for the four-flavor extended linear sigma model. In this case, all nonets are enlarged to $4\times 4$ matrices, hence 16-th multiples.

To this end, we introduce new mesons including the charm quark. For the case of pseudoscalars, these are the charmed states $D^{0,\pm}$, the open strange-charmed states $D^\pm_s$, and a hidden charmed ground
 state $\eta_c(1S)$, which are collected within the matrix $P$ given below:
\begin{equation}
P=\frac{1}{\sqrt{2}}\left(
\begin{array}
[c]{cccc}%
\bar{u} i\gamma^5 u & \bar{d} i\gamma^5 u & \bar{s} i\gamma^5 u & \bar{c} i\gamma^5 u\\
\bar{u} i\gamma^5 d & \bar{d} i\gamma^5 d & \bar{s} i\gamma^5 d & \bar{c} i\gamma^5 d\\
\bar{u} i\gamma^5 s & \bar{d} i\gamma^5 s & \bar{s} i\gamma^5 s & \bar{c} i\gamma^5 s\\
\bar{u} i\gamma^5 c & \bar{d} i\gamma^5 c & \bar{s} i\gamma^5 c & \bar{c} i\gamma^5 c
\end{array}
\right) =\frac{1}{\sqrt{2}}\left(
\begin{array}
[c]{cccc}%
\frac{1}{\sqrt{2}}(\eta_{N}+\pi^{0}) & \pi^{+} & K^{+} & D^{0}\\
\pi^{-} & \frac{1}{\sqrt{2}}(\eta_{N}-\pi^{0}) & K^{0} & D^{-}\\
K^{-} & \overline{K}^{0} & \eta_{S} & D_{s}^{-}\\
\overline{D}^{0} & D^{+} & D_{s}^{+} & \eta_{c}%
\end{array}
\right)  \text{ .} \label{p}%
\end{equation}
For scalar fields, we include the open charmed
 $D^{\ast\,0,\pm}_0$, the strange-charmed meson $D^{\ast\,\pm}_{s0}$, and the hidden
charmed resonance $\chi_{c0}\equiv \chi_{c0}(1P) $, which are
 assigned to $D^\ast_0(2400)^{0,\pm}$, $D_{s0}^\ast(2317)^\pm$, and to the $\eta_c$ meson.
The matrix has the following form:
\begin{equation}
S=\frac{1}{\sqrt{2}}\left(
\begin{array}
[c]{cccc}%
\frac{1}{\sqrt{2}}(\sigma_{N}+a_{0}^{0}) & a_{0}^{+} &
K_{0}^{\ast+} &
D_{0}^{\ast0}\\
a_{0}^{-} & \frac{1}{\sqrt{2}}(\sigma_{N}-a_{0}^{0}) &
K_{0}^{\ast0} &
D_{0}^{\ast-}\\
K_{0}^{\ast-} & \overline{K}_{0}^{\ast0} & \sigma_{S} & D_{s0}^{\ast-}\\
\overline{D}_{0}^{\ast0} & D_{0}^{\ast+} & D_{s0}^{\ast+} & \chi_{c0}%
\end{array}
\right)  \text{ ,}%
\end{equation}
where also the explicit quark content is displayed. As for the three-flavor case, we construct a chiral field containing both of the above-mentioned nonets $\Phi=S+iP$, which has analogous transformation rules under parity, charge conjugation, and chiral symmetry (here under $SU(4)_L\times SU(4)_R$) presented in Table \ref{tab:chiral-transformations}.  
Of course, a large explicit breaking is present due to the large $c$-quark mass, but the idea is that this breaking can be relegated to the linear and quadratic flavor-symmetry-breaking mass terms.
For vector mesons, one  includes the nonstrange-charmed fields $D^{\ast 0},\,D^{\ast\pm }$, that
corresponds to $D^{\ast }(2007)^{0}$ and
$D^{\ast }(2010)^{\pm }$, the strange-charmed
$D_{0}^{\ast\pm}$ is assigned to the resonance $D_{0}^{\ast\pm}$
(with mass $m_{D_{0}^{\ast\pm}}$), and the $J/\psi$ as the well-known
charmonium state $J/\psi (1S)$. All of them are part of the  matrix $V^{\mu}$:%
\begin{equation}
V^{\mu}=\frac{1}{\sqrt{2}}\left(
\begin{array}
[c]{cccc}%
\frac{1}{\sqrt{2}}(\omega_{N}+\rho^{0}) & \rho^{+} &
K^{\ast}(892)^{+} &
D^{\ast0}\\
\rho^{-} & \frac{1}{\sqrt{2}}(\omega_{N}-\rho^{0}) &
K^{\ast}(892)^{0} &
D^{\ast-}\\
K^{\ast}(892)^{-} & \bar{K}^{\ast}(892)^{0} & \omega_{S} & D_{s}^{\ast-}\\
\overline{D}^{\ast0} & D^{\ast+} & D_{s}^{\ast+} & J/\psi
\end{array}
\right)  ^{\mu}\text{.}%
\end{equation}
%%%%%%%%%%%%%%%%%%%%
For axial-vector fields, one adds the open charmed mesons $D_{1}$ and $D_{s1}$, assigned to
$D_{1}(2420)$ and $D_{s1}(2536)$, and the 
$c\overline{c}$ state $\chi _{c1}$ that corresponds to the well-known
resonance $\chi _{c1}(1P)$. The matrix is:
\begin{equation}
A_1^{\mu}=\frac{1}{\sqrt{2}}\left(
\begin{array}
[c]{cccc}%
\frac{1}{\sqrt{2}}(f_{1N}+a_{1}^{0}) & a_{1}^{+} & K_{1}^{+} & D_{1}^{0}\\
a_{1}^{-} & \frac{1}{\sqrt{2}}(f_{1N}-a_{1}^{0}) & K_{1}^{0} & D_{1}^{-}\\
K_{1}^{-} & \bar{K}_{1}^{0} & f_{1S} & D_{s1}^{-}\\
\bar{D}_{1}^{0} & D_{1}^{+} & D_{s1}^{+} & \chi_{c1}%
\end{array}
\right)  ^{\mu}\,.
\end{equation}
Homochiral (axial-)vector fields are  $L^{\mu}=V^{\mu}+A_1^{\mu}$ and
$R^{\mu}=V^{\mu }-A_1^{\mu}$ and follow the corresponding transformation rules given in Table \ref{tab:chiral-transformations} for three-flavor fields. 
%%%%%%%%%%%%%%%%%%
\begin{table}[ht!]
\centering
\begin{tabular}
[c]{|c|c|c|c|c|c|}\hline
Field & PDG & Quark content & $I$ & $J^{\mathcal{P}\mathcal{C}}$ &  Mass (MeV)\\\hline\hline
 $\bar{D}^{0}$, $D^{0}$, $D^{\pm}$ & $D$  & $u\bar{c}$, $\bar{u}c$,$d\bar{c}$, $\bar{d}c$ & $\frac{1}{2}$ & $0^{-+}$ &$1864.86 \pm 0.13$\\\hline
%%%%%%%%%%%%
$D_S^{\pm}$ & $D_S$  & $s\bar{c}$, $\bar{s}c$ & $0$& $0^{-+}$ &$1968.50 \pm 0.32 $\\\hline
%%%%%%%%%%%%
$\eta_c$ & $\eta_c(1s)$  & $c\bar{c}$&  $0$ & $0^{-+}$ &$2673 \pm 118$\\\hline\hline
%%%%%%%%%%%%
$\bar{D}_0^\ast$, $D_0^\ast$ , $D_0^{\ast\pm}$  & $D_0^\ast(2400)^{0}$ & $u\bar{c}$, $\bar{u}c$,$d\bar{c}$, $\bar{d}c$ & $\frac{1}{2}$ & $0^{++}$&$2414 \pm 77 $\\\hline
%%%%%%%%%%%%
 $D_{s0}^{\ast\pm}$ & $D_{s0}^\ast(2317)^{\pm}$  & $s\bar{c}$, $\bar{s}c$ & 0 & $0^{++}$&$2467 \pm 76 $\\\hline
%%%%%%%%%%%%
$\chi_{c0}$& $\chi_{c0}(1P)$  & $c\bar{c}$ & 0 & $0^{++}$ &$3144 \pm128 $\\\hline\hline
%%%%%%%%%%%%
$\bar{D}^{\ast0}$, $D^{\ast0}$, $D^{\ast\pm}$ & $D^\ast(2007)^{0}$ & $u\bar{c}$, $\bar{u}c$,$d\bar{c}$, $\bar{d}c$ & $\frac{1}{2}$  & $1^{--}$& $2168 \pm 70  $\\\hline
%%%%%%%%%%%%
$D_{s}^{\ast\pm}$& $D^{\ast\pm}_S(2010)$  & $s\bar{c}$, $\bar{s}c$ & 0& $1^{--}$&$2203 \pm 69 $\\\hline
%%%%%%%%%%%%
$J/\psi$ & $J/\psi(1s)$  & $c\bar{c}$& 0 & $1^{--}$ &$2947 \pm 109$\\\hline\hline
%%%%%%%%%%%%
 $\bar{D}_1^{0}$, $D_1^{0}$, $D_1^{\pm}$& $D_1(2420)^{0}$  & $u\bar{c}$, $\bar{u}c$,$d\bar{c}$, $\bar{d}c$ & $\frac{1}{2}$ & $1^{++}$ &$2429 \pm 63$\\\hline
%%%%%%%%%%%%
$D_{s1}^{\pm}$& $D_{s1}^{\pm}$  & $s\bar{c}$, $\bar{s}c$ & 0& $1^{++}$ &$2480 \pm 63 $\\\hline
%%%%%%%%%%%%
$\chi_{c1}$& $\chi_{c1}(1P)$  & $c\bar{c}$& 0 & $1^{++}$ &$3239 \pm 101  $\\\hline
%%%%%%%%%%%%
\end{tabular}
\caption{Charmed meson fields within the eLSM, PDG correspondence, quark content, quantum numbers, and PDG mass values from \cite{ParticleDataGroup:2012pjm}.}
\label{Tab:charmed_masses_PDG}
\end{table}
The Lagrangian of the eLSM presented in Sec. \ref{Ssec:elsm_Lagr} retains its form when the matrices are $4 \times 4$. A special attention is devoted to the symmetry breaking terms, which are a straightforward extensions of those in \ref{Ssec:elsm_Lagr} but are now important (and not only a correction) to describe the large charm mass: 
\begin{align}   \mathcal{L}_{\text{sym-brk}}^{N_f=4}=\Tilde{c}\Big(\det{\Phi}-\det{\Phi}^\dagger\Big)^2 +\mathrm{Tr}[H(\Phi^{\dag}%
+\Phi)] -2\,\mathrm{Tr}%
[\hat{\epsilon}\Phi^{\dagger }\Phi]+\mathrm{Tr}[\Delta(L_{\mu\nu}^2+R_{\mu\nu}^2)]
\text{ ,}
\end{align}
where $\Tilde{c}\equiv 2c_2/\phi_C^2$ and explicit forms of the matrices (only diagonal elements are non-zero) are
\begin{equation}
    H  = \text{diag}\Big(\frac{h_{0N}}{2},\frac{h_{0N}}{2},\frac{h_{0S}}{2}, \frac{h_{0C}}{\sqrt{2}}\Big)\,,\, 
\Delta = \text{diag}\Big(\delta_{N},\delta_{N}, \delta_{S},\delta_{C}\sim m_c^2\Big) \text{ , } \hat{\epsilon}= \text{diag}\Big(\varepsilon_N,\varepsilon_N,\varepsilon_S,\varepsilon_C\Big)\,.
\end{equation}
Above: $h_{0,N,S,C} \propto m_{n,s,c}$, $\delta_{N,S,C} \propto m_{n,s,c}^2$, and $\epsilon_{N,C,S} \propto m_{n,s,c}^2$. As previously, we set $\delta_N = \epsilon_N =0$.
SSB implies that also the scalar charmed meson condenses, hence we need to perform a shift on the field $\chi _{C0}$ similar to Eq. \eqref{eq:scalar-shift} by its 
vacuum expectation value $\phi_C$ as
\begin{align}
\chi_{c0} \rightarrow \chi_{c0}+\phi_C\,.
\end{align}
We present the expression for the masses of (pseudo)scalar and (axial-)vector charmed mesons within the eLSM in Tables \ref{tab:mass-spin0} and \ref{tab:mass-spin1}. In particular, for chiral partners one has:
\begin{align}    m_{D_1}^2&=m_{D^\star}^2+\sqrt{2}(g_1^2-h_3)\phi_N\phi_C\,,\\    m_{\chi_{c1}}^2&=m_{J/\psi}^2+2(g_1^2-h_3)\phi_C^2\,,\\
m_{D_{s1}}^2&=m_{D_S^\star}^2+2(g_1^2-h_3)\phi_S\phi_C\, \text{ ,}
    \end{align}
thus the mass differences do not depend on the charm mass, but only on the chiral condensates, as in the light sector.
For the coupling constants of the model ($g_1$, $\lambda_{1,2}$ and $h_{1,2,3}$), we use the values determined in the $N_f=3$ case, see Table \ref{Tab:params_eLSM_2013}. Using the PDG masses for 12 mesonic states given in Table \ref{Tab:charmed_masses_PDG}, we determine the following three parameters:
\begin{align*}
    \phi_C =178\pm 28\, \text{ MeV}\,,\quad \delta_C= (3.91\pm 0.36)\cdot 10^6\, \text{ MeV}^2 \,,\quad \varepsilon_C= (2.23\pm 0.71)\cdot 10^6\, \text{ MeV}^2\,.
\end{align*}
It is worth noting that the charm condensate is comparable to the non-strange and strange quark condensates. The masses of open charmed mesons closely align with PDG values within theoretical errors, while the masses of charmonia, excluding $J/\psi$, are underestimated by about $10\%$.
%------------------
\begin{table}[ht!]
\centering
\renewcommand{\arraystretch}{1.3} 
\begin{tabular}{|c|c|c|}
\hline
Mass  & $\text{Equations of Mass Square}$ & eLSM (MeV)  \\ \hline\hline	
$m_{\eta_C}$ &  $Z_{\eta_C}^{2}\left[  m_{0}^{2}+\lambda_1(\phi_N^2+\phi_S^2)+(\lambda_1+\lambda_2)\phi_C^2+\frac{c}{8}\phi_N^4\phi_S^2\right]
$ & $2673 \pm 118$  \\
\hline
$m_{D}$ &  $Z_{D}^{2}\left[  m_{0}^{2}+\left(  \lambda_{1}%
+\frac{\lambda_{2}}{2}\right)  \phi_{N}^{2}+\lambda_{1}\phi_{S}^{2}-\frac{\lambda_2}{\sqrt{2}}\phi_N\phi_C+(\lambda_1+\lambda_2)\phi_C^2
\right]$ & $1981 \pm 73$  \\
\hline
$m_{D_{s}}$ &  $Z_{D_S}^{2}\left[  m_{0}^{2}+\lambda_1\phi_N^2+\left(  \lambda_{1}%
+\lambda_{2}\right)  \phi_{S}^{2}-\lambda_2\phi_S\phi_C+(\lambda_1+\lambda_2)\phi_C^2
\right]$ & $2004 \pm 74$  \\
\hline
$m_{\chi_{C0}}$ &  $m_{0}^{2}+\lambda_1(\phi_N^2+\phi_S^2)+3(\lambda_1+\lambda_2)\phi_C^2$ & $3144 \pm 128$  \\
\hline
$m_{D_0^\star}$  &  $Z_{D_0^\star}^{2}\left[  m_{0}^{2}+\left(  \lambda_{1}%
+\frac{\lambda_{2}}{2}\right)  \phi_{N}^{2}+\lambda_{1}\phi_{S}^{2}+\frac{\lambda_2}{\sqrt{2}}\phi_N\phi_C+(\lambda_1+\lambda_2)\phi_C^2
\right]$ & $2414 \pm 77$  \\\hline
$m_{D_{s0}^\star}$  & $Z_{D_{s0}}^{2}\left[  m_{0}^{2}+\lambda_1\phi_N^2+\left(  \lambda_{1}%
+\lambda_{2}\right)  \phi_{S}^{2}+\lambda_2\phi_S\phi_C+(\lambda_1+\lambda_2)\phi_C^2
\right]$ & $ 2467 \pm 76 $  \\\hline
	\end{tabular}%
\caption{ Masses of (pseudo)scalar charmed mesons within the eLSM.  }\label{tab:mass-spin0}
\end{table}
%---------------
\begin{table}[ht!]
\centering
\renewcommand{\arraystretch}{1.3} 
\begin{tabular}{|c|c|c|}
\hline
Mass & $\text{Equations of Mass Square}$  & eLSM (MeV)\\ \hline\hline	
 $m_{D^\star}$  &  $m_{1}^{2}+\delta_N+\delta_C+\frac{\phi_{N}^{2}}{2}(\frac{g_1^2}{2}+h_{1}+\frac{h_{2}}{2})%
+\frac{\phi_N\phi_C}{\sqrt{2}}(h_3-g_1^2)+\frac{\phi_C^2}{2}
(g_1^2+h_1+h_2)+\frac{h_1}{2}\phi_S^2$ & $2168 \pm 70$  \\ \hline
$m_{J/\psi}$  &  $m_{1}^{2}+2\delta_{C}+\frac{h_1}{2}(\phi_N^2+\phi_S^2)+(\frac{h_1}{2}+h_2+h_3)\phi_C^2$ & $2947 \pm 109$  \\\hline
$m_{D_{s}^\star}$  &  $m_{1}^{2}+\delta_S+\delta_C+\frac{\phi_{S}^{2}}{2}(g_1^2+h_{1}+h_{2})%
+\phi_S\phi_C(h_3-g_1^2)+\frac{\phi_C^2}{2}
(g_1^2+h_1+h_2)+\frac{h_1}{2}\phi_N^2$ & $2203 \pm  69$  \\\hline
$m_{D_{s1}}$  &  $m_{1}^{2}+\delta_S+\delta_C+\frac{\phi_S^2}{2}(g_1^2+h_1+h_2)+\phi_S\phi_C(g_1-h_3^2)+\frac{\phi_C^2}{2}(g_1^2+h_1+h_2)+\frac{h_1}{2}\phi_N^2$ & $2480 \pm 63$  \\\hline
$m_{D_1}$  &  $m_{1}^{2}+\delta_N+\delta_C+\frac{\phi_N^2}{2}(\frac{g_1^2}{2}+h_1+\frac{h_2}{2})+\frac{\phi_N\phi_C}{\sqrt{2}}(g_1^2-h_3)+\frac{\phi_C^2}{2}
(g_1^2+h_1+h_2)+\frac{h_1}{2}\phi_S^2$ & $2429 \pm 63$   \\\hline
$m_{\chi_{C1}}$  &  $m_{1}^{2}+2\delta_{C}+\frac{h_1}{2}(\phi_N^2+\phi_S^2)+2g_1\phi_C^2+\phi_C^2(\frac{h_1}{2}+h_2-h_3)$ & $3239 \pm 101$  \\\hline
\end{tabular}%
\caption{ Masses of (axial-)vector charmed mesons within the eLSM.  }\label{tab:mass-spin1}
\end{table}
%--------
Table \ref{tab:decay-nf=4} illustrates decay channels in eLSM versus the PDG, see Ref. \cite{Eshraim:2014eka} for details. 
Of course, a chiral model, being intrinsically a low-energy QCD approach, cannot be as precise as approaches constructed for heavy quarks (e.g., based on heavy quark symmetry \cite{Brambilla:2019esw}). Yet the very fact that many decay widths can be correctly reproduced (especially the ones that are allowed by large-$N_c$ dominant terms such as the first entry in Table \ref{tab:decay-nf=4}) by using the parameters that were determined in the low-energy sector, implies that a `remnant' of chiral symmetry is still active for charmed mesons as well, even if the explicit breaking is large. 
The extension of the eLSM to $N_f=4$ also predicts an interesting decay channel of $\eta_c$ into pseudoscalar glueball $\Gamma_{\eta_c\rightarrow P_G\pi\pi} = 0.124 \,\text{MeV}$ (using $2.6$ GeV for the pseudoscalar glueball mass). Other decay channels of $\eta_c$ within the eLSM can be found in Ref. \cite{Eshraim:2018jkt}.

The somewhat unexpected result that some interactions still fulfill chiral symmetry opens also interesting outlooks, such as the inclusion of (axial-)tensor chiral multiples ($\mathbf{L}^{\mu\nu}$ and $\mathbf{R}^{\mu\nu}$) by (including the resonances $\chi_{c2}(1P)$ and $\psi_2(3823)$) and the study of the charm condensate at nonzero temperature.
\begin{table}[ht!]
\centering
   \begin{tabular}
[c]{|c|c|c|}\hline
Decay Channel & eLSM [MeV] & PDG [MeV]\\\hline
$D_{0}^{\ast}(2400)^{0}\rightarrow D\pi$ &
$139_{-114}^{+243}$ & $\Gamma_{\text{tot}}=267\pm40$\\\hline
$D_{0}^{\ast}(2400)^{+}\rightarrow D\pi$ &
$51_{-51}^{+182}$ &   $\Gamma_{\text{tot}}=283\pm24\pm34$\\\hline
$D^{\ast}(2007)^{0}\rightarrow D^{0}\pi^{0}$ & $0.025\pm0.003$ & $<1.3$\\\hline
$D^{\ast}(2010)^{+}\rightarrow D^{+}\pi^{0}$ & $0.018_{-0.003}^{+0.002}$ &
$0.029\pm0.008$\\\hline
$D^{\ast}(2010)^{+}\rightarrow D^{0}\pi^{+}$ & $0.038_{-0.004}^{+0.005}$ & $0.065\pm0.017$\\\hline
$D_{1}(2420)^{0}\rightarrow D^{\ast}\pi$ &
$65_{-37}^{+51}$ &   $\Gamma_{\text{tot}}=27.4\pm
2.5$\\\hline
$D_{1}(2420)^{0}\rightarrow D^{0}\pi\pi$ & $0.59\pm0.02$ & seen\\\hline
$D_{1}(2420)^{0}\rightarrow D^{+}\pi^{-}\pi^{0}$ & $0.21_{-0.015}^{+0.01}$ &
seen\\\hline
$D_{1}(2420)^{+}\rightarrow D^{\ast}\pi$ &
$65_{-36}^{+51}$ &  $\Gamma_{\text{tot}}=25\pm
6$\\\hline
$D_{1}(2420)^{+}\rightarrow D^{+}\pi\pi$ & $0.56\pm0.02$ & seen\\\hline
$D_{1}(2420)^{+}\rightarrow D^{0}\pi^{0}\pi^{+}$ & $0.22\pm0.01$ &
seen\\\hline
\end{tabular}
\caption{Decay widths of charmed
mesons.}
\label{tab:decay-nf=4}
\end{table}

\section{The eLSM at nonzero temperature and density}
\label{Sec:eLSM_finite_T}

\subsection{General features and the coupling of the eLSM to quark d.o.f.}
\begin{figure}[ht!]
    \centering    
    \includegraphics[width=0.6\textwidth]{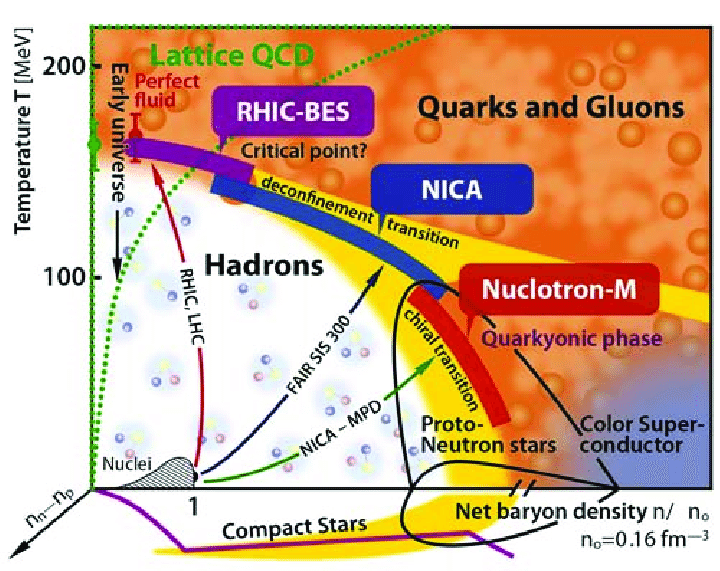}
    \caption{The envisioned phase diagram of strongly interacting matter, as presented in \cite{RodriguezCahuantzi:2017upc}, is shown along with the regions of exploration for various experiments.}
    \label{Fig:envis_QCD_phase_diag}
\end{figure}

The study of the strong interaction in the medium represents one of the most significant research areas in high-energy physics in recent decades. This is due to the fact that matter produced in hadron colliders (CERN, RHIC) heats up and becomes very dense during the collision. Subsequently, as it expands and cools, it undergoes a trajectory in the $T$-$\rho_B$ or $T$-$\mu_B$ plane, where $T$ is the temperature, while $\rho_B$ and $\mu_B$ are the baryonic density and chemical potential, respectively. The envisioned image of the phase diagram of the strongly interacting matter can be seen in Fig.~\ref{Fig:envis_QCD_phase_diag}.
According to the figure, it is commonly believed that at large temperature and/or baryon chemical potential/density, the normal hadronic matter transforms into the quasi-free gas of quarks and gluons. There is a phase boundary between the two phases. The transition itself is the deconfinement phase transition. When the matter is deconfined, the chiral symmetry is also restored, which is called the chiral phase transition. The two phase boundaries, i.e. deconfinement and chiral, go hand in hand on some part of the $T$-$\mu_B$ plane, but there can be regions where they are separated. From lattice QCD it is known that at $\mu_B=0$ the chiral phase transition is continuous or crossover type \cite{Aoki:2006we}, while other parts of the phase diagram are less known, since they cannot be directly accessed. For example, lattice QCD cannot be used directly at finite $\mu_B$, due to the notorious sign problem \cite{Gattringer:2016kco}. However, most effective models of QCD -- with which finite $\mu_B$ regions can be reached -- predict that the phase transition as a function of $\mu_B$ at $T=0$ is of first order. Accordingly, there should be a so-called critical end point (CEP) along the chiral phase boundary separating the crossover region from the first-order one. Different models predict the location of the CEP in a wide range of $T$ and $\mu_B$. On the other hand, current and future experiments address different parts of the phase diagram, but the CEP has not yet been found. Another interesting part of the phase diagram is the small temperature $T\approx 0$ and large $\mu_B$ region where the so-called neutron stars or more generally compact stars live. Neutron stars, which are extremely dense stable astrophysical objects, cool down to $T\sim 100$~eV within a few years after their formation due to neutrino cooling, while their core density is estimated to be around $2-5 \rho_0$, where $\rho_0=0.16 \text{\,fm}^{-3}$ is the nuclear saturation density. In the last decade, neutron star observations have regained momentum with new experiments such as the NICER experiment \cite{NICER:2016nasa}, which can simultaneously measure the radii and masses of neutron stars.
The advantage of effective models of QCD such as the eLSM is that all parts of the phase diagram can be explored. Thus, the eLSM can be used to determine the chiral phase boundary, or even to compute mass-radius curves for compact stars that can be directly compared with experimental astrophysical data. 
In this section we present the eLSM in the medium by following the treatment presented in Ref. \cite{Kovacs:2016juc}. For additional eLSM works in the medium using the mean-field approximation, see Refs.~\cite{Schaefer:2007pw,Schaefer:2009ui,Kovacs:2022zcl,Gupta:2009fg,Tawfik:2014gga,Tawfik:2016edq,Tawfik:2019tkp}, while for using the renormalization group approach, see e.g. \cite{,Herbst:2013ail} and references therein.
It is also worth noting that the phase structure is thought to be more complicated in the low temperature, high density regime, where a color superconducting phase is expected \cite{Alford:RevModPhys801455,Anglani:RevModPhys86509}. However, as far as we know, this phase has not been studied in the eLSM.          

Theoretically, the in-medium properties of strongly interacting matter can be described within the framework of finite temperature field theory \cite{Kapusta:2023eix,Bellac:2011kqa,Das:2023abc}. To go to finite temperature, the standard procedure is to perform an analytic continuation to imaginary times, i.e. $t\to -i\tau$. Then the grand potential $\Omega(T,\mu_q)$ can be calculated from the partition function $\mathcal{Z}$ for a spatially uniform system of volume $V$ in thermal equilibrium for temperature $T=1/\beta$. Any thermodynamic quantity can be derived from the grand potential. So far we have only considered mesons, but to get to finite baryon density/baryon chemical potential, particles carrying baryon number should be included. In practice, instead of baryons, the so-called constituent quarks ($q_i,\; i\in(u,d,s)$) are included in the eLSM by adding a Yukawa term to the Lagrangian as follows,
\begin{equation}
    \bar{q}\left[i \gamma_{\mu}D^{\mu}-\mathcal{M}\right]q\,, 
\end{equation}
with
\begin{equation}
    q = \left(\begin{array}[c]{c}
         q_u  \\
         q_d  \\
         q_s
    \end{array}\right)\; \text{ , } D^{\mu}q = \partial^{\mu}q - iG^{\mu}q \label{Eq:cov_deriv_const_q} \text{ , }
    \mathcal{M} = g_F\left(\mathds{1}_{4\times 4} S + i\gamma_5 P\right)\; .
    %\\    M_{S} &\equiv \sum_{a=0}^{8}S_{a} t_{a}\; ,\quad M_{PS} \equiv \sum_{a=0}^{8} iP_{a}t_{a}\; \text{ .}
\end{equation}
Here, $S$ and $P$ are the scalar and pseudoscalar nonets (see Eq.~\eqref{Eq:J0_sc_ps_nonet}), and, as before, $G^{\mu}$ is the gluon field. Since the eLSM is an effective approach where the fields are physical meson fields, the gluons are taken into account via the so-called Polyakov loop variable, which mimics some properties of confinement (see below). Returning to the grand potential and the partition function, $\mathcal{Z}$ can be written according to \cite{Kapusta:2023eix},
\begin{align}
{\mathcal{Z}} &= e^{-\beta V \Omega(T,\mu_q)}=\textnormal {Tr}\exp
\bigg[ -\beta \bigg( \hat {\mathcal{H}} - \sum_{i=u,d,s}\mu_i \hat
{\mathcal{Q}}_i \bigg)\bigg]\nonumber \\
&=\int\limits_\textnormal{PBC} \prod_a {\mathcal{D}} \varphi_a
\int\limits_\textnormal{APBC} \prod_i {\mathcal{D} q_i} {\mathcal{D} q^\dagger_i}  \exp\bigg[-\int_0^\beta d\tau \int_V d^3 x \bigg({\mathcal{L}} + \mu_q\sum_{i} q^\dagger_i q_i\bigg)\bigg]
\text{ ,}
\label{Eq:Z_finite_T}
\end{align}
where PBC and APBC refer to the periodic and antiperiodic boundary conditions, $\hat{\mathcal{Q}}_i$ is a conserved charge operator, and $\varphi_a$ denotes all the mesonic fields. There is no integration over the gluons, since they are only used as a constant background field to generate the mean-field level Polyakov loop variable. Moreover, here we assumed symmetric quark matter, i.e. $\mu_u = \mu_d = \mu_s \equiv \mu_q = \mu_B/3$, where $\mu_q$ is the quark and $\mu_B$ is the baryon chemical potential. Unfortunately, even in the case of eLSM, the partition function in Eq.~\eqref{Eq:Z_finite_T} cannot be calculated exactly, so some approximation is needed. In \cite{Kovacs:2016juc} a so-called hybrid approach was used, where the meson potential is at the tree-level, the fermion determinant obtained after performing the functional integration over the quark fields is evaluated for vanishing mesonic fluctuating fields (but the integration itself is exact), and the lightest mesonic thermal fluctuations ($\pi$, $K$, $f_0^L$) are taken into account in the pressure. This approach considers fermion vacuum and thermal fluctuations, neglects meson vacuum fluctuations, and considers only the lightest meson fluctuations in the pressure. Considering the mesons, this approach is not self-consistent. Self-consistent approaches at different loop levels in the eLSM are outlined in \cite{Kovacs:2021kas}. The explicit form of the grand potential can be given in this hybrid approach by implementing the Polyakov loop variables, see the next subsection. 

Before proceeding, it is worth discussing the relationship between quarks and mesons in the case of the NJL and eLSM approaches. In our approach, the mesons are ` de facto' treated as elementary fields rather than emerging from integrating out quark degrees of freedom as in NJL-type models. 
Formally, the eLSM version implemented in this chapter is analogous to the quark-level LSM that arises as an intermediate step in the bosonization processes from quark to hadronic d.o.f. Yet, we will not attempt to link the parameters of NJL-like models to ours.
Nevertheless, the explicit coupling between mesons and constituent quarks via a Yukawa interaction ensures that the dynamics of the quark sector is tightly bound to the mesonic observables. All model parameters are constrained by a detailed fit to PDG data, yielding good agreement with experimental meson masses and decay widths, while the symmetry constraints mirror those of QCD. Thus, while compositeness is not explicitly modeled as in NJL-based LSM approaches, it is implicitly embedded through parameter choices resulting from the fit to PDG data and symmetry constraints that are the same as the global symmetries of QCD. Moreover, meson masses consistently incorporate contributions from quark loops at the one-loop level, maintaining a dynamical connection between quark and meson degrees of freedom. With respect to the Mott transition (see e.g. \cite{Ropke:1982ino}), the underlying mechanism can be explored analogously to the NJL model; for example, the Mott condition of the pion, $m_{\pi}(T,\mu_B) = 2m_{u,d}^{\star}(T,\mu_B)$, where both the pion and the constituent quark masses are medium dependent, remains a meaningful criterion (for a study of the Mott transition in NJL, see \cite{Mao:2019avr}, while in the eLSM, see \cite{Blaschke:2016sqn}).  Finally, the resulting chiral phase diagram shows qualitative similarities with those obtained in the NJL and PQM models, underscoring the reliability of our finite-temperature and density investigations within the eLSM framework. 

\subsection{Polyakov loop and Polyakov loop variables}
\label{Ssec:Polyakov}

The Polyakov loop operator, which is a special Wilson loop of the gauge field in the temporal direction is defined after going to imaginary times ($G_0(t, \mathbf{x}) \to -iG_4(\tau,\mathbf{x})$) as
\begin{equation}
L = \mathcal{P}\exp\left(i\int_{0}^{\beta}d\tau G_4(\tau,\mathbf{x}) \right),
\end{equation}
where we assumed that the spatial components of the gluon field vanish (for more details see e.g. \cite{Fukushima:2003fw, Ratti:2005jh, Hansen:2006ee, Chatterjee:2011jd}). $L$ and $L^{\dagger}$ are matrices in the fundamental representation of the color gauge group $SU(3)$. Next we define the color traced Polyakov loop as,
\begin{equation}
    \Phi_{P}(\mathbf{x}) = \frac{1}{3}\Tr_c L(\mathbf{x}),
    \quad \bar\Phi_{P}(\mathbf{x}) = \frac{1}{3} \Tr_c L^{\dagger} (\mathbf{x})\,,
\label{Eq:Phi_barPhi}
\end{equation}
while the Polyakov loop variables are the thermal expectation values of  $\Phi_{P}(\mathbf{x})$ and $\bar\Phi_{P}(\mathbf{x})$,
\begin{equation}
    \braket{\Phi_{P}}_T = \frac{1}{3}\braket{\Tr_c L(\mathbf{x})}_T,\;  \braket{\bar\Phi_{P}}_T = \frac{1}{3}\braket{\Tr_c L^{\dagger}(\mathbf{x})}_T\;.
\end{equation}
In the pure gauge case, $\braket{\Phi_{P}}_T$ and $\braket{\bar\Phi_{P}}_T$ are related to the free energy of the infinitely heavy static quark and antiquark, respectively. As a further simplification, we go to the Polyakov gauge in which $G_4(\tau,\mathbf{x}) = G_4(\mathbf{x})$ is time-independent and diagonal in color space, and we also assume that $G_4(\mathbf{x})=G_4$ is homogeneous, so it can be written as
\begin{equation}
G_4 = \varphi_3\lambda_3+\varphi_8\lambda_8 \label{eq:G4_form}\;,\text{ with } \varphi_3,\varphi_8 \in \mathbb{R}\;.
\end{equation}
This can be substituted into the covariant derivative expression for the quarks \eqref{Eq:cov_deriv_const_q} and calculate the propagation of the constituent quarks on a constant gluon background. The role of the gluons in this approximation will be a color dependent imaginary chemical potential \cite{Kovacs:2016juc}. 

The Polyakov loop variables also have a Polyakov loop potential that drives the deconfinement phase transition with increasing temperature. The potential comes from pure gauge theory, and its form is constructed to reproduce some thermodynamic properties calculated on the lattice. There are still several possibilities for the form of the potential, but a commonly used one can be given as
\begin{equation}
    \beta^4U_{\text{Pol}}(\Phi_{P},\bar{\Phi}_{P}) = -\frac{1}{2}a(T)\Phi_{P}\bar{\Phi}_{P} + b(T)\ln\left(1 - 6\Phi_{P}\bar{\Phi}_{P} + 4(\Phi_{P}^3 + \bar{\Phi}_{P}^3) - 3(\Phi_{P}\bar{\Phi}_{P})^2\right), \label{Eq:Ulog}
\end{equation}
with
\begin{equation}
a(T) = a_0 + a_1\left(\frac{T_0}{T}\right) + a_2\left(\frac{T_0}{T}\right)^2, \quad b(T) = b_3\left(\frac{T_0}{T}\right)^3,
\end{equation} 
where for simplicity we used $\Phi_{P} \equiv  \braket{\Phi_{P}}_T, \bar\Phi_{P} \equiv \braket{\bar{\Phi}_{P}}_T$ and the values of the constants are $a_0=3.51$, $a_1=-2.47$, $a_2=15.22$, and $b_3=-1.75$.  

Returning to the grand potential in the hybrid approximation outlined above, it can be written as
\begin{equation}
\Omega_\textnormal{H}(T,\mu_q) = U(\left<\Phi\right>) +  U_{\text{Pol}}(\Phi_{P},\bar{\Phi}_{P})
+ \Omega_{\bar q q}^{(0)}(T,\mu_q),
\label{Eq:grand_pot_H}
\end{equation}
where $U(\left<\Phi\right>)$ represents the tree-level meson potential, $U(\Phi_{P},\bar\Phi_{P})$ is the Polyakov loop potential, and $\Omega_{\bar q q}^{(0)}$ is the fermion contribution for non-vanishing scalar backgrounds $\phi_N$ and $\phi_S$ and vanishing mesonic fluctuating fields. The tree-level classical potential is given by
\begin{equation}
 U(\left<\Phi\right>) =\frac{m_0^2}{2}\big(\phi_N^2+\phi_S^2\big) -\frac{c_1}{2\sqrt{2}}\phi_N^2\phi_S - h_S\phi_S - h_N\phi_N +\frac{\lambda_1}{4}\big(\phi_N^2+\phi_S^2\big)^2+\frac{\lambda_2}{8}\big(\phi_N^4+2\phi_S^4\big)\;,
\label{Eq:meson_pot}
\end{equation}
which, a part from the anomaly term, coincides with Eq. $V_{G \Phi}(G_0,\phi_N,\phi_S)$ discussed in Eq. \eqref{vgphi}. Skipping some details of the calculation of the fermion part -- see \cite{Kovacs:2016juc} -- one can arrive at 
\begin{align} 
  \Omega_{\bar q q}^{(0)}(T,\mu_q) &= \Omega_{\bar q q}^{(0)\textnormal{v}} + \Omega_{\bar q q}^{(0)\textnormal{T}}(T,\mu_q), \label{Eq:grandp_quark_tot}\\
  \Omega_{\bar q q;R}^{(0)\textnormal{v}} &= -\frac{3}{8\pi^2}\sum_{i=u,d,s} m_i^4 \ln\frac{m_i}{M_0} \label{Eq:grandp_quark_vac_R}\\
   \Omega_{\bar q q}^{(0)\textnormal{T}}(T,\mu_q) &= -2 T \sum_i\int \frac{d^3 p}{(2\pi)^3}\big[\ln g_i^+(p) + \ln g_i^-(p)\big]\;, \label{Eq:grandp_quark_T}
\end{align}
where
\begin{align}
    g_i^{+}(p) &= 1 + 3 \bar\Phi_{P}\, e^{-\beta E_i^+(p)} + 3\Phi_{P}\, e^{-2\beta E_i^+(p)} + e^{-3\beta  E_i^+(p)}\;, \\
    g_i^{-}(p) &= 1 + 3 \Phi_{P}\, e^{-\beta E_i^-(p)} + 3\bar\Phi_{P}\, e^{-2\beta E_i^-(p)} + e^{-3\beta  E_i^-(p)}\;, \\
    E_i^{\pm}(p) &= E_i(p)\mp\mu_i,\; E_i(p)=\sqrt{\mathbf{p}^2+m_i^2}\;, \\
    m_{u,d} &= \frac{g_F}{2}\phi_N \quad \textrm{and} \quad m_s = \frac{g_F}{\sqrt{2}} \phi_S\;. \label{Eq:const_quark_mass}
\end{align}
In \eqref{Eq:grandp_quark_tot} the contribution consists of two parts, a vacuum and a thermal part. The vacuum part in \eqref{Eq:grandp_quark_vac_R} is already renormalized. In the thermal part \eqref{Eq:grandp_quark_T} the appearing $\ln g_f^{\pm}$ factors contain the Polyakov loop variables and are related to modified Fermi-Dirac factors (see next section). Eq~\eqref{Eq:const_quark_mass} are the tree-level constituent quark masses. It is worth noting that if we consider the PCAC relations, then the $\phi_{N/S}$ condensates are fixed by the pion and kaon decay constants. Consequently, if one wanted to use baryons instead of the constituent quarks, then $m_p = g_F \phi_N/2$ would be the proton mass, which would lead to a very large value for the $g_F$ Yukawa coupling. However, since $g_F$ enters the expressions for the meson masses through the fermion vacuum fluctuations, this would lead to an unacceptable meson spectrum. Therefore, the baryons cannot be used instead of the constituent quarks in the way described above.   

\subsection{Field equations in the eLSM}
\label{Ssec:FE_finite_T}

The field equations, which determines the temperature and baryon chemical potential dependence of the order parameters of the model, which are the $\phi_N$, $\phi_S$ scalar condensates and the $\Phi_{P}$ and $\bar \Phi_{P}$ Polyakov loop variables, are given by the stationary point of the grand potential, 
\begin{equation}
    \frac{\partial\Omega_\textnormal{H}}{\partial \phi_N} =
    \frac{\partial\Omega_\textnormal{H}}{\partial \phi_S} =
    \frac{\partial\Omega_\textnormal{H}}{\partial \Phi} =
    \frac{\partial\Omega_\textnormal{H}}{\partial \bar\Phi} = 0,
\end{equation}
which leads in the current hybrid approximation to
\begin{align}
  &-\frac{d}{d \Phi_{P}}\left( \frac{U_{\text{Pol}}(\Phi_{P},\bar\Phi_{P})}{T^4}\right) + \frac{6}{T^3}\sum_{i=u,d,s} 
  \int \frac{d^3p}{(2\pi)^3} \left(\frac{e^{-\beta E_i^-(p)}}{g_i^-(p)}
    + \frac{e^{-2\beta E_i^+(p)}}{g_i^+(p)} \right) = 0, \label{Eq:FE_Phi}\\
  &-\frac{d}{d \bar\Phi_{P}}\left( \frac{U_{\text{Pol}}(\Phi_{P},\bar\Phi_{P})}{T^4}\right) + \frac{6}{T^3}\sum_{i=u,d,s}
  \int \frac{d^3p}{(2\pi)^3} \left(\frac{e^{-\beta E_i^+(p)}}{g_i^+(p)}
    + \frac{e^{-2\beta E_i^-(p)}}{g_f^-(p)} \right) = 0, \label{Eq:FE_Phibar}\\
  &m_0^2 \phi_N + \left(\lambda_1 + \frac{1}{2} \lambda_2 \right)
  \phi_N^3 + \lambda_1 \phi_N \phi_S^2 -\frac{1}{\sqrt{2}} c_1\phi_N\phi_S - h_{0N}
  +\frac{3}{2}g_F\left(\langle \bar q_u q_u\rangle_{_{T}} + \langle\bar q_d q_d\rangle_{_{T}} \right) = 0,\label{Eq:FE_phiN}\\
  &m_0^2 \phi_S + \left(\lambda_1 + \lambda_2 \right) \phi_S^3 +
  \lambda_1 \phi_N^2 \phi_S -\frac{\sqrt{2}}{4}c_1\phi_N^2 - h_{0S} 
  +\frac{3}{\sqrt{2}}g_F \langle \bar q_s q_s\rangle_{_{T}} = 0,\label{Eq:FE_phiS}
\end{align}
where the renormalized quark tadpole integral is given by
\begin{align}
     \langle \bar q_i q_i\rangle_{_{T}} &= 4 m_i\left[ -\frac{m_i^2}{16\pi^2}\left(\frac{1}{2} + \ln\frac{m_i^2}{M_0^2} \right) + T_i\right], \\
       T_i &= \int \frac{d^3p}{(2\pi)^3}\frac{1}{2E_i(p)}\big(f^-_i(p) + f^{+}_{i}(p)\big) \,,\\
   f^+_i(p) &=\frac{ \bar\Phi_{P} e^{-\beta E_i^+(p)} + 2\Phi_{P} e^{-2 \beta E_i^+(p)} + e^{-3\beta E_i^+p)} } {g^+_i(p)}, \\
    f^-_i(p) &=\frac{ \Phi_{P} e^{-\beta E_i^-(p)} + 2\bar\Phi_{P} e^{-2 \beta E_i^-(p)} + e^{-3\beta E_i^-(p)} } {g^-_i(p)}\; . 
\end{align}
Here $f^{\pm}_i(p)$ are the modified Fermi-Dirac distribution functions. To solve the set of field equations Eqs.~\eqref{Eq:FE_Phi}-\eqref{Eq:FE_phiS} a set of parameters is needed, which are usually determined at $T=\mu_B=0$. The parameterization procedure is described in Section~\ref{Ssec:parameterization}. One could choose as an example the parameters presented in Section~\ref{Ssec:parameterization}, but at finite temperature there are some new aspects to consider. The most important one is that from the lattice it is known that the pseudocritical temperature $T_c$ at $\mu_B=0$ is $T_c = 151$~MeV \cite{Aoki:2006we}. However, once we have fixed the parameters, the set of field equations can be solved for a given value of $T,\mu_B$. In practice, one usually starts with a solution at $T=\mu_B=0$ and increases either $T$ or $\mu_B$, using the previous solution as the initial condition. Since the set of equations is highly nonlinear, it is very difficult to find a solution without a good initial condition. In this way we can determine the order parameters ($\phi_N$, $\phi_S$, $\Phi_{P}$, $\bar\Phi_{P}$) either as a function of $T$ for a fixed $\mu_B=\mu_{B,\text{fix}}$ or as a function of $\mu_B$ for a fixed $T=T_{\text{fix}}$. As an example a parameterization result from \cite{Kovacs:2016juc} is shown, in Table~\ref{Tab:fit_res_eLSM_fin_T}, together with the values for the parameters in Table~\ref{Tab:param_fin_T}. 
\begin{table}[ht!]
\centering
\begin{tabular}{|c|c|c|c|c|c|}\hline
Observable & Fit $\left[  \text{MeV}\right]  $ & Experiment $\left[\text{MeV}\right]  $ &Observable & Fit $\left[  \text{MeV}\right]  $ & Experiment $\left[\text{MeV}\right]  $\\\hline
$f_{\pi}$ & $95.5$ & $92.2\pm0.2$ &$f_{K}$ & $109.4\pm0.6$ & $110.5\pm0.8$ \\\hline
$m_{\pi}$ & $140.5$ & $138\pm3.0$& $m_{K}$ & $499.5$ & $495.6\pm2.0$ \\\hline
$m_{\eta}$ & $542.1$ & $547.9\pm0.02$ &$m_{\eta^{\prime}}$ & $964.3$ & $957.8\pm0.06$\\\hline
$m_{\rho}$ & $806.4$ & $775.3\pm0.3$ & $m_{K^{\star}}$ & $915.2$ & $894.7\pm0.3$\\\hline
$m_{\phi}$ & $990.1$ & $1019.5\pm0.02$ &$m_{a_{1}}$ & $1076.6$ & $1230\pm40$\\\hline
$m_{f_{1}\left(1420\right)  }$ & $1416.0$ & $1426.4\pm0.9$ &$m_{a_{0}(980)}$ & $720.8$ & $980\pm20$\\\hline
$m_{K_{0}^{\star}\left(800\right)}$ & $752.9$ & $682\pm29$ & $m_{f_0^L(500)}$ & $283.7$ & $475\pm75$ \\\hline 
$m_{f_0^H(980)}$ & $737.6$ & $990\pm20$ & $m_{u,d}$ & $322.4$ & $308\pm31$ \\\hline
$m_{s}$ & $457.7$ & $483\pm49$ & $\Gamma_{\rho\rightarrow\pi\pi}$ & $151.5$ & $149.1\pm1.1$ \\\hline
$\Gamma_{K^{\star}\rightarrow K\pi}$ & $47.8$ & $48\pm1.3$ & $\Gamma_{\phi\rightarrow\bar{K}K}$ & $3.53$ & $3.54\pm0.02$\\\hline
$\Gamma_{a_{1}\rightarrow\rho\pi}$ & $199.4$ & $425\pm175$ &$\Gamma_{a_{1}\rightarrow\pi\gamma}$ & $0.37$ & $0.64\pm0.25$\\\hline
$\Gamma_{f_{1}\left(1420\right)\rightarrow K^{\star}K}$ & $44.5$ & $44.5\pm2.1$ &$\Gamma_{a_{0}(980)}$ & $68.3$ & $75\pm25$\\\hline
$\Gamma_{K_{0}^{\star}(800)\rightarrow K\pi}$ & $600.1$ & $547\pm24$ & $\Gamma_{f_0^L(500)\rightarrow\pi\pi}$ & $554.2$ & $550\pm 150$\\\hline
$\Gamma_{f_0^L(500)\rightarrow KK}$ & $0.0$ & $0.0\pm 100$ & $\Gamma_{f_0^H(980)\rightarrow\pi\pi}$ & $81.7$ & $70\pm 30$ \\\hline
$\Gamma_{f_0^H(980)\rightarrow KK}$ & $0.0$ & $0.0\pm 20$ & $T_c(\mu_B=0)$ & $170.4$ & $151\pm15.1$ \\\hline
\end{tabular}
\caption{An example of fit results from \cite{Kovacs:2016juc}, together with the experimental values taken from PDG (2012) \cite{ParticleDataGroup:2012pjm}. \label{Tab:fit_res_eLSM_fin_T}}
\end{table}
\begin{table}[t]
\centering
\begin{tabular}[c]{|c|c||c|c|}\hline
Parameter & Value & Parameter & Value \\\hline\hline
$\phi_{N}$ [GeV]& $0.141$ & $g_{1}$ & $5.62$\\\hline
$\phi_{S}$ [GeV]& $0.142$ & $g_{2}$ & $3.05$\\\hline
$m_{0}^2$ [GeV$^2$] & $2.39\times 10^{-4}$ & $h_{1}$ & $27.46$\\\hline
$m_{1}^2$ [GeV$^2$] & $6.33\times 10^{-8}$ & $h_{2}$ & $4.23$\\\hline
$\lambda_{1}$ & $-1.67$ & $h_{3}$ & $5.98$\\\hline
$\lambda_{2}$ & $23.51$ & $g_{F}$ & $4.57$\\\hline
$c_{1}$ [GeV]& $1.31$ & $M_{0}$ [GeV]& $0.35$\\\hline
$\delta_{S}$ [GeV$^2$] & $0.11$ & \multicolumn{1}{c}{} & \multicolumn{1}{c}{} \\\cline{1-2}
\end{tabular}
\caption{Parameter values for the fit shown in Table~\ref{Tab:fit_res_eLSM_fin_T}. \label{Tab:param_fin_T}}
\end{table}
Some notes are in order, in this particular parameterization the scalar-isoscalar $f_0^L$ and $f_0^H$ masses and their decay widths are also fitted. In addition, to be compatible with lattice results, the pseudocritical temperature $T_c$ at $\mu_B=0$ is also part of the fit. Note that scalar states below $1$~GeV are used here simply because they gave better chi-squared values than other particle assignments. In this respect, a tension between the vacuum's results and the present ones is visible. 
Namely, the role of the $f_0^L$ (or $\sigma$) mass is paramount, since if its value is too high the value of $T_c$ is dramatically increased, e.g. for $m_{f_0^L} > 1$~GeV, $T_c > 500$~MeV and the phase transition along the $\mu_B$ axis would also be crossover instead of first order, so there would be no CEP. The existence of the CEP is a common belief, but in principle it is not excluded that the chiral or deconfinement phase boundary is a crossover all along.

\subsection{Selected results}
\label{Ssec:FiniteT_results}

\begin{figure}[ht!]
  \centering
  \includegraphics[width=0.48\textwidth]{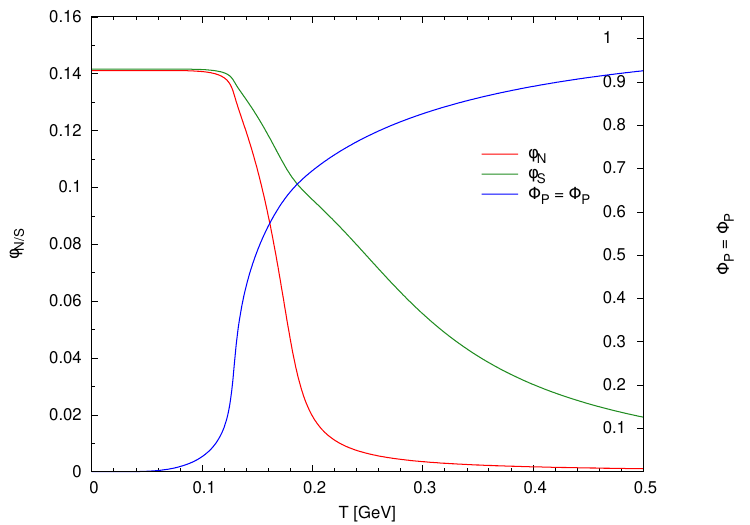}
  \includegraphics[width=0.48\textwidth]{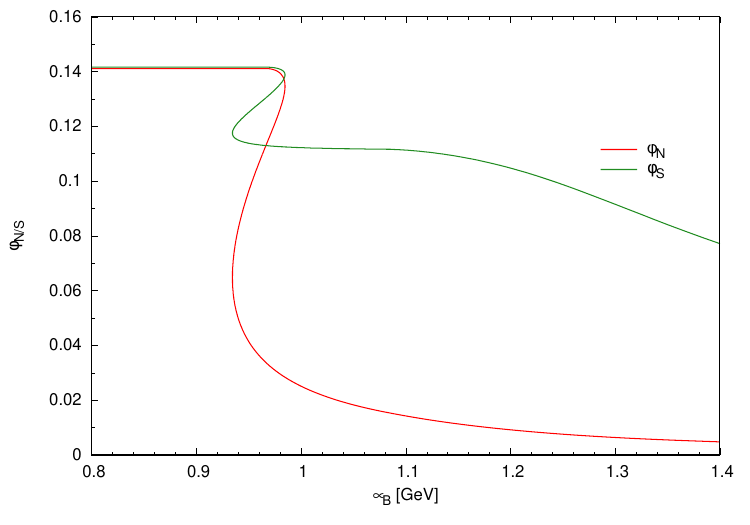}
  \caption{\label{Fig:phiNS_T} Temperature (left) and baryon chemical potential (right) dependence of the order parameters ($\phi_N$, $\phi_S$, $\Phi_P$, $\bar\Phi_P$). On the left $\mu_B=0$, on the right $T=0$.}
\end{figure}

In this section we show some results using the solution of Eqs.~\eqref{Eq:FE_Phi}-\eqref{Eq:FE_phiS} with the parameter set of Table~\ref{Tab:param_fin_T}. First, the left figure of Fig.~\ref{Fig:phiNS_T} shows the temperature dependence of the $\phi_N(T)$ and $\phi_S(T)$ non-strange and strange scalar condensates together with the Polyakov loop variable $\Phi_P(T)$ at $\mu_B=0$. 
At $\mu_B$ the two Polyakov loop variables are the same $\Phi_P(T,\mu_B=0) = \bar\Phi_P(T,\mu_B=0)$. It can be seen that both $\phi_N$ and $\phi_S$ decrease smoothly from some initial value at $T=0$ and tend to zero at $T\to\infty$. There are various definitions for the pseudocritical temperature $T_c$ in the case of a crossover transition. One possibility is the value of the inflection point of $\phi_n(T)$, i.e. $d\phi_N/dT = 0$ and $d^2\phi_N/dT^2 = 0$ for some $T=T_c$. Another definition can be the value of $T$ for which $\phi_N(T)/phi_N(T=0) = 0.5$, i.e. when the non/strange condensates drop to half of their original value. While one definition, which is also used on the \cite{Cheng:2007jq} lattice, is based on the so-called subtracted condensate, which is defined as
\begin{equation}
    \Delta_{\text{sub}}(T) = \frac{(\phi_N - \frac{h_{0N}}{h_{0S}} \phi_S)\vert_{T}}{(\phi_N - \frac{h_{0N}}{h_{0S}} \phi_S)
    \vert_{T=0}},
\end{equation}
and which takes values between $1$ and $0$, while the definition of $T_c$ is $\Delta_{\text{sub}}(T_c) = 0.5$. In this particular case, both the inflection point of $\phi_N$ and the half value of the subtracted condensate give $T_c=172$~MeV. The right figure in Fig.~\ref{Fig:phiNS_T} shows the $\phi_{N/S}$ condensates as a function of $\mu_B$ at $T=0$. At $T=0$ the Polyakov loop variables are identically zero, $\Phi_P(\mu_B,T=0) = \bar\Phi_P(\mu_B,T=0) = 0$. The phase transition is first order along the $\mu_B$ axis, the transition temperature can again be defined as the infection point in the unstable region, while the two spinodals are given by the condition $d\phi_N/dT = \infty$ (or $dT/d\phi_N = 0$). 

\begin{figure}[ht!]
  \centering
  \includegraphics[width=0.48\textwidth]{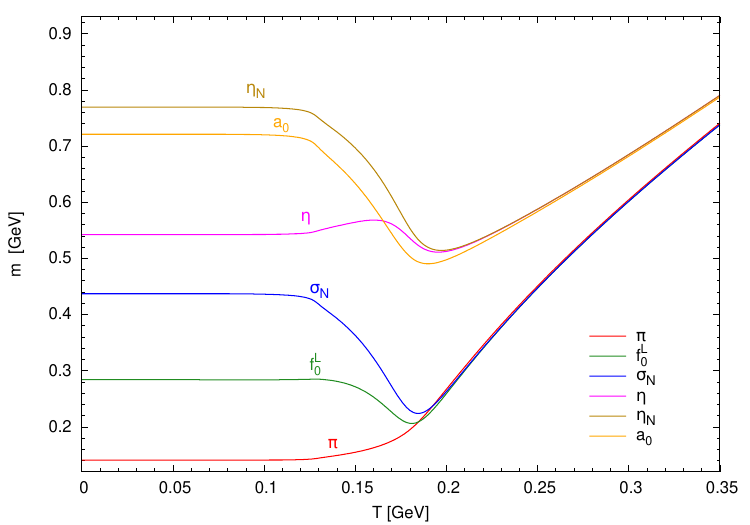}
  \includegraphics[width=0.48\textwidth]{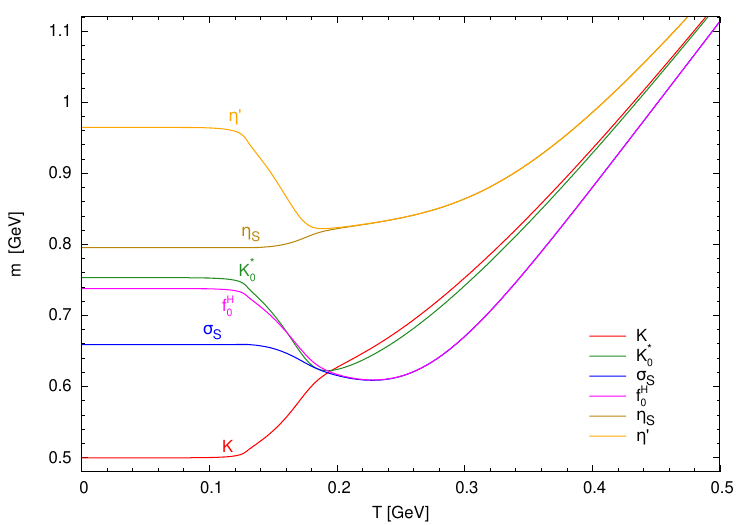}
  \caption{\label{Fig:masses_T}Temperature dependence of the scalar and pseudoscalar meson masses.}
\end{figure}

The so-called curvature masses are defined as the second derivative of the grand potential (or effective potential) at the minimum, which is the solution of the field equations, 
\begin{equation}
  m^2_{ab} = \frac{\partial^2 \Omega_H(T,\mu_q )}{\partial
    \varphi_{a} \partial \varphi_{b}}
  \bigg|_\textnormal{min}=m^2_{\text{tree},ab}+\delta m^2_{\text{vac},ab}+\delta
  m^2_{T,ab}, 
\label{Eq:M_curve_s_ps}
\end{equation}
where $\varphi_{a}$ can be any scalar or pseudoscalar field. The curvature mass has a tree-level part, a fermion vacuum part, and a fermion thermal part coming from the zero and finite temperature fermion loop corrections to the grand potential. The details and the explicit expression for the scalar and pseudoscalar curvature masses can be found in \cite{Kovacs:2016juc}. The temperature dependence of the masses can be seen in Fig.~\ref{Fig:masses_T}.
It can be seen that the masses degenerate around $T_c$ as expected, but not all of them. This is due to the fact that the $U(1)_A$ axial symmetry is not fully restored, at least up to temperatures of $350-400$~MeV, where the model is no longer reliable. In the case of mesons heavier than the pion or kaon, there is a drop down around $T_c$ to reach the degenerate state with their lighter partners. After this region, the thermal contributions become dominant in all masses and all masses increase with increasing $T$. Recently, the temperature dependence of the curvature masses of the vector and axial vectors in \cite{Kovacs:2021kas} has also been investigated within the eLSM.  

\begin{figure}[ht!]
  \centering
  \includegraphics[width=0.7\textwidth]{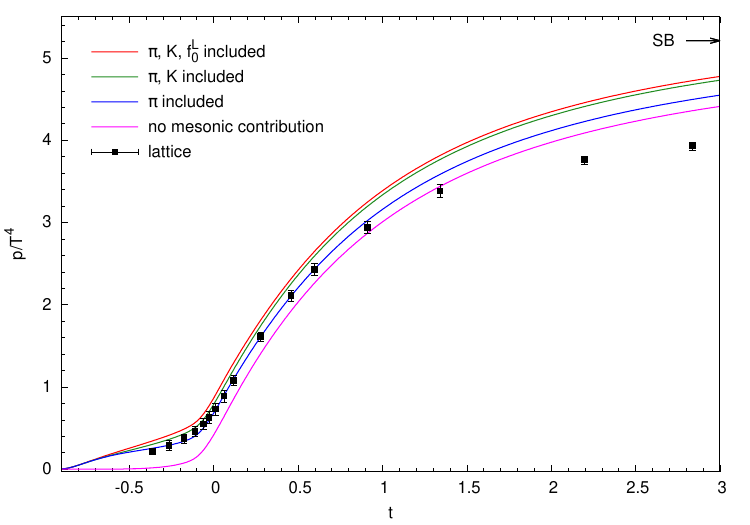}
  \caption{\label{Fig:pressure_T} The normalized pressure as a function of the reduced temperature ($t=(T-T_c)/T_c$). A continuum lattice result is also shown. Thermal meson fluctuations are also added for the lowest mass mesons ($\pi$, $K$, $f_0^L$). The arrow on the right indicates the Stefan-Boltzmann limit of QCD: $p_{SB}/T^4 = 5.2$.}
\end{figure}

After solving the field equations and substituting the solution at each $T$ and $\mu_B$ back into the grand potential, it can be used to determine any thermodynamic quantity. The pressure is given by
\begin{equation}
p (T,\mu_q)=\Omega_\textnormal{H}(T=0,\mu_q) - \Omega_\textnormal{H}(T,\mu_q), 
\end{equation}
while the entropy, baryon number density, baryon number susceptibility and energy density are given by
\begin{align}
    s &= \frac{\partial p}{\partial T}\;,\quad \rho_B=\frac{\partial p}{\partial \mu_B}\; , \\
    \chi_B &= \frac{\partial^2 p}{\partial \mu^2_B}\;,\quad \varepsilon = -p + T s + \mu_B\rho_B\; .  
\end{align}
The normalized pressure $p/T^4$ is shown in Fig.~\ref{Fig:pressure_T} together with a continuum lattice result taken from \cite{Borsanyi:2010cj} for comparison as a function of the reduced temperature $t=(T-T_c)/T_c$.  
As already mentioned, meson fluctuations are only considered in the pressure with the following contribution
\begin{equation}
\Delta p_x(T) = - N_{x}T\int\frac{d^3 p}{(2\pi)^3}\ln(1-e^{-\beta E_x(p)}),\quad E_x(p)=\sqrt{p^2 + m_x^2}, \quad x\in(\pi,K,f_0^L), 
\end{equation}
where $N_x$ is the meson multiplicity factor with $N_{\pi} =3$, $N_{K} =4$ and $N_{f_0^L} =1$. This treatment of the meson thermal fluctuation is not self-consistent, since it is not introduced at the level of the grand potential, so there is no meson fluctuation in the field equation. The introduction of a self-consistent meson contribution beside the fermion one is more complicated, as described in \cite{Kovacs:2021kas}. It can be seen that the meson contributions, especially for the pion, have a huge effect at lower temperatures and with their help the pressure curve is much closer to the lattice result. On the other hand, as $T$ increases, the additional thermal meson fluctuations overshoot the lattice results more and more. However, it should be noted again that the validity of the model decreases at higher temperatures.

\begin{figure}[ht!]
  \centering
  \includegraphics[width=0.7\textwidth]{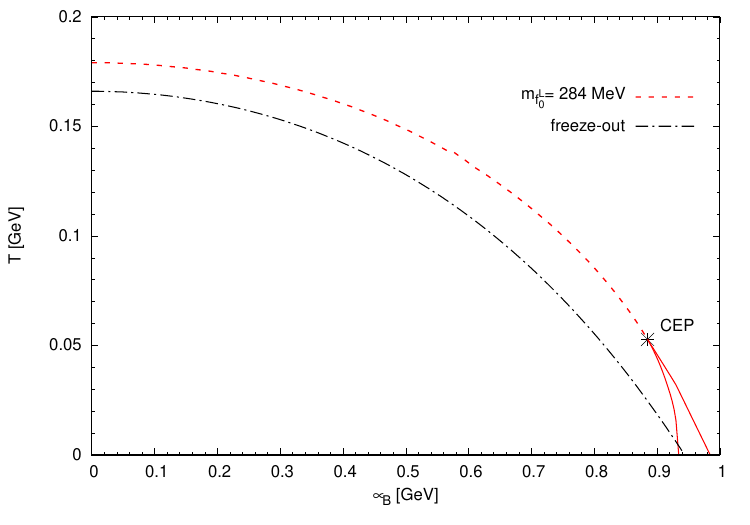}
  \caption{\label{Fig:CEP} Chiral phase boundary in eLSM for a given set of parameters from Table~\ref{Tab:param_fin_T}. The dashed curve shows the crossover, while the solid curve shows the first order phase transition. The dashed-dotted curve is the chemical freeze-out curve from \cite{Cleymans:2006qe}. The coordinates of the CEP are $(T,\mu_B)=(885,52.7)$ MeV.}
\end{figure}

Determining the pseudocritical temperatures for increasing values of $\mu_B$ draws the line of the phase boundary on the $T$-$\mu_B$ plane. A particular example is shown in Fig.~\ref{Fig:CEP} for the parameter set presented in the previous section. 
As $\mu_B$ increases, the phase transition as a function of $T$ becomes stronger and stronger, while for a certain value the derivative of $d\phi_N/dT$ at the infection point becomes infinite. At this point, the phase transition is second order, which we call the critical endpoint or CEP. As $\mu_B$ continues to increase, the phase transition becomes first order and there is an unstable region bounded by the two spinodal lines as shown in the figure. In this particular example the coordinates of the CEP are $(T,\mu_B)=(885,52.7)$ MeV. With different parameterizations and different approximations the location of the CEP can change, it can also happen, especially for parameterizations with large $f_0^L$ mass ($m_{f_0^L}\gtrapprox 300$~MeV), that the CEP vanishes and the phase boundary is only crossover. In many other models, the CEP also exists, but it could be in a very different place, so finding the CEP experimentally is not easy because there is no clear indication of where to find it.  

If we introduce in addition to the scalar condensates also condensates of different vector fields in the temporal direction, one can calculate $p(\mu_B)$ and $\varepsilon(\mu_B)$ at $T=0$ and determine the so-called equation of state, i.e. the $p(\varepsilon)$ curve. This can then be used to study hybrid stars in the eLSM (for more details see \cite{Kovacs:2021ger,Takatsy:2023xzf}).

Chiral symmetry restoration at finite temperature and density leads to parity doubling, and at sufficiently high densities and temperatures, hadrons undergo Mott dissociation as the quarks become effectively massless. Although the eLSM does not explicitly incorporate hadron compositeness, the Mott transition can still be analyzed within this framework in a manner similar to NJL-based models. In particular, mesons do not simply disappear beyond the chiral transition, but change into unbound quark-antiquark correlations. The lack of a fully dynamical treatment of hadron dissociation into the continuum is a limitation of the current formulation, as reflected in Figures 8.3-8.5, where such effects are not explicitly included. However, it should be noted that no existing effective model provides a fully consistent description of this transition, as a complete understanding of hadron dissolution in hot and dense QCD matter remains an open problem. Finally, we list here some other approaches using somewhat different models, such as a hybrid quark-meson-nucleon model \cite{Marczenko:2018jui}, a quark-meson-nucleon model with dilaton and statistical confinement \cite{Benic:2015pia}, and a parity doublet model \cite{Dexheimer:2012eu}.
 
%\newpage
\section{Summary and outlook}\label{Sec:sum}
In this review, the genesis, motivation, and main results of the extended Linear Sigma Model (eLSM) have been presented \cite{Parganlija:2012fy,Janowski:2014ppa,Kovacs:2016juc,Eshraim:2020ucw,Jafarzade:2022uqo}.  The eLSM is a chiral model for low-energy hadrons that contains on the same level chiral partners, which are further classified as heterochiral and homochiral multiplets \cite{Giacosa:2017pos,Giacosa:2023fdz}. We have shown that the eLSM for $N_f=3$ can be used to successfully describe the most salient features of mesonic phenomenology up to 2.5 GeV. 
Below, we summarize some of them and then we discuss possible future tasks.

 \begin{itemize}
     \item Besides (pseudo)scalar fields, which are present in every LSM, the eLSM contains (axial-)vector mesons from the beginning. For instance, the broad resonance $a_1(1230)$ is regarded as the chiral partner of the $\rho$ meson \cite{Parganlija:2012fy,Parganlija:2010fz}. 

     \item The chiral partners of pseudoscalar fields, the famous scalar mesons, turn out to be heavier than $1$~GeV. In this respect, the chiral partner of the pion is the resonance $f_{0}(1370)$ \cite{Parganlija:2012fy,Janowski:2014ppa}. 
     However, when the eLSM is studied in the medium \cite{Kovacs:2016juc,Kovacs:2022zcl}, light scalar states are required to correctly reproduce the chiral phase transition. A way out of this conundrum could be the inclusion at the same time of a nonet of light four-quark states (scalar resonances below 1 GeV) into the eLSM.

     \item Other $J^{\mathcal{P}\mathcal{C}}$ quark-antiquark mesons have been coupled to the eLSM as well. An interesting example is the inclusion of tensor and their chiral partners, the axial-tensor mesons \cite{Jafarzade:2022uqo}. The former states are very well known experimentally, the latter are still mostly unknown. The eLSM could make prediction for the not-yet discovered axial-tensor states, finding that they are broad.

    \item Within the eLSM, hybrid mesons can be easily added, since they form chiral multiplets which, under chiral transformations, behave just as regular quark-antiquark multiplets \cite{Eshraim:2020ucw}. The lowest hybrid nonet involves the exotic quantum numbers $J^{\mathcal{P}\mathcal{C}} = 1^{-+}$. In particular, the state $\pi_1(1600)$ and the recently discovered resonance $\eta_1(1855)$ fit well in this picture \cite{Shastry:2022mhk,Shastry:2023ths}. Two additional predicted states, $\eta_1(1660)$ and $K_1(1750)$, are still missing. 

     \item  The eLSM contains the dilaton/glueball from its very onset, since the model is built (in the chiral limit and neglecting the chiral anomaly) as being dilatation and chirally invariant, the former being broken by one dimensionful parameter present in the dilaton potential. The decays and mixing patterns of the scalar glueballs are then naturally described within the eLSM. The resonance $f_0(1710)$ turns out to be mostly gluonic \cite{Janowski:2014ppa,Janowski:2011gt}. 

     \item The tensor glueball can be easily coupled to the eLSM \cite{Vereijken:2023jor}: the main result is a strong coupling to (axial-)vector mesons, thus the channel $T_G \rightarrow \rho \rho \rightarrow 4\pi$ is expected to dominate. At present, the resonance $f_2(1950)$  can be regarded as a good glueball candidate, but further studies are needed.

     \item The pseudoscalar glueball is coupled to regular $\bar{q}q$ states of the eLSM via an $U(1)_A$ anomalous term \cite{Eshraim:2012jv}. The main decay channels are $KK\pi$ as well as $\eta \pi \pi$ and $\eta' \pi \pi$. The recently confirmed pseudoscalar states $X(2370)$ and $X(2600)$ (both measured by BESIII in the channel $\eta' \pi \pi$ \cite{BESIII:2010gmv,BESIII:2023wfi,BESIIICollaboration:2022kwh}) are good candidates for containing a sizable pseudoscalar glueball amount in their wave function. Note, as discussed in Ref. \cite{Giacosa:2023fdz} also the intensity of the interaction as driven by an instanton gas agrees with the experimentally measured widths of these states. An important task for the future is the exact allocation of the glueball amount in the two resonances mentioned above.

     \item The vector glueball has been also coupled to the eLSM \cite{Giacosa:2016hrm}. This is an example in which a relatively heavy glueball (about 3.8 GeV) is considered: the main outcome is that the decay channel $\omega \pi \pi$ is dominant.

     \item Further topics have been considered: (i) a successful description of isospin breaking is achieved, that is able to describe e.g. the small mass difference of the neutral vs the charged  pions \cite{Kovacs:2024cdb}; (ii) the inclusion of radially excited (pseudo)scalar fields allow to describe various resonances such as $\pi(1300)$ and $f_0(1790)$ \cite{Parganlija:2016yxq}; (iii) the extension to the four-flavor case, $N_f =4$, leads to the correct description of certain decay widths, such as $D^{*}(2010)^{+} \rightarrow D^{+} \pi^{0}$ \cite{Eshraim:2014eka}.

     \item The eLSM, together with Polyakov loop and quarks d.o.f., allows to study at the same time the chiral and the deconfinement phase transitions and offers a phenomenological description in agreement with other models and lattice QCD. Moreover, it offers a consistent framework to investigate the location of the CEP \cite{Kovacs:2016juc}. It should be stressed that mesons are treated as elementary particles in the eLSM, i.e. at high temperature and/or density, the validity of the model is lost when mesons dissociate into quarks.
     
\end{itemize}

Besides what it has been achieved up to now, it is important to mention interesting topics that can be studied in the future. The list below is not intended to be complete, but it offers an idea of the topics that can be addressed using the eLSM.

 \begin{enumerate}
     \item The addition and/or a study in depth of quark-antiquark multiplets is a possible straightforward application of the eLSM. For instance, the heterochiral multiplet involving the pseudovector mesons and the orbitally excited vector mesons has already been introduced in the eLSM but has not yet been studied in detail. Also, the novel inclusion of the pseudotensor mesons and the mesons with $J^{\mathcal{P}\mathcal{C}}=3^{--}$ \cite{Jafarzade:2021vhh} together with their -yet unknown- chiral partners is doable, eventually shedding light on the latter states. Moreover, further radially excited states can be added, such as the rather well known radially excited vector mesons together with their chiral partners, the radially excited axial-vector mesons (for which candidates do exist in the PDG). 

     \item The tension linked to the scalar meson assignment shows that a nonet of light scalar mesons in form of four-quark states should be added to the eLSM, thus completing the approaches of Refs. \cite{Fariborz:2007ai,Napsuciale:2004au,Giacosa:2006tf}. In particular, in the scalar-isoscalar sector, five scalar states need to be present: two mainly four-quark objects, two quarkonia, and one glueball. Preliminary two-flavor works that involve light four-quark mesonic field(s) have been discussed in Refs. \cite{Heinz:2008cv,Gallas:2011qp,Lakaschus:2018rki}.

     \item A particularly interesting topic is the systematic study of terms that break the chiral anomaly and involve other multiplets than the (pseudo)scalar one. In this respect, a novel and quite promising mathematical object (denoted as `GPKJ $\epsilon$-product', see Eq. \eqref{epsilon} and Refs. \cite{Giacosa:2017pos,Giacosa:2023fdz}) has been introduced for this purpose and affects all the heterochiral multiplets.

     \item A related topic to the previous one is a systematic study of isoscalar mixing in each homochiral nonet (such as the (axial-)vector one), for which there are no axial anomalous terms. To this end, some terms that were at first neglected in the eLSM because large-$N_c$ suppressed can be added and have a small but significant contribution. Such terms may play also an important role when isospin breaking is taken into account. 

     \item The behavior of the model when changing certain parameters can be studied. While the quark masses are fixed in Nature, it is quite easy to modify them in models and on the lattice. The behavior of masses and decays as function of the bare quark masses is in this respect promising.
     
     \item Two further and speculative directions are possible: one is the study of the $N_f =1$ case. In this limit, most mesons are stable and, quite interestingly, it is possible to study such a system on the lattice \cite{DellaMorte:2023ylq,Verbaarschot:2014upa}. An eLSM study could help to understand the physics of the dilaton and the lightest scalar meson. The other limit consists in analyzing the case $N_f=5$: even if the explicit breaking of chiral symmetry is huge, some specific decay terms could still show that some interaction terms retain chiral symmetry, even in such a different energy domain. 

     \item There is a full tower of glueballs predicted by lattice QCD and other approaches. Up to now, only four of them (the 3 lightest ones and the vector glueball) have been coupled to the eLSM. On the one hand, open questions concerning the glueballs already coupled to the eLSM still exist. On the other hand, in view of the interest of the community on such states, and also in the perspective of future experimental searches, their complete study within the eLSM can be useful to isolate their dominant decay widths and to predict various decay ratios.

     \item Another important direction is the development of the model in the baryonic sector, in particular the addition of the decuplet and its chiral partners, extending the work of Ref. \cite{Kovacs:2013gxa}.

     \item Novel studies of the QCD phase diagrams are interesting, such as the investigation of neutron stars, in connection with processes emitting gravitational waves. Also the detailed investigation of quarkyonic matter belongs to the outlook of the eLSM approach.

     \item The application of deep learning techniques in the study of parameterization could be useful, since this problem is similar to finding a global minimum of the loss function in deep learning.
     
 \end{enumerate}

In conclusion, due to its versatility and generality, the eLSM was applied in various areas of low-energy QCD phenomenology, but at the same time many potential studies -both in the vacuum and in the medium- can be realized in the future. It is also expected that many experimental data from ongoing and planned experiments will enlarge and complete the PDG tables of mesonic (as well as baryonic) states, requiring a useful, general, and adequate modeling. Together with other approaches and with computer simulations, the eLSM offers its valuable contribution to the understanding of old and new resonances, in particular relating chiral partners and isolating states that cannot be described as (predominantly) quark-antiquark objects.

	%end of the core of the manuscript
	
	\newpage
	\section*{Acknowledgments}
It is a great pleasure to thank our numerous collaborators and colleagues with whom we extensively discussed the eLSM over the past 15 years: F. Divotgey, W. Eshraim, C. Fischer, A. Heinz, S. Janowski, A. Koenigstein, Gy. Kovács, P. Lakaschus, L. Olbrich, G. Pagliara, D. Parganlija, M. Piotrowska, R. Pisarski, C. Reisinger, D. H. Rischke, J. Sammet, J. Schaffner-Bielich, V. Shastry, S. Strueber, Zs. Szép, J. Takátsy, A. Vereijken, Gy. Wolf, N. Weickgenannt, Th. Wolkanowski-Gans, and M. Zetenyi. F.~G. acknowledges support from the Polish National Science Centre (NCN) through the OPUS 440 Project No. 2019/33/B/ST2/00613 during the preparation of this paper (up to 16/2/2024). S.~J. has partially been supported by the Polish National Science Centre (NCN) Grant SONATA project No. 2020/39/D/ST2/02054 and the U.S. Department of Energy through the ExoHad Topical Collaboration, Contract DE-SC0023598. P.~K. acknowledges support by the Hungarian National Research, Development and Innovation Fund under Project numbers FK 131982 (until 31/05/2024) and K 138277.

	%\section*{Author's contributions \textit{(optional section)}} Detailing here the contributions of the authors of the review.

 \newpage
\appendix
%\renewcommand*{\thesection}{\Alph{section}}
	
%\section{Appendices}\label{appendix}
  
\section{Brief recall of the \texorpdfstring{$U(N)$}{UN} and \texorpdfstring{$SU(N)$}{SUN} groups}
\label{Ssec:UN_group}

Elements $U$ of the unitary $U(N)$ group are (in the fundamental representation) $N\times N$ complex matrices satisfying the following conditions
\begin{align}    
UU^\dagger=U^\dagger U=\mathbb{1}_{N\times N}
\text{ .}
\end{align}
Each matrix $U$ can be written as (see e.g. Ref. \cite{Zee:2016fuk}):
\begin{align}
U=\exp{-i\sum_{a=0}^{N^2-1}\theta_at_a}\, \text{ ,}
\label{groupun}
\end{align}
where the $t^a$ form a basis of Hermitian matrices. As usual, we set $t_0=\frac{1}{\sqrt{2N}}\mathbb{1}_{N\times N}$ and for $(a,b)\in [0,\cdots,N^2-1]$:
 \begin{align}     \mathrm{Tr}\Big(t_at_b\Big)=\frac{1}{2}\delta^{ab}\, \text{ .}
 \end{align}
Then, for $a\in [1,\cdots,N^2-1]$ one has $\mathrm{Tr}(t_a)=0$.

Adding the additional condition $\det U=1$ to the elements of $U(N)$ we obtain the special unitary group $SU(N)$. It can be expressed as in Eq. (\ref{groupun}) upon setting $\theta_0 =0$, thus there are $N^2-1$ generators that satisfy the following commutation relation
\begin{align}    [t_a,t_b]=if_{abc}t_c
\,,\text{ where } (a,b,c)\in [1,\cdots , N^2-1]\,,
\end{align}
where $f_{abc}$ are the antisymmetric structure constants of $SU(N)$.  The Jacobi identity is also fulfilled
\begin{align}
     [[t_a,t_b],t_c]+[[t_b,t_c],t_a]+[[t_c,t_a],t_b]=0\, \text { .}
\end{align}
Moreover, the symmetric structure constants $d_{abc}$ are given by
\begin{align}
    \{t_a,t_b\}=\frac{\delta_{ab}}{N}+d_{abc}t_c
    \text{ .}
\end{align}

Another important subgroup of $SU(N)$ is the discrete center group $Z_N$, defined as the diagonal elements $Z_n = e^{2 \pi n/N}\mathbb{1}_{N\times N}$ for $n=0,1,...,N-1$. The symmetry under $Z_N$ is spontaneously broken at high $T$, since one of the $N$ possible choice is picked up, but is fulfilled in the QCD vacuum. This is correctly implemented by the Polyakov loop variable, see chapter~\ref{Sec:eLSM_finite_T} for the eLSM application to the case $N= N_c = 3$ and Ref. \cite{Kovacs:2022zcl} for the arbitrary case $N =N_c$.

The dimensions of the generators, as well as the structure constants, are dependent on the form of the representation. Throughout the paper, we are interested in the fundamental and the adjoint representation. 
For the generic representation $\mathcal{R}$, $SU(N)$ generators satisfy
\begin{align}    \mathrm{Tr}\Big(t^{\mathcal{R}}_at_b^{\mathcal{R}}\Big)=C_1^{\mathcal{R}}\delta^{ab}\,,\qquad\text{and}\,,\qquad\sum_{a=1}^{N^2-1}\Big(t^{\mathcal{R}}_a\Big)_{ij}\Big(t^{\mathcal{R}}_a\Big)_{jk}=C_2^{\mathcal{R}}\delta_{ik}
\end{align}
Denoting the dimension of the representation $d^{\mathcal{R}}$ and the number of the independent parameters $n_{G}$ of the group $G$, one has the general relation $d^{\mathcal{R}}C_{2}^{\mathcal{R}}=n_{G}C_{1}^{\mathcal{R}}$.
In the case of the fundamental representation $\mathcal{F}$, its dimension $d^{\mathcal{F}}=N$, $n_{SU(N)}=N^2-1$, $C_1^{\mathcal{F}}=\frac{1}{2}$ and $C_2^{\mathcal{F}}=\frac{N^2-1}{2N}$.
For the adjoint representation $\mathcal{A}$, the dimension of the representation is equal to the independent parameters of the group: $d^{\mathcal{A}}=n_{G}=N^2-1$ and consequently, $C_1^{\mathcal{A}}=C_2^{\mathcal{A}}=N$. In the later, a generic transformation of $\phi^a$ under the adjoint representation reads $\phi^a \rightarrow (U_{\mathcal{A}})_{a,b} \phi^b$. Upon introducing the $N \times N$ matrix $\phi = \sum_{a=1}^{N^2-1}\phi^a t^a$, the matrix $\phi$ transforms as $\phi \rightarrow U \phi U^{\dagger}$, where $U$ belongs to the fundamental representation. The latter equation is applied allover the manuscript.

In the case $N=2$, one sets $t^a=\sigma^a/2$, where $\sigma^a$ are the Pauli matrices defined in Eq. (\ref{pauli}). In this case, $f_{abc}=\epsilon_{abc}$, $d_{abc}=0$, and the $SU(2)$ group elements take the simple form 
\begin{equation} \exp{\Big(i\vec{\sigma}\cdot\vec{n}\theta\Big)}=\cos{\theta}+i\vec{\sigma}\cdot\vec{n}\sin{\theta} \text{ .}
\end{equation}

For $N=3$, the generators are represented in terms of the 8 Gell-Mann matrices $\lambda_a$ as $t_a=\frac{\lambda_a}{2}$ given in Eq. \eqref{eq:Gell-Mann matr}. 
\begin{align} \label{eq:Gell-Mann matr}
    \lambda^1 = \begin{pmatrix}
        {0}&{1}&{0}&\\
        {1}&{0}&{0}\\
        {0}&{0}&{0}
    \end{pmatrix},\qquad\lambda^2 =\begin{pmatrix}  {0}&{-i}&{0}\\
    {i}&{0}&{0}\\
    {0}&{0}&{0}\end{pmatrix},\qquad \lambda^3 = \begin{pmatrix}   {1}&{0}&{0}\\
    {0}&{-1}&{0}\\
    {0}&{0}&{0}\end{pmatrix},\qquad
    \lambda^4 =\begin{pmatrix}
        {0}&{0}&{1}\\
        {0}&{0}&{0}\\
        {1}&{0}&{0}
    \end{pmatrix} \\\nonumber
\lambda^5 =\begin{pmatrix}
     {0}&{0}&{-i}\\
     {0}&{0}&{0}\\
     {i}&{0}&{0}
\end{pmatrix} ,\quad
\lambda^6 = \begin{pmatrix}
    {0}&{0}&{0}\\
    {0}&{0}&{1}\\
    {0}&{1}&{0}
\end{pmatrix} ,\qquad
\lambda^7 =\begin{pmatrix}
    {0}&{0}&{0}\\
    {0}&{0}&{-i}\\
    {0}&{i}&{0}
\end{pmatrix} ,\quad
\lambda^8 = \frac{1}{\sqrt{3}} \begin{pmatrix}
    {1}&{0}&{0}\\
    {0}&{1}&{0}\\
    {0}&{0}&{-2}
\end{pmatrix} \, 
\text{ ,}
\end{align}
 and $\lambda_0 = \sqrt{2/3} \mathbb{1}_{3 \times 3}$. The structure constants $f^{abc}$ read \cite{Zee:2016fuk}
 \begin{align}
f_{abc}=-2i\text{Tr}\Big( [t_a,t_b]t_c\Big)\quad&\rightarrow \quad  f^{147}=f^{165}=f^{246}=f^{257}=f^{345}=f^{376}={\frac {1}{2}}\,,\\\nonumber
&f^{123}=1\ ,\quad f^{458}=f^{678}={\frac {\sqrt {3}}{2}}\,,
 \end{align}
 while $d^{abc}$ are: 
 \begin{align}
     d_{abc}=2\text{Tr}\Big( \{t_a,t_b\}t_c\Big)\,\quad\rightarrow &\, \qquad d^{146}=d^{157}=d^{256}=d^{344}=d^{355}=-d^{247}=-d^{366 }=-d^{377}=\frac{1}{2}\,,\\\nonumber
    d^{118 }&=d^{228 } =d^{338 }=-d^{888}=\frac{1}{\sqrt{3}}\,,\qquad d^{448}=d^{558}=d^{668}=d^{778}=-\frac{1}{2\sqrt{3}}\,.
 \end{align}

\section{Brief recall of the large-\texorpdfstring{$N_c$}{Nc} limit in QCD}
\label{App:Large-N}

\begin{figure}[htb]
    \centering
    %\begin{subfigure}
    %    \centering
        \includegraphics[width=0.18\textwidth]{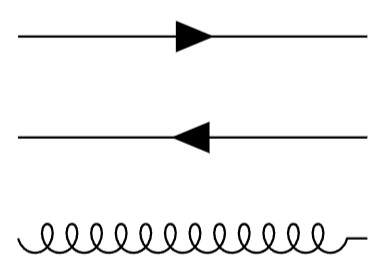}
        \includegraphics[width=0.22\textwidth]{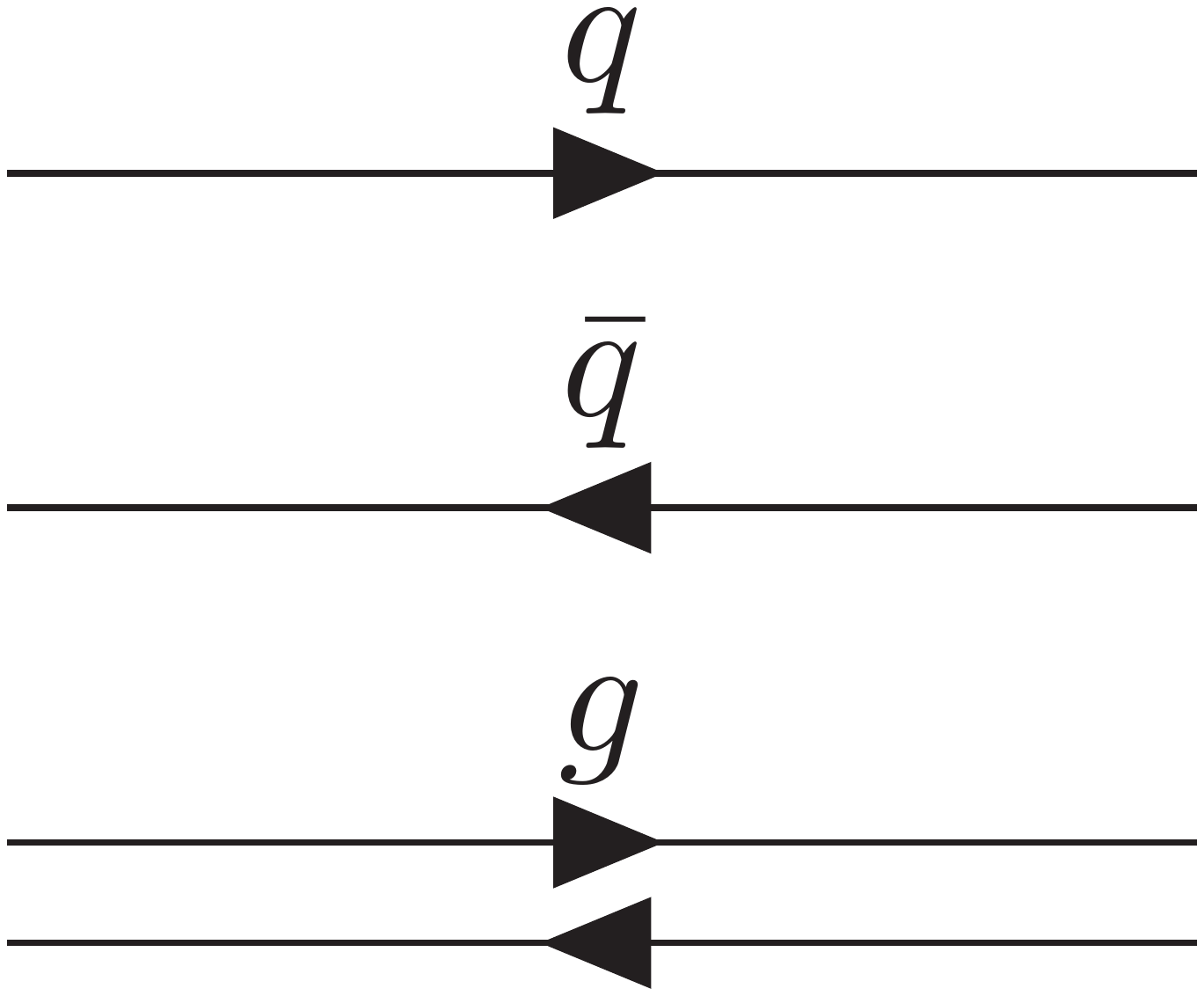} \\
    %\end{subfigure} \\
    %\begin{subfigure}
    %    \centering
        \includegraphics[width=0.22\textwidth]{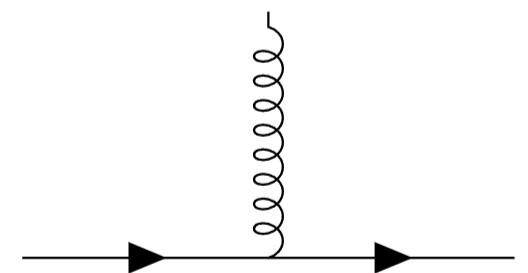}
        \includegraphics[width=0.22\textwidth]{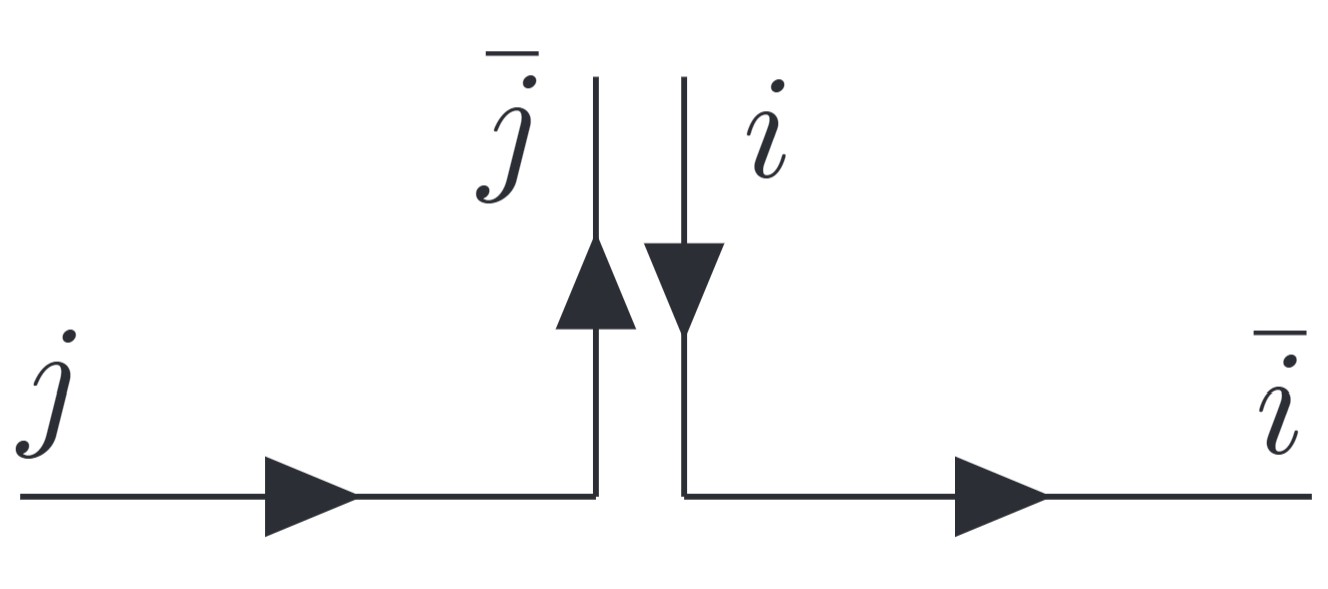} \\
    %\end{subfigure} \\
    \includegraphics[width=0.18\textwidth]{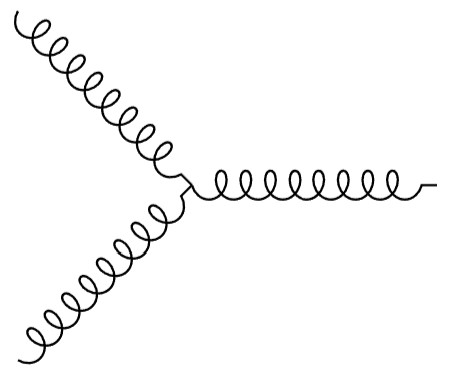}
    \includegraphics[width=0.22\textwidth]{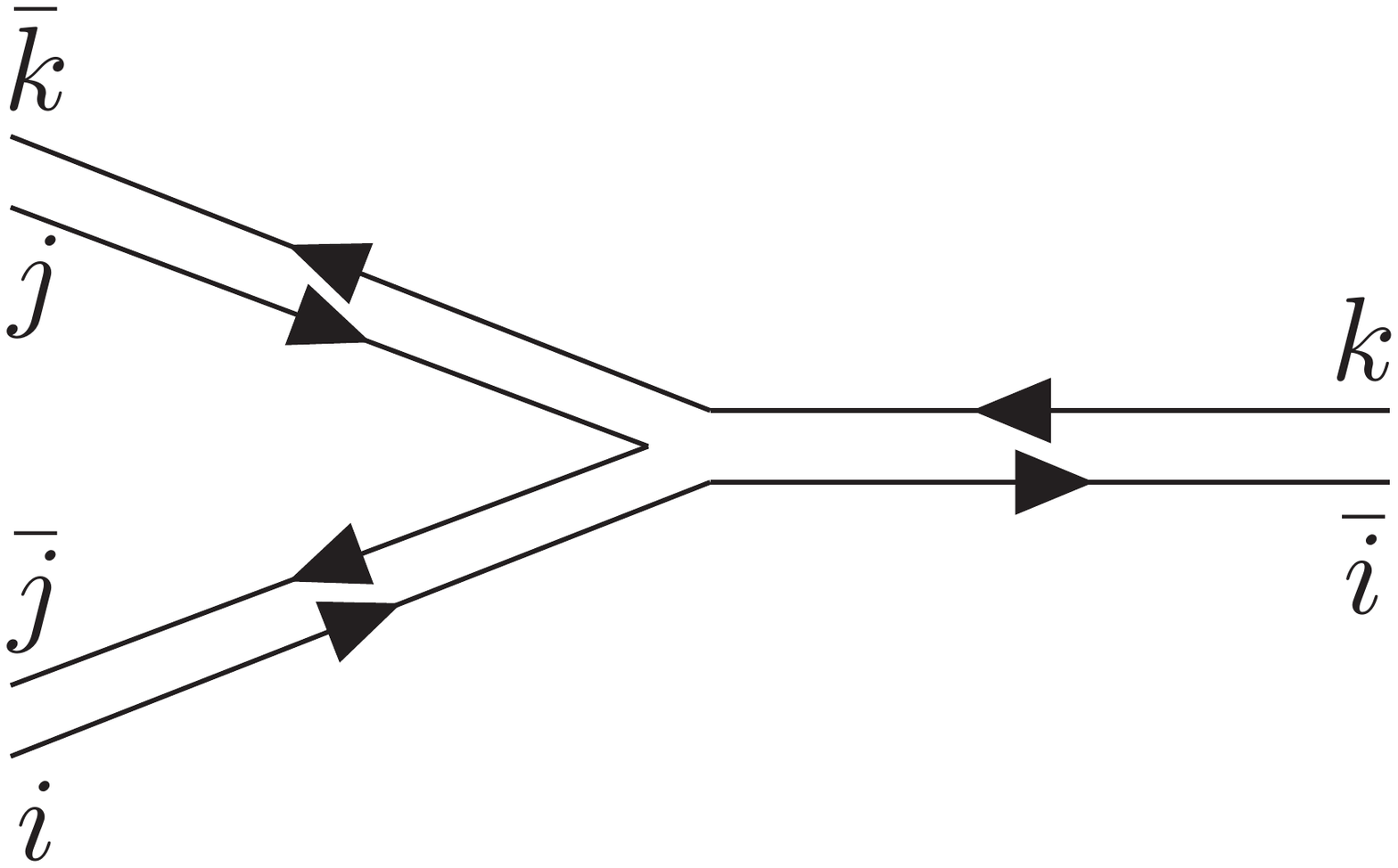} \\
    \includegraphics[width=0.16\textwidth]{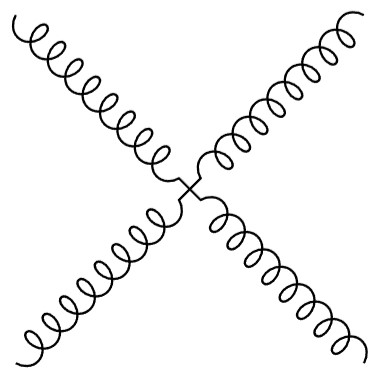} 
    \includegraphics[width=0.22\textwidth]{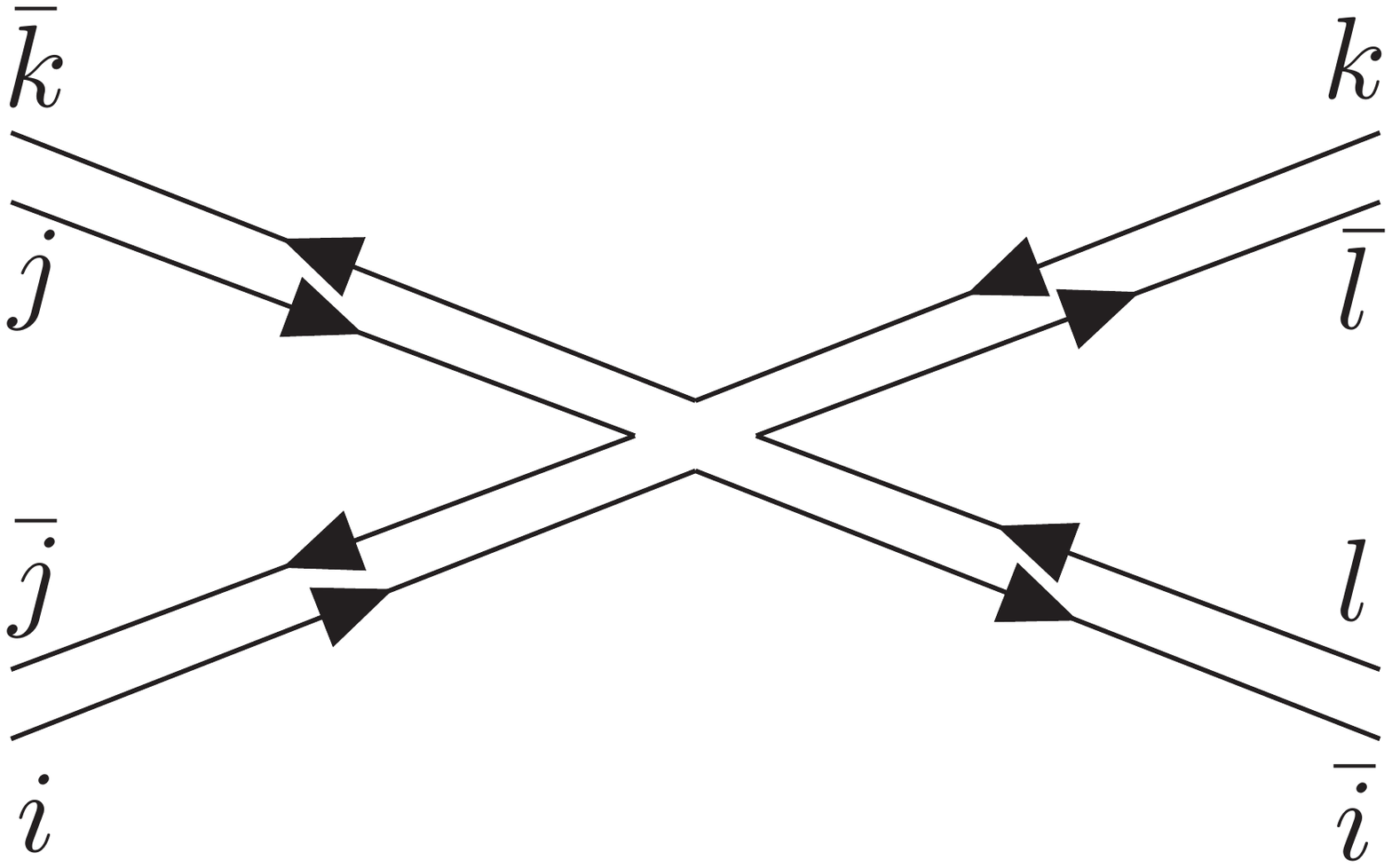} 
    \caption{The normal (left) and the double line (right) notation for the propagators and vertices of QCD. The indices on the right are color indices.}
    \label{Fig:Double_line_notation}
\end{figure}

As well known, QCD is nonperturbative in its low-energy domain since the coupling constant is large. Yet, a different kind of expansion as function of the inverse of the color numbers $1/N_c$ is possible \cite{tHooft:1973alw}. This expansion,  analogous to the semi-classical expansion, allows to explain many non-perturbative features of QCD \cite{Witten:1979kh}. To this end, one needs to assume the scaling $g_{\text{QCD}} \propto N_c^{-1/2}$, which is a natural consequence of $\Lambda_{\text{QCD}}$ being taken as $N_c$-independent, see Sec. \ref{Ssec:symmetries}.
Moreover, it is convenient to introduce the double-line representation, since it greatly simplifies the $1/N_c$ counting. In this notation, gluons can be seen as color-anticolor objects, as it is shown in Fig. \ref{Fig:Double_line_notation}. 

\begin{figure}[htb]
    \centering
    \includegraphics[width=0.3\textwidth]{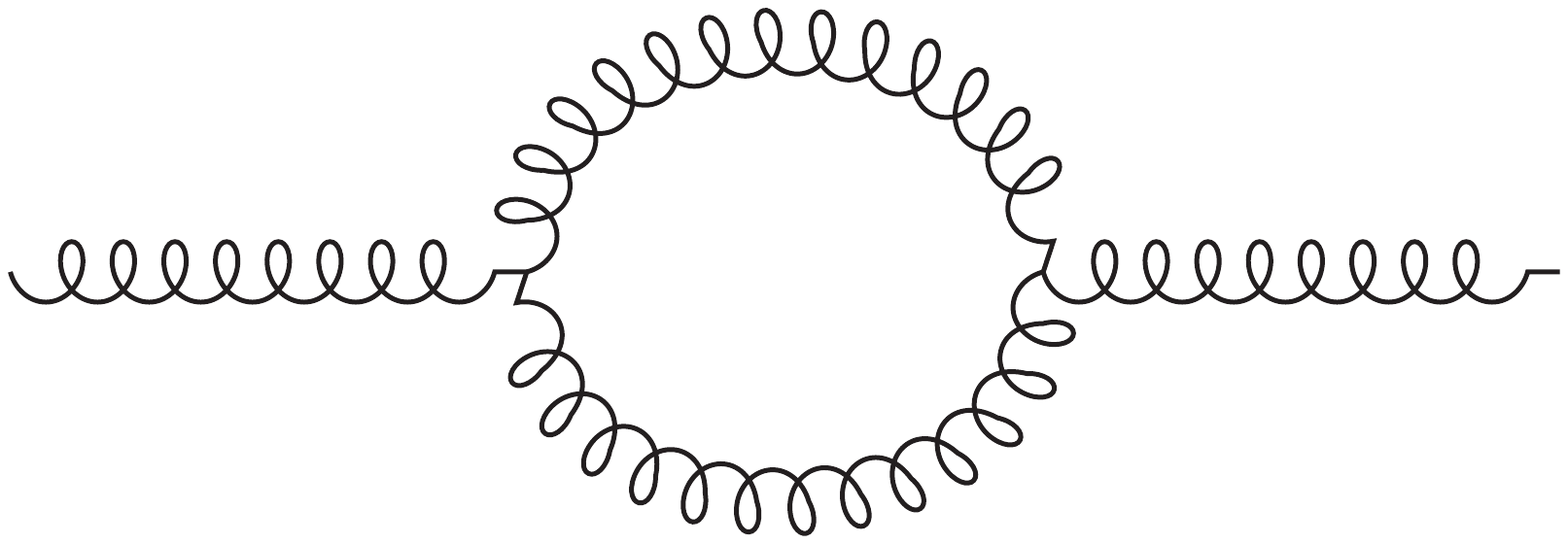}
    \includegraphics[width=0.3\textwidth]{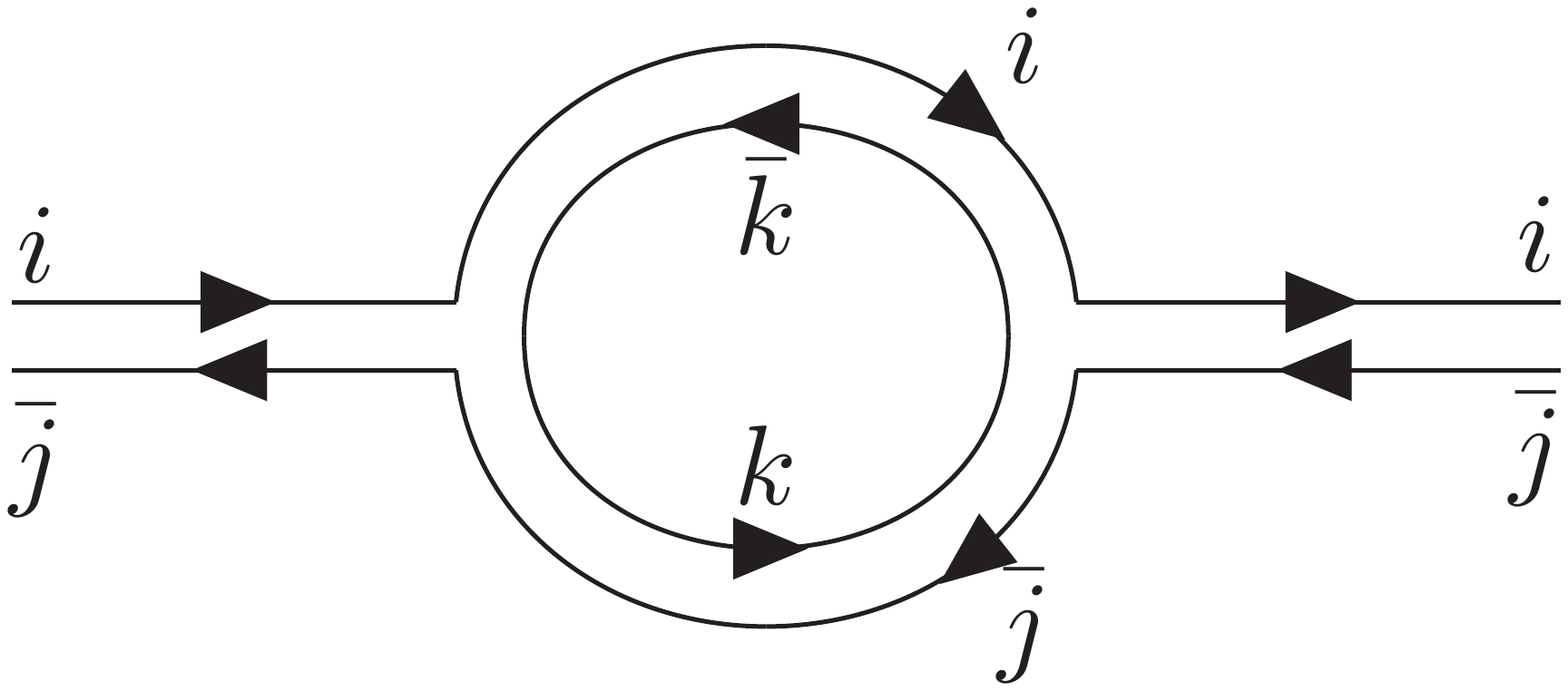}\\
    \includegraphics[width=0.3\textwidth]{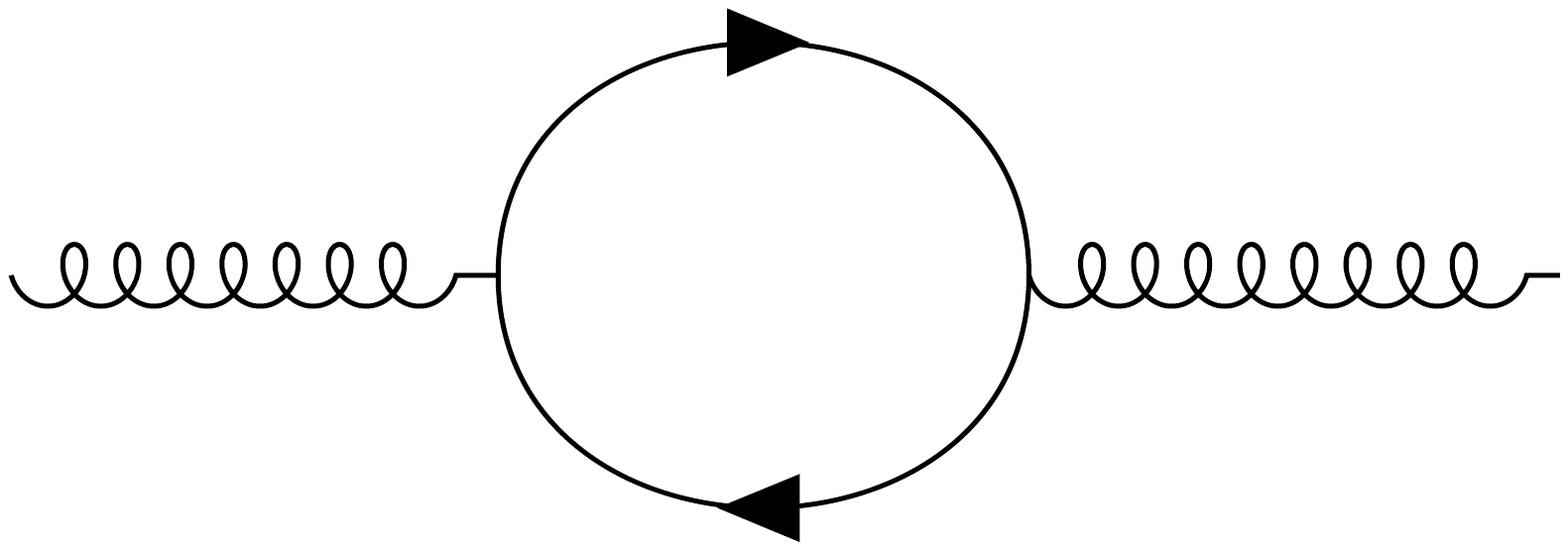}
    \includegraphics[width=0.3\textwidth]{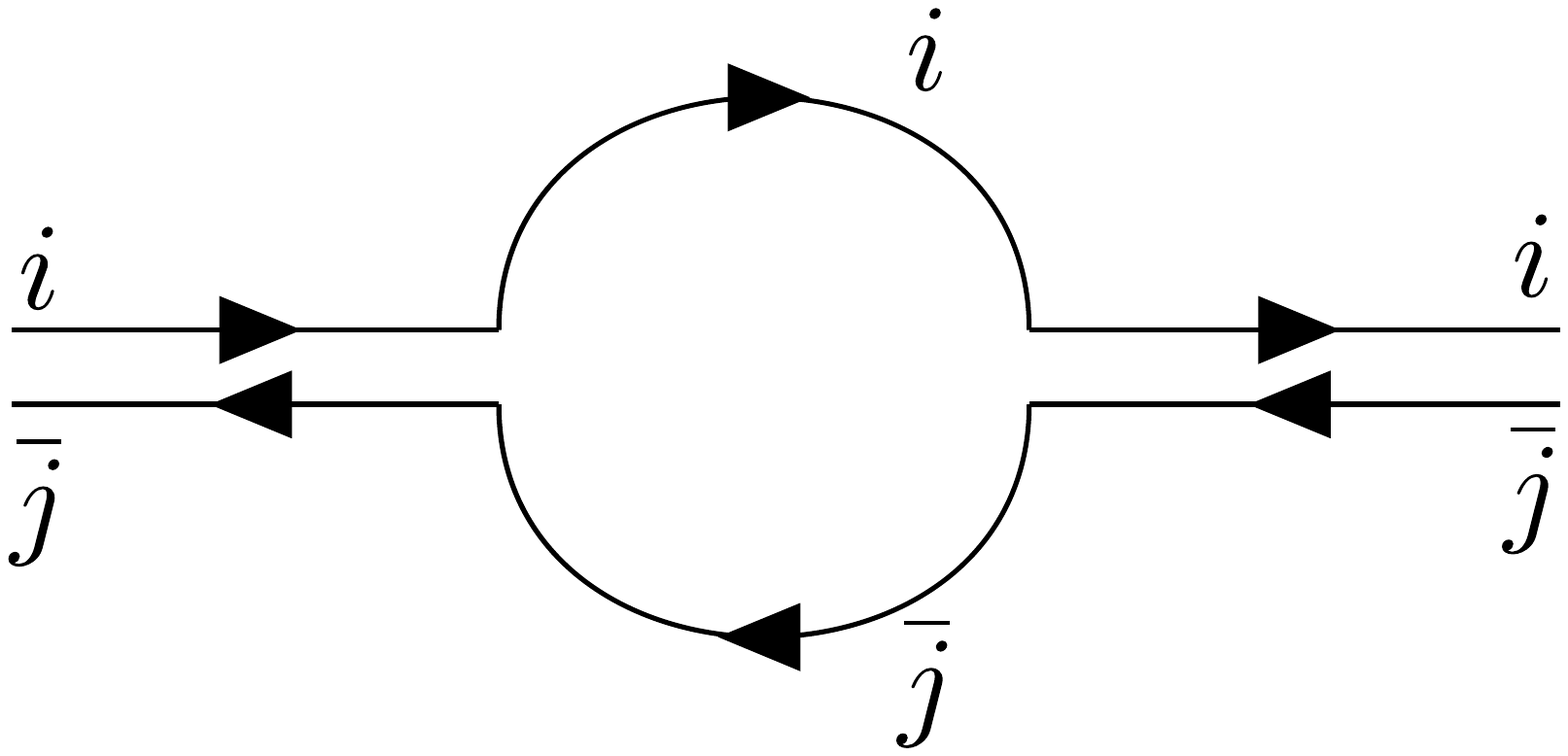}
    \caption{The one-loop gluon (upper part) and fermion (lower part) contribution to the gluon self-energy in the usual (left) and the double line notation (right). The upper diagram scales as $g^2 N_c \propto N_c^0$, while the lower one as $g^2 \propto N_c^{-1}$.}
    \label{Fig:SE_doubleline}
\end{figure}

According to this counting, the dominant Feynman graphs are the planar ones with the minimal possible number of quark loops (diagrams on the plane without any crossing lines), see details in Ref. \cite{Witten:1979kh}. Also, gluon loops dominate w.r.t. quark loops, see for instance Fig. \ref{Fig:SE_doubleline}.
%-----------------

Following the recent review of Ref. \cite{Giacosa:2024scx}, the consequences of large-$N_c$ phenomenology are summarized below:

\begin{enumerate}

\item The masses of $\bar{q}q$ states $Q$, glueballs
$G\equiv gg$, and hybrids $H\equiv\overline{q}qg$ are $N_c$-independent at large $N_c$:
\begin{equation}
M_{Q}\propto N_{c}^{0}\text{ , }M_{G}\propto N_{c}^{0}\text{ , }M_{H}\propto
N_{c}^{0} \text{ .}
\end{equation}

\item The interaction between $n_{Q}$ quarkonia scales as
\begin{equation}
A_{n_{Q}Q}\propto\frac{N_{c}}{N_{c}^{n_{Q}/2}}\text{ for }n_{Q}\geq1\text{ ,}%
\end{equation}
implying that the scattering amplitudes vanish at large $N_c$. In particular, the decay process corresponds to $n_Q=3$, thus $A_{\text{decay}}\propto N_{c}^{-1/2}$. Then, a generic decay width scales as $\Gamma\propto1/N_{c}$. Conventional quarkonia are stable at large $N_{c}.$

\item The scattering amplitude between $n_{G}$ glueballs goes as
\begin{equation}
A_{n_{G}G}\propto\frac{N_{c}^{2}}{N_{c}^{n_{G}}}\text{ for }n_{G}\geq1\text{ ,}%
\end{equation}
even smaller than the previous case.

\item The scattering amplitude between $n_{Q}$ quarkonia and $n_{G}$
glueballs scales as
\begin{equation}
A_{\left(  n_{Q}Q\right)  \left(  n_{G}G\right)  }\propto\frac{N_{c}}%
{N_{c}^{n_{Q}/2}N_{c}^{n_{G}}}\text{ for }n_{Q}\geq1\text{ .}%
\end{equation}
The glueball-quarkonium mixing corresponds to $n_{G}=n_{Q}=1$ and scales as $A_{\text{mixing}}\propto
N_{c}^{-1/2}$, hence suppressed as well.  

\item The scattering amplitude for $n_Q$ quarkonia and $n_H$ hybrids behaves as
\begin{equation}
A_{\left(  n_{Q}Q\right)  \left(  n_{H}H\right)  }\propto\frac{N_{c}}%
{N_{c}^{n_{Q}/2}N_{c}^{n_{H}/2}}\text{ for }n_{Q}+n_{H}\geq1\text{ .}%
\end{equation}
For $n_Q = n_H =1$ one sees that the quarkonium-hybrid mixing scales as $N_c^0$, implying that quarkonia and hybrids are indistinguishable at large $N_c$.

\item The general scattering amplitude with $n_{H}$ quarkonia, $n_{G}$ glueballs, and
$n_{H}$ hybrids reads:%
\begin{equation}
A_{\left(  n_{Q}Q\right)  \left(  n_{G}G\right)  \left(  n_{H}H\right)
}\propto\frac{N_{c}}{N_{c}^{n_{Q}/2}N_{c}^{n_{G}}N_{c}^{n_{H}/2}}\text{ for
}n_{Q}+n_{H}\geq1\text{ .}%
\end{equation}

\item Four-quark states (both  molecular and diquark-anti-diquark) tend to disappear at large $N_c$, see however the discussion in Refs. \cite{Lucha:2021mwx,Giacosa:2024scx}.

\end{enumerate}

\section{The non-relativistic limit as a justification for the currents}\label{Ssec:non-relimit}

As illustrative examples, we consider the cases of pseudoscalar, scalar, and
vector mesons with quark content $u\bar{d}$, that correspond to the mesons
$\pi^{+}$, $a_{0}^{+},$ and $\rho^{+}$, respectively. All the other cases can
be derived similarly.

The quark field `$i$' (with $i=u,d,s$) can be expressed as%
\begin{equation}
q_{i}(x)=\sum_{s=\pm1/2}\int\frac{d^{3}p}{(2\pi)^{3/2}}\sqrt{\frac{m_{i}%
}{E_{i}}}\left(  b_{i}^{(s)}(\mathbf{p})u_{i}^{(s)}(\mathbf{p})e^{-ipx}%
+d_{i}^{(s)\dagger}(\mathbf{p})v_{i}^{(s)}(\mathbf{p})e^{ipx}\right)
\text{ ,}
\end{equation}
where $u^{(s)}(\mathbf{p})$ and $v^{(s)}(\mathbf{p})$ are spinors, while $b_i^{(s)}$ and $d_i^{(s) \dagger}$ are the annihilation and creation operators of quarks and antiquarks.
Choosing
the Dirac representation, suited for the non-relativistic limit, one has:
\begin{equation}
u_{i}^{(s)}(\mathbf{p})= \sqrt{\frac{E_i + m_i}{2m_i}} \left(
\begin{array}
[c]{l}%
\chi^{(s)}\\
\frac{\mathbf{\sigma\cdot p}}{E_{i}+m_{i}}\chi^{(s)}%
\end{array}
\right)  \text{ , } v_{i}^{(s)}(\mathbf{p})=\sqrt{\frac{E_i + m_i}{2m_i}} \left(
\begin{array}
[c]{l}%
\frac{\mathbf{\sigma\cdot p}}{E_{i}+m_{i}}\varepsilon\chi^{(s)}\\
\varepsilon\chi^{(s)}%
\end{array}
\right)
\text{ ,} 
\label{diracuv}
\end{equation}
where the matrix
$\varepsilon$ is such that $\left(  \varepsilon\right)  ^{ij}=\varepsilon
^{ij}$ (Levi-Civita), and 
with
\begin{equation}
\chi^{(1/2)}=\left(
\begin{array}
[c]{c}%
1\\
0
\end{array}
\right)  \text{ , }\chi^{(-1/2)}=\left(
\begin{array}
[c]{c}%
0\\
1
\end{array}
\right)  \text{ , }p\cdot x=E_it-\mathbf{p\cdot x}\text{ , }E_i=\sqrt
{\mathbf{p}^{2}+m_{i}^{2}}\text{ .}%
\end{equation}
The vector $\chi^{(s)}$ describes the spin of the quark.
%the upper solution corresponds to an quark, the lower one to an antiquark
The anti-commutation relations read
\begin{equation}
\left\{  b_{i}^{(s)}(\mathbf{p}_{1}),b_{j}^{(r)\dagger}(\mathbf{p}%
_{2})\right\}  =\delta_{ij}\delta_{rs}\delta(\mathbf{p}_{1}-\mathbf{p}%
_{2})\text{ ; }\left\{  d_{i}^{(s)}(\mathbf{p}_{1}),d_{j}^{(r)\dagger
}(\mathbf{p}_{2})\right\}  =\delta_{ij}\delta_{rs}\delta(\mathbf{p}%
_{1}-\mathbf{p}_{2})\text{ .}%
\end{equation}
The generic pseudoscalar current is $P_{ij}\sim\bar{q}_{j}(x)i\gamma^{5}%
q_{i}(x),$ hence the state $\left\vert \pi^{+}\right\rangle $
\begin{equation}
\left\vert \pi^{+}\right\rangle \sim\int d^{4}xe^{-ipx}\bar{q}_{u}%
(x)i\gamma^{5}q_{d}(x)\left\vert 0\right\rangle
\end{equation}
corresponds to a pion-like structure with four-momentum $p=(p^{0}%
,\mathbf{p}).$ Namely,  when acting on the vacuum $\bar{q}_{u}$ creates a quark $u$ and $q_{d}$ an
antiquark $\bar{d}$. A straightforward calculation (by plugging Eq. (\ref{diracuv}) into the one above),
upon choosing $\mathbf{p}=\mathbf{0}$, leads to (neglecting an overall
constant):%
\begin{equation}
\left\vert \pi^{+}\right\rangle \sim\sum_{s_{1},s_{2}=\pm1/2}\int d^{3}%
q\sqrt{\frac{m_{u}}{E_{u}}}\sqrt{\frac{m_{d}}{E_{d}}}\bar{u}_{u}^{(s_{1}%
)}(\mathbf{q})i\gamma^{5}v_{d}^{(s_{2})}(-\mathbf{q})b_{u}%
^{(s_{1})\dagger}(\mathbf{q})d_{d}^{(s_{1})\dagger}(-\mathbf{q}%
)\left\vert 0\right\rangle \text{ .}%
\end{equation}
Thus, upon using %
\begin{equation}
\bar{u}_{u}^{(s_{1})}(\mathbf{q})i\gamma^{5}v_{d}^{(s_{2})}(-%
\mathbf{q})=-i\sqrt{\frac{E_{u}+m_{u}}{2m_{u}}}\sqrt{\frac{E_{d}+m_{d}}%
{2m_{d}}}\left(  1+\frac{\mathbf{q}^{2}}{\left(  E_{u}+m_{u}\right)  \left(
E_{d}+m_{d}\right)  }\right)  \varepsilon^{s_{1}s_{2}}\sim\varepsilon
^{s_{1}s_{2}}\text{ ,}%
\end{equation}
one arrives at:
\begin{equation}
\left\vert \pi^{+}\right\rangle \sim\int d^{3}qA(\mathbf{q}^{2})\left(
b_{u}^{(1/2)\dagger}(\mathbf{q})d_{d}^{(-1/2)\dagger}(-\mathbf{q}%
)-b_{u}^{(-1/2)\dagger}(\mathbf{q})d_{d}^{(1/2)\dagger}(-%
\mathbf{q})\right)  \left\vert 0\right\rangle 
\text{ ,}
\end{equation}
where the spin singlet $\left\vert \uparrow\downarrow\right\rangle -\left\vert
\downarrow\uparrow\right\rangle $ emerges automatically, showing that $S=0$. In general, the
function $A(\mathbf{q}^{2})$ is the momentum wave function of the pion, which
here cannot be determined since we used local currents (for nonlocal extension, see e.g. Refs. \cite{Faessler:2003yf,Giacosa:2004ug}).
%(the expression would be of the type $1+$relativistic corrections). 
Yet, the generic expression
$A(\mathbf{q}^{2})$, being rotational invariant, assures that $L=0.$ Finally, a study of parity and charge
conjugation confirms the previously shown assignment. Under parity,
\begin{equation}
U_{\mathcal{P}}\left\vert \pi^{+}\right\rangle =-\left\vert \pi^{+}\right\rangle
\end{equation}
follows from $U_{\mathcal{P}}b_{u}^{(s)\dagger}(\mathbf{q})U_{\mathcal{P}}^{\dagger}%
=b_{u}^{(s)\dagger}(-\mathbf{q})$ and $U_{\mathcal{P}}d_{d}^{(1/2)\dagger
}(-\mathbf{q})U_{\mathcal{P}}^{\dagger}=-d_{d}^{(1/2)\dagger}(\mathbf{q}),$
while under charge conjugation
\begin{equation}
U_{\mathcal{C}}\left\vert \pi^{+}\right\rangle =\left\vert \pi^{-}\right\rangle
\end{equation}
is a consequence of $U_{\mathcal{C}}b_{u}^{(s)\dagger}(\mathbf{q})U_{\mathcal{C}}^{\dagger}%
=d_{u}^{(s)\dagger}(\mathbf{q})$ and $U_{\mathcal{C}}d_{d}^{(1/2)\dagger}(\mathbf{q}%
)U_{\mathcal{C}}^{\dagger}=u_{d}^{(1/2)\dagger}(\mathbf{q})$. 

The scalar case is, at first glance simple, but indeed the result is quite
instructive. Upon using the scalar current $S_{ij}\sim\bar{q}_{j}(x)q_{i}(x),$ for the $\bar{d}u$ object
one studies
\begin{equation}
\left\vert a_{0}^{+}\right\rangle \sim\sum_{s_{1},s_{2}=\pm1/2}\int
d^{3}q\sqrt{\frac{m_{u}}{E_{u}}}\sqrt{\frac{m_{d}}{E_{d}}}\bar{u}_{u}%
^{(s_{1})}(\mathbf{q})v_{d}^{(s_{2})}(-\mathbf{q})b_{u}^{(s_{1}%
)\dagger}(\mathbf{q})d_{d}^{(s_{1})\dagger}(-\mathbf{q})\left\vert
0\right\rangle
\text{ ,}
\end{equation}
where $\gamma^{5}$ has been replaced by the identity matrix. Hence:%
\begin{equation}
\bar{u}_{u}^{(s_{1})}(\mathbf{q})v_{d}^{(s_{2})}(-\mathbf{q})\sim
\chi^{(s_{1})\dagger}\mathbf{\sigma}\cdot\mathbf{q}\chi^{(s_{2})\dagger}\text{
.}%
\end{equation}
Let us look, for instance, at the specific direction $q_{z},$ leading to the
following component of the wave function:
\begin{equation}
\left\vert a_{0}^{+}\right\rangle \sim\int d^{3}qq_{z}A(\mathbf{q}^{2})\left(
b_{u}^{(1/2)\dagger}(\mathbf{q})d_{d}^{(1/2)\dagger}(-\mathbf{q}%
)-b_{u}^{(-1/2)\dagger}(\mathbf{q})d_{d}^{(-1/2)\dagger}(-%
\mathbf{q})\right)  \left\vert 0\right\rangle +...
\end{equation}
where dots refer to the terms proportional to $q_{x}$ and $q_{y},$
respectively. It is evident that the spin is $S=1,$ since the structure
$\left\vert \uparrow\uparrow\right\rangle -\left\vert \downarrow
\downarrow\right\rangle $ arises.  Moreover, the momentum wave function
$q_{z}A(\mathbf{q}^{2})$ belongs to the $L=1$ multiplet. Thus, this exercise
shows that scalar current corresponds to $L=S=1$, as expected. Moreover, it is a scalar $J=0$ since the original current is such. 
A direct
application of parity and charge conjugation confirms that%
\begin{equation}
U_{\mathcal{P}}\left\vert a_{0}^{+}\right\rangle =\left\vert a_{0}^{+}\right\rangle
\text{ },\text{ }U_{\mathcal{C}}\left\vert a_{0}^{+}\right\rangle =\left\vert a_{0}%
^{-}\right\rangle \text{ }.
\end{equation}
As a last example we may consider vector mesons with currents $V_{ij}^{\mu
}\sim\bar{q}_{j}(x)\gamma^{\mu}q_{i}(x).$ Thus, upon choosing $\mu=3$ for
simplicity, the meson $\rho^{+}$ corresponds to
\begin{equation}
\left\vert \rho^{+}\right\rangle \sim\int d^{4}xe^{-ipx}\bar{q}%
_{u}(x)\gamma^{3}q_{d}(x)\left\vert 0\right\rangle
\text{ .}
\end{equation}
Since $\bar{u}_{u}^{(s_{1})}(\mathbf{q})\gamma^{3}v_{d}^{(s_{2})}%
(-\mathbf{q})\sim\chi^{(s_{1})\dagger}\mathbf{\sigma}_{z}%
\varepsilon\chi^{(s_{2})\dagger}$, one arrives to
\begin{equation}
\left\vert \rho^{+}\right\rangle \sim\int d^{3}qA(\mathbf{q}^{2})\left(
b_{u}^{(1/2)\dagger}(\mathbf{q})d_{d}^{(-1/2)\dagger}(-\mathbf{q}%
)+b_{u}^{(-1/2)\dagger}(\mathbf{q})d_{d}^{(-1/2)\dagger}({ -}%
\mathbf{q})\right)  \left\vert 0\right\rangle \text{ ,}
\end{equation}
which corresponds to $L=0$ and $S=1$ (the latter in the form $\left\vert
\uparrow\downarrow\right\rangle +\left\vert \downarrow\uparrow\right\rangle $,
referring to $S=1$ and $S_{z}=0).$

These arguments can be repeated for any current introduced in Sec. \ref{ssec:qq and hybr} and show the correctness of
the field assignments performed in this work in particular and in all these
types of model in general. 

	\newpage
	
	\bibliography{mybibfile}
	%Please use Bib\TeX\ to generate your bibliography and include DOIs whenever available. Example of bib file: 
	
	%%%%%%%%%%%%%%%%%%%%%%%%%%%%%%%%%%%%%%%%%%%%%%%%%%%%%%%%%%%%%%%%%%%
	% Encoding: ISO-8859-1

	%@Article{Eichmann:2016yit,
	%author        = {Eichmann, Gernot and Sanchis-Alepuz, Helios and Williams, Richard and Alkofer, Reinhard and Fischer, Christian S.},
	%title         = {{Baryons as relativistic three-quark bound states}},
	%journal       = {Prog. Part. Nucl. Phys.},
	%year          = {2016},
	%volume        = {91},
	%pages         = {1-100},
	%archiveprefix = {arXiv},
	%doi           = {10.1016/j.ppnp.2016.07.001},
	%eprint        = {1606.09602},
	%owner         = {chfi},
	%primaryclass  = {hep-ph},
	%slaccitation  = {%%CITATION = ARXIV:1606.09602;%%},
	%timestamp     = {2018.08.02},
	%}

	%@Comment{jabref-meta: databaseType:bibtex;}
	%%%%%%%%%%%%%%%%%%%%%%%%%%%%%%%%%%%%%%%%%%%%%%%%%%%%%%%%%%%%%%%%%%%

 %\bibliography{ELSM}
	
\end{document}